\documentclass[a4paper,11pt,numbered,oneside,timestimes, custommargin]{Classes/PhDThesisPSnPDF} 

\input{Preamble/preamble}

\title{Ab-initio Quantum Monte Carlo study of ultracold atomic mixtures}

\author{Viktor Cikojević}

\dept{}

\university{Universitat Politècnica de Catalunya \\ University of Zagreb}
\crest{\includegraphics[width=0.7\textwidth]{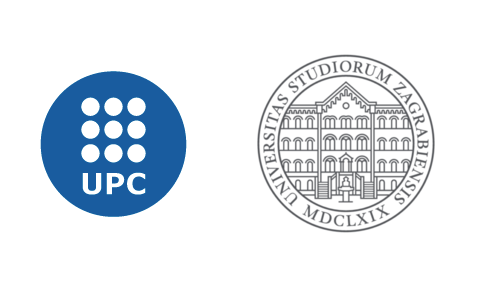}}

\supervisor{Prof. Leandra Vranješ Markić \newline Prof. Jordi Boronat}

 \supervisorrole{\textbf{Supervisors: }}

\supervisorlinewidth{0.6\textwidth}

\degreetitle{Doctor in Physics}

\degreedate{September 2020}

\ifdefineAbstract
\fi

\ifdefineChapter
\fi

\usepackage{soul}
\usepackage{multicol}

\begin{document}

	\begin{titlepage}

		\newgeometry{left=30mm,right=30mm,top=30mm,bottom=30mm}
		
		\begin{figure}[H]
			\centering
			\includegraphics[width=0.2\textwidth]{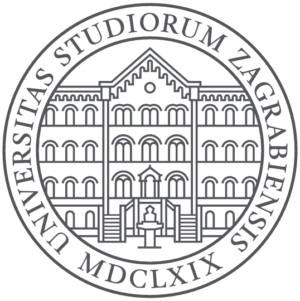} 
			\hspace{7cm}
			\includegraphics[width=0.2\textwidth]{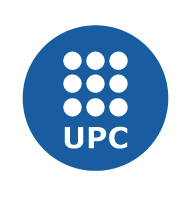}
		\end{figure}\vspace{-2ex}

		{\large \hspace{-0.65cm} University of Zagreb \hspace{5 cm} Universitat Politècnica de Catalunya}\\[3ex]
		{\large Faculty of Science \hspace{7 cm} Department of Physics}\\[3ex]
		{\large{\fontfamily{Arial} Deparment of Physics }}\\[2.5cm]

%
		
		\begin{center}

			{\Large\fontfamily{phv}\selectfont Viktor Cikojević}\\[2 cm]
			
			{ \Huge \bfseries \fontfamily{phv}\selectfont Ab-initio Quantum Monte Carlo study of ultracold atomic mixtures}\\[2.5cm]
			{\Large\fontfamily{phv}\selectfont INTERNATIONAL DUAL DOCTORATE} \\[2.5cm]
			{\Large\fontfamily{phv}\selectfont Supervisors:} \\
			
			{\Large\fontfamily{phv}\selectfont prof. dr. sc. Leandra Vranješ Markić} \\
			{\Large\fontfamily{phv}\selectfont prof. dr. sc. Jordi Boronat}

			\vfill
			{\large Split/Zagreb/Barcelona, 2020.}
		\end{center}
	
		\restoregeometry

	\end{titlepage}

	\cleardoublepage
	
	\begin{titlepage}
		
		\newgeometry{left=30mm,right=30mm,top=30mm,bottom=30mm}
		
		\begin{figure}[H]
			\centering
			\includegraphics[width=0.2\textwidth]{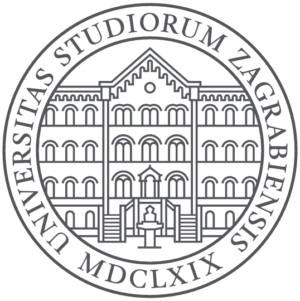} 
			\hspace{7cm}
			\includegraphics[width=0.2\textwidth]{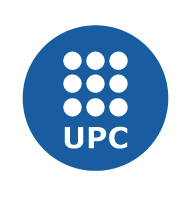}
		\end{figure}\vspace{-2ex}

%

		{\large \hspace{-0.65cm} Sveučilište u Zagrebu \hspace{5 cm} Universitat Politècnica de Catalunya}\\[3ex]
		{\large Prirodoslovno-matematički fakultet \hspace{4 cm} Department of Physics}\\[3ex]
		{\large{\fontfamily{Arial} Fizički odsjek }}\\[2.5cm]
		
		\begin{center}

			{\Large\fontfamily{phv}\selectfont Viktor Cikojević}\\[2 cm]
			
			{ \Huge \bfseries \fontfamily{phv}\selectfont Ultrahladne atomske mješavine istražene ab-initio kvantnom Monte Carlo metodom}\\[2.5cm]
			{\Large\fontfamily{phv}\selectfont MEÐUNARODNI DVOJNI DOKTORAT ZNANOSTI} \\[2.5cm]
			{\Large\fontfamily{phv}\selectfont Mentori:} \\
			
			{\Large\fontfamily{phv}\selectfont prof. dr. sc. Leandra Vranješ Markić} \\
			{\Large\fontfamily{phv}\selectfont prof. dr. sc. Jordi Boronat}

			\vfill
			{\large Split/Zagreb/Barcelona, 2020.}
		\end{center}
		
		\restoregeometry
	\end{titlepage}

	
	\thispagestyle{empty}
	\section*{INFORMATION ABOUT THE MENTORS}
	\hspace{5ex} This doctoral thesis was made at the Faculty of Science in Split and the Universitat Politècnica de Catalunya, under the mentorship of prof. dr. sc. Leandre Vranješ Markić (leandra@pmfst.hr, ORCID number 0000-0002-4912-3840) and prof. dr. sc. Jordi Boronat (jordi.boronat@upc.edu, ORCID number 0000-0002-0273-3457). 
	
	So far, under the mentorship of prof. dr. sc. Leandra Vranješ Markić, 4 doctoral theses were written, during which 5 co-authored CC papers were published with dr. sc. Ivana Bešlić (defended her doctoral thesis in 2010), 4 CC papers with doc. dr. sc. Petar Stipanović (defended his doctoral thesis in 2015), 4 CC papers with Krešimir Dželalija and 5 CC papers with Viktor Cikojević. Leandra Vranješ Markić, together
	with collaborators, has published so far 50 papers registered by Current Contents.
	
	Under the mentorship of prof. dr. sc. Jordi Boronat Medico, 12 doctoral theses were written, 5 of them in the last 5 years: Adrián Macia, Guillem Ferré, Raúl Bombín, Juan Sánchez, and Grecia Guijarro. The papers published as a result of these Thesis (last 5 years) were: Macia: 6 articles; Ferré: 5 articles; Bombín: 6 articles; Sánchez: 5 articles, and Guijarro: 2 articles (+ 1 submitted and 2 in preparation). Jordi Boronat Medico, together with collaborators, has published so far 200 papers registered by Current Contents.

	\newpage


\thispagestyle{empty}

\chapter*{Acknowledgements}
\addcontentsline{toc}{chapter}{Acknowledgements}

Many people contributed to this Thesis.

First and foremost, I give big thanks to two outstandingly warm people and excellent scientists who accepted to be my supervisors, Professors Leandra Vranje\v{s} Marki\'c and Jordi Boronat. During this journey I was lucky to gain experience and knowledge from you, both from personal and scientific aspect. This work would not be made without you. I hope this Thesis is just a beginning, and that we will continue the journey of research in the years to come. 

I was lucky to have worked with Petar Stipanović, a "numeric king". Petar, thank you for your friendly help and advises with building the DMC code.

It was a pleasure to have shared my office with such friendly people: Hrvoje Vrcan, Krešimir Dželalija, Josipa Šćurla and Raúl Bomb\'in Escudero. I cherished your company as you've made the office a comfortable and fun place.

A special place in my heart hold the people with whom I spend time at the UPC: Raúl Bombín Escudero, Juan Sanchez, Grecia Guijarro, Giulia De Rosi, Huixia Lu. Thank you for all the support and for making my stay in Barcelona exciting. I truly appreciated many many lunches, coffees, dinners, vermuts  and other things we did together. Guyz, good luck with everything.

Other than my supervisors, I am lucky to have worked or discussed with other brilliant people who influenced me: Gregory Astrakharchik, Pietro Massignan and Ferran Mazzanti, thanks for all the great conversations and new ideas. Thanks to density-functional experts Mart\'i Pi, Manuel Barranco and Francesco Ancilotto who were patient with my questions about the DFT codes and who showed me the beauty of this theory. Thanks to the ICFO people who gave me a great insights on the physics of droplets: Leticia Taruell, Julio Sanz, Cesar R. Cabrera, Luca Tanzi. Thanks, Albert Gallem\'i, for your time explaining me how you simulated the collisions.

Many thanks to Marko Hum. You've made the mesh of cotutelle bureaucracy procedure easy.

Finally, a huge thanks to all my family and friends. Thanks Ante, Roje and other musicians for the great music you're doing, it served as a soundtrack for this thesis. Andrea and Vedran, my longest friends, for a reason. Kristian, Luce, Ivana, I remember with happiness all our wonderful meetings. I very grateful to my dear brother Frane and mother Zvjezdana, for being positive and a huge support. Others that I omitted, you will not be omitted a celebration party, for sure.

\thispagestyle{empty}

	\newpage
	\thispagestyle{empty}
	\begin{center}
		Abstract\vspace{3ex}
		
		{\large{\bf  Ab-initio Quantum Monte Carlo study of ultracold atomic mixtures}}\vspace{2ex}
		
		VIKTOR CIKOJEVIĆ\\
		University of Split, Faculty of Science\\
		Ruđera Boškovića 33, 21 000 Split
	\end{center}

	The properties of mixtures of Bose-Einstein condensates at $T=0$ have been investigated using quantum Monte Carlo (QMC) methods and Density Functional Theory (DFT) with the aim of understanding physics beyond the mean-field theory in Bose-Bose mixtures. In particular, quantum liquid droplets with attractive intraspecies and repulsive interspecies attraction were studied, for which we observed significant contributions beyond Lee Huang Yang (LHY) theory that affect the energy, saturation density, and surface tension. The critical atom number in droplets in free space for total number of atoms $N$ between $N=30$ and $N=2000$ was obtained. Results of the surface tension for three values of the attractive interspecies interactions are presented. For a homogeneous system, extensive calculations of the equations of state were performed and we report the influence of finite-range effects in beyond-Bogoliubov theory. In systems interacting with a small (large) effective range, we observe repulsive (attractive) beyond-LHY contributions to the energy. For the droplets in a mixture of $^{39}$K atoms, which were observed experimentally for the first time, the calculations of equations of state were performed. Combining QMC-built functionals with DFT, the discrepancy in the estimation of critical atom number between the mean-field theory and experimental results was explained by the proper inclusion of the effective range in inter-particle interaction models. The influence of finite-range effects on breathing and quadrupole modes in $^{39}$K quantum droplets was investigated. We predicted a significant deviation in the excitation frequencies when entering a more correlated regime. Finally, the phase diagram of repulsive Bose-Bose mixtures in a spherical harmonic trap using Quantum Monte Carlo calculations was studied. Density profiles were obtained reported and we found the occurrence of three phases: separation of condensates in two blobs, fully mixed and shell-separated phase. A comparison with the Gross-Pitaevskii solutions showed a large deviation in the regime of large mass imbalance and strong interactions. We showed the universality in the density profiles with respect to the $s$-wave scattering length and found numerical evidence for Gross-Pitaevskii scaling present beyond the regime of applicability of Gross-Pitaevskii equations. \vspace{2ex}

	\noindent{\bf Original in:} English
	
	\noindent{\bf Keywords:} Quantum Monte Carlo methods, Diffusion Monte Carlo, Density Functional Theory, Gross-Pitaevskii equation, Bose-Bose mixtures, quantum liquids, beyond-Bogoliubov calculations, finite-range effects, excitation modes
	
	\noindent{\bf Supervisor:} Professor Leandra Vranješ Markić, PhD, Full Professor
	
	\noindent{\bf Supervisor:} Professor Jordi Boronat, PhD, Full Professor

	\newpage
	\thispagestyle{empty}	
	\begin{center}
		Sažetak \vspace{3ex}
		
		{\large{\bf Ultrahladne atomske mješavine istražene ab-initio kvantnom Monte Carlo metodom}}\vspace{2ex}
		
		VIKTOR CIKOJEVIĆ \\
		Sveučilište u Splitu, Prirodoslovno-matematički fakultet\\
		Ruđera Boškovića 33, 21 000 Split
	\end{center}

	Svojstva smjesa Bose-Einsteinovih kondenzata pri $T = 0$ istražena su korištenjem metoda kvantnog Monte Carla (QMC) i teorije funkcionala gustoće (DFT) s ciljem proučavanja fizike izvan teorije srednjeg polja u bozonskim mješavinama. Proučili smo kvantne kapljice s jednakim i odbojnim interakcijama između atoma istovrsne komponente te privlačnim interakcijama atoma različitih komponenti u interakciji i opazili smo značajne doprinose povrh Lee Huang Yang (LHY) teorije koji utječu na energiju, saturacijsku gustoću i površinsku napetost. Odredili smo kritični broj atoma za kapljice u slobodnom prostoru za broj atoma u kapljici $ N $ između $ N = 30 $ i $ N = 2000 $. Izračunali smo površinsku napetost za tri vrijednosti privlačnih međuatomskih interakcija. Izvršili smo opsežne proračune jednadžbi stanja iznimno rijetke tekućine bozonske mješavine i uočili utjecaj efekata konačnog dosega koji nije predviđen Bogoliubovljevom teorijom. U sustavima koji interagiraju s malim (velikim) efektivnim dosegom, opaženi su odbojni (privlačni) doprinosi koje ne predviđa LHY teorija. Izračunali smo jednadžbe stanja za kapljice bozonskih mješavina koje su po prvi put eksperimentalno uočene u smjesi $^{39}$K atoma. Kombinirajući funkcionale gustoće izgrađene pomoću kvantnog Monte Carla s DFT-om, neslaganje u procjeni kritičnog broja atoma između teorije srednjeg polja i eksperimentalnih rezultata je objašnjeno preko pravilnog uključivanja efektivnog dosega u modele međudjelovanja čestica. Istražen je utjecaj efektivnog dosega na pobuđenja kapljice $^{39}$K, i to na mod disanja i kvadrupolni mod. Dobiveni rezultati prikazuju značajno odstupanje frekvencija pobude pri ulasku u korelirani režim. Detaljno smo proučili fazni dijagram odbojnih Bose-Bose mješavina u sfernoj harmonijskoj zamci koristeći kvantne Monte Carlo račune. Dobiveni su profili gustoće koji pokazuju pojavu tri faze: separacija kondenzata u dvije nakupine, potpuno miješanje i separacije u obliku ljuske. Usporedba s riješenjima Gross-Pitaevskii jednadžbi pokazuje veliko odstupanje u režimu velike masene neravnoteže i jakih interakcija. Pokazali smo univerzalnost profila gustoće s obzirom na $ s $-valnu duljinu raspršenja te postojanje Gross-Pitaevskii skaliranja prisutnog izvan dosega primjenjivosti Gross-Pitaevskii jednadžbi. \vspace{2ex}

	\noindent{\bf Jezik izvornika:} engleski
	
	\noindent{\bf Ključne riječi:} Kvantne Monte Carlo metode, Difuzijski Monte Carlo, teorija funkcionala gustoće, Gross-Pitaevskii jednadžba, bozonske mješavine, kvantne tekućine, povrh-Bogoliubov računi, efekti konačnog dosega, modovi pobuđenja
	
	\noindent{\bf Mentorica:} Prof. dr. sc. Leandra Vranješ Markić

	\noindent{\bf Mentor:} Prof. dr. sc.  Jordi Boronat

	\newpage


	
	
	
	
	
	\tableofcontents
	
	\listoffigures
	
	\listoftables
	
	
	\printnomenclature
	
	\mainmatter
	
	\renewcommand{\vec}[1]{{\boldsymbol{#1}}} 
	\newcommand{\abs}[1]{\left| #1 \right|} 
	\newcommand{\ket}[1]{\left| #1 \right>} 
	\newcommand{\bra}[1]{\left< #1 \right|} 
	\newcommand{\ave}[1]{\left< #1 \right>}
	\newcommand{\braket}[2]{\left< #1 \vphantom{#2} \right|
		\left. #2 \vphantom{#1} \right>} 
	\newcommand{\vecR}{\vec{\mathrm{R}}}
	\newcommand{\vecF}{\vec{\mathrm{F}}}
	\newcommand{\vecr}{\vec{\mathrm{r}}}
	\newcommand{\vecx}{\vec{\mathrm{x}}}


\chapter{Introduction}



States of matter in Nature such as liquids, gases and solids are characterized by the interparticle correlations. Solids on one side, and liquids and gases on the other, can mutually be distinguished by the degree of periodicity of the total density $\rho(\vec{r}) = \ave{\hat{\Psi}^\dagger(\vec{r}) \hat{ \Psi}(\vec{r})}$, that is, the diagonal long-range order. For solids, it is manifested through the occurrence of peaks in the Fourier transform of $\rho$, whereas for gases and liquids the periodicity is absent. Liquids stand in between gases and solids, as there is no spatial long-range order, yet they are self-bound.
Physical intuition between different classical states of matter dates back to van der Waals in the 19th century, who won the Nobel prize in 1910 for introducing an equation of state for liquids and gases. 
The basic idea is that ordinary liquids or solids occur due to two features of interatomic potentials: long-range interparticle attraction and a short-range repulsion. For weakly interacting atoms, attractive long-range interaction is of a van der Waals type, coming from dipole-dipole interaction between neutral atoms. On the other hand, at small distances, the Pauli exclusion principle acts as a repulsive force, so the balance between these two effects ultimately defines the system properties, together with external parameters such as the pressure, geometry, and temperature. 


After the 1910s, experimental techniques allowed cooling the matter down to very low temperatures, initiating the journey of ultracold physics and chemistry. With the decrease of temperature, the effect of thermal fluctuations is reduced, allowing for the study of quantum effects in the many-body system. In the low-temperature domain, two new phenomena emerged which defied the laws of classical statistical mechanics: superfluidity of $^4$He and superconductivity of mercury, both manifesting resistless flow of its constituents \cite{london1938lambda,allen1938flow} below a critical temperature, indicating that a new state of matter occurs at very low temperatures. Both phenomena are captured with the phenomenological two-fluid model, first introduced by Laszlo Tisza in 1938 \cite{tisza1938transport}, where the total density is decomposed in superfluid and normal components. 

London made a connection between a superfluid state of $^4$He with the condensation phenomena in a Bose-Einstein gas \cite{london1938lambda}. Lowering the temperature, transition from normal to Bose condensed state in a system obeying Bose-Einstein statistics is accompanied by a spike in the heat capacity, which bears clear resemblance to that of $^4$He, and is called the $\lambda$-transition due to its peculiar shape. Additionally, similar values of the measured critical temperature and entropy in a superfluid $^4$He state with those in a corresponding Bose gas gave support to this hypothesis.

By postulating symmetric (Bose-Einstein) statistics on the interchange between the two bosonic particles, a homogeneous ensemble of ideal gas undergoes a Bose-Einstein condensation \cite{satyendranath1976beginning,einstein1924quantum}  below a critical temperature $T_c$, defined as the macroscopic occupation of $\vec{\mathrm{k}}=0$ state. To generalize the concept of a Bose-Einstein condensation to a system of interacting particles, Penrose and Onsager \cite{penrose1956bose} considered the long-range behaviour of the non-diagonal density matrix $\rho(\vec{r}, \vec{r}') = \ave{\hat{\Psi}^\dagger(\vec{r}) \hat{ \Psi}(\vec{r}')}$. It was proven that in a BEC, a density matrix has a non-zero value at large distances, i.e., $|\vec{r} - \vec{r}'| \rightarrow \infty$, which is equal to the density of Bose-Einstein condensed system. 

With the concept of macroscopic wavefunction present in both the superfluid and BEC theories, the first attempt to explain the superfluidity came from the theory of Bose-Einstein condensation in an ideal gas. Approximating liquid Helium as an ideal Bose gas proved to be a crude assumption, giving predictions significatly off the measurements. The reason for these disagreements is that atoms in liquid Helium are far from being non-interacting, leading to a significant condensate depletion, which leads to a condensate fraction of around 8$\%$ \cite{moroni1997momentum}.

An extended period of development of experimental techniques had to pass before the first observation of a pure BEC, achieved in alkali atoms \cite{anderson1995observation, davis1995bose, bradley1995evidence}. Typical density in which the experiments with cold Bose gases are performed today is of the order of $10^{14} \mathrm{cm}^{-3}$, meaning that the temperature to achieve quantum degeneracy is of the order $10^{-5}$K \cite{pethick2008bose}. Experimental control of various system parameters is supreme, making the field of ultracold atoms a fertile playground for the manifestation of different phases of matter \cite{bloch2008many}. It is now possible to tune rather easily the strength and sign of interactions, geometry of external traps, or dimensionality \cite{bloch2008many}. Various accessible system properties involve the density profile, static structure factor, momentum distribution, collective excitations, or release energy.

A simple theoretical understanding of cold Bose gases is the Gross-Pitaevskii equation. It is a mean-field approach, where all the particles occupy the same single-particle quantum state ($\vec{k} = 0$), whereas the interactions are incorporated through the average external field due to other particles. This approach works well for very dilute systems. Increasing the density, the energy of a Bose gas can be calculated perturbatively, where the first energy correction term to the mean-field one is called the Lee-Huang-Yang (LHY) energy. This term is known for both single-component \cite{lee1957eigenvalues,huang1957quantum} and two-component Bose mixtures \cite{larsen1963binary}. The next energy correction is called the Wu term \cite{wu1959ground} and is known only for single-component systems, but it proved to deviate significantly from the quantum Monte Carlo energies \cite{giorgini1999ground}. The latter values are taken as a reference, because quantum Monte Carlo methods give, within statistical errorbars, exact values for bosonic systems.

Other than the mean-field approach, non-perturbative numerical treatments such as the variational hypernetted chain method \cite{ripka1979practical} or the quantum Monte Carlo techniques \cite{boronat2002microscopic, ceperley1980ground, ceperley1995path} are favorable when the system under study enters a strongly correlated regime. Many-body quantum properties of a Bose system at zero temperature can be directly investigated with the exact diffusion Monte Carlo (DMC) method \cite{casulleras1995unbiased}. In this Thesis, we have implemented and exploited the DMC method to benchmark and investigate the predictions of commonly used mean-field theories of Bose condensed mixtures.

In a single-component Bose system, beyond mean-field effects are very small, as it has been confirmed with first principles numerical calculations, namely the diffusion Monte Carlo \cite{giorgini1999ground} and Path Integral Ground State methods \cite{rossi2013path}. However, a dramatic manifestation of beyond mean-field physics was predicted by Petrov \cite{petrov2015quantum}, manifested in the mixture of two Bose-Einstein condensates with interspecies attraction and intraspecies repulsion. In that particular system, the repulsive beyond mean-field repulsion stabilizes the mean-field collapse, resulting in a possibility of quantum droplet formation. This phenomenon was first studied in three-dimensional systems, and later extended to one and two dimensions \cite{petrov2016ultradilute,parisi2019liquid, parisi2020quantum} and at a dimensional crossover \cite{zin2018quantum}. Very soon after this theoretical prediction, the first Bose-Bose quantum droplet was observed in a homonuclear mixture of two hyperfine states of $^{39}$K \cite{cabrera2018quantum,semeghini2018self} and later in a  heteronuclear mixture of $^{41}$K - $^{87}$Rb atoms \cite{derrico2019observation}. The first efforts in understanding this system were done within the local density approximation \cite{ancilotto2018self}, where the general thermodynamic conditions for droplet stability were discussed. At very small densities, droplet properties are well reproduced with Petrov's theory. However, already in the first experiment \cite{cabrera2018quantum}, beyond-LHY terms appeared to play a role at magnetic fields producing more correlated droplets. Microscopic understanding of beyond-LHY physics was first made in numerical studies using diffusion Monte Carlo (see Chapters \ref{chapter:ultradilute_liquid_drops}, \ref{ch:symmetric_liquids} and \ref{ch:finite_range_effects}) and the hypernetted chain method\cite{staudinger2018self}, where it was shown that the details of interaction potential play an important role. These non-universal effects were already studied theoretically in the single-component scenario \cite{tononi2019zero,tononi2018condensation,salasnich2017nonuniversal, cappellaro2017thermal}, but they were not yet extended to two-component systems.
Overall, beyond mean-field effects are naturally incorporated in the diffusion Monte Carlo calculations, and it is of interest to benchmark the Petrov functional, or even generate a correction to it at higher densities. This functional can be used in dynamical studies, for example the calculation of collective excitation modes, which in the past proved to be a powerful technique for exploring microscopic theories in a many-body system \cite{tylutki2020collective, cappellaro2018collective, jin1996collective,altmeyer2007precision,dalfovo1999theory,pi1986time}.

These mixed quantum droplets resemble in some aspects the classical droplets, and already there have been dynamical studies observing their liquid-like properties, such as collisions between quantum droplets in 3D \cite{ferioli2019collisions} or in 1D \cite{astrakharchik2018dynamics}. Interestingly, in Ref. \cite{ferioli2019collisions}, the authors observed, for the first time, a compressible regime of the liquid phase, when the number of particles is so small that the droplet has a surface size comparable to its radius. Interestingly, experimental results show disagreement with the prediction of merging vs. separation in the collision outcome using the conventional Petrov theory. Dynamical properties of quantum droplets were studied in lower dimensions as well, where it was shown that they can support vortex states in 2D \cite{tengstrand2019rotating}, as in 3D \cite{kartashov2018three}, and that they can host exotic metastable phases in 2D \cite{kartashov2019metastability}.

Another advancement in the field was made by producing gases with dipolar interactions \cite{ferrier2016observation}. In contrast to quantum droplets in a Bose-Bose mixture, where the interactions are isotropic and short-ranged, dipolar interactions are anisotropic and long-ranged \cite{gadway2016strongly}. A similar methodology of the stabilization mechanism in dipolar droplet formation can be made following Petrov's work \cite{petrov2015quantum}. However, the failure of this approach has been observed in dilute dipolar quantum droplets \cite{bottcher2019dilute}, by measuring radically different density profiles and critical atom numbers, all of which were recovered with a full quantum Monte Carlo approach \cite{bottcher2019dilute}. Lack of consistent theory proves the necessity to treat these systems with a full many-body approach. A system with dipolar interactions is unique because of the exotic feature of supersolidity, a phenomenon where periodically structured matter exhibits superfluid behavior, first predicted in $^4$He \cite{andreev1969quantum}, but never experimentally confirmed. By measuring the excitation spectra \cite{natale2019excitation} and performing time-of-flight experiments \cite{chomaz2019long}, the coexistence of phase coherence and periodicity in dipolar gases has been observed, indicating the existence of a supersolid of droplets.

Nowadays, it is possible to routinely produce mixtures of same-species atoms \cite{myatt1997production,hall1998measurements}, different isotopes \cite{papp2008tunable,stellmer2013production} or different elements \cite{modugno2002two, thalhammer2008double}. In a properly tuned magnetic field, a Bose mixture can have the repulsive interactions in all of three channels. For example, in a harmonically trapped mixture \cite{lee2016phase}, there is a variety of spatial configurations that two-component species can have. They can form a mixed phase, where the two species completely overlap and they can be in a two-blob structure, where the overlap is minimized, meaning that the two condensates physically separate. Alternatively, they can form a shell structure, such that one species forms a shell structure around an inner one. These different regimes depend on the atom number \cite{wen2020effects} and are quite sensitive to the interaction strength \cite{jezek2002interaction}, making the phase diagram much richer than the usual single-component trapped BEC. Occurrence of these phases happens due to mean-field instability. Thus, this system is a promising candidate for exploring beyond mean-field effects. Additionally, at ultracold temperatures, these mixtures can enter in a superfluid regime \cite{fava2018observation}, making it possible to study and improve our understanding of physics of the two interpenetrating superfluids. With an increase of temperature, these mixtures exhibit interesting thermodynamic properties. For example, recently, a counter-intuitive prediction of a phase separation with temperature has been proposed, in the regime where zero-temperature mean-field theory predicts mixing \cite{ota2019magnetic}. Usually, at very small densities, the description of these mixtures is well reproduced with the Gross-Pitaevskii equation \cite{lee2018time}. However, it is not clear how the phase space changes when the densities or correlations become larger, thus it is of interest to study these systems with ab-initio quantum Monte Carlo methods.

\section{Outline}

This Thesis is devoted to the computational study of quantum Bose-Bose mixtures. The outline of a Thesis is as follows.

\vspace{0.5 cm}
\textbf{Chapter 2: Overview of ultracold gases.} In chapter 2, we discuss the basics of ultracold Bose gases. This chapter is relevant for the whole thesis, as the main physical quantities and terminology are introduced. We start the discussion with the physics of scattering in ultracold gases, where the two most important parameters are introduced: the $s$-wave scattering length and the effective range. The numerical method to calculate those parameters is outlined. Next, the overview of the main results of the Bogoliubov theory for a Bose gas is given. This problem can be formulated in a density-functional manner, which we present with the numerical algorithm specialized to solve it. Finally, basic mean-field theory and its extension, the LHY term, are introduced.

\vspace{0.5 cm}
\textbf{Chapter 3: QMC methods.} In chapter 3 are discussed the quantum Monte Carlo methods used in this Thesis. Since the goal is to study ultracold systems, it is an excellent approximation to tackle the problem at $T = 0$, thus making it possible to use the power of the variational and diffusion Monte Carlo methods, suitable for zero temperature quantum many-body studies. Variational Monte Carlo offers a variational solution, and it is used both to sample and to optimize the trial wavefunction. Improvement of this method can be made by performing the imaginary-time propagation, allowing to reach the ground-state, and therefore extract the ground-state averages.

\vspace{0.5 cm}
\textbf{Chapter 4: Ultradilute quantum liquid drops.} In chapter 4, we present the study of dilute symmetric Bose-Bose liquid droplets using quantum Monte Carlo methods with attractive interspecies interaction in the limit of zero temperature. The calculations are exact within some statistical noise and thus go beyond previous perturbative estimations. By tuning the intensity of the attraction, we observe an evolution from a gas to a self-bound liquid drop in an $ N $- particle system. This observation agrees with recent experimental findings and allows for the study of an ultradilute liquid never observed before in Nature. 

\vspace{0.5 cm}
\textbf{Chapter 5: Universality in ultradilute liquid Bose-Bose mixtures.} In chapter 5, we present the study of dilute symmetric Bose-Bose liquid mixtures of atoms with attractive interspecies and repulsive intraspecies interactions using quantum Monte Carlo methods at $T=0$, in the thermodynamic limit. Using several models for interactions, we determine the range of validity of the universal equation of state of the symmetric liquid mixture as a function of two parameters: the $s$-wave scattering length and the effective range of the interaction potential. It is shown that the Lee-Huang-Yang correction is sufficient only for extremely dilute liquids, with the additional restriction that the range of the potential is small enough. Based on the quantum Monte Carlo equation of state, we develop a new density functional which goes beyond the Lee-Huang-Yang term and use it, together with local density approximation, to determine density profiles of realistic self-bound drops.

\vspace{0.5 cm}
\textbf{Chapter 6: Finite-range effects in ultradilute quantum drops.} In chapter 6, we present the study of bulk properties of two hyperfine states of $^{39}$K utilizing the quantum Monte Carlo technique and introduce an improved density functional based on two scattering parameters: the $s$-wave scattering length and the effective range. In the first experimental realization of dilute Bose-Bose liquid drops in the same hyperfine states of $^{39}$K, the prediction of the critical numbers using the Lee-Huang-Yang extended mean-field theory \cite{petrov2015quantum} was significantly off the experimental measurements. Using a new functional, based on quantum Monte Carlo results of the bulk phase incorporating finite-range effects, we can explain the origin of the discrepancies in the critical number. This result proves the necessity of including finite-range corrections to deal with the observed properties in this setup. The controversy on the radial size is reasoned in terms of the departure from the optimal concentration ratio between the two species of the mixture. 

\vspace{0.5 cm}
\textbf{Chapter 7: Finite range effects on the excitation modes of a $^{39}$K quantum droplet.} In chapter 7, we present the study of the influence of finite-range effects on the monopole and quadrupole excitation spectrum of extremely dilute quantum droplets, in the mixture of $^{39}$K. We present a density functional, built from first-principles quantum Monte Carlo calculations, which can be easily introduced in the existing Gross-Pitaevskii numerical solvers. Our results show differences of up to $20\%$ with those obtained within the extended Gross-Pitaevskii theory, likely providing another way to  observe finite-range effects in  mixed quantum droplets by measuring their lowest excitation frequencies.

\vspace{0.5 cm}
\textbf{Chapter 8: Harmonically trapped Bose-Bose mixtures with repulsive interactions.} Going from self-bound to gaseous systems, in Chapter 8 we present the study of a phase diagram of a harmonically confined repulsive Bose-Bose mixture using quantum Monte Carlo methods, and we find emergence of three phases: two-blob, mixed, and separated. Our results for the density profiles are systematically compared with mean-field predictions derived through the Gross-Pitaevskii equation in the same conditions. We observe significant differences between the mean-field results and the Monte Carlo ones that magnify when the asymmetry in mass and interaction strength increases. In the analyzed interaction regime, we observe universality of our results, which extends beyond the applicability regime for the Gross-Pitaevskii equation.

	

\chapter{{\label{ch:overview}}Overview of ultracold atomic gases}

\section{Introduction}

In Nature, elementary particles manifest in two statistics: bosons and fermions. They are distinguished by spin: bosons have integer spin, while fermions have a half-integer spin. Statistics of composite particles, such as atoms or molecules, is defined by the total spin: atoms with odd (even) atomic number are fermions (bosons). The importance of statistics in a many-body system is generally enhanced as the temperature is cooled down, as both Bose and Fermi particles are guided towards a quantum degeneracy regime. Lowering the temperature increases the de Broglie wavelength associated with each particle, and a quantum degenerate regime can be recognized when the de Broglie wavelength is comparable to the mean interparticle spacing. For Bose particles at sufficiently low temperatures, wave packets mutually overlap in a way that there can occurr a macroscopic occupation of a single-particle $\vec{k} = 0$ state. Almost 100\% Bose-Einstein condensates were first achieved in Ref. \cite{anderson1995observation}, earning Eric Cornell, Wolfgang Ketterle, and Carl Wieman a Nobel prize in 2001, for the achievement of Bose-Einstein condensate in alkaline dilute vapors. Ever since that discovery, the field of ultracold atomic gases has been one of the most active avenues of research in contemporary physics.

The focus of this Thesis is on ultracold Bose-Bose mixtures. To present the relevance of the conducted research, in Sec. (\ref{sec:bogoliubov_theory}) we will first explore the basic concepts of Bogoliubov and density-functional theory relevant to the field of study, resulting in the famous Gross-Pitaevskii equation. Additionally, an efficient numerical scheme for dealing with real- and imaginary-time properties of general density-functional theories, based on the Trotter decomposition, is presented. The mean-field formalism of Bose-Bose mixtures is covered in Sec. (\ref{sec:two_comp_bose}), where the stability and formation of two-component Bose droplets are discussed. Crucial to the connection with real-world experiments, it is relevant to study the interaction between ultracold atoms, which is covered in Sec. (\ref{sec:scattering_theory}), where the basic scattering theory is presented. Finally, the nomenclature and methodology to calculate the relevant parameters used in the Thesis, such as the $s$-wave scattering length and the effective range, is discussed in Sec. (\ref{sec:calculation_swave_reff}).

\section{{\label{sec:bogoliubov_theory}}Single-component Bose gas}

\subsection{Bogoliubov theory}

Having more than two particles and entering into the many-body physics,  $N$-particle system obeys the Schr\"odinger equation
\begin{equation}
    i\hbar \dfrac{\Psi(\vec{r}_1, \ldots, \vec{r}_N)}{\partial t} = \hat{H} \psi(\vec{r}_1, \ldots, \vec{r}_N),
\end{equation}
with the Hamiltonian given by
\begin{equation}
    \hat{H} = \sum_{i=1}^{N} \left\{-\dfrac{\hbar^2}{2m} \nabla_i^2 + V_{\rm ext}(\vec{r}_i)  \right\} + \sum_{i  < j}^{N} V_{\rm pair}(\vec{r}_i - \vec{r}_j),
\end{equation} 
where $V_{\rm ext}(\vec{r})$ is the external potential and $V_{\rm pair}$ is the interaction between the two particles. It is convenient to write the Hamiltonian in the second-quantization formalism \cite{stoof2009ultracold}
\begin{eqnarray}
\hat{H} = \displaystyle \int d^3 r \hat{ \Psi}^{\dagger}(\vec{r}) \left\{ -\dfrac{\hbar^2}{2m} \nabla^2 + V_{\rm ext}(\vec{r})    \right\} \hat{\Psi}(\vec{r}) \nonumber \\ + \dfrac{1}{2}  \displaystyle \int d^3 r \displaystyle \int d^3 r' \hspace{0.1cm}  \hat{\Psi}^{\dagger}(\vec{r}) \hat{\Psi}^{\dagger}(\vec{r}')  V(\vec{r} - \vec{r}')   \hat{\Psi}(\vec{r}') \hat{\Psi}(\vec{r}) ,
\end{eqnarray}
where $\hat{\Psi}$ is the field operator 
\begin{equation}
    \hat{\Psi}(\vec{r}) = \sum_{i} \psi_i(\vec{r}) \hat{a}_i,
\end{equation}
with $\psi_i(\vec{r})$ being the $i$-th single-particle state, and $\hat{a}_i$ ($\hat{a}_i^{\dagger}$) is the annihilation (creation) operator which incorporates the statistics. For Bose particles, commutation relations are $\left[\hat{a}_i, \hat{a}_j^\dagger\right] = \delta_{ij}$ and $\left[\hat{a}_i, \hat{a}_j\right] = \left[\hat{a}_i^\dagger, \hat{a}_j^\dagger\right] = 0$, and their expectation values are \cite{lifshitz2013statistical}
\begin{equation}
	 \ave{\hat{a}_k^\dagger \hat{a}_k} = N_k,
\end{equation}
with $N_k$ being the occupation number of the $k$-th state. At ultracold temperatures and weak interactions, excited states are scarcely populated in a Bose system, thus to a first approximation the field operator can be replaced by its average value \cite{lifshitz2013statistical,bogoliubov1947theory}
\begin{equation}
    \label{eq:approximation_bogoliubov}
    \hat{ \Psi}  (\vec{r}) \rightarrow \psi = \ave{\hat{ \Psi}} \approx \psi_0(\vec{r}) ,
\end{equation}
where $\psi_0 = \sqrt{\rho_0}$ is called the wavefunction of the condensate, with $\rho_0$ being the number density of the condensate. It plays the role of the order parameter of the Bose-Einstein condensate. In reality, not all the particles are in the condensate. This is because the zero-momentum state is mixed with higher excited states due to the two-body interaction. It can be show \cite{pethick2008bose} that the depletion of the condensate is equal to
\begin{equation}
\dfrac{\rho_{0}}{\rho} = 1 - \dfrac{8}{3 \sqrt{\pi}} \sqrt{\rho a^3},
\end{equation}
where $\rho$ is a total density and $a$ is the $s$-wave scattering length (see Sec. \ref{sec:scattering_theory}). Consequently, the gas parameter $\rho a^3$ plays a role of a perturbative parameter of the theory. In the experiments, usually, this parameter is of the order of $10^{-5}$, making the depletion of the condensate around one percent. However, this is not true for atomic gases in the vicinity of a Feshbach resonance or in the unitary limit where $|a| \rightarrow \infty$ \cite{eigen2017universal}.

Finally, from the equation of motion for $\hat{ \Psi}$
\begin{equation}
    i\hbar \dfrac{\partial \hat{ \Psi}}{\partial t} = \left[\hat{\Psi}, \hat{H}\right],
\end{equation}
and the Bogoliubov replacement (Eq. \ref{eq:approximation_bogoliubov}), the equation of motion for $\psi$ reads
\begin{eqnarray}
    \label{eq:before_gp}
    i\hbar \dfrac{\partial \psi(\vec{r}, t)}{\partial t} = \left\{-\dfrac{\hbar^2}{2m} \nabla^2 + V_{\rm ext}(r)   \right\} \psi(\vec{r}, t)  \nonumber \\ +  \displaystyle \int d^3 r' \hspace{0.1cm} \psi^*(\vec{r}', t)  V_{\rm int} (\vec{r} - \vec{r}') \psi(\vec{r}', t) \psi(\vec{r}, t).
\end{eqnarray}
Interaction potential between atoms usually has a complicated form (see for example the Aziz potential for $^4$He \cite{aziz1979accurate}), but generally it consists of a short-range hard-core and a long-range Van der Waals tail which varies as $\propto r^{-n}$ at large distances. When the collision between particles are weak, i.e., the incident wave-vector $k$ is small, and the mean interparticle distance is much larger than the typical range of the potential, realistic inter-particle potentials can be replaced with the effective potential \cite{pethick2008bose}
\begin{equation}
    \label{eq:effective_interaction}
    V_{\rm int}(\vec{r} - \vec{r}') \rightarrow V_{\rm eff}(\vec{r} - \vec{r}') = \dfrac{4\pi \hbar^2 a}{m} \delta(\vec{r} - \vec{r}'),
\end{equation}
which is characterized solely by the $s$-wave scattering length $a$. When $V_{\rm int}$ in Eq. (\ref{eq:before_gp}) is replaced with the effective interaction $V_{\rm eff}$ (Eq. \ref{eq:effective_interaction}), we get the famous Gross-Pitaevskii equation
\begin{equation}
    i\hbar \dfrac{\partial \psi(\vec{r}, t)}{\partial t} = \left\{-\dfrac{\hbar^2}{2m} \nabla^2 + V_{\rm ext}(r)   \right\} \psi(\vec{r}, t)  +  \dfrac{4\pi\hbar^2a}{m}|\psi(\vec{r}, t)|^2 \psi(\vec{r}, t).
\end{equation}
The lowest eigenvalue for the Gross-Pitaevskii equation corresponds to the chemical potential, because the condensate wavefunction is $\psi(t) = \ave{\hat{ \Psi} (\vec{r}, t)} = \bra{N} \hat{ \Psi} \ket{N + 1}  \propto \exp\left[-(E_N - E_{N-1}) it / \hbar\right]$, so the stationary Gross-Pitaevskii equation reads
\begin{equation}
    \mu \psi(\vec{r}) = \left\{-\dfrac{\hbar^2}{2m} \nabla^2 + V_{\rm ext}(r)   \right\} \psi(\vec{r})  +  \dfrac{4\pi\hbar^2a}{m}|\psi(\vec{r})|^2 \psi(\vec{r}).
\end{equation}

\subsection{LHY energy}

Replacement of the realistic interaction with the effective one $V_{\rm eff}(r) = 4 \pi \hbar^2 a  \delta(r) / m$ is correct only for very small values of the gas parameter $\rho a^3$. This is because, as the density increases, the condensate depletion starts to be non-negligible, i.e., higher-order momentum contributions to the effective interactions start playing a role \cite{pethick2008bose}. First correction to the mean-field energy is derived by Lee and Yang \cite{lee1957many}, followed by Huang \cite{huang1957quantum}, thus the correction is called the Lee-Huang-Yang (LHY) term. Ground-state energy of the single-component Bose gas, with the included effect of LHY term, is given by
\begin{equation}
\label{eq:energy_per_particle}
    \dfrac{E}{N} = \dfrac{2 \pi \hbar^2 a}{m}\rho \left\{   1 + \dfrac{128}{15 \sqrt{\pi}} \sqrt{\rho a^3} \right\}.
\end{equation}
The energy per particle (Eq. \ref{eq:energy_per_particle}) was observed experimentally \cite{navon2011dynamics}, in the measurement of the equation of state. A theoretical recipe for correcting the mean-field energy with the LHY term is shown to be correct up to $\rho a^3 \approx 10^{-3}$, by comparing the predictions of Eq. (\ref{eq:energy_per_particle}) with a QMC calculation performed in \cite{giorgini1999ground}, showing only the weak dependence on the shape of the potential. For higher densities however, more scattering parameters are required because the universality in terms of the $s$-wave scattering length is ended \cite{braaten2001nonuniversal,giorgini1999ground}. The same form of the correction was also derived for the inhomogeneous Bose gas \cite{braaten1997quantum}, and confirmed to be correct up to $\rho a^3 \approx 10^{-3}$ in a QMC calculation \cite{blume2001quantum}.

\subsection{Density-functional formulation}

An equivalent formulation of a many-body problem is the density-functional theory \cite{hohenberg1964inhomogeneous,eberhard2011density}. The starting point of a density-functional theory is the ground-state density, defined as
\begin{equation}
\rho(\vec{r}) = \displaystyle \int d^3 r_2 \ldots d^3 r_N \Psi^*(\vec{r},\vec{r}_2, \ldots, \vec{r}_N)  \Psi(\vec{r},  \vec{r}_2, \ldots, \vec{r}_N),
\end{equation}
where $\Psi$ is the full many-body solution to the Schr\"odinger equation. This density is obtained by minimizing the energy functional 
\begin{eqnarray}
    \label{eq:energy_functional}
    E[\rho(\vec{r})]  & =  & T[\rho(\vec{r})] + V_{\rm ext}[\rho(\vec{r})] + V_{\rm int}[\rho(\vec{r})] \nonumber \\
    & = & \bra{\Psi} -\dfrac{\hbar^2}{2m}\nabla^2\ket{\Psi} + \bra{\Psi} V_{\rm ext}\ket{\Psi} + \bra{\Psi} V_{\rm int}\ket{\Psi},
\end{eqnarray}
where $\nabla^2 = \sum_{i=1}^{N}  \nabla_i^2$, $V_{\rm ext} = \sum_{i=1}^{N} V_{\rm ext}(\vec{r}_i)$ and $V_{\rm int} = \sum_{i<j}^{N} V_{\rm int}(|\vec{r}_i - \vec{r}_j|)$.
Minimization of the energy Eq. (\ref{eq:energy_functional}) yields the ground-state density and other ground-state observables. In principle this theory is exact \cite{eberhard2011density} since there is a one-to-one correspondence between the full Hamiltonian of the system and the ground-state density profile $\rho(r)$, as proved in a seminal paper by Hohenberg and Kohn \cite{hohenberg1964inhomogeneous}. However, the interaction energy-density $V_{\rm int}[\rho(\vec{r})]$ is unknown and needs to be approximated, meaning that the Density Functional Theory does not necessarily produce exact results. Highly accurate density functionals can be derived employing quantum Monte Carlo data. This is in fact one of the goals of this Thesis, with the specific application to the quantum self-bound Bose-Bose liquid. In the past, density functionals of superfluid liquid helium \cite{ancilotto2017density} were obtained to reproduce various static and dynamic properties from either the experiments, Hartree-Fock calculations, or first-principle QMC calculations \cite{stringari1987surface,barranco2006helium,dalfovo1985hartree}.

For the fully condensed Bose gas, we work in the Hartree approximation, i.e., we assume that all the atoms are in the same single-particle orbital
\begin{equation}
\Psi_{\rm H}(\vec{r}_1, \ldots, \vec{r}_N) = \prod_{i=1}^N \psi(\vec{r}_i),
\end{equation}
We further assume that the interaction part of the functional can be written as
\begin{equation}
    \bra{\Psi} V_{\rm int}\ket{\Psi} = \displaystyle \int d^3 r \mathcal{E}_{\rm int}[\rho],
\end{equation}
where all the interaction effects are incorporated in the energy-density term $\mathcal{E}_{\rm int}(\vec{r})$. The time evolution can thus be obtained starting from the Lagrangian $L$
\begin{eqnarray}
    L & = & \displaystyle \int d^3 r \mathcal{L}, \\
    & = & \displaystyle \int d^3 r\left[ \dfrac{i\hbar}{2} \left(\psi^* \dfrac{\partial \psi}{\partial t} - \psi \dfrac{\partial \psi^*}{\partial t}\right) - \dfrac{\hbar^2}{2m} |\nabla \psi|^2 - V_{\rm ext}(\vec{r})|\psi|^2  - \mathcal{E}_{\rm int}(\vec{r})\right], \nonumber
\end{eqnarray}
whereupon the imposed action principle $\delta \displaystyle \int_{t_1}^{t_2} L dt = 0$ leads to a Schr\"odinger-like equation. For simple, local density functionals behaving as a power-law of the density $\rho$, the equation of motion for $\psi$ reads
\begin{equation}
    \label{eq:dft_equation_of_motion}
    i\hbar \dfrac{\partial \psi}{\partial t} = \hat{\mathcal{H}}_{\rm DFT}\psi = \left\{-\dfrac{\hbar^2}{2m} \nabla^2 \psi + V_{\rm ext} + \dfrac{\partial \mathcal{E}_{\rm int}}{\partial \rho} \right\} \psi.
\end{equation}
To make a connection of density-functional theory with QMC calculations, bulk properties can be exactly obtained by fitting the numerically accesible energy per particle $E/N$, so that the interaction term $\mathcal{E}_{\rm int}$ can be written as
\begin{equation}
    \mathcal{E}_{\rm int} = \dfrac{E}{N} \rho .
\end{equation}
For a density functional to satisfy other than the bulk properties, more terms need to be included. 
To name one example, we give the OT (Orsay-Trento) functional of $^4$He, which, on top of the bulk energy, properly accounts for the static response function and the phonon-roton dispersion in the uniform liquid \cite{dalfovo1995structural}
\begin{equation}\begin{aligned}
\mathcal{E}_{\rm int}^{\text {Full } \mathrm{OT}}[\rho]=& \frac{1}{2} \int \mathrm{d} \mathbf{r}^{\prime} \rho(\mathbf{r}) V_{\mathrm{LJ}}\left(\left|\mathbf{r}-\mathbf{r}^{\prime}\right|\right) \rho\left(\mathbf{r}^{\prime}\right) \\
&+\frac{1}{2} c_{2} \rho(\mathbf{r})[\bar{\rho}(\mathbf{r})]^{2}+\frac{1}{3} c_{3} \rho(\mathbf{r})[\bar{\rho}(\mathbf{r})]^{3} \\
&-\frac{\hbar^{2}}{4 m} \alpha_{s} \int \mathrm{d} \mathbf{r}^{\prime} F\left(\left|\mathbf{r}-\mathbf{r}^{\prime}\right|\right)\left[1-\frac{\tilde{\rho}(\mathbf{r})}{\rho_{0 s}}\right] \nabla \rho(\mathbf{r}) \cdot \nabla^{\prime} \rho\left(\mathbf{r}^{\prime}\right)\left[1-\frac{\tilde{\rho}\left(\mathbf{r}^{\prime}\right)}{\rho_{0 s}}\right] \\
&-\frac{m}{4} \int d \mathbf{r}^{\prime} V_{J}\left(\left|\mathbf{r}-\mathbf{r}^{\prime}\right|\right) \rho(\mathbf{r}) \rho\left(\mathbf{r}^{\prime}\right)\left[\mathbf{v}(\mathbf{r})-\mathbf{v}\left(\mathbf{r}^{\prime}\right)\right]^{2}.
\end{aligned}\end{equation}

\subsection{\label{numerical_solution_gpe_equations}Numerical solution of GPE-like equations}

The numerical solution to the Eq. (\ref{eq:dft_equation_of_motion}) is obtained by mapping the wavefunction $\psi$ on a three-dimensional grid of equally spaced grid points \cite{barrancozero}. Time evolution is performed by applying the time-evolution operator at each iteration
\begin{equation}
    \psi(t + \Delta t) = \mathcal{T}(\Delta t) \psi(t),
\end{equation}
where $  \mathcal{T}(\Delta t)  =  \exp \left\{ -i \hat{ \mathcal{H}}_{\rm DFT}  \Delta t  / \hbar \right\}$. The evolution operator $\mathcal{T}$ is not known analytically. However, it can be decomposed by means of the Trotter formula. The simplest approximation we use in this thesis reads 
\begin{eqnarray}
    \label{eq:decomposition_trotter}
    \mathcal{T}(\Delta t) = e^{-i \Delta t  V(\vec{r}') / 2\hbar } e^{-i \Delta t \hbar \nabla^2 / (2m) }  e^{-i \Delta t  V(\vec{r}) / 2\hbar}   + \mathcal{O}(\Delta t^3).
\end{eqnarray}
Potential propagators are trivially evaluated in real space, and the kinetic propagator is evaluated in $k$-space. Higher order decomposition schemes are also available \cite{chin2009arbitrary, chin2009any, auer2001fourth}, but in this Thesis we mantain the second-order scheme. The methodology following the Eq. (\ref{eq:decomposition_trotter}) inherently conserves the number of particles during real-time evolution since the time-evolution operator is still unitary. In order to obtain ground-state properties, the time evolution is performed in imaginary time $\tau = -it / \hbar$. For a sufficiently large imaginary-time propagation, the wavefunction reaches a ground state, since it becomes the dominant mode due to exponentially decaying contributions of excited states in large imaginary-time propagation. This is a general feature of imaginary-time propagation, used also in a diffusion Monte Carlo method (see Sec. \ref{sec:dmc}). The algorithm is presented below as a pseudocode 
\begin{enumerate}
    \item  $\psi_1(\vec{r})= G_1 * \psi(\vec{r}, t)$, where $G_1 = \exp\left\{ -\dfrac{1}{2} i \Delta t  V / \hbar  \right\} $, $V = V_{\rm ext}(\vec{r}) + \dfrac{\partial \mathcal{E}_{\rm int}}{\partial \rho} $ and $\rho = |\psi(\vec{r}, t)|^2$. $*$ stands for element-wise multiplication, where a result tensor has the same dimension as the two tensors to be multiplied, with the value at the grid point $(i,j,k)$ being the product of elements $(i,j,k)$ of the two tensors.
    \item $\psi_2(\vec{r}) = {\rm FFT^{-1}}\left[G_2 * {\rm FFT[\psi_1]} \right]$, where $G_2 = \exp\left\{ -i\Delta t \hbar k^2 / (2m) \right\}$ and FFT stands for Fourier transform.
    \item $\psi(\vec{r}, t + \Delta t)= G_1 * \psi_2(\vec{r})$, where $G_1 = \exp\left\{ -\dfrac{1}{2} i \Delta t  V / \hbar  \right\} $, $V = V_{\rm ext}(\vec{r}) + \dfrac{\partial \mathcal{E}_{\rm int}}{\partial \rho} $ and $\rho = |\psi_2(\vec{r})|^2$.
\end{enumerate}

\section{{\label{sec:two_comp_bose}}Two-component Bose system}

The formalism of two-component Bose-Einstein condensates is readily extended from the single-component one. Assuming the Hartree wavefunction
\begin{equation}
    \Psi(\vec{r}_1, \ldots, \vec{r}_N) = \prod_{i=1}^{N_1} \psi_1(\vec{r}_i) \prod_{j=N_1 + 1}^{N} \psi_2(\vec{r}_j),
\end{equation}
where $N_1$ ($N_2$) is the number of atoms of the first (second) component, and $N = N_1 + N_2$ is the total atom number. A general density functional describing two-component quantum Bose-Bose mixture reads
\begin{eqnarray}
    \mathcal{E}[\rho_1, \rho_2]&  = & \dfrac{\hbar^2}{2m_1} |\nabla \psi_1(\vec{r})|^2 +V_{\rm ext}^{(1)}(\vec{r}) |\psi_1(\vec{r})|^2 + \nonumber \\ 
    & &  \dfrac{\hbar^2}{2m_2} |\nabla \psi_2(\vec{r})|^2  + V_{\rm ext}^{(2)}(\vec{r}) |\psi_2(\vec{r})|^2 \nonumber+ \\
    & &   \mathcal{E}_{\rm int}[\rho_1, \rho_2],
\end{eqnarray}
where $m_i$ is the mass of atom of $i$-th species and $V_{\rm ext}^{i}(\vec{r})$ is the external potential subjected to the $i$-th component. In a mixture of two Bose-Einstein condensates, the mean-field energy-functional with point interactions reads
\begin{equation}
    \mathcal{E}_{\rm int}^{\rm MF} = \dfrac{1}{2} g_{11} \rho_1^2 + \dfrac{1}{2} g_{22} \rho_2^2 + g_{12} \rho_1 \rho_2,
\end{equation}
where $g_{ii} = 4\pi\hbar^2 a_{ii} / m_i$ ($i = 1, 2$), $g_{12}= 2\pi\hbar^2 a_{12} / \mu$ and $\mu = (1 / m_1 + 1 / m_2)^{-1}$.  The corresponding coupled equations for $\psi_1$ and $\psi_2$ are named Gross-Pitaevskii equations, and they read
\begin{equation}
    i\hbar \dfrac{\partial \psi_i}{\partial t} =  \left\{ -\dfrac{\hbar^2}{2m_i} \nabla^2 + V_i(\rho_1, \rho_2)   \right\} \psi_i  \hspace{0.5cm} i=1, 2
\end{equation}
where
\begin{equation}
    V_i = g_{ii} \rho_i + g_{12} \rho_j \hspace{0.5cm} (j \neq i=1, 2),
\end{equation}
and $\rho_i = |\psi_i|^2$, for $i=1,2$.

\subsection{\label{sec:stability_bose_mixture}Stability of a Bose-Bose mixture}

Here, we consider a stability criteria of a homogeneous Bose-Bose mixture with contact interaction, on a mean-field level. For a mixture to be stable, the total energy must increase for small deviations of each density $\delta \rho_1$ and $\delta \rho_2$ \cite{ao1998binary}. For a very small density variation, kinetic energy terms can be neglected, so the total energy reads \cite{pethick2008bose}
\begin{equation}
    E = \displaystyle \int d^3 r \mathcal{E}_{\rm int}=  \int d^3 r  \left\{\dfrac{1}{2} g_{11} \rho_1^2 + \dfrac{1}{2} g_{22} \rho_2^2 + g_{12}  \rho_1 \rho_2  \right\}.
\end{equation}
A first-order variation $\delta E$ reads
\begin{equation}
    \delta E = \displaystyle \int d^3 r \left\{ \dfrac{\partial \mathcal{E}_{\rm int}}{\partial \rho_1} \delta \rho_1 + \dfrac{\mathcal{E}_{\rm int}}{\partial \rho_2} \delta \rho_2  \right\},
\end{equation}
which vanishes because the number of atoms is conserved $\int d^3r \delta \rho_i = 0$, for $i=1, 2$. Second-order variation reads
\begin{equation}
    \delta^2 E = \dfrac{1}{2} \displaystyle \int d^3 r \left\{  \dfrac{\partial^2 \mathcal{E}_{\rm int}}{\partial \rho_1^2} (\delta \rho_1)^2 + \dfrac{\partial^2 \mathcal{E}_{\rm int}}{\partial \rho_2^2} (\delta \rho_2)^2 + 2 \dfrac{\partial^2 \mathcal{E}_{\rm int}}{\partial \rho_1 \partial \rho_2} \delta \rho_1 \delta \rho_2    \right\}.
\end{equation}
Expanding the integrand to a positive definite square, we get
\begin{eqnarray}
     \left( \sqrt{\dfrac{\partial \mu_1 }{\partial \rho_1}} \delta \rho_1 \pm \sqrt{\dfrac{\partial \mu_2}{\partial \rho_2}}  \delta \rho_2\right)^2 & = & \dfrac{\partial \mu_1}{\partial \rho_1} (\delta \rho_1)^2 + \dfrac{\partial \mu_2}{\partial \rho_2} (\delta \rho_2)^2 +   2 \delta \rho_1 \delta \rho_2  \dfrac{\partial^2 \mathcal{E}_{\rm int}}{\partial \rho_1 \partial \rho_2}    + \nonumber \\  & + &    2 \delta \rho_1 \delta \rho_2 \left\{   \pm \sqrt{\dfrac{\partial \mu_1}{\partial \rho_1} \dfrac{\partial \mu_2}{\partial \rho_2} }  - \dfrac{\partial^2 \mathcal{E}_{\rm int}}{\partial \rho_1 \partial \rho_2}\right\} 
\end{eqnarray}
where the chemical potential is $\mu_i = \partial \mathcal{E}_{\rm int} / \partial \rho_i$, for $i=1, 2$. A requierement $\delta^2E > 0$ leads to the conditions
\begin{equation}
    \dfrac{\partial \mu_i}{\partial \rho_i} = g_{ii} \ge 0 \hspace{0.5 cm} i = 1, 2,
\end{equation}
\begin{equation}
    \label{eq:stability}
    \pm \sqrt{\dfrac{\partial \mu_1}{\partial \rho_1} \dfrac{\partial \mu_2}{\partial \rho_2}} - \sqrt{\dfrac{\partial \mu_1}{\partial \rho_2} \dfrac{\partial \mu_2}{\partial \rho_1}} = \pm \sqrt{g_{11} g_{22}} - g_{12} \ge  0.
\end{equation}
First conditions, $g_{ii} > 0$, for $i=1, 2$, are due to the requirement of stability against collapse of each of the components. In case of attractive interactions (negative $g_{ii}$), the system is unstable under variations of density, so that the ground-state of a component $i$ with a negative $g_{ii}$ has an non-uniform density, i.e., it is a self-bound cluster. For the repulsive interspecies interactions, $g_{12} >  0$, the requirement for stability occurs for a positive sign in Eq. (\ref{eq:stability}), finally reading
\begin{equation}
    \label{eq:repulsive}
    g_{12} \le \sqrt{g_{11} g_{22}}.
\end{equation}
This condition ensures that the density disturbance in any of the two components does not lead to their phase separation. When $g_{12} = \sqrt{g_{11} g_{22}}$ is satisfied, deviations to the ground-state energy scale as $(\sqrt{g_{11}} \rho_1 + \sqrt{g_{22}} \rho_2)^2$. For attractive interspecies interactions, i.e., for a negative sign in Eq. (\ref{eq:stability}), the stability condition reads
\begin{equation}
    g_{12} \le - \sqrt{g_{11} g_{22}},
\end{equation}
preventing the formation of a self-bound cluster due to the stronger attractive interspecies interaction over geometric average of intraspecies repulsion. When $g_{12} = -\sqrt{g_{11} g_{22}}$, the optimal configuration satisfies $\rho_2 / \rho_1 = \sqrt{g_{11} / g_{22}}$, according to Eq. (\ref{eq:stability}). For $g_{12} > - \sqrt{g_{11} g_{22}}$, the ground-state is either a dense droplet or a collapsed state, depending on the particle number \cite{pethick2008bose}.

\subsection{LHY energy for Bose-Bose mixtures}

At higher densities and stronger interactions, the quantum corrections to the mean-field energy due to the condensate depletion start to play a role. Energy correction to the mean-field is called the Lee-Huang-Yang (LHY) term, and was first derived for quantum mixtures in Ref. \cite{larsen1963binary}. The energy density term is given by \cite{ancilotto2018self}
\begin{equation}
    \label{eq:lhy_term}
    \mathcal{E}_{\rm LHY} = \dfrac{8}{15 \pi^2} \left(\dfrac{m_1}{\hbar^2}\right)^{3 / 2} \left(g_{11} \rho_1\right)^{3 / 2}f\left(\dfrac{m_2}{m_1}, \dfrac{g_{12}^2}{g_{11} g_{22}}, \dfrac{g_{22} \rho_2}{g_{11} \rho_1}\right),
\end{equation}
where $f(z, u, x)$ is given in \cite{ancilotto2018self}. For a homonuclear case $z = 1$, the expression reads \cite{petrov2015quantum}
\begin{equation}
    f(1, u, x) = \dfrac{1}{4 \sqrt{2}} \sum_{\pm} \left(1 + x \pm \sqrt{(1 - x)^2 + 4 u x}\right)^{5 / 2}.
\end{equation}
Note that for $u > 1$, i.e., $g_{12}^2 > g_{11} g_{22}$, the LHY term is a complex valued number. This feature is also present in the heteronuclear scenario \cite{minardi2019effective}. This is usually omitted by taking an \textit{ad hoc} approximation $u = 1$ valid only in the vicinity of $g_{12}^2 \approx g_{11} g_{22}$, resulting in $f(1, 1, x) = (1 + x)^{5/2}$. For a heterogenous mixture with $g_{12} \neq g_{22} \neq g_{11}$, the LHY term in Eq. (\ref{eq:lhy_term}) is given in terms of an non-analytic integral and thus it has to be obtained numerically. However, an effective and accurate expression for this term was deduced in Ref. \cite{minardi2019effective}, so that finally the LHY term for a general mixture, approximating $g_{12}^2 = g_{11} g_{22}$, reads
\begin{equation}
    \mathcal{E}_{\rm LHY} \simeq \dfrac{8 m_1^{3 / 2} (g_{11} \rho_1)^{5 / 2)}}{15 \pi^2 \hbar^3} \left[1 + \left(\dfrac{m_2}{m_1}\right)^{3 / 5} \dfrac{g_{22} \rho_2}{g_{11} \rho_1}\right]^{5 / 2}.
\end{equation}

\subsection{Quantum droplets}

For a single-component Bose gas, the LHY term is small and usually included only when the density is high and the influence of interactions is important. This stands for the repulsive Bose mixtures as well. However, when $g_{12} \lesssim -\sqrt{g_{11} g_{22}}$, the mean-field term starts to be of the same order as the LHY term, first noted in a seminal paper by D. Petrov \cite{petrov2015quantum}. A mixture with residual attractive mean-field energy can be balanced with the LHY term, resulting in a droplet having a density proportional to $\propto \delta g^2 / g^5$, where $g$ is the coupling constant and $\delta g = g_{12} + \sqrt{g_{11} g_{22}}$. Thus, it is theoretically possible to create a droplet of arbitrarily low density. The prediction was confirmed by experimental observation of ultradilute droplets in homonuclear \cite{cabrera2018quantum,semeghini2018self}, and heteronuclear mixtures \cite{derrico2019observation}.

\begin{figure}[!htb]
    \centering
    \includegraphics[width=0.8\linewidth]{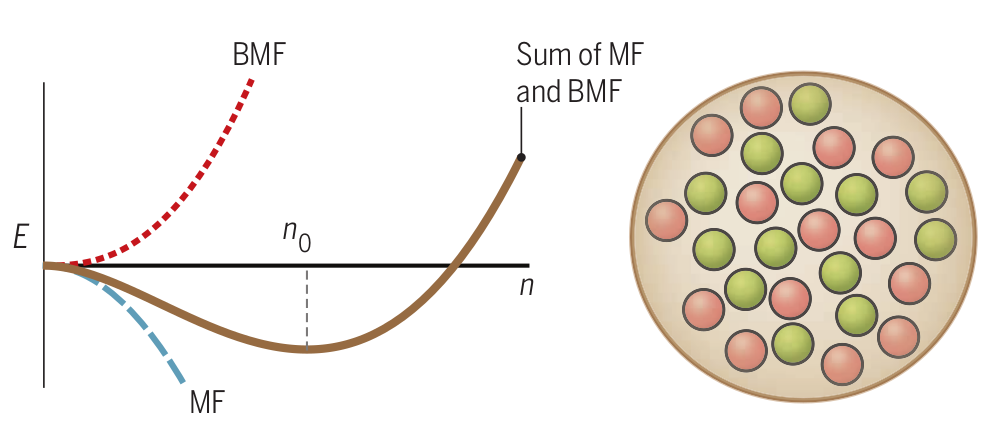}
    \caption[Graphical illustration of the balance between mean-field (MF) and Lee-Huang-Yang (LHY) energy, resulting in a formation of a quantum droplet.]{Graphical illustration of the balance between mean-field (MF) and Lee-Huang-Yang (LHY) energy, resulting in a formation of a quantum droplet. Figure taken from \cite{ferrier2018quantum}.}
    \label{fig:twocomp}
\end{figure}

\section{\label{sec:scattering_theory}Scattering theory  of ultracold collisions}

Interaction plays a crucial role in the complexity of the many-body problem, otherwise, the problem could be separated into $N$ one-body problems. In a dilute many-body system, most of the interactions occurs as binary collisions. Therefore, a natural starting point for constructing the many-body wavefunction is the two-particle problem. 

Let us consider a collision between two particles in vacuum, of mass $m_1$ and $m_2$, each having coordinates $\vec{r}_1$ and $\vec{r}_2$ and interacting through the pair-potential $V_{\rm int}(|\vec{r}_i - \vec{r}_j|)$. The Schr\"oedinger equation for the two-body problem reads \cite{landau2013quantum}
\begin{equation}
\label{eq:two_body}
\left\{ \frac{-\hbar^2}{2m_1} \nabla^2_{\vec{r}_1} + \frac{-\hbar^2}{2m_2} \nabla^2_{\vec{r}_2} + V_{\rm int}(|\vec{r}_i - \vec{r}_j|)\right\}
\Psi(\vec{r}_1, \vec{r}_2) = \mathcal{E} \Psi(\vec{r}_1, \vec{r}_2).
\end{equation}
The problem (\ref{eq:two_body}) is translationary invariant, so it can be reduced to a one-body problem after making a transformation to the relative coordinate $\vec{r}=\vec{r}_1 - \vec{r}_2$, and the center-of-mass coordinate $\vec{r}_{\rm cm}=\sum_i m_i\vec{r}_i/(m_1+m_2)$. Thus, the wavefunction is decomposed as $\Psi(\vec{r}_1, \vec{r}_2) = \psi_{\rm cm}(\vec{r}_{\rm cm})\psi(\vec{r})$, as well as the energy $\mathcal{E} = E_{\rm cm} + E$, with $E_{\rm cm}$ being the center-of-mass energy and $E$ the incident energy of the relative motion. The stationary equation for $\psi$ reads
\begin{equation}
\label{eq:two_body_pair}
\left\{ \frac{-\hbar^2}{2\mu}\nabla^2 + V_{\rm int}(r)\right \}
\psi(\vec{r}) = E \psi(\vec{r}),
\end{equation}
where $\mu=m_1m_2/(m_1+m_2)$ is the reduced mass. 
\begin{figure}[H]
    \centering    
    \includegraphics[width=0.7\textwidth]{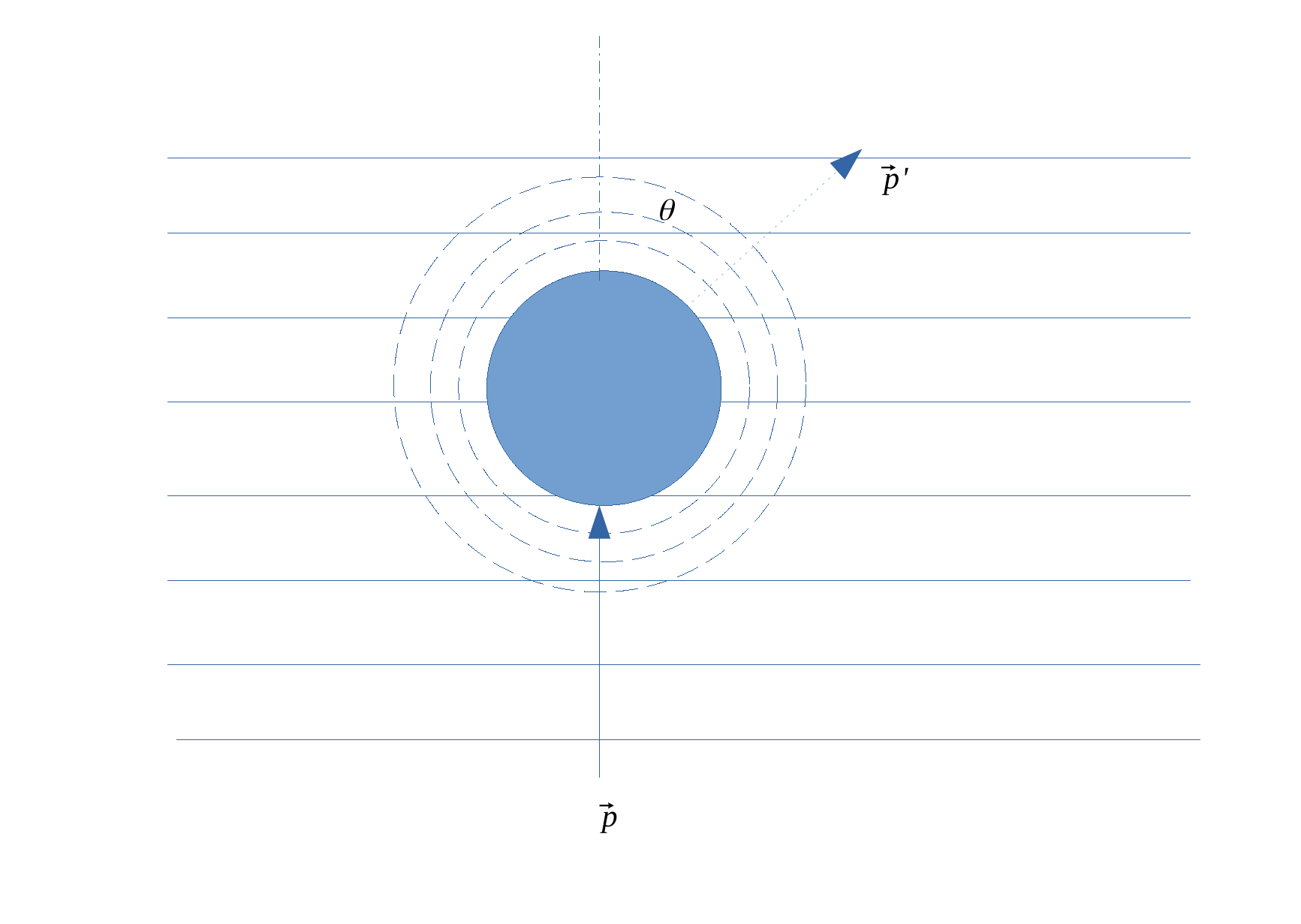}
    \caption[Schematic representation of the three-dimensional scattering]{Schematic representation of the three-dimensional scattering. Incoming free particle with the momentum $\vec{p}$, presented with solid lines, enter into the region where an interaction potential $V_{\rm pair}$ is non-negligible. An outcome of the collision is a spherical wave and an outgoing particle with momentum $\vec{p}'$, denoted by dashed and dotted lines, respectively. A scattering angle of collision is denoted by $\theta$.}
    \label{fig:scattering_figure}
\end{figure}
Thinking of the scattering problem, the solution of Eq. (\ref{eq:two_body_pair}) can be expanded into partial waves \cite{blume2012few}
\begin{equation}
\psi_k(\vec{r}) = \sum_{k} R_{kl}(r) P_l(\theta),
\end{equation}
where $P_l$ is the Legendre polynomial of degree $l$, $\theta$ is the scattering angle (see Fig. \ref{fig:scattering_figure}) and $k$ is the incident wavevector $k = \sqrt{2 \mu E} / \hbar$. In the asymptotic regime $r \rightarrow \infty$ \cite{blume2012few,landau2013quantum,newton2013scattering}, the wavefuction behaves as 
\begin{equation}
\label{eq:rkl_first}
R_{kl}(r) \displaystyle \propto j_l(kr) - \tan\left[\delta_l(k)\right] n_l(kr),
\end{equation}
where $j_l$ and $n_l$ are the Bessel and Neumann functions, respectively. $\delta_l(k)$ is called the phase shift for the $l$-th partial wave and it is equal to zero in the absence of interaction, i.e., when $V_{\rm int} = 0$. Functions $j_l$ and $n_l$ correspond to homogeneous and particular solution of Eq. (\ref{eq:two_body_pair}), respectively, meaning that the non-zero interaction can be seen as a unitary transformation which changes a phase of $l$-th angular momentum eigenstate. This phase accumulates in the region where the potential is non-negligible, and a small positive (negative) value of $\delta_l(k)$ can be interpreted as an effective attractive (repulsive) interaction, which can be directly seen by rewriting Eq. (\ref{eq:rkl_first}) as \cite{blume2012few}
\begin{equation}
R_{kl}(r) \displaystyle \propto \sin(kr + \delta_l(k)).
\end{equation}
It is useful to define the general energy-dependent scattering length \cite{taylor2006scattering} as
\begin{equation}
a_l^{(2l + 1)}(k) = - \dfrac{\tan \left[\delta_l(k)\right]}{k^{2l + 1}},
\end{equation}
where we name $a_0(k)$ the energy-dependent $s$-wave scattering length, $a_1(k)$ the energy-dependent $p$-wave scattering length, and so on. Generally, for a given interaction potentail, \textit{all} the scattering lengths are non-zero. However, in the field of ultracold atomic gases, experiments are performed at very low temperature and low density, meaning that collisions are occurring at very low incident wavevector $k$. It can be shown that this allows a large number of phase shifts to be neglected. This is because the order of importance of $l$-th phase shift is smaller for larger values of $l$, since $\delta_l(k)$ goes to zero as $k^{2l +1}$ for small $k$, provided that $l < (n - 3)/ 2$, and that the potential varies as $\sim r^{-n}$ at large distances \cite{landau2013quantum,pethick2008bose}. To obtain the low-energy properties of a binary collision, we expand the $a_0(k)$ around $k=0$ \cite{landau2013quantum}
\begin{equation}
\label{eq:tan_expansion}
\dfrac{\tan \left[\delta_0(k)\right]}{k} = a  + \dfrac{1}{2} r_{\rm eff} a^2 k^2 + \mathcal{O}(k^4),
\end{equation}
where $a$ is the zero-momentum $s$-wave scattering length and $r_{\rm eff}$ is the effective range \cite{taylor2006scattering}. These two very useful quantities are used throughout the Thesis for describing the interactions between atoms. In the field of ultracold atomic gases, the second term in Eq. (\ref{eq:tan_expansion}) is usually omitted, allowing for an interaction to be described solely by one parameter, the $s$-wave scattering length. Usually, the potential between atoms is not known precisely, except for the knowledge of $a$, so the real atom-atom interaction is usually replaced by the $\delta$-pseudopotential \cite{huang1957quantum}
\begin{equation}
V(r) = \dfrac{4\pi\hbar^2 a}{m} \delta(r),
\end{equation}
which is characterized solely by $a$. For this potential, there are no higher-order scattering lengths than the $s$-wave. The $s$-wave effective range is zero for a contact interaction because it is a constant in $k$-space. 

\section{\label{sec:calculation_swave_reff}Calculation of $s$-wave scattering length and the effective range}

Two most important parameters we use to describe the interactions in this Thesis are the $s$-wave scattering length $a$ and the effective range $r_{\rm eff}$. Because the $s$-wave scattering affects the $l= 0$ state, the wavefunction is spherically symmetric, and the two-body Schr\"odinger equation describing binary collision reads
\begin{equation}
\label{eq:radial_schroedinger}
-\dfrac{d^2}{dr^2} \left[r\psi(r)\right] + \left[\dfrac{2\mu V_{\rm int}(r)}{\hbar^2} - k^2 \right] \left[r\psi(r)\right] = 0.
\end{equation}
For any short-range potential which does not support a two-body bound state \cite{landau2013quantum}, the long-range behavior of the wavefunction reads
\begin{equation}
\label{eq:asymptotics_a}
\psi \propto 1 - \dfrac{a}{r}.
\end{equation}
It is readily seen that positive (negative) values of $a$ can be interpreted as an effective repulsive (attractive) interaction. Therefore, one way of obtaining $a$ is by fitting a solution of the two-body problem (Eq. \ref{eq:radial_schroedinger}) to the form of Eq. (\ref{eq:asymptotics_a}) at large $r$. A second way to calculate $a$ is to solve Eq. (\ref{eq:radial_schroedinger}) for several values of small $k$, and then to fit the long-range part of each of the solutions with
\begin{equation}
    \psi \propto \sin(kr + \delta_0(k)),
\end{equation}
to obtain the dependence of a $l=0$ phase shift $\delta_0(k)$ on $k$. Then, for a sufficiently small $k$, both $a$ and the effective range $r_{\rm eff}$ can be obtained by fitting to the form
\begin{equation}
    \dfrac{\tan \left[\delta_0(k)\right]}{k} = a  + \dfrac{1}{2} r_{\rm eff} a^2 k^2.
\end{equation}
Alternatively, the effective range can be calculated by performing the integration 
\begin{equation}
r_{\rm eff} = \dfrac{2}{a^2}  \int_{0}^{\infty} \left\{ (r - a)^2  - \dfrac{1}{C_0^2} [r \psi(r)]^2  \right\} dr,
\end{equation} 
where $C_0 = \lim\limits_{r\rightarrow\infty} \psi(r)$. Practically, in a numerical estimation of the effective range, the upper limit is replaced by a sufficiently large value so that the true value is reached below a given error threshold.

\subsection{Numerov algorithm}
\label{subsec:numerov_algorithm}

Scattering parameters such as the $s$-wave scattering length $a$ and the effective range $r_{\rm eff}$ can be obtained analytically \cite{stoof2009ultracold,flambaum1999analytical} for some potentials. However, in most cases, they must be determined numerically. We perform the numerical integration of the Schr\"odinger equation utilizing the Numerov algorithm \cite{allison1972calculation,zettili2003quantum}.

The two-body Schr\"odinger equation
\begin{equation}
\label{eq:radial_schroedinger_again}
-\dfrac{d^2}{dr^2} \left[r\psi(r)\right] + \left[\dfrac{2\mu V_{\rm int}(r)}{\hbar^2} - k^2 \right] \left[r\psi(r)\right] = 0,
\end{equation}
can be written as 
\begin{equation}
    -\chi(r)'' + u(r)\chi(r) = 0,
\end{equation}
where
\begin{equation}
    \chi(r) = r\psi,
\end{equation}
and
\begin{equation}
    u(r) = \dfrac{2\mu V_{\rm int}(r)}{\hbar^2} - k^2.
\end{equation}
The problem is mapped to a discrete, equally spaced one-dimensional grid in the radial coordinate with the grid size $h = r_{i+1} - r_{i}$. Since the equation is second-order, two initial conditions are required. The first condition is $\chi(0) = 0$. The next point $\chi (\Delta r)$ can be chosen to have an arbitrary value, since this is equivalent to a normalization condition. If the potential is not finite at $r = 0$, then the solution is started at small but finite $0^+$, which helps to stabilize the algorithm. In that case, additional checks need to be made in order to reach a physical result. Finally, a numerical solution of Eq. (\ref{eq:radial_schroedinger_again}) is obtained by performing forward iteration Numerov algorithm \cite{jamieson2003elastic,petar}
\begin{equation}
    \chi_{i + 1} = \dfrac{2\left(1 - \dfrac{5}{12} h^2 u_n^2\right)\chi_i - \left(1 + \dfrac{1}{12} h^2 u_{i-1}^2\right)\chi_{i-1}}{1 + \dfrac{1}{12} h^2 u_{n+1}^2},
\end{equation}
where $\chi_i = \chi(r_i)$ and $u_i = u(r_i) $. This algorithm has a $\mathcal{O}(h^6)$ local error and serves as a usual technique for calculating scattering properties of interatomic potentials. Once the exact solution to the two-body problem is solved, it can be used either to calculate the scattering properties of a interaction potential (see Sec. \ref{sec:calculation_swave_reff}) or as a basis for the construction of the Jastrow trial wavefunction (see Sec. \ref{sec:vmc}).

	

\chapter{{\label{ch:qmc_methods}}Quantum Monte Carlo methods}

\section{Introduction}

The field of many-body physics goes beyond ultracold atomic gases. It is a general framework to be used whenever there are \textit{many} and, more importantly, \textit{interacting} particles which are put together. Any real physical system that we encounter consists of interacting particles: nuclei, atoms, molecules and so on.  They are all described by the Schr\"odinger equation
\begin{equation}
\label{eq:hat_psi_e_psi}
\hat{H} \psi = E \psi,
\end{equation}
where $\psi$ is the wavefunction and $E$ is the corresponding energy. In this Thesis, we are interested in understanding $N$-particle systems, where the particles interact through the two-body pair-wise potential $V_{\rm pair}$, and are possibly in the external potential defined by the $V_{\rm ext}(\vec{r})$. Thus, the Hamiltonian $\hat{H}$ generally reads
\begin{equation}
\label{eq:hamiltonian_many_body}
\hat{H} = \sum_{i=i}^{N} -\dfrac{\hbar^2}{2m} \nabla_i^2 +  \sum_{i=i}^{N} V_{\rm ext}(\vec{r}_i) + \sum_{i<j}^{N} V_{\rm pair}(|\vec{r}_i - \vec{r}_j|),
\end{equation}
There are numerous approaches to solving Eq. (\ref{eq:hat_psi_e_psi}) \cite{pethick2008bose}. A first approximation is the Hartree approach, in which the motion of every particle is considered to be independent of other particles, and they rather ``feel'' the forces in a mean-field manner. This approach is motivated by the non-interacting picture, where the problem is trivially separable into $N$ one-body problems. Another analytical approach is the perturbation theory, where a theory is developed through the introduction of a small perturbation parameter, \`a la Bogoliubov \cite{bogoliubov1947theory}. Nowadays, the method of quantum field theory is a general way of theoretically understanding the ultracold atomic quantum gases \cite{stoof2009ultracold}. Still, out of the applicability domain of a given theory, numerical approaches provide an important, and sometimes the only tool in understanding highly correlated systems.

Quantum Monte Carlo methods offer an elegant \textit{ab initio} numerical solution for the quantum many-body problem. Quantum nature emerges at ultracold temperatures, where the many-body system is in the ground-state. The suitable Monte Carlo technique for studying ground-state properties in an \textit{exact} manner is the Diffusion Monte Carlo method. In this chapter, we introduce the quantum Monte Carlo methods used in the Thesis. The Monte Carlo method is a broad and general technique that relies on (pseudo)random numbers to study a given problem. In this sense, its applicability goes beyond studying quantum many-body physics. In order to perform high-dimensional integrals, we utilize the Metropolis algorithm, presented in Sec. (\ref{sec:metropolis}). Quantum Monte Carlo techniques are exact within some statistical noise, and the method for estimating the associated statistical error is presented in Sec. (\ref{sec:error_analysis}). The Variational Monte Carlo method, a method for obtaining good trial wavefunctions and variational estimates of ground-state properties, is presented in Sec. (\ref{sec:vmc}). Finally, in Sec. (\ref{sec:dmc}) we present the Diffusion Monte Carlo method, suitable for exploring quantum many-body systems at zero temperature in the ground state.

\section{\label{sec:metropolis}Metropolis algorithm}

Integral formulation of the quantum many-body problem containing $N$ particles leads to highly multidimensional integrals of the following type
\begin{equation}
\label{eq:init_problem}
\int d\vecR f(\vecR),
\end{equation}
where $\vecR = \left\{\vec{r}_1, \ldots, \vec{r}_N\right\}$ is a high-dimensional vector in the space $\mathbb{R}^{DN}$, with $D$ being the dimension. Often these integral are non-trivial and without analytical solution, so the only way to solve them is by employing a numerical technique. Numerical finite-grid integration techniques are not a suitable approach for this particular problem because of memory and processor requirements to deal with exponentially large number of integration points. A general Metropolis algorithm \cite{metropolis1949monte}, a method for sampling a given distribution function is used instead. Monte Carlo methods applied in this Thesis all rely on this approach. The trick is to rewrite integral in the following form
\begin{equation}
\label{eq:integral}
\int d\vecR f(\vecR) = \int d\vecR \hspace{0.05cm}  g(\vecR) \rho(\vecR),
\end{equation}
where $\rho(\vecR)$ is interpreted as the normalized probability distribution function, whereas the integral in Eq. (\ref{eq:init_problem}) is recognized as the expectation value of the function $g(\vecR)$.

In the Metropolis algorithm, the simulation is performed as a Markov process which starts at a given point $\vecR$, with the update from state $\vecR$ to state $\vecR'$ defined with a transition probablity $P(\vecR \rightarrow \vecR')$, satisfying the detailed balance condition \cite{metropolis1949monte,hammond1994monte}
\begin{equation}
\label{eq:detailed_balance}
\rho(\vecR) P(\vecR \rightarrow \vecR') = \rho(\vecR') P(\vecR' \rightarrow \vecR).
\end{equation}
The transition probability $P$ is decomposed as $P(\vecR \rightarrow \vecR') = \mathcal{G}\left(\mathbf{R} \rightarrow \mathbf{R}^{\prime}\right) \mathcal{A}\left(\mathbf{R} \rightarrow \mathbf{R}^{\prime}\right)$ \cite{hammond1994monte}, where $\mathcal{A}$ is the acceptance distribution of a move $\mathbf{R} \rightarrow \mathbf{R}^\prime$ and $\mathcal{G}$ is a proposal distribution, chosen in such a way that the phase space of the problem is sampled efficiently. Metropolis algorithm then accepts the move with probability
\begin{equation}
\label{eq:metropolis_algorithm}
\mathcal{A}\left(\mathbf{R} \rightarrow \mathbf{R}^{\prime}\right) = \text{min}\left(1, \dfrac{\rho(\vecR') \mathcal{G}(\vecR' \rightarrow \vecR) }{\rho(\vecR) \mathcal{G}(\vecR \rightarrow \vecR')} \right).
\end{equation}
Eq. (\ref{eq:metropolis_algorithm}) satisfies the detailed balance condition (\ref{eq:detailed_balance}), which finally ensures that the distribution $\rho(\vecR)$ is sampled ergodically, thus making possible to express the integral (\ref{eq:integral}) as the average during the stochastic walk
\begin{equation}
\label{sec:metrAlg_integral}
\int d\vecR \, g(\vecR) \rho(\vecR) = \lim\limits_{N \rightarrow \infty} \frac{1}{N}\sum_{i=1}^{N} g(\vecR_i),
\end{equation}
where $\vecR_i$ are stochastically generated high-dimensional vectors, drawn from the given distribution $\rho(\vecR)$. In Quantum Monte Carlo techniques, the average is computed over a set of \textit{walkers}, each one being a point in a phase space $\vecR = \left(\vec{r}_1, \ldots, \vec{r}_N\right)$, with $N$ being the number of particles. 

There are multiple ways in which the proposal distribution $\mathcal{G}$ can be chosen. In this Thesis we have mainly used the uniform distribution
\begin{equation}
\mathcal{G}_{\rm uniform}(\vec{r} \rightarrow \vec{r}') =
\begin{cases}
\textrm{const}, & \vec{r}' = [\vec{r} - \Delta \vec{r}, \vec{r} + \Delta \vec{r}] \\
0, & \textrm{otherwise}.
\end{cases}
\end{equation}
The parameter $\Delta \vec{r}$ ultimately defines the rate of step acceptance. For this distribution, a proposed new particle coordinate is chosen on a symmetric interval defined by $\Delta \vec{r}$. Note that this distribution satisfied the detailed balance condition only when the interval is symmetric because otherwise it would not be possible for a random walk to occur in the opposite direction, $\vec{r}' \rightarrow \vec{r}$.  

In case of $P$ being a free-particle-like Gaussian (see Sec. \ref{sec:dmc}), then a move can be generated according to the Box-Muller algorithm \cite{marsaglia2000ziggurat}, and we numerically implemented a $\mathtt{gasdev}$ numerical function \cite{numerical_recipes_press1988} to propose new particle coordinates.

General Metropolis algorithm is given below:
\begin{enumerate}
	\item Generate a new configuration $\vecR'$ from $\vecR$ according to the function $\mathcal{G}(\vecR \rightarrow \vecR')$.
	\item \label{first} Calculate transition probability $T=\dfrac{\rho(\vecR') \mathcal{G}(\vecR' \rightarrow \vecR) }{\rho(\vecR) \mathcal{G}(\vecR \rightarrow \vecR')}$.
	\item if $T \ge 1$ , step \ref{first} is accepted, i.e., we set $\vecR= \vecR'$ and we proceed to step \ref{fin}.
	\item if $T<1$, a proposed step is accepted and we set $\vecR = \vecR'$ only if $r \le T$, where $r$ is a random number chosen from the uniform distribution in interval $[0,1]$.
	\item \label{fin} Calculate the averages evaluated at $\vecR$.
	\item In case the proposal distribution $P$ is a uniform transition probability distribution; update the size of the step: if the acceptance rate is greater (less) than the wanted acceptance rate, then decrease (increase) the step size.
\end{enumerate}
Iterations of the Metropolis algorithm are performed until the desired statistical accuracy is achieved. The rate of exploration of the phase space is determined by the walker's step size when the uniform distribution is chosen for proposal distribution $P$. The size of a step must be variable during simulation, such that the target percentage of accepted steps is in the interval $50\%-70\%$ to ensure that the walker moves through the phase space as efficiently as possible \cite{hammond1994monte}. If the proposed moves are large, it is improbable for a move to be accepted. For example, large steps can result in two particles being at a very small relative distance where it is expected that the wavefunction is negligible due to the hard core. On the other hand, if the proposed step is too small, then accepting a move is very likely because the distribution function will not change too much from the original value. A high acceptance rate is a problem because two successive steps are then be highly correlated, i.e., statistically dependent.

\section{Error estimation in Monte Carlo calculations: data blocking\label{sec:error_analysis}}

\begin{figure}
	\centering
	\includegraphics[width=0.75\linewidth]{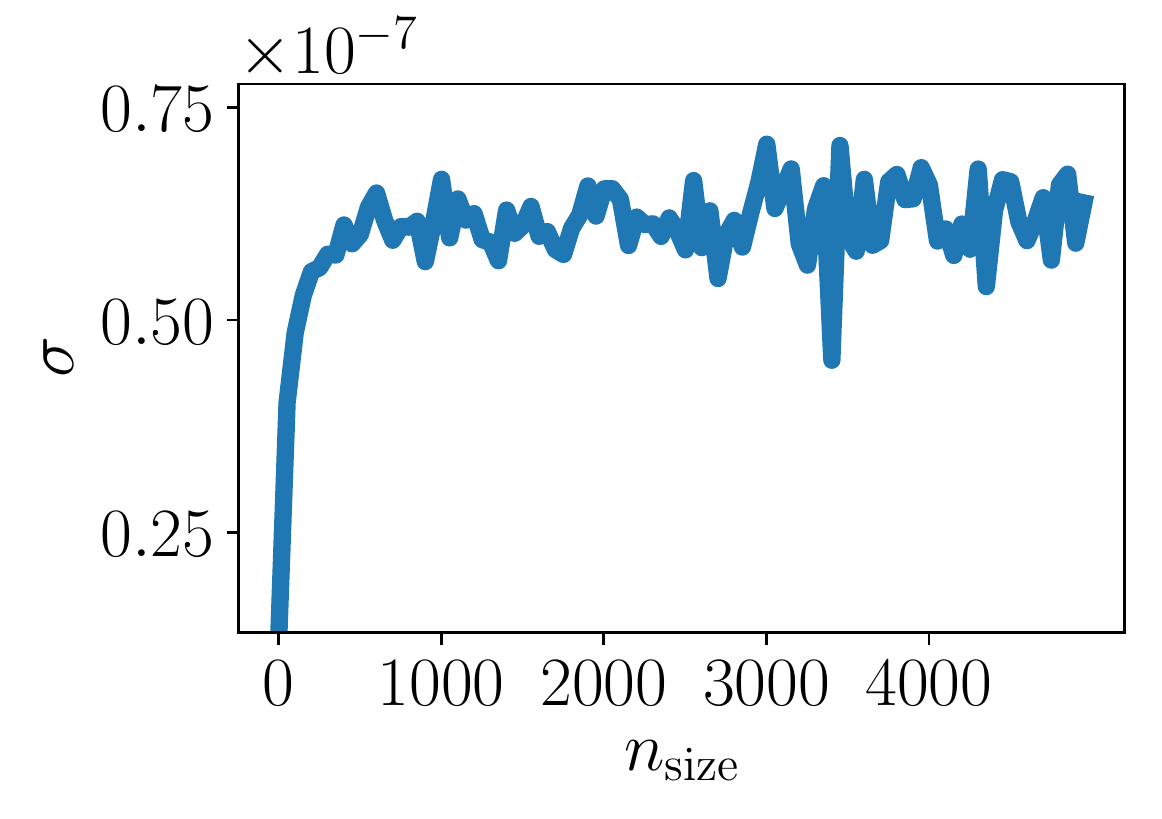}
	\caption[Dependence of the error estimation $\sigma$ as the function of the number of data in a single block, $n_{\rm size}$.]{Dependence of the error estimation $\sigma$ as the function of the number of data in a single block, $n_{\rm size}$.}
	\label{fig:sigmansns}
\end{figure}

Numerical error of Monte Carlo integration technique comes from the stochastic nature of the Markov process and it is determined by the number of accumulated data. The usual definition of the statistical error works only when the data points are statistically independent. For uncorrelated samples, the measure of uncertainty is determined from standard deviation \cite{hammond1994monte}
\begin{equation}
\label{eq:stdev_uncorrelated}
\sigma_f = \sqrt{\dfrac{1}{n-1} \sum_{k}^{n} \left(f_k - \ave{f}\right)^2},
\end{equation}
with $n$ being the total number of data points, and $f$ the function to be averaged. In Monte Carlo calculations there is always an inherent statistical dependence of two successive steps, and this problem can be solved by data blocking technique (see Fig. {\ref{fig:sigmansns}}). The basic idea of data blocking is to divide the data into large enough blocks to assume that they are mutually uncorrelated, where the value of each block is the average of the data within a  block. If all the blocks have a specific size $n_{\rm size}$, than there is $n_{\rm blocks} = n / n_{\rm size}$ number of blocks. Each block carries a value corresponding to an average of measured quantities during that block. This leads to the correct estimation of the standard deviation of function $f$
\begin{equation}
\label{eq:correct_stdev}
\sigma = \sqrt{\dfrac{1}{n_{\rm blocks}-1} \sum_{k}^{n_{\rm blocks}} \left(\tilde{f}_k^2 - \ave{f}^2\right)},
\end{equation}
where $\tilde{f}_k = (1 / n_{\rm size}) \sum_{i=k \times {\rm size}}^{(k+1) \times {\rm size}} f_i$ is an average during $k$-th block. Illustration of the convergence of $\sigma$ with respect to the size of each block $n_{\rm size}$ is shown in Fig. (\ref{fig:sigmansns}).

\section{\label{sec:vmc}Variational Monte Carlo}

Variational Monte Carlo (VMC) is a method for obtaining variational estimations in a stochastic sense. We are interested in finding a variational many-body ground state $\psi$ (Eq. \ref{eq:hamiltonian_many_body}). In this Thesis we are only interested in continuous Hamiltonians, but the VMC methodology applies to discrete systems as well as problems involving three-body forces, or many-body forces in general. Also, is is possible to use the same methodology on the problems in $2^{\rm nd}$ quantization \cite{booth2009fermion}. Finally, VMC method is applicable both to Bose and Fermi systems because it is possible to interpret the $|\psi_{\rm T}|^2$ as the probability density function for particles of both symmetries.

In VMC, the wavefunction is written in the coordinate representation and the expectation values are integrated out by performing Markov chains over the states defined by $\vecR = \left\{\vec{r}_1, \ldots, \vec{r}_N\right\}$. The central object in the VMC is a trial wavefunction $\psi_{\rm T}(\vecR)$, written in a physically motivated way, which is a function of all the coordinates of the system.

The idea of the VMC method is to reformulate the many-body problem as an optimization problem, where the goal is to find the best wavefunction $\psi$, according to the minimization criteria of the energy (functional)
\begin{equation}
\label{eq:many_body_problem_density}
E[\psi] = \dfrac{\bra{\psi} \hat{H}\ket{\psi}}{\braket{\psi}{\psi}}.
\end{equation}
It is also possible to extend the methodology to seek for excited states \cite{boronat1998diffusion,blunt2015excited}, but this is outside the scope of this Thesis.

Physical observables, of which the most important one is the energy, are estimated as the expectation values using the trial wavefunction $\psi_{\rm T}$. Generally, the expectation value of a generic operator $\hat{O}$ is calculated as
\begin{equation}
\label{eq:definition_ave_O}
\ave{\hat{O}} = \dfrac{\bra{\psi_{\rm T}} \hat{O}\ket{\psi_{\rm T}}}{\braket{\psi_{\rm T}}{\psi_{\rm T}}} = \dfrac{\displaystyle \int d\vec{r}_1  \dots \int  d\vec{r}_N |\psi_{\rm T}(\vecR)|^2 O_L(\vecR)}{\displaystyle \int d\vec{r}_1  \dots \int  d\vec{r}_N |\psi_{\rm T}(\vecR)|^2}.
\end{equation}
It is customary in all of quantum Monte Carlo techniques to define a local quantity $O_L$ of the operator $\hat{O}$ as
\begin{equation}
O_L(\vecR) = \dfrac{\hat{O} \psi_{\rm T}}{\psi_{\rm T}}.
\end{equation}
Integration in Eq. (\ref{eq:definition_ave_O}) is readily recognized as a problem which is possible to solve with the Metropolis algorithm (Sec. \ref{sec:metropolis}). Note that the normalization constant of the wavefunction does not play any role in the Metropolis sampling because the move $\vecR \rightarrow \vecR'$ is always accepted with a probability $|\psi_{\rm T}(\vecR')|^2 / |\psi_{\rm T}(\vecR)|^2$, thus canceling the normalization constant. Therefore the expectation value of the operator $\hat{O}$ is calculated as
\begin{equation}
\ave{\hat{O}} = \lim\limits_{n \rightarrow \infty} \dfrac{1}{n}\sum_{i}^{n} O_L(\vecR_i),
\end{equation}
where the set of states $\{\vecR_1, \ldots, \vecR_n\}$ is sampled according to the $|\psi_{\rm T}|^2$. Of course, in practice $n$ is always finite, meaning that the stochastic estimation has an intrinsic statistical error. Consequently, the error analysis is always performed as described in Sec. \ref{sec:error_analysis}. Starting from the initial configuration, Metropolis algorithm ensures that after a certain amount of iterations, the particle coordinates will reach the desired distribution.

The VMC method is primarily used to optimize the trial wavefunction, which is then used in the Diffusion Monte Carlo method (see Sec. \ref{sec:dmc}) in order to obtain the exact ground-state averages. Variational energy $E_{\rm VMC}$ gives an upper bound to the exact ground state energy $E_0$ according to the variational principle of Quantum Mechanics
\begin{equation}
\label{eq:var_principle_qmc_method}
E_{\rm VMC} = \dfrac{    \bra{\psi_{\rm T}} \hat{H}\ket{\psi_{\rm T}} }{\braket{\psi_{\rm T}}{\psi_{\rm T}}} \ge E_0,
\end{equation} 
meaning that the quality of a wavefunction can be judged by its relation with the DMC energy. Note that Eq. (\ref{eq:var_principle_qmc_method}) is true whenever $\braket{\psi_{\rm T}}{\psi_0} \neq 0$. To physically guide the shape of the wavefunction, it is optimal to follow the Bijl-Dingle-Jastrow-Feenberg expansion \cite{bijl1940lowest,dingle1949li,jastrow1955many,feenberg2012theory} of the many-body wavefunction
\begin{equation}
\Psi(\vecR) = \exp\left\{  \sum_{i}^N f_1(\vec{r}_i) +  \sum_{i<j}^N f_2(\vec{r}_i, \vec{r}_j)  + \sum_{i<j<k}^{N}f_3(\vec{r}_i, \vec{r}_j, \vec{r}_k)  + \cdots  \right\},
\end{equation}
which is a rapidly converging series usually truncated to the first two terms, due to exponential increase of computational requirements. It is expected that in the weakly interacting limit the wavefunction is well described with one- and two-body functions, so we usually write the wavefunction as
\begin{equation}
\label{eq:psi_t_simple_two_body}
\psi_{\rm T} = \prod_{i}^{N} f_{\rm 1b} (\vec{r}_i) \prod_{i<j}^{N} f_{\rm 2b} (|\vec{r}_i - \vec{r}_j|).
\end{equation}
$\psi_{\rm T}$ is not an eigenstate, and it is usually parameterized in a physical manner with a set of parameters $\vec{\alpha}$, with the purpose of improving the wavefunction. The choice of these parameters is determined by minimizing either the energy or the energy variance \cite{hammond1994monte, landau2013quantum}. Those parameters with the minimal energy, or minimal energy variance \cite{drummond2005variance}, are then chosen to give the optimal wavefunction. Minimal energy criteria comes on from the energy-functional formulation of the quantum many-body problem (Eq. \ref{eq:many_body_problem_density}), and the energy variance criterion is due to the zero variance principle for an eigenstate of the Hamiltonian. One should be careful when the optimization with respect to the energy variance is being carried out because the energy variance vanishes for excited states as well. In our calculations, we usually use only a few parameters, allowing the optimization process to be performed by carrying out many energy calculations and then picking one set of parameters which gives the lowest energy.

Improvements over the form written as Eq. (\ref{eq:psi_t_simple_two_body}) can be made by including effective three-body correlations \cite{schmidt1980variational}. Recently, there has been a great advancement in the development of a VMC method by writing it in the form of a neural network
\cite{ruggeri2018nonlinear}, allowing for study of precise real-time dynamics \cite{carleo2011spectral,carleo2017unitary}, and allowing for the improvement of knowledge of the nodal surface in Fermi systems \cite{umrigar2007alleviation,carleo2019machine,carleo2017solving}.

\section{\label{sec:dmc} Diffusion Monte Carlo}

The Diffusion Monte Carlo method is a widely used technique suitable for the study of strongly-correlated quantum many-body systems at zero temperature. It is an improvement of the Variational Monte Carlo method and allows the exact evaluation of ground-state observables for bosonic systems. This method can be applied to fermions as well \cite{ceperley1980ground}, but because the nodal surface is unknown, it can provide only variational energies and other properties. In contrast with mean-field theory or conventional perturbative approaches commonly used in the field of ultracold atomic gases, Diffusion Monte Carlo is a first-principles method, meaning that it operates directly with the many-body wave function. The Diffusion Monte Carlo method belongs to a more general class of projection methods \cite{umrigar2015observations}, for which the starting point is the Schr\"odinger equation in imaginary time.  We focus on continuous Hamiltonians and view the problem in the first quantization, but the generality of the DMC method allows for the study in second quantization as well \cite{booth2009fermion}. Diffusion Monte Carlo method scales as $N^2$, which is due to $N(N-1)/2$ pairs appearing in the calculation of the potential energy.  We have implemented a DMC code for quantum Bose-Bose mixtures by following the works in Ref. \cite{boronat2002microscopic,chin1990quadratic}.

\subsection{Schr\"oedinger equation in imaginary time}

The fundamental equation of motion for any quantum system is given by the Schr\"odinger equation
\begin{equation}
i\hbar \dfrac{\partial \psi}{\partial t} = \hat{H} \psi,
\end{equation}
where $\hat{H}$ is the Hamiltonian of the system. To search for the ground-state properties it is common to introduce the concept of imaginary-time $\tau = -i t / \hbar$, in which the Schr\"odinger equation reads
\begin{equation}
-\dfrac{\partial \psi}{\partial \tau} = \hat{H} \psi.
\end{equation}
Primary motivation for formulating the problem in imaginary time is the fact that the imaginary-time evolution of any given wavefunction $\psi$ with a given symmetry leads to the ground-state. To see closely what happens with the wavefunction $\psi$ in an imaginary-time evolution, we expand it in Hamiltonian eigenstate wavefunctions $\phi_n$ with increasing order of energies 
\begin{equation}
\psi(\vecR, \tau)  = \sum_{n=0}^{\infty} c_n \phi_n (\vecR, 0)e^{-E \tau},
\end{equation}
where $c_n$ are components of $\psi$ in the basis $\{\phi_n\}$. For the methodology to work, i. e. to ensure that we reach the ground-state in $\tau \rightarrow \infty$ limit, we must assume $c_0 \neq 0$, meaning that the overlap of $\psi$ with a ground state wavefunction $\phi_0$ is non-zero. In the limit $\tau \rightarrow \infty$, excited states decay to zero exponentially faster than the ground-state. Since the normalization of $\psi$ falls to zero as well, it is convenient to introduce a constant shift to the Hamiltonian $\hat{H} \rightarrow \hat{H} - E_0$, thus making the $\psi$ to finally arrive to the many-body ground state wave function
\begin{equation}
\psi(\vecR, \tau \rightarrow \infty) = c_0 \phi_0(\vecR).
\end{equation}

\subsection{Naive implementation}

Because $\vecR$ is often a highly-dimensional vector, the representation of the many-body wavefunction $\psi$ on the finite-grid is not feasible due to current computer memory and processor limitations. This limitation is alleviated by providing the solution in a stochastic sense. Numerically, a stochastic process which solves the Schr\"odinger equation must follow a probability distribution function. When dealing with bosonic systems, the wavefunction is positive definite everywhere, meaning that it can be interpreted as the probability distribution function. In what follows, the naive approach to the solution for a bosonic many-body problem is presented \cite{kosztin1996introduction}, setting the stage for a final algorithm (\ref{subsec:short-time_approximation_green_function}) implemented to study problems presented in the Thesis.

In coordinate representation, the wavefunction is numerically represented as a set of \textit{walkers}, which are points in $\vecR$ space
\begin{equation}
\psi (\vecR(\tau)) = \mathcal{N} \sum_{i}^{n_{\rm w}} \delta \left(\vecR(\tau)- \vecR_i\right),
\end{equation}
with $\mathcal{N}$ representing the normalization of the wavefunction and $n_{\rm w}$ is the number of walkers. Quantum mechanically, coordinate representation of the wavefunction requires infinite number of walkers, but in practice the series is truncated due to numerical limitations. Particle coordinates $\{\vecR_i\}$ are evolved according to the Schr\"odinger equation
\begin{equation}
- \dfrac{\partial \psi}{\partial \tau} = -D \nabla_{\vecR}^2 \psi + V(\vecR) \psi - E_{\rm ref} \psi(\vecR),
\end{equation}
where $D = \hbar^2 / (2m)$, with $m$ being the mass of particle in the system. The total potential $V(\vecR)$ reads 
\begin{equation}
V(\vecR) = \sum_{i=i}^{N} V_{\rm ext}(\vec{r}_i) + \sum_{i<j}^{N} V_{\rm pair}(|\vec{r}_i - \vec{r}_j|) ,
\end{equation}
and $E_{\rm ref}$ is the referent energy used to stabilize the number of walkers in the simulation. Particle coordinates are evolved according to the many-body Green function $G$ as
\begin{equation}
\psi(\vecR', \tau + \Delta \tau) = \displaystyle \int G(\vecR', \vecR, \Delta \tau) \psi(\vecR, \tau),
\end{equation}
where $G$ is the approximation to the full many-body Green's function, accurate up to the second-order
\begin{eqnarray}
 G(\vecR', \vecR, \Delta \tau)   & = &  \left(\dfrac{m}{2\pi \Delta \tau}\right)^{3 / 2} \underbrace{\exp\left\{-\dfrac{m(\vecR - \vecR')^2}{2 \hbar^2\Delta \tau}\right\}}_{G_{\rm K}}
\underbrace{\exp\left\{-(V(\vecR) - E_{\rm ref})\Delta \tau\right\} }_{G_{\rm R}} \nonumber  \\ &+  &\mathcal{O}(\Delta \tau^2). \label{eq:gree_function_naive}
\end{eqnarray}
Eq. \ref{eq:gree_function_naive} sets the basic ingredients for the evolution of particle coordinates:
\begin{enumerate}
	\item For each particle, $\vec{r} \rightarrow \vec{r} + \xi \sqrt{2 \hbar^2\Delta \tau / m} $, where $\xi$ is a pseudo-random number drawn from the Gaussian distribution.
	\item Replicate the walker according to $G_{\rm R} = \exp\left\{-(V(\vecR_i) - E_{\rm ref})\Delta \tau\right\}$, where $\vecR_i$ are particle coordinates of the $i$-th walker. This is numerically achieved by rounding $G_{\rm R}$ to the closest integer, erasing the walker if $G_{\rm R} = 0$, and making $G_{\rm R} - 1$ new copies of a given walker to be used in the next iteration.
	\item $E_{\rm ref}$ is adjusted in order to achieve and maintain the number of walkers around target value.
\end{enumerate}
After a long enough imaginary-time propagation, distribution of particles evolves according to the ground-state wavefunction $\phi_0$. In the $\tau\rightarrow\infty$ limit, average of the reference energy $E_{\rm ref}$ over a set of walkers converges to the ground-state energy $E_0$, as the population of walkers reaches a steady distribution \cite{binder2012monte}. 

The fundamental problem of the previous algorithm is that a probability distribution function is not the quantum-mechanical probability density $|\psi|^2$, but rather a probability amplitude $\phi_0$ \cite{landau2013quantum}. Therefore, expectation values cannot be calculated, except for the ground-state energy $E_0$. Practically, this approach is never used for the study of many-body systems because of substantial fluctuations of the bare potential $V(\vecR)$, resulting in a high variance in the estimation of energy, especially when the potentials have a hard-core \cite{boronat2002microscopic}. Finally, only bosonic systems can be treated in this way, since fermions have to obey the Fermi principle. This algorithm can be significantly improved when a physically motivated trial wavefunction is introduced, leading to the importance sampling.

\subsection{\label{sec:importance_sampling}Importance sampling}

Importance sampling is a commonly used variance reduction Monte Carlo technique for efficient sampling of the distribution function \cite{kalos2009monte, boronat2002microscopic,binder2012monte}. The basic physical idea is that certain phase space regions play a much more important role than others in the estimation of statistical averages. Hence, it is desired that these regions are sampled more frequently. For example, in a system interacting with a pair-wise realistic Lennard-Jones-like potential, it is expected that the wavefunction is small in the hard-core region, large around the potential minimum, and approaching a constant at large distances \cite{pade2007exact}. Importance sampling is introduced in the Diffusion Monte Carlo method by exploiting the knowledge of an approximate ground-state wavefunction $\psi_{\rm T}(\vecR)$ obtained from Variational Monte Carlo. This leads to the biased estimation of the averages, and a practitioner must ensure that the results are unbiased (see Sec. \ref{systematic_errors_dmc}). The probability distribution function $f$ which is to be sampled is defined as
\begin{equation}
f(\vecR, \tau) = \psi_{\rm T}(\vecR) \phi(\vecR, \tau),
\end{equation} 
where $\psi_{\rm T}(\vecR)$ is the trial wavefunction and $\phi(\vecR, \tau)$ is a function with an initial condition set to $\phi(\vecR, 0) = \psi_{\rm T}(\vecR)$. $\phi$ has the desired infinite projection time limit $\phi(\vecR, \tau \rightarrow \infty) = \phi_0$, with $\phi_0$ being the ground-state wavefunction. 

Choosing a good trial wavefunction is crucial for the efficient and unbiased Diffusion Monte Carlo calculation. Benefits of importance sampling with a good trial wavefunction $\psi_{\rm T}$ coming from the Variational Monte Carlo are twofold: variance-reduction leads to lower computation times, and finally to a weaker dependence on the imaginary time-step and the number of walkers.

Probability distribution function $f$ is represented as a finite set of particle coordinates $\{\vecR  \}$, i.e., walkers, at each instant $\tau$
\begin{equation}
f(\vecR, \tau) = \mathcal{N} \sum_{i}^{n_{\rm w}} \delta \left(\vecR(\tau) - \vecR_i(\tau)\right).
\end{equation}
The time evolution equation of $f(\vecR, \tau)$ is given by 
\begin{equation}
\label{eq:time_evolution_f}
-\dfrac{\partial f(\vecR, \tau)}{\partial \tau} = -D\nabla_{\vecR}^2 f(\vecR, \tau) + D\nabla_{\vecR}\left(\vecF(\vecR) f(\vecR, \tau)\right) + \left(E_L(\vecR) - E_{\rm ref}\right) f(\vecR, \tau),
\end{equation}
where $\vecF(\vecR)$ is defined as \textit{quantum force}
\begin{equation}
\vecF(\vecR) = 2 \dfrac{\nabla_{\vecR} \psi_{\rm T}(\vecR)}{\psi_{\rm T}(\vecR)},
\end{equation}
acting in the direction of the fastest increase of the wavefunction. The local energy $E_L(\vecR)$ is
\begin{equation}
E_L(\vecR) = \dfrac{\hat{H} \psi_{\rm T}(\vecR)}{\psi_{\rm T}(\vecR)} = -D\dfrac{\nabla_{\vecR}^2 \psi_{\rm T}(\vecR)}{\psi_{\rm T}} + V(\vecR),
\end{equation}
and finally $E_{\rm ref}$ is the reference energy used to stabilize the number of walkers in the simulation. Derivation of the quantum force $\vecF(\vecR)$ and local energy $E_L(\vecR)$ is presented in the Appendix \ref{appendix:local_energy}. Eq. (\ref{eq:time_evolution_f}) can be written as  
\begin{equation}
-\dfrac{\partial f(\vecR, \tau)}{\partial \tau} = \left(\hat{O}_{\rm K} + \hat{O}_{\rm D} + \hat{O}_{\rm B}\right) f(\vecR, \tau),
\end{equation}
where the kinetic term is $\hat{O}_{\rm K} = -D\nabla_{\vecR}^2$, the drift term is $\hat{O}_{\rm D}(\cdot) = D\nabla_{\vecR}(\vecF(\vecR) \cdot)$ and finally a branching term is $\hat{O}_{\rm B} = E_L(\vecR) - E_{\rm ref}$. The formal solution to the Eq. (\ref{eq:time_evolution_f}) is given by
\begin{equation}
f(\vecR, \tau ) = \displaystyle \int G(\vecR', \vecR, \tau) f(\vecR', 0) d\vecR',
\end{equation}
where $G(\vecR', \vecR, \tau)$ is the full many-body Green's function
\begin{equation}
G(\vecR', \vecR, \tau) = \exp\left\{ - (\hat{O}_{\rm K} + \hat{O}_{\rm D} + \hat{O}_{\rm B}) \tau \right\}.
\end{equation}
For the full many-body problem, Green's function is not known analytically. However, each of the Green's functions $\hat{G}_{\rm K}$, $\hat{G}_{\rm D}$ and $\hat{G}_{\rm B}$, corresponding to the operators $\hat{O}_{\rm K}$, $\hat{O}_{\rm D}$ and $\hat{O}_{\rm B}$, are known analytically. Since the operators $\hat{O}_{\rm K}$, $\hat{O}_{\rm D}$ and $\hat{O}_{\rm B}$ do not mutually commute, it is not possible to decompose $G(\vecR', \vecR, \tau)$ into a product ${G}_{\rm K} {G}_{\rm D} {G}_{\rm B}$, due to the
Baker–Campbell–Hausdorff theorem \cite{landau2013quantum}. Therefore, we resort to the short-time approximation of the full Green's function.

\subsection{Short-time approximation of the Green's function \label{subsec:short-time_approximation_green_function}}

In the limit of short-time propagation, the full Green's function $G(\vecR', \vecR, \tau)$ can be expressed perturbatively as a power series in the time-step $\Delta \tau$ \cite{boronat2002microscopic}. Therefore, the idea is to divide the propagation time into discrete timesteps of length $\Delta \tau$, and at each iteration solve the equation
\begin{equation}
f(\vecR, \tau + \Delta \tau) = \displaystyle \int d\vecR' G(\vecR', \vecR, \Delta \tau) f(\vecR', \tau).
\end{equation}
There are known linear, quadratic \cite{chin1990quadratic,boronat2002microscopic} and fourth-order \cite{forbert2001fourth} expansions of the $G$ into a product of analytically solvable Green's functions which can be numerically implemented. In this Thesis, we use second-order expansion in the timestep
\begin{eqnarray}
\label{eq:decomposition_G}
G(\vecR', \vecR, \Delta \tau) = G_{\rm B}\left(\vecR, \vecR_4, \dfrac{\Delta \tau}{2}\right) G_{\rm D}\left(\vecR_, \vecR_3, \dfrac{\Delta \tau}{2}\right) G_{\rm K}\left(\vecR_3, \vecR_2, \Delta \tau\right) \nonumber \\ \times G_{\rm D}\left(\vecR_2, \vecR_1, \dfrac{\Delta \tau}{2}\right) G_{\rm B}\left(\vecR_1, \vecR', \dfrac{\Delta \tau}{2}\right) + \mathcal{O}( \Delta \tau^3),
\end{eqnarray}
which produces a quadratic dependence of energy on the timestep $\Delta \tau$ \cite{chin1990quadratic}. In Eq. (\ref{eq:decomposition_G}), there are three different particle updates. First update is due to Green's function $G_{\rm K}$, which corresponds to the free-particle diffusion term $\hat{O}_{\rm K} = -D\nabla_{\vecR}^2$, and is given by
\begin{equation}
\label{eq:gk_short_time}
G_{\rm K} (\vecR', \vecR, \Delta \tau) = \left( 4 \pi D  \Delta  \tau \right)^{-3N / 2} \exp \left[-\dfrac{\left(\vecR - \vecR'\right)^2}{4 D  \Delta  \tau}\right].
\end{equation}
Numerically, the action of this operator, i.e., $\displaystyle \int d\vecR' G_{\rm K}(\vecR', \vecR, \Delta  \tau) f(\vecR', \Delta  \tau)$, corresponds to the isotropic Gaussian displacement of variance $2D \Delta  \tau$ for each particle. The second particle update corresponds to the drift force movement due to the operator $\hat{O}_{\rm D} = D \nabla_{\vecR}(\vecF(\vecR) \cdot)$, whose Green's function is given by
\begin{equation}
\label{eq:gd_short_time}
G_{\rm D} (\vecR', \vecR, \Delta \tau)= \delta \left(\vecR' - \vecR(\Delta \tau)\right),
\end{equation}
where 
\begin{equation}
\label{eq:equation_of_motion}
\dfrac{d\vecR}{d\tau} =\vecF(\vecR).
\end{equation}
The particle move update according to the $\hat{O}_{\rm D}$ operator is hence defined deterministically by integrating the equation of motion (Eq. \ref{eq:equation_of_motion}) for time $\Delta \tau$, with the initial condition $\vecR(0) = \vecR$. Since the initial decomposition of the full Green's function is quadratic (Eq. \ref{eq:decomposition_G}), the equation of motion Eq. (\ref{eq:equation_of_motion}) must be solved with the same accuracy. Particulary, we have implemented second-order Runge-Kutta integrator to solve Eq. (\ref{eq:equation_of_motion}). Finally, the crucial term for the Diffusion Monte Carlo method is the branching term $G_{\rm B}$, which selects regions of the phase space according to the value of local energy
\begin{equation}
\label{eq:gb_short_time}
G_{\rm B} (\vecR', \vecR, \Delta \tau) = \exp \left[ - \left(E_L(\vecR) - E_{\rm ref}\right)\Delta \tau\right] \delta\left(\vecR' - \vecR\right).
\end{equation}
This term does not change particle coordinates, but the walker population, according to the exponential part of the Green's function $G_{\rm B}$. The statistical weight of each walker $w = \exp \left[ - \left(E_L(\vecR) - E_{\rm ref}\right)\Delta \tau\right]$ determines the probability that a walker would be passed on to the next iteration. Numerically, the branching term is implemented by generating a random number from uniform distribution in the range $r \in (0, 1)$, and making $[r + w]$ replications of the walker, where the brackets denote integer rounding of $r + w$. 

Finally, pseudo-code steps for a $2^{\rm nd}$ order DMC we have implemented, are given below:
\begin{enumerate}
	\item Gaussian drift. For each walker, and for each particle coordinate $\vec{r}$:
	\subitem $\vec{r} \rightarrow \vec{r} + \sqrt{2D\Delta \tau} \times \mathtt{gasdev()}$, where $\mathtt{gasdev()}$ is the random number generator according to the Gaussian distribution with zero mean and unity variance.
	\item Drift move. For each walker:
	\subitem $\vecR_1 = \vecR + \vecF(\vecR) D\Delta \tau / 2.$
	\subitem $\vecR_2 = \vecR + (\vecF(\vecR) + \vecF(\vecR_1)) D \Delta \tau / 4$. Observables are calculated at this intermediate step.
	\subitem $\vecR_3 = \vecR + \vecF(\vecR_2) D \Delta \tau.$
	\item  Branching step. For each walker:
	\subitem Calculate the weight $w = \exp \left[ - \left( (E_L(\vecR) + E_L(\vecR_2))/2 - E_{\rm ref}\right)\Delta \tau\right]$.
	\subitem Replicate the walker for $[ w + r]$ times, where $r$ is a random number from uniform distribution in the range $r \in (0, 1)$, and brackets stand for integer rounding of $r + w$. If $r = 0$, a walker is destroyed.
\end{enumerate}

Summarizing, after a long enough application of the short-time operators defined above, a steady-state distribution of walkers is generated according to the distribution $\psi_{\rm T} \phi$. The observables are collected at each iteration of the simulation only after the convergence is achieved.

\subsection{ \label{systematic_errors_dmc}Systematic errors in Diffusion Monte Carlo}

\begin{figure}
	\centering
	\includegraphics[width=0.8\linewidth]{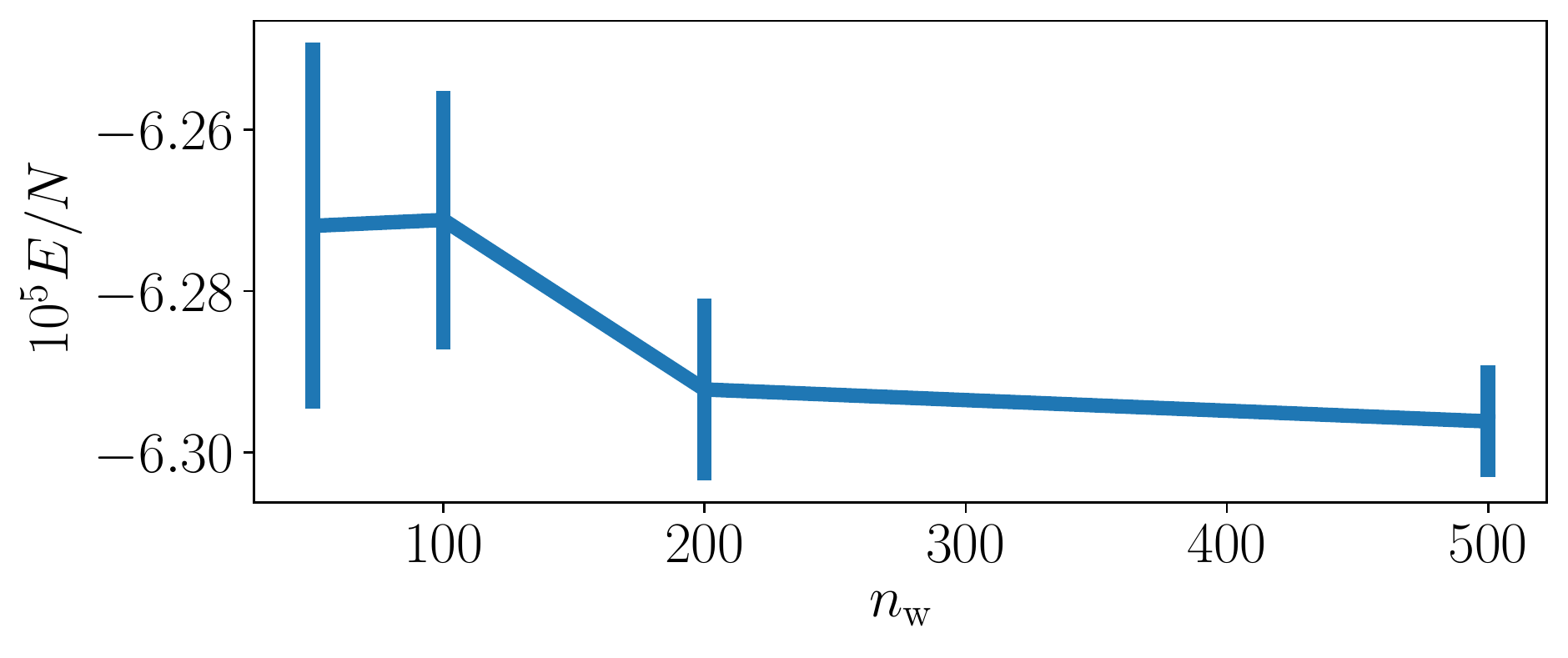}
	\caption[Dependence of energy per particle with the number of walkers, for the Bose-Bose symmetric mixture.]{Dependence of energy per particle with the number of walkers, for the Bose-Bose symmetric mixture. Bias is well eliminated for $n_{\rm w} \approx 200$.}
	\label{fig:nwanalysisplot}
\end{figure}

Two main biases appear in the Diffusion Monte Carlo calculation: the finite population size and the time-step error. They can be eliminated by studying the convergence of energy in the limit of small time-steps and large population sizes. A typical dependence of the energy per particle on the population size is shown in Fig. \ref{fig:nwanalysisplot}, where the convergence of energy at large $n_{\rm w}$ is shown. The system under study is the Bose-Bose liquid with symmetric interactions.

\begin{figure}
	\centering
	\includegraphics[width=0.8\linewidth]{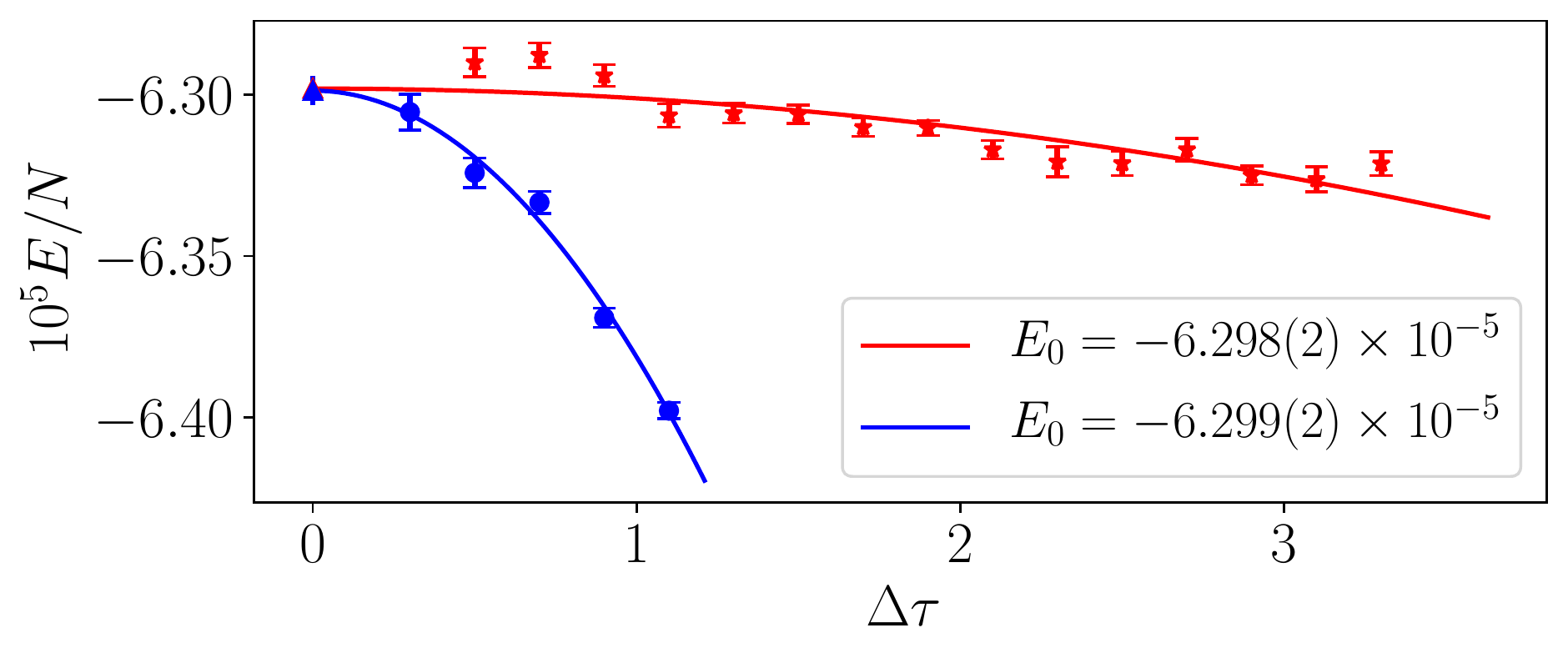}
	\caption[Dependence of energy per particle with the time-step, for the Bose-Bose symmetric mixture.]{Dependence of energy per particle with the time-step, for the Bose-Bose symmetric mixture. Two set of results correspond to different guiding wavefunctions, with the upper one being the best available wavefunction.}
	\label{fig:dtanalysisplot}
\end{figure}

For the same system, the dependence of energy on the time-step is shown in Fig. \ref{fig:dtanalysisplot}. To finally confirm the robustness of the DMC method, we usually perform a third set of calculations with purposely worsened wavefunction. In Fig. \ref{fig:dtanalysisplot}, the predictions of energy per particle in the limit $\Delta \tau \rightarrow 0$ are indistinguishable with two wavefunctions, up to the statistical uncertainty, showing finally that the results are unbiased. It is noted that for a trial wave function which has a greater overlap with the true ground state, larger time-steps can be used. In addition, the efficiency of DMC increases when that overlap is large. In particular, the number of walkers necessary to avoid population control bias is reduced.

\section{\label{sec:observables}Observables}

The ground-state expectation value of a given observable $\hat{O}$ according to the equilibrium distribution $f = \psi_{\rm T} \phi$ is 
\begin{equation}
\label{eq:mixed_estimators}
\ave{\hat{O}}_{\rm mixed} = \dfrac{\bra{\phi}  \hat{O}  \ket{\psi_{\rm T}}}{\braket{\phi}{\psi_{\rm T}}}.
\end{equation}
Numerically, this is calculated by evaluating the statistical average over all walkers and iteration steps
\begin{equation}
\ave{\hat{O}}_{\rm mixed} \approx \dfrac{1}{n_s} \sum_{i_{\rm s}}^{n_{\rm s}} \left(\dfrac{1}{n_{\rm w}^{(i_{\rm s})}}  \sum_{i_{\rm w}}^{n_{\rm w}^{(i_{\rm s})}}    O_L(\vecR_{i_{\rm w}}(i_{\rm s}))\right),
\end{equation}
where $n_{\rm s}$ is the number of simulation steps, $n_{\rm w}^{(i_{\rm s})}$ is the number of walkers at the timestep $i_{\rm s}$, and $O_L(\vecR) = \psi_{\rm T}^{-1} \hat{O} \psi_{\rm T}$. In general, statistical averages calculated according to Eq. (\ref{eq:mixed_estimators}) are biased from the choice of the trial wavefunction $\psi_{\rm T}$. Therefore, $\ave{\hat{O}}_{\rm mixed}$ is called the $\textit{mixed}$ estimator. When $\hat{O}$ commutes with the Hamiltonian $\hat{H}$, then the statistical average is exact. This is readily shown by noting that $\hat{O}$ and $\hat{H}$ share the same basis of the eigenfunctions, thus
\begin{equation}
\dfrac{\bra{\phi}  \hat{H}  \ket{\psi_{\rm T}}}{\braket{\phi}{\psi_{\rm T}}} = \dfrac{\bra{\phi}  \hat{H}  \left(\ket{\phi} + \sum_{n\neq 0} c_n\ket{\phi_n}\right)}{\bra{\phi} \left(\ket{\phi} + \sum_{n\neq 0} c_n\ket{\phi_n}\right)  } = \dfrac{\bra{\phi}  \hat{H}  \ket{\phi}}{\braket{\phi}{\phi}},
\end{equation} 
where $\phi$ converges to a ground state $\phi_0$ in the long projection limit. When $\hat{O}$ does not commute with the Hamiltonian, then the average according to Eq. (\ref{eq:mixed_estimators}) is not exact. When a trial wavefunction is close to the ground state, a first-order extrapolation towards an exact value $\ave{\hat{O}}_{\rm extr.} $ can be used to  
\begin{equation}
\ave{\hat{O}}_{\rm extr.} =  2 \ave{\hat{O}}_{\rm mixed} - \ave{\hat{O}}_{\rm VMC}, 
\end{equation}
where $\ave{\hat{O}}_{\rm VMC}$ is the VMC estimate. Another way of extrapolating to the ground state value, which preserves a sign of the estimate is
\begin{equation}
\ave{\hat{O}}_{\rm extr.} = \dfrac{\ave{\hat{O}}_{\rm mixed}^2}{\ave{\hat{O}}_{\rm VMC}}.
\end{equation}
Additionally, in order to access to an exact estimation of an observable in a ground-state, a forward walking technique \cite{liu1974quantum,casulleras1995unbiased} can be used.

\subsection{Unbiased estimators in Diffusion Monte Carlo}

The exact estimation of an expectation value in a ground state is called a \textit{pure} estimator, and it is defined as
\begin{equation}
\label{eq:pure_estimator}
\ave{\hat{O}}_{\rm pure} = \dfrac{\bra{\phi}  \hat{O}  \ket{\phi}}{\braket{\phi}{\phi}}.
\end{equation}
In this subsection, we lay out the forward walking technique used to evaluate ground-state averages of operators that do not commute with the Hamiltonian. In the Diffusion Monte Carlo calculation, it is not possible to directly access the value of a ground-state wavefunction $\phi$. However, it is possible rewrite Eq. (\ref{eq:pure_estimator}) so that the density distribution function is $f = \psi_{\rm T} \phi$
\begin{equation}
\label{eq:pure_mixed}
\ave{\hat{O}}_{\rm pure} = \dfrac{\bra{\phi}  O_L \dfrac{\phi}{\psi_{\rm T}} \ket{\psi_{\rm T}}}{\bra{\phi}  \dfrac{\phi}{\psi_{\rm T}}   \ket{\psi_{\rm T}}} = \dfrac{\ave{O_L(\vecR) W(\vecR)}_{\rm mixed}}{\ave{W(\vecR)}_{\rm mixed}},
\end{equation}
where $W(\vecR) = \phi / \psi_{\rm T}$. Computationally, Eq. (\ref{eq:pure_mixed}) reads
\begin{equation}
\ave{\hat{O}}_{\rm pure} = \dfrac{\sum_{i} O_L (\vecR_i) W(\vecR_i)  }{W(\vecR_i)},
\end{equation}
where the sum is over walkers and times following the course of the DMC calculation. It is shown \cite{liu1974quantum} that $W(\vecR)$ represents the number of descendants for a walker $\vecR$ in the $\tau \rightarrow \infty$ limit. Thus, the estimation of a pure estimator can be performed by means of a tagging algorithm \cite{barnett1991monte}: a local quantity $O_L(\vecR)$ is computed at time $t$, but the statistical average is performed at $t + T$, where $T$ is large enough. In this way, $O_L(\vecR)$ is aposteriori weighted with the number of walker descendants. 

A more simple version for the pure estimator algorithm we have implemented following references \cite{casulleras1995unbiased, sarsa2002quadratic, petar}, which alleviates the need to evaluate the summation (Eq. \ref{eq:pure_mixed}) ``from the distance'', making it more practical and more memory efficient. Instead of performing average of $O_L$ at each instant over all walkers, which is the way to obtain mixed estimates, this method introduces an individual auxiliary variable $P_i$ associated with the walker at the index $i$.  As the walkers die or replicate, the local estimate of each walker is evolved together with a walker such that
\begin{equation}
P_i(\tau + \Delta \tau) = O_L (\vecR_i(\tau + \Delta \tau)) + P_i (\tau),
\end{equation}
where at the beginning of each block the $P_i$ is initialized to zero. After the block of $M$ steps is finished, the average is performed over $N_{\rm f}$ values of $P_i$
\begin{equation}
\ave{\hat{O}}_{\rm pure} = \dfrac{1}{M \times N_{\rm f}} \sum_{i}^{N_{\rm f}} P_i.
\end{equation}
In a new simulation block, the revision of  calculated values of $P_i$ are being propagated as $P_i(\tau + \Delta \tau) = P_i(\tau)$ for each walker, in order to ensure the asymptotic behavior $W(\vecR)$ in the $\tau \rightarrow \infty$ limit. Thus, for a simulation involving $N_{\rm b}$ blocks, we end up having $N_{\rm b}  - 1$ block averages of a pure estimator. Finally, with block averages, the error associated with pure estimators is calculated with the standard data blocking technique (Sec. \ref{sec:error_analysis}). The described pure estimator methodology works only in the limit of large enough block sizes. Therefore, a value of a block size is chosen for which a convergence is obtained.



\chapter{\label{chapter:ultradilute_liquid_drops}Ultradilute quantum liquid drops}

\section{Introduction}

The high tunability of interactions in ultracold Bose and Fermi gases is allowing for exploration of regimes and phases difficult to find in other condensed-matter systems~\cite{pethick2008bose}. By adjusting the applied magnetic field properly, Bose and Fermi gases are driven to Feshbach resonances with an increase of interaction practically at will, and with the possibility of turning the system from repulsive to attractive and vice-versa. This level of controllability is not possible in conventional condensed matter where interactions are generally not tunable at this level. A significant example of this versatility has been the clean experimental realization of the unitary limit for fermions \cite{ohara2002observation,zwerger2011bcs} and the precise characterization of the BCS-BEC crossover~\cite{carlson2003superfluid,astrakharchik2004equation}, which up to that moment was only a theoretical scenario.

Recently, it has been possible to explore the formation of liquid/solid patterns in dilute gases by modifying the strength of the short-range interatomic interactions. Probably the most dramatic example of this progress has been the observation of the Rosensweig instability in a confined system of $^{164}$Dy atoms with a significant magnetic dipolar moment~\cite{kadau2016observing}.  By tuning the short-range interaction, Kadau \textit{et al.}~\cite{kadau2016observing} observed the spontaneous formation of an array of self-bound droplets remembering the characteristics of a classical ferrofluid. The observation of solid-like arrangements in dilute gases have also been possible working with highly-excited Rydberg atoms~\cite{schauss2012observation}. By direct imaging, Schauss \textit{et al.}~\cite{schauss2012observation} obtained ordered excitation patterns with a geometry close to the well-known arrangements obtained in few-body confined Coulomb particles.

In the line of obtaining other \textit{dense} systems starting from extremely dilute Bose and Fermi gases, the mechanism suggested by Petrov relying on Bose-Bose mixtures~\cite{petrov2015quantum} is noteworthy. According to this proposal, it is possible to stabilize a mixture with attractive interspecies interaction in such a way that the resulting system is self-bound, i.e., a liquid. Whereas a mean-field (MF) treatment of the mixture predicts a collapsed state, the first beyond mean field correction, the Lee-Huang-Yang (LHY) term, can stabilize the system by properly selecting the interspecies $s$-wave scattering length. In what follows, we call the perturbative LHY-corrected mean-field theory MF+LHY. Further work has shown that reducing the dimensionality of the setup to two or quasi-two dimensions may help to stabilize the liquid phase~\cite{petrov2016ultradilute}. The LHY correction has also been used to account for the formation of dipolar drops~\cite{wachtler2016quantum} and then confirmed by full first-principles quantum Monte Carlo (QMC) simulations~\cite{saito2016path,macia2016droplets}.

The exciting idea of producing self-bound liquid drops by using interspecies attractive interaction acting as a glue of the entire Bose-Bose mixture has been put forward by Tarruell and collaborators~\cite{cabrera2018quantum}, followed by Semeghini \textit{et al.} \cite{semeghini2018self} and most recently by D'Errico \textit{et. al} \cite{derrico2019observation}. Therefore, the theoretical prediction seems confirmed and thus a new window for exploring matter in unprecedented situations is open. On one side, it proves the way of forming liquid drops with high density in the world of cold gases and, on the other, makes possible the study of a liquid state of matter with an extremely low density, lower than any other existing liquid.

In the present chapter, we study the formation of liquid drops in a Bose-Bose mixture using the diffusion Monte Carlo (DMC) method, which solves the $N$-body Schr\"odinger equation exactly within some inherent statistical uncertainties due to stochastic sampling (see Chapter \ref{ch:qmc_methods}). The DMC method (see Chapter \ref{ch:qmc_methods}) had been extensively used in the past for determining the structure and energy properties of liquid drops of $^4$He~\cite{pandharipande1983calculations,chin1992structure}, $^3$He~\cite{guardiola2005minimal,sola2006ground}, H$_2$~\cite{sindzingre1991superfluidity}, and spin-polarized tritium~\cite{beslic2009quantum}. At difference with previous perturbative estimates, DMC allows for an exact study of the system's quantum properties relying only on its Hamiltonian. Our results confirm the LHY prediction on the stability of self-bound mixtures and determine quantitatively the conditions under which liquid drops are stable and how they evolve when the attractive interaction is increased. Within the regime here explored, we do not observe a full collapse of the drop but an increase of the density and reduction of the size, which is rather progressive.

\section{Hamiltonian and the trial wavefunction}

The Bose-Bose mixture under study is composed by $N_1$ bosons of mass $m_1$ and $N_2$ bosons of mass $m_2$ with Hamiltonian
\begin{eqnarray}
H & = &  -\frac{\hbar^2}{2m_1} \sum_{i=1}^{N_1} {\rm \nabla}_i^2 
-\frac{\hbar^2}{2 m_2} \sum_{j=1}^{N_2} {\rm \nabla}_j^2 + \frac{1}{2} 
\sum_{\alpha,\beta=1}^{2} \sum_{i_\alpha,j_\beta=1}^{N_\alpha,N_\beta} 
V^{(\alpha,\beta)}(r_{i_\alpha j_\beta}) \ ,
\label{hamiltonian_liquid_drops}
\end{eqnarray}
with $V^{(\alpha,\beta)}(r)$ the interatomic interaction between species $\alpha$ and $\beta$. Our interest is focused on a mixture of intraspecies repulsive interaction, i.e., positive $s$-wave scattering lengths $a_{11}>0$ and $a_{22}>0$, and interspecies attractive potential, $a_{12}<0$. To set up this regime, we use a hard-sphere potential of diameter $a_{\alpha \alpha}$ for potentials $V^{(\alpha,\alpha)}(r)$ and an attractive square well of depth $-V_0$ and range $R$ for $V^{(\alpha,\beta)}(r)$. In the latter case, we fix $R$ and change $V_0$ to reproduce the desired negative scattering length; notice that we work with negative  $a_{\alpha \beta}$ values and we impose that the attractive potential does not support a pair bound state.

The DMC method uses a guiding wave function as importance sampling to reduce the variance to a manageable level (see Sec. \ref{sec:importance_sampling}). We adopt a Jastrow wave function in the form
\begin{equation}
\Psi(\vecR)= \prod_{1=i<j}^{N_1} f^{(1,1)}(r_{ij}) \prod_{1=i<j}^{N_2 
}f^{(2,2)}(r_{ij}) \prod_{i,j=1}^{N_1,N_2} f^{(1,2)}(r_{ij}) \ ,
\label{trialwf}
\end{equation}
with $\vecR=\{\vecr_1,\ldots,\vecr_{N}\}$. In the case of equal particles 
the Jastrow factor is taken as the scattering solution

\begin{equation}
f^{(\alpha,\alpha)}(r)=
\begin{cases}
1-a_{\alpha \alpha}/r  &  r \ge a_{\alpha \alpha} \\
0 & \mathrm{otherwise}.
\end{cases}
\end{equation}
If the pair is composed of different particles then we take 
\begin{equation}
f^{(\alpha,\beta)}(r)= \exp(-r/r_0),
\end{equation}
with $r_0$ a variational parameter.

In order to reduce the number of variables of the problem, keeping the essentials, we consider $m_1=m_2=m$, $N_1=N_2$, and $a_{11}=a_{22}$. In this way, our study explores the stability and formation of liquid drops as a function of $a_{12}$ and the number of particles $N$ ($N_1=N_2=N/2$). The $s$-wave scattering length $a_{12}$ of an attractive well is analytically known
\begin{equation}
\label{eq:definition_a12}
a_{12}=R \left[ 1- \tan(KR)/(KR) \right],
\end{equation}
with $K^2=m_1 V_0/\hbar^2$.  We take $a_{12}<0$ which correspond to $KR < \pi/2$. In practice, we fix the range of the well $R$ and vary the depth $V_0$ to reproduce a desired value of $a_{12}$. As it is obvious from the Eq. (\ref{eq:definition_a12}) for $a_{12}$, its value depends on the product $R V_0^{1/2}$ and then decreasing $R$ means to increase $V_0$. If for a fixed $a_{12}$ value we want 
to approach the limit $R \to 0$ then $V_0 \to \infty$, our calculations become extremely demanding in terms of accuracy and number of particles required to observe saturation. 

\begin{figure}[H]
	\centering
	\includegraphics[width=0.7\linewidth]{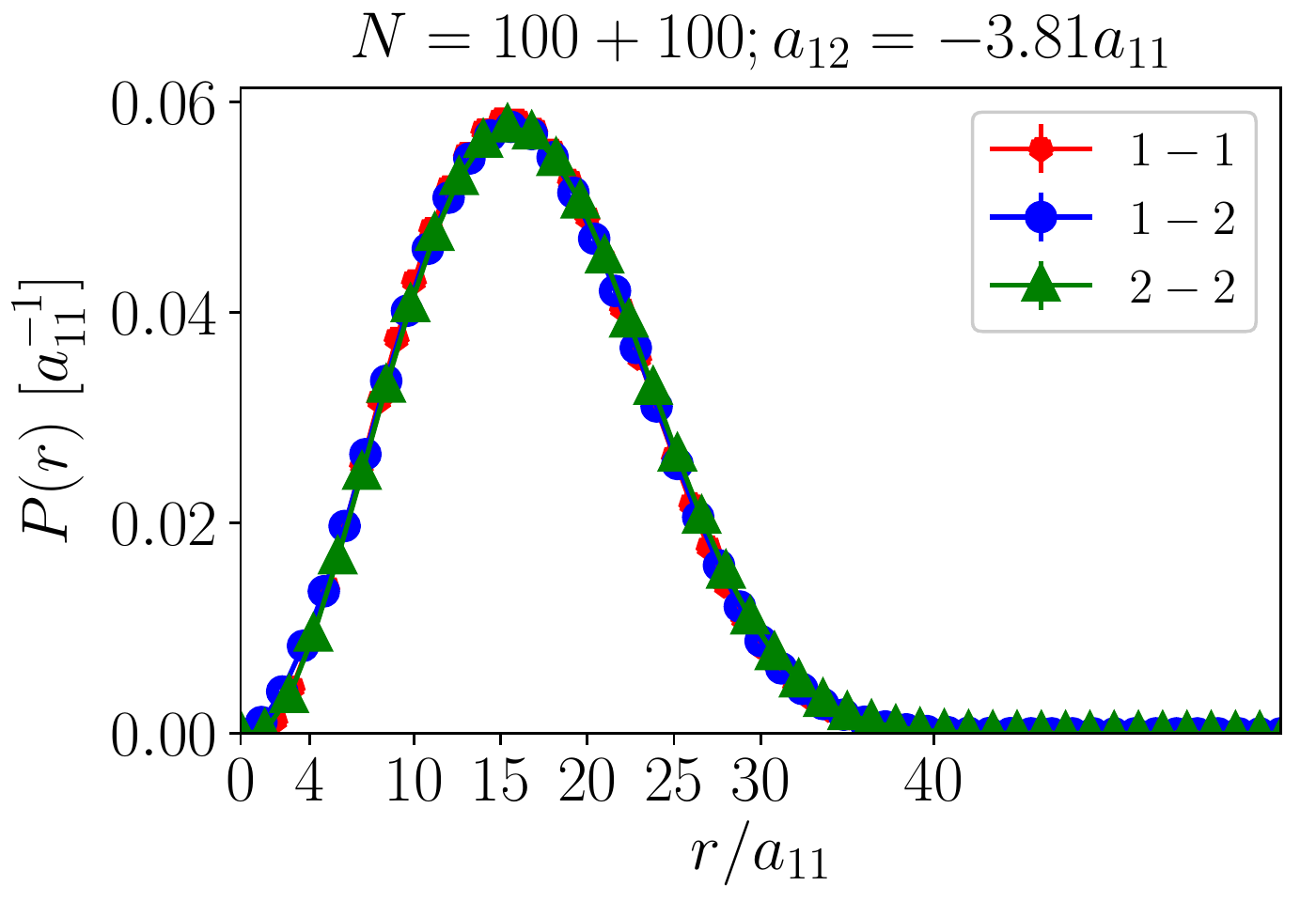}
	
	\caption[Pair distribution function $P(r)$ for three types of pairs in a Bose-Bose quantum droplet]{Pair distribution function $P(r)$ for three types of pairs in a Bose-Bose quantum droplet. The probablity of two particle interacting within $r=4 a_{11}$ is $\int_0^{4a_{11}} P(r) dr \approx 0.01$.}
	\label{fig:distrdistance3}
\end{figure}

After preliminary studies, we determined that $R=4 a_{11}$ is a good compromise between accuracy and reliability, and thus, the major part of our results is obtained with that value of $R$. To better illustrate this point, in Fig. \ref{fig:distrdistance3} we report the results for the pair distribution function $P(r)$ in a typical drop, in particular $N=100+100$, normalized to have integral one
\begin{equation}
	\int_0^\infty \mathrm{d}r  P(r) = 1 .
\end{equation}
As one can see, the most probable distance between particles is $\sim 15 a_{11}$, nearly four times larger than the potential range. If we integrate the distribution probability between 0 and 4, we get 0.01. Only $1\%$ of the particles are, on average, at distances below the range of the well. On the other hand, the repulsive potential between equal species is given by a hard-sphere potential with a core at $a_{11}=a_{22}$, which, in reality, fix our length units. Indeed, both potentials are not delta potentials (with zero range) because the contact interaction is problematic in 3D simulations due to the singularity that it introduces in the wave function at the origin. Still, we think that our model can be quite close to reality, taking into account the large values of the effective ranges of potassium interactions that can play a relevant role.

The trial wave function $\Psi(\vecR)$ (Eq. \ref{trialwf}) depends on a single parameter $r_0$. This parameter is previously optimized using the variational Monte Carlo method. Its value increases with the total number of particles $N$; for instance, when $R=4$ (in $a_{11}$ units) and $V_0=0.166$ (in $\hbar^2/(2 m_1 a_{11}^2)$ units), $r_0$ increases monotonously from $106$ up to $622$ when $N$ grows from $100$ to $2000$. 

To obtain ground-state averages, we resort to the Diffusion Monte Carlo (DMC) algorithm, as explained in Sec. \ref{sec:dmc}. Our DMC algorithm is accurate up to second order in the imaginary-time step~\cite{boronat1994monte} and uses forward walking to remove any bias of the trial wave function in the estimation of diagonal operators, which do not commute with the Hamiltonian~\cite{casulleras1995unbiased}. Any systematic bias derived from the use of a finite time step and a finite number of walkers in the diffusion process is kept smaller than the statistical noise.

\section{\label{sec:critical_atom_number_finite_drops_symmetric}Critical atom number}

\begin{figure}[t]
	\begin{center}
		\includegraphics[width=\linewidth]{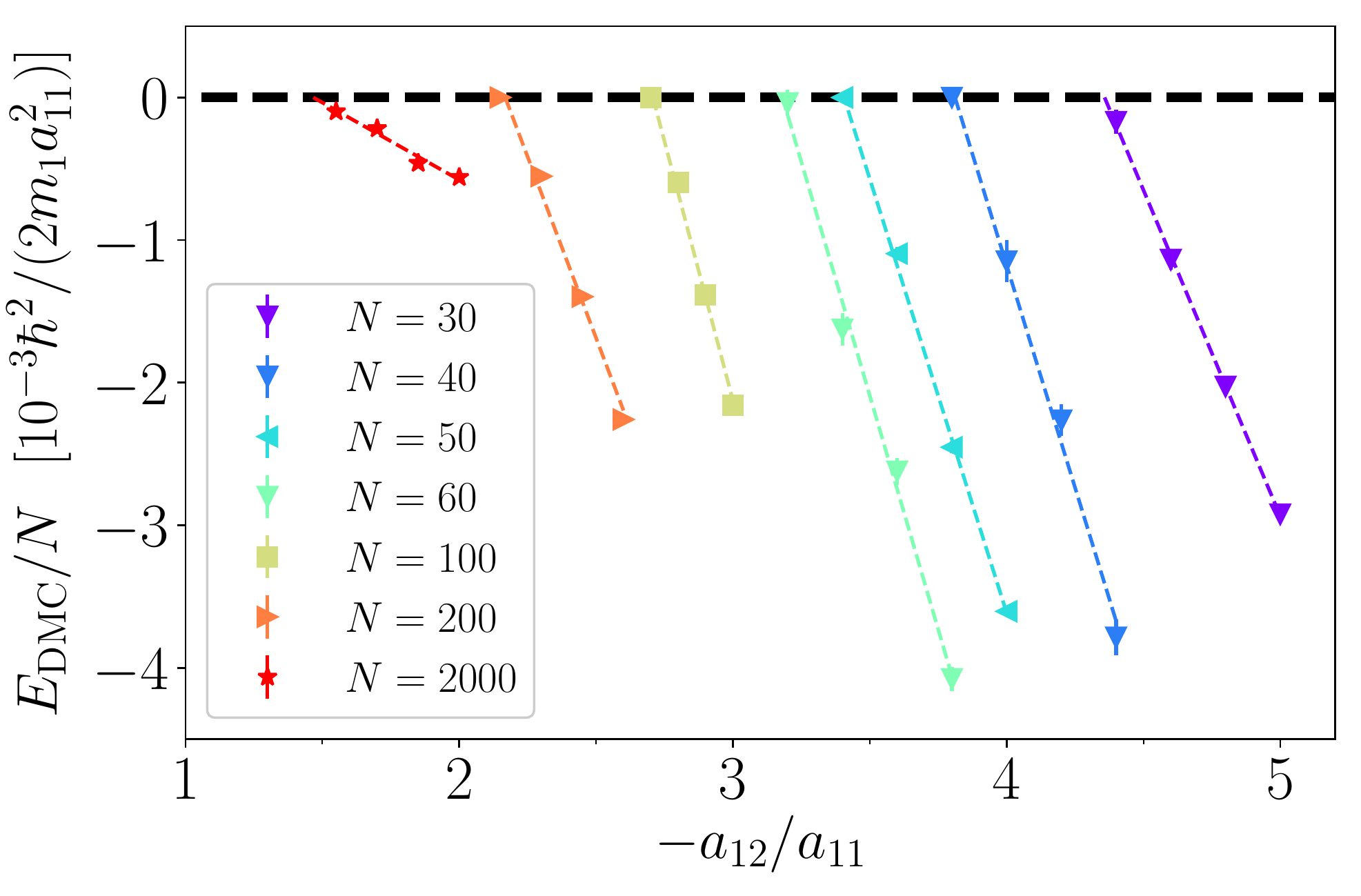}
		\caption[Energy per particle of the Bose-Bose symmetric liquid droplet as a 
		function of the scattering length $a_{12}/a_{11}$]{Energy per particle of the Bose-Bose symmetric liquid droplet as a 
			function of the scattering length $a_{12}/a_{11}$. Different symbols and lines 
			correspond to DMC calculations with different number of particles.}
		\label{fig1_droplet}
	\end{center}
\end{figure}

In Fig. \ref{fig1_droplet}, we report results for the energy per particle of the symmetric Bose-Bose liquid droplet, for a different number of particles and as a function of the scattering length $a_{12}$. For each $N$, we observe a similar behavior when we tune $a_{12}$. There is a critical value that separates systems with positive and negative energies. When the energy is positive, the system is in a gas phase, and, by increasing $|a_{12}|$, the $N$ system condenses into a self-bound system, that is,  a liquid drop. Around the critical value, the energy decreases linearly. Our results show a clear dependence of the critical scattering length for binding on the number of particles: smaller drops require more 
attraction (larger $V_0$) than larger ones.  

\begin{figure}[t]
	\begin{center}
		\includegraphics[width=\linewidth]{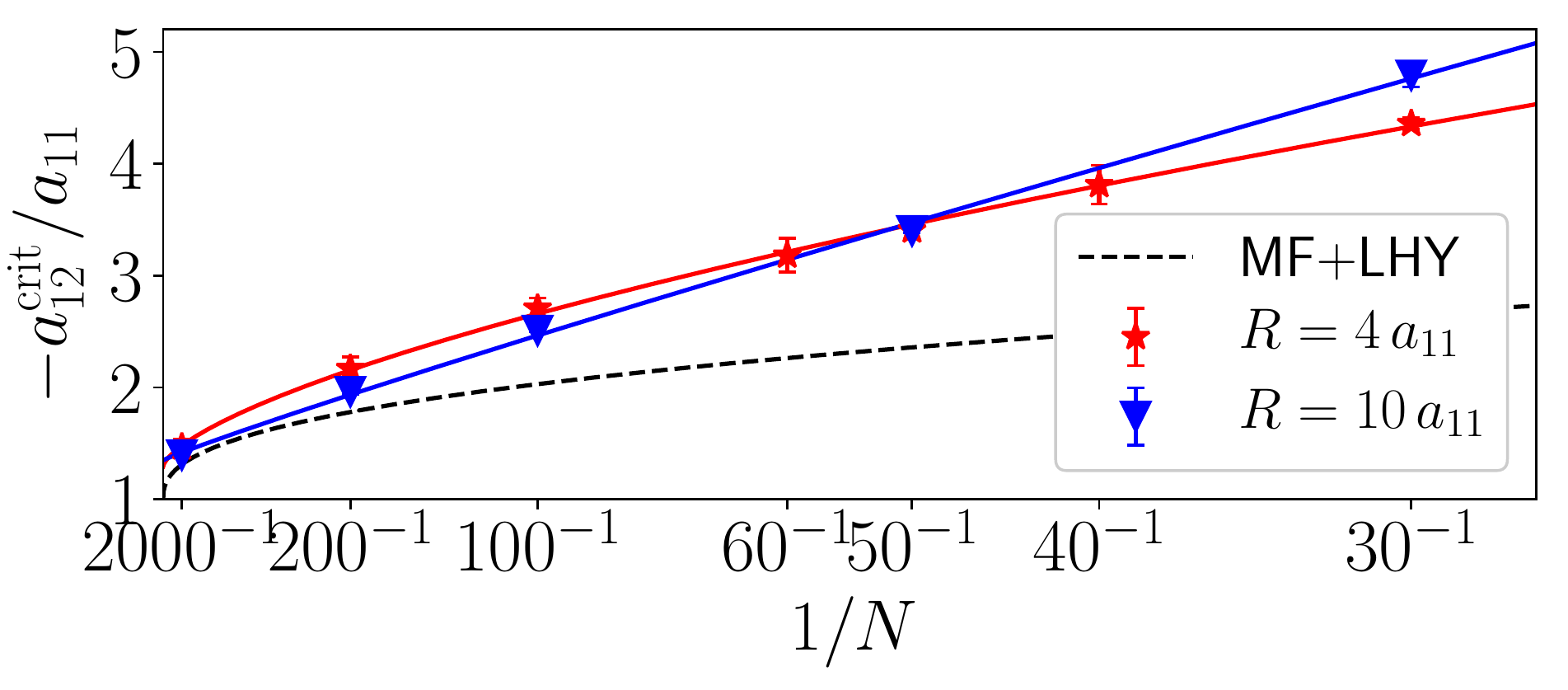}
		\caption[Critical values $a_{12}^{\rm{crit}}$ for symmetric liquid drop 
		formation as a function of $1/N$.]{Critical values $a_{12}^{\rm{crit}}$ for symmetric liquid drop 
			formation as a function of $1/N$. Red and blue points stand for $R=4 a_{11}$ 
			and $R=10 a_{11}$, respectively. The full lines correspond to fits to the DMC 
			results, and the dashed line is the prediction according to the MF+LHY theory \cite{petrov2015quantum}.}
		\label{fig2_droplet}
	\end{center}
\end{figure}

The dependence of the critical scattering length for self-binding, $a_{12}^{\rm{crit}}$, on the number of particles is shown in Fig. \ref{fig2_droplet}. Plotted as a function of $1/N$, we observe a linear decrease of $a_{12}^{\rm{crit}}$, reaching in the thermodynamic limit ($N \to \infty$) a value slightly larger than one. MF+LHY theory has been applied to the formation of Bose-Bose drops around this value $|a_{12}|\sim a_{11}$ corresponding to drops with a very large number of particles~\cite{petrov2015quantum}. In the same figure, we show results derived using a different range $R=10 a_{11}$ of the attractive well. As we can see, the results are slightly different, with an extrapolation to the thermodynamic limit a bit closer to one. The influence of the effective range has been confirmed in Chapter \ref{ch:symmetric_liquids} by performing extensive calculations of the equations of state. Together with the finite-range effects appearing in a mixture of $^{39}$K for which $a_{11} \neq a_{22}$ (presented in Chapter \ref{ch:finite_range_effects}), this gives strong indications that effects due to effective range are a general feature emerging in correlated quantum drops. Finally, a prediction of MF+LHY theory predicts systematically lower values of $-a_{12}^{\rm crit} / a_{11}$ for a given $N$, signaling that we observe, for the first time, repulsive LHY contributions to the energy not accounted for by the Petrov theory \cite{hu2020consistent, ota2020beyond}.

\section{\label{sec:liquid_state_calculation_finite_drops_symmetric}Liquid state calculation}

\begin{figure}[H]
	\centering
	\includegraphics[width=0.7\linewidth]{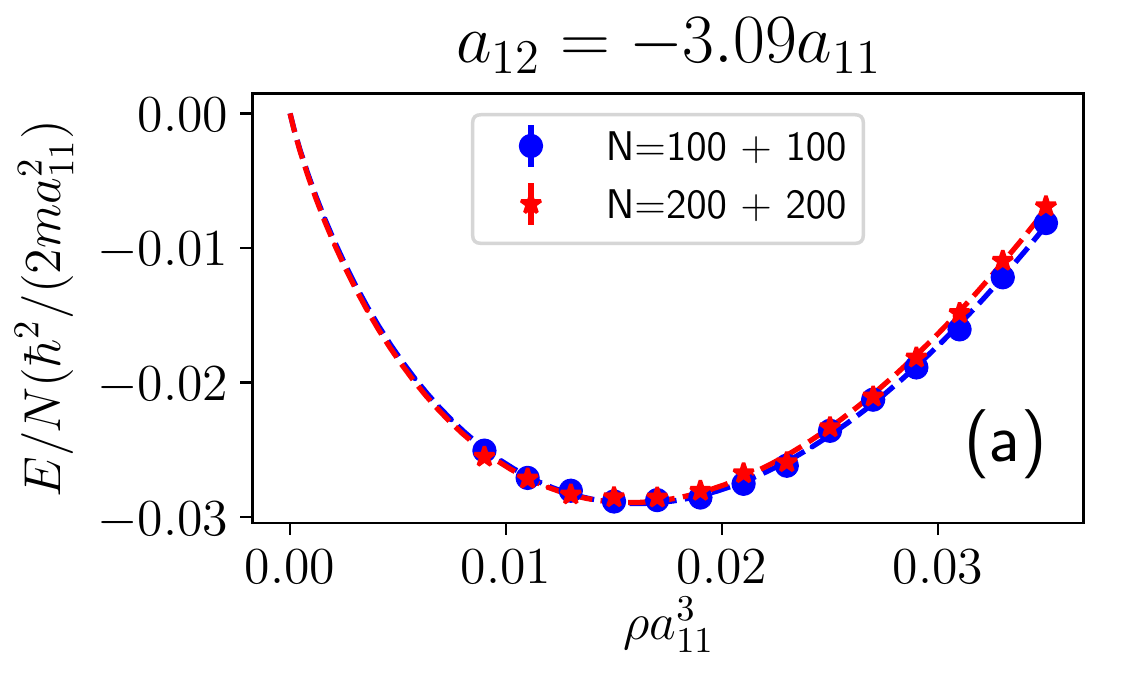}
	\includegraphics[width=0.7\linewidth]{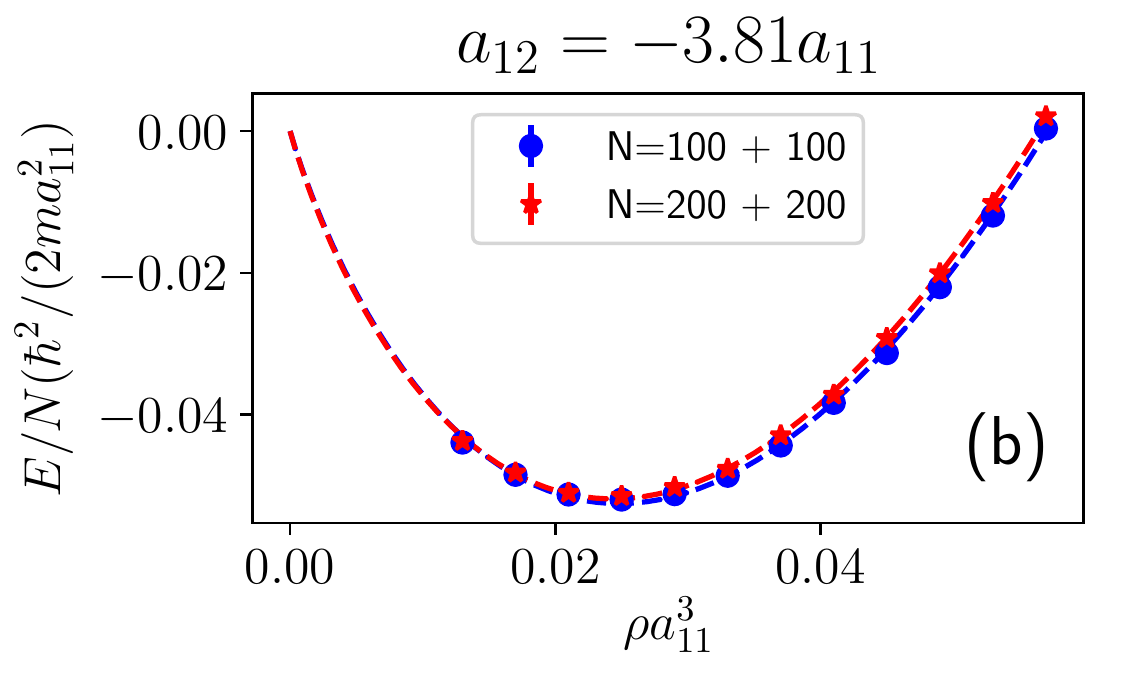}
	\includegraphics[width=0.7\linewidth]{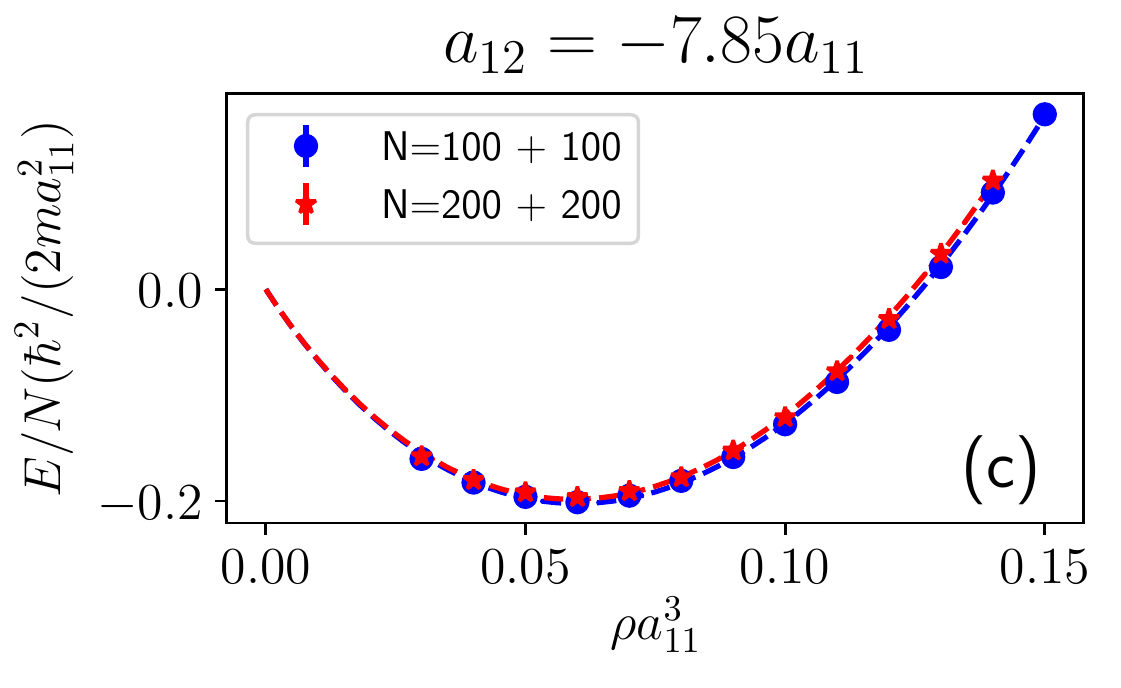}
	\caption[DMC energy per particle in a symmetric liquid as a function of density for three values of the scattering length $a_{12}$]{DMC energy per particle in a symmetric liquid as a function of density for three values of the scattering length $a_{12}$. In each figure, a set of two calculations with increasing particle numbers in a box is presented as an illustration that the finite-size effects are well eliminated.}
	\label{fig:combinedeossymmetricdrops}
\end{figure}

To compare our results of finite-$N$ drops with the thermodynamic limit, we have carried out the calculations in the homogeneous phase, where we can obtain the equilibrium density and energy per particle. These two parameters served as a consistency check in the determination of a density profile and surface tension. Our calculations of the homogeneous phase are performed by imposing periodic boundary condition on particle coordinates in a box of size $L=\sqrt[3]{\rho / N}$, where $\rho$ is the total number density and $N$ is the total number of particles. We used a trial wavefunction built as a product of Jastrow factors 
\cite{reatto1967phonons},
\begin{equation}
\Psi(\vec{\mathrm{R}}) = \prod_{i<j}^{N_1} f^{(1,1)}(r_{ij}) 
\prod_{i<j}^{N_2} f^{(2,2)}(r_{ij}) \prod_{i,j}^{N_1, N_2} f^{(1,2)}(r_{ij}) \ ,
\end{equation} 
with $N_1=N_2 = N/2$, and where the two-particle correlation functions $f(r)$ are
\begin{equation}
\label{eq:trial_wf_2_liquid_drops}
f^{\alpha, \beta}(r)=
\begin{cases}
f_{\rm 2b}(r) &  r < R_0 \\
B\exp(-\frac{C}{r} + \frac{D}{r^2})       ,& R_0 <r < L/2 \\
1      ,& r > L/2 \ . \\
\end{cases}
\end{equation}
Coefficients B, C and D are adjusted to match the continuity condition of the wavefunction and its first derivative. $R_0$ is a variational parameter, which is set to $R_0 = 0.45L$ in all calculations. Function $f_{\rm 2b}(r)$ is a solution to the two-body problem, and it is connected to a long-range phononic wavefunction \cite{reatto1967phonons}.  By performing two set of calculations with increasing particle numbers, we observe that already at $N \approx 100$ the thermodynamic limit is achieved. Results of the energy per particle as a function of density are shown in Fig. (\ref{fig:combinedeossymmetricdrops}). By fitting the DMC energy per particle to the functional form given by
\begin{equation}
\dfrac{E}{N} = \alpha \rho + \beta \rho^\gamma,
\end{equation}
we can obtain the equilibrium energy per particle and density as
\begin{eqnarray}
\label{eq:symmetric_equilibrium_en_rho}
\rho_0  & = & \left(\dfrac{-\alpha}{\beta \gamma} \right)^{1 / (\gamma - 1)} \\
\dfrac{E_0}{N}  & = & \alpha \rho_0 + \beta \rho_0^\gamma.
\end{eqnarray}

\section{\label{sec:density_saturation_finite_drops_symmetric}Density saturation}

\begin{figure}[H]
	\begin{center}
		\includegraphics[width=\linewidth,angle=0]{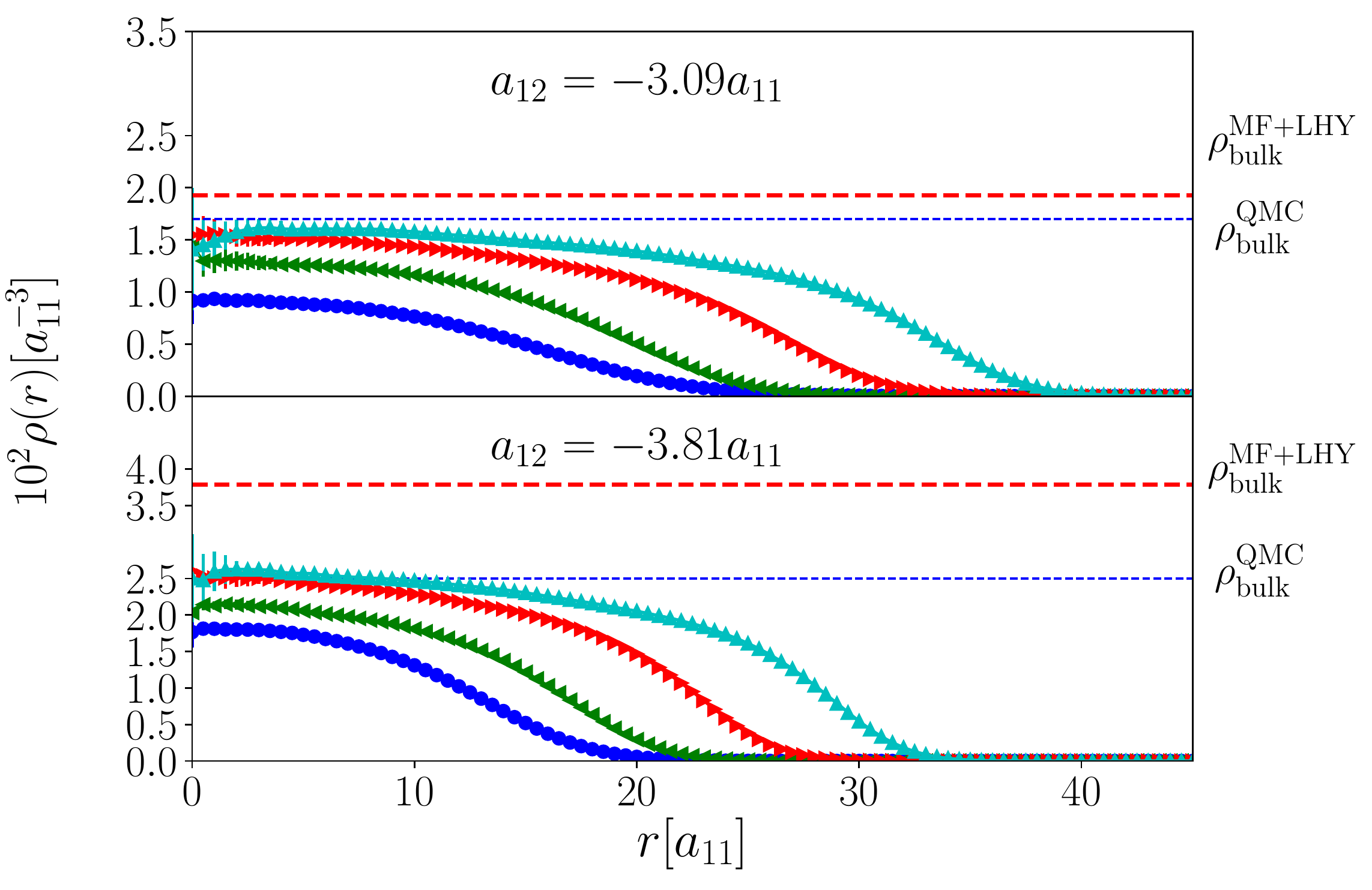}
		\caption[Density profiles of the symmetric Bose-Bose liquid drops for different number of particles]{Density profiles of the symmetric Bose-Bose liquid drops for different number of particles. Top and bottom panels correspond to $V_0=0.150$, $a_{12}=-3.09$ and $V_0=0.166$, $a_{12}=-3.81$, respectively. From small to large drops, $N=200$, 400, 1000, and 2000. Red dashed lines denote the equilibrium density in the bulk phase within the MF+LHY theory \cite{petrov2015quantum}, whereas blue dashed lines correspond to the equilibrium density of the bulk phase estimated from QMC calculations in the homogeneous setup (see Eq. \ref{eq:symmetric_equilibrium_en_rho} and Fig. \ref{fig:combinedeossymmetricdrops}).}
		\label{fig3_droplet}
	\end{center}
\end{figure}

The calculation of the density profiles $\rho(r)$ allows for a better knowledge of the shape and size of the formed drops. In Fig. (\ref{fig3_droplet}) we report DMC results on the density profiles of the obtained drops. Notice that there is not any difference between the partial density profiles due to our election of interactions and masses, $\rho^{(1)}(r)=\rho^{(2)}(r)=\rho(r)/2$. The two cases shown in Fig. (\ref{fig3_droplet}) correspond to scattering lengths $a_{12}=-3.09$ (top) and $a_{12}=-3.81$ (bottom). When the number of particles increases, one observes that both the central density and radius of the drop grow. Progressive increase of density is expected to happen until the central density reaches the equilibrium density of the bulk phase. Once the drop saturates, only the radius increases with the addition of more particles. The density profiles, shown in Fig. (\ref{fig3_droplet}) for two illustrative examples, correspond to very dilute liquids because we need $\sim 2000$ particles to reach saturation. In the figures, we have also shown the equilibrium densities that we have obtained for the same potentials in the bulk phase (see Fig. \ref{fig:combinedeossymmetricdrops}).

By increasing the scattering length, i.e., by making the system more attractive, we observe that the central density increases and the size of the drop squeezes. Apart from the central 
density one can also extract from the density profiles the surface width, 
usually measured as the length $W$ over which the density decrease from 90 to 
10\% of the inner density. It is expected that $W$ increases with $N$ for unsaturated drops, and then it stabilizes when saturation is reached. Our 
results show also this trend: for $a_{12}=-3.09$, $W=15$ for the smallest drop 
and stabilizes then to $W\simeq 20$; for $a_{12}=-3.81$, these values are 
$W=11$ and $18$. 

\begin{figure}[H]
	\begin{center}
		\includegraphics[width=\linewidth,angle=0]{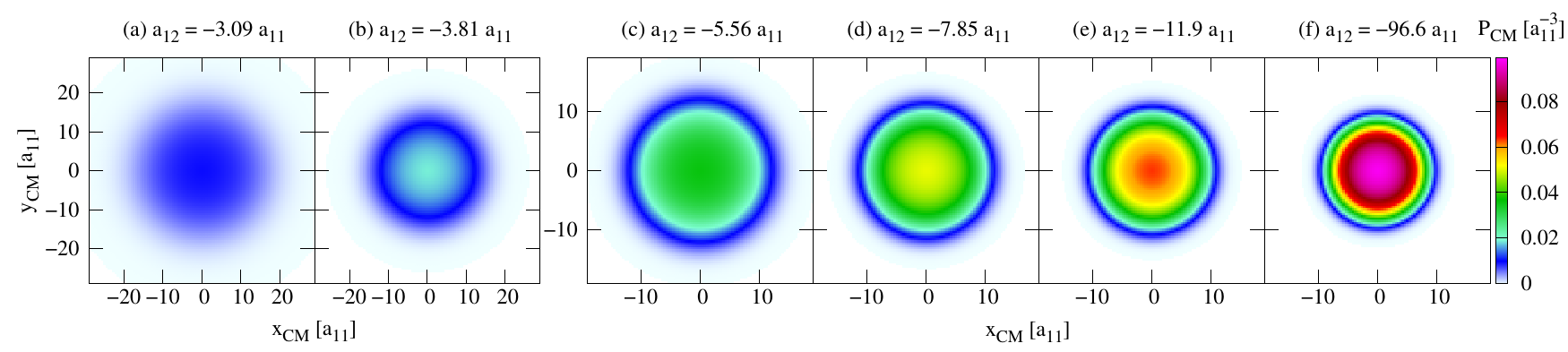}
		\caption[Contour plots of the density profiles of a liquid drop 
		with $N=200$ as a function of $a_{12}$.]{Contour plots of the density profiles of a liquid drop 
			with $N=200$ as a function of $a_{12}$. }
		\label{fig4_droplet}
	\end{center}
\end{figure}

DMC allows for the study of the drops around the gas to liquid transition but also can show how the evolution towards a collapsed state happens. By increasing the depth of the attractive well $V_0$, we can see the change in the shape and size of a 
given drop. In Fig. \ref{fig4_droplet}, we report this evolution as a contour plot of the density profiles for a particular liquid drop with $N=200$ particles. The range of $a_{12}$ values starts close to $a_{12}^{\rm{crit}}$, for this $N$ value, and ends quite deep into the Feshbach resonance at a scattering length $a_{12} \simeq 30  \, a_{12}^{\rm{crit}}$. Following this ramp, we observe an increase of an order of magnitude in the inner density and a shrinking of the size, with a three-fold reduction of the radius. Therefore, the drop becomes denser, but it is still a fully stable object which is not at all collapsed.

\section{\label{sec:surface_calculation_finite_drops_symmetric}Surface tension calculations}

\begin{figure}
	\begin{center}
		\includegraphics[width=0.8\linewidth,angle=0]{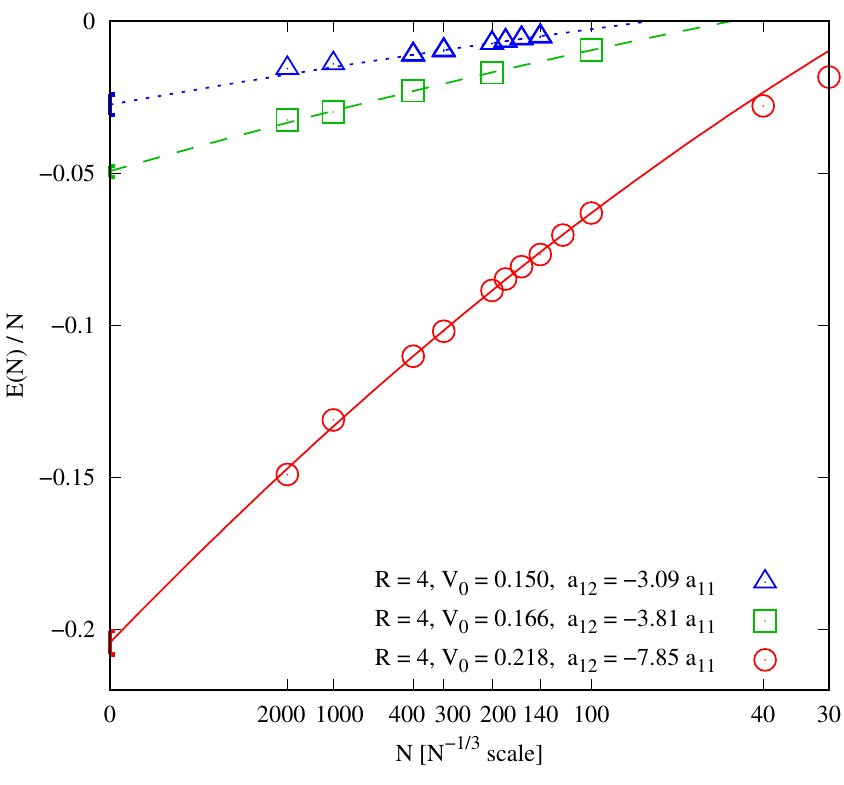}
		\caption[Energy per particle of symmetric Bose-Bose drops as a function of $N^{-1/3}$.]{Energy per particle of symmetric Bose-Bose drops as a function of $N^{-1/3}$. The open symbols are the DMC results and the lines are 
			fits according to the liquid drop model (\ref{dropmodel}). The errorbars are 
			smaller than the size of the symbols. Different sets correspond to different 
			values of the interparticle scattering length $a_{12}$. The points at zero 
			$x$-axis with error bars correspond to bulk calculations (see Fig. \ref{fig:combinedeossymmetricdrops}).}
		\label{fig5_droplet}
	\end{center}
\end{figure}

The microscopic characterization of the Bose-Bose liquid drops is not complete without the knowledge of the energy. As we commented before, the result of the energy determines if an $ N $-particle system is in a gas or liquid state. Once in the liquid phase, it is important to calculate the dependence of the energy on the number of particles. In Fig. (\ref{fig5_droplet}), we report the DMC 
energies as a function of $N$ and for three different $a_{12}$ values. From 
intensive calculations carried out in the past on liquid $^4$He 
drops~\cite{pandharipande1983calculations,chin1992structure}, we know that the energy of the drops is well accounted for by a liquid-drop model. 
According to this, the energy per particle is
\begin{equation}
E(N)/N = E_v + E_s \, x + E_c \, x^2 \ , 
\label{dropmodel}
\end{equation}
with $x\equiv N^{-1/3}$. The coefficients in Eq. (\ref{dropmodel}), $E_v$, 
$E_s$, and $E_c$ are termed volume, surface, and curvature energies, 
respectively. The term $E_v$ corresponds to the energy of an infinitely large 
drop or, in other words, to the energy per particle of the bulk. The second 
term $E_s$ is important because, from it, we can estimate the surface tension 
of the liquid $t$ as
\begin{equation} 
t = \frac{E_s}{4 \pi r_0^2} \ .
\label{tension}
\end{equation}
The parameter $r_0$ is the unit radius of the liquid, and can be estimated from 
the relation 
\begin{equation}
4 \pi/3 \, r_0^3 \rho_0 =1,
\end{equation}
with $\rho_0$ the equilibrium 
density of the liquid.

In Fig. (\ref{fig5_droplet}), we plot as lines the results of the liquid-drop model obtained as least-squares fit to the DMC energies. In the three cases studied, we obtain a high-fidelity fit. In Fig. (\ref{fig5_droplet}) we plot in the zero $x$-axis the energies of the bulk liquid with the same Hamiltonian as in the drops. These results are not included in the fit (\ref{dropmodel}), but they are completely coincident with the energies $E_v$ obtained solely from the drop energies. Matching of finite-N calculations with those in a homogenous phase is, in fact, a stringent test of accuracy on the calculations of the liquid drops. In the figure, we see the effect of the potential on the energy of the drops for three selected cases. The binding energy of a given $N$ drop increases with the magnitude of $V_0$, and thus with $a_{12}$. We have verified that the energy grows linearly with $V_0$ close to the critical value for self-binding 
but, for larger potential depths, increases faster. From relation 
(\ref{tension}) and the values obtained for $E_s$ from the fits using the liquid-drop model, we estimate that the surface tension for the three cases shown in Fig. (\ref{fig5_droplet}) are $0.18 \cdot 10^{-3}$, $0.37\cdot 10^{-3}$, and $2.41 
\cdot 10^{-3}$ (in units $\hbar^2/(2 m_1 a_{11}^4)$) when $a_{12}=-3.09$, 
$-3.81$, and $-7.85$, respectively.   

We think that a comparison between the Bose-Bose drops here studied and the 
well-known properties of stable superfluid $^4$He drops can help to better 
visualize their extraordinary properties. We can consider a typical value for 
$a_{11}$ used in the experiments with ultracold mixtures of $^{39}$K-$^{41}$K, 
say $a_{11}=50 \, a_0$, with $a_0$ the Bohr radius. Then, the saturation 
densities of the drops shown in Fig. \ref{fig3_droplet} are $\sim 1.0 \cdot 10^{-6}$ and 
$ 1.4 \cdot 10^{-6}\ \text{\AA}^{-3}$. Near the critical scattering length for a given size, the drops are even more dilute, e.g., for $N=2000$ and $a_{12} = -1.75$, the central density is about $3\cdot 10^{-8}\text{\AA}^{-3}$. The saturation density of liquid $^4$He 
is $2.2 \cdot 10^{-2} \ \text{\AA}^{-3}$ implying that the Bose-Bose drops can 
be as dilute as $\sim 10^4$ times the $^4$He ones (a similar ratio happens when 
compared with water, with density $3.3\cdot10^{-2}\ \text{\AA}^{-3}$). For the 
same number of atoms, the Bose-Bose drop is much larger than the $^4$He one: 
$9.8 \cdot 10^{-2} \mu m$ for $V_0=0.150$ and $3\cdot 10^{-3}  \mu m$ for $^4$He with $N=2000$ \cite{barranco2006helium}. The surface of dilute drop for this $N$ is $\sim 50\%$ of the total size, much larger than the $20\%$ value in $^{4}$He.

\section{Summary and conclusions}

Summarizing, we have carried out a DMC calculation of Bose-Bose mixtures with attractive interspecies interaction. Relying only on the Hamiltonian, we describe the system without further approximations. As announced by Petrov using LHY approximate theory~\cite{petrov2015quantum}, it is possible to get self-bound systems by a proper selection of the interactions between equal and different species. Our results clearly show the transition from gas, with positive energy, to a self-bound system (liquid), and accurately determine the critical scattering lengths for the transition as a function of the number of particles. In the range of parameters here studied, we do not observe universality in the sense that the results depend only on the $s$-wave scattering lengths. For the same $a_{12}$ value, we observe dependence on the range of the potential $R$. This motivated us to study whether the inclusion of the $s$-wave effective range to the model potentials provides with a more precise description, which is discussed in Chapters \ref{ch:symmetric_liquids} and \ref{ch:finite_range_effects}.


The experimental realization of Bose-Bose liquid drops~\cite{cabrera2018quantum} opens the possibility of accessing to denser systems than the usual trapped ultracold gases where quantum correlations can be much more relevant. The point of view from the liquid state is different: the liquid that emerges from these 
mixtures is ultradilute, much less dense than any other stable liquid in 
Nature. Therefore, the liquid phase realm extends to unexpected regimes never achieved before.



\chapter{\label{ch:symmetric_liquids}Universality in ultradilute liquid Bose-Bose mixtures}

\section{Introduction}

For more than two decades, most of the experiments in ultracold atoms were done in the low-density gas phase, in the universal regime fixed solely by the gas parameter $\rho a^3$, with $a$ the $s$-wave scattering length and $\rho$ the density. The range of universality of the homogeneous single-component Bose gas was established using different model potentials and solving the $N$-body problem in an exact way with quantum Monte Carlo (QMC) methods~\cite{giorgini1999ground}. One of the most important advances in the field of ultracold atoms in the last years is the recent creation of ultradilute quantum droplets in Bose-Bose mixtures. Petrov~\cite{petrov2015quantum} pointed out that liquid drops can be created in a setup composed by a two-component mixture of bosons with short-ranged attractive interspecies and repulsive intraspecies interactions. However, the perturbative technique employed by Petrov is valid only very close to the mean-field (MF) instability limit, that is for extremely dilute liquids. The collapse predicted on the MF level is avoided by stabilization due to the quantum fluctuations described by the Lee-Huang-Yang (LHY) correction to the energy. Two experimental groups recently managed to obtain self-bound liquid drops~\cite{cabrera2018quantum,semeghini2018self}, which, upon releasing the trap, did not expand. The drops required a certain critical number of atoms to be bound. Importantly, measurements of the critical number and size of the smallest droplets could not be fully accounted for by the MF+LHY 
term~\cite{cabrera2018quantum}.

We have studied self-bound Bose-Bose droplets using 
the diffusion Monte Carlo (DMC) method, thus solving \textit{exactly} the 
full many-body problem for a given Hamiltonian at zero 
temperature, discussed in Chapter \ref{chapter:ultradilute_liquid_drops} of this Thesis. Our results have confirmed the transition from gas, with positive energy, to a self-bound droplet with negative energy. We have determined the critical number of atoms needed to form a liquid droplet as a function of the intraspecies scattering length. Using two different models for the attractive interaction, we did not get quantitatively the same results for the range of scattering lengths studied, which points to the lack of universality in terms of $\rho a^3$. Thus, it is of fundamental interest to find whether there is a range of densities and scattering lengths where such universality exists. In the case of homogeneous Bose gases, departures from universality start to appear around $\rho a^3 \gtrsim 10^{-3}$~\cite{giorgini1999ground}. In that case, adding the LHY correction allowed for a good approximation of the equation of state up to higher densities. Recently, a variational hypernetted-chain Euler-Lagrange calculation~\cite{staudinger2018self} of unbalanced mixtures showed that the drops could only be stable in a very narrow range an optimal ratio of partial densities and near the energy minimum. Moreover, Ref.~\cite{staudinger2018self} found dependence on the effective range, even at low densities.

In this chapter, we use the DMC method to address the question of the 
universality in the equation of state of dilute Bose-Bose mixtures with symmetric interactions, i.e. $a_{11} = a_{22}$ and $m_1 = m_2$. The second question we pose here is whether there exists a regime where instead of using only one parameter ($s$-wave scattering length) inclusion of an additional parameter (effective range~\cite{newton2013scattering}) extends the validity of the universal description. To answer these questions directly for finite-size droplets would require enormous computational resources, as at least thousands of atoms are needed to achieve a self-bound state close to the mean-field limit~\cite{cabrera2018quantum}. In order to eliminate the finite-size effects caused by the surface tension and simplify the analysis, we study here bulk properties corresponding to the interior of large saturated droplets. From the obtained equation of state, we construct a new density functional and use it to predict the profiles of the drops, discussing the effects of the potential range.

\section{Hamiltonian and the methods} We rely on the DMC method, as described in Chapter \ref{ch:qmc_methods}. The Hamiltonian of our system is given by 
\begin{equation}
H = -\sum_{\alpha=1}^{2}\frac{\hbar^2}{2 m_{\alpha}} \sum_{i=1}^{N_{\alpha}} \nabla_{i\alpha}^2
+\frac{1}{2}\sum_{\alpha, \beta=1}^{2} \sum_{i_{\alpha},
	j_{\beta=1}}^{N_{\alpha}, N_{\beta}} V^{(\alpha, \beta)}
(r_{i_{\alpha}j_{\beta}}) \ ,
\label{eq:hamiltonian_symmetric_liquid}
\end{equation}
where $V^{(\alpha,\beta)}(r)$ is the interatomic interaction between species $\alpha$ and $\beta$. The intraspecies interactions with positive $s$-wave scattering length are modeled either by a hard-core potential of diameter $a_{ii}$
\begin{equation}
V(r) =
\begin{cases}
\infty & r \le a_{ii} \\
0      & \mathrm{otherwise},
\end{cases}
\end{equation}
or by a 10-6 potential~\cite{pade2007exact} that does not support a two-body bound state
\begin{equation}
V(r) =  V_0 \left[ \left(\frac{r_0}{r}\right)^{10} -
\left(\frac{r_0}{r}\right)^{6} \right].
\end{equation}
The latter model has an analytic scattering length given in  Ref.~\cite{pade2007exact}. The interspecies interactions with negative scattering length, $a_{12} < 0$, are modeled by a square-well potential of range $R$ and depth $-V_0$
\begin{equation}
V(r) =
\begin{cases}
-V_0 & r\le R_0 \\
0   & \mathrm{otherwise},
\end{cases}
\end{equation}
or by a 10-6 potential with no bound states. 

We resort to a second-order DMC method and use a guiding wave function to reduce the variance, as described in  Ref.~\cite{boronat1994monte} and Sec. (\ref{sec:importance_sampling}). Our calculations in the homogeneous phase are performed by imposing periodic boundary condition on particle coordinates in a box of size $L=\sqrt[3]{\rho / N}$, where $\rho$ is the total number density and $N$ is the total number of particles. We construct the trial wave function a product of Jastrow factors~\cite{reatto1967phonons}
\begin{equation}
\Psi({\bf R})= \prod_{1=i<j}^{N_1} f^{(1,1)}(r_{ij}) \prod_{1=i<j}^{N_2 
}f^{(2,2)}(r_{ij}) \prod_{i,j=1}^{N_1,N_2} f^{(1,2)}(r_{ij}) \ .
\label{eq:trialwf_symmetric_liquid}
\end{equation}
The particular form of the two-particle correlation function depends on the model of the interaction potential. For the hard-core potential we used
\begin{equation}
f^{\alpha, \alpha}(r)=
\begin{cases}
1 - a_{\alpha,\alpha}/r , &  r < \tilde{R} \\
B\exp(-\frac{C}{r} + \frac{D}{r^2})       ,& \tilde{R} <r < L/2 \\
1      ,& r > L/2, \\
\end{cases}
\end{equation} 
The parameter $\tilde{R}$ can be optimized, but energies do not change 
drastically when $\tilde{R} \approx L/2$, so we set $\tilde{R} 
= 0.9 L/2$. Other parameters were obtained by continuity conditions for the function and its first derivative. For the square well potential we used 
\begin{equation}
f^{\alpha, \beta}(r)=
\begin{cases}
\sin(k r) /r, &  r < R \\
A(1 - \tilde{a}_{\alpha, \beta}/r) &   R  < r < \tilde{R}\\
B\exp(-\frac{C}{r} + \frac{D}{r^2})       ,& \tilde{R} <r < L/2 \\
1      ,& r > L/2, \\
\end{cases}
\end{equation} 
where $\tilde{a}_{\alpha, \beta}$ is a variational parameter. We set $\tilde{R} = 0.9 L/2$, while other parameters were obtained by continuity conditions.
Finally, for the 10-6 potential we used
\begin{equation}
f^{\alpha, \beta}(r)=
\begin{cases}
h(r,\tilde{a}_{\alpha, \beta} ) &  r < R_0 \\
B\exp(-\frac{C}{r} + \frac{D}{r^2})       ,& R_0 <r < L/2 \\
1      ,& r > L/2, \\
\end{cases}
\end{equation}
where $h(r, \tilde{a}_{\alpha, \beta})$ is the two-body scattering solution 
given in \cite{pade2007exact}. This function has a variational parameter 
$\tilde{a}_{\alpha, \beta}$, and we set $R_0 = 0.9 L/2$. Zero derivative was imposed at the half size of the simulation box $L$ for all three Jastrow factors.

We consider a mixture with equal masses of particles $m_1=m_2=m$. Such a situation is typical in experiments where different hyperfine states of the same atomic species are used to create two components~\cite{cabrera2018quantum}. Furthermore, to reduce the number of degrees of freedom, we choose to study the symmetric mixture with $a_{11}=a_{22}$ resulting in the ground-state concentration ratio of $N_1=N_2$. 
The calculations are performed in a box with periodic boundary conditions. 

\begin{figure}[]
	\centering
	\includegraphics[width=0.7\textwidth]{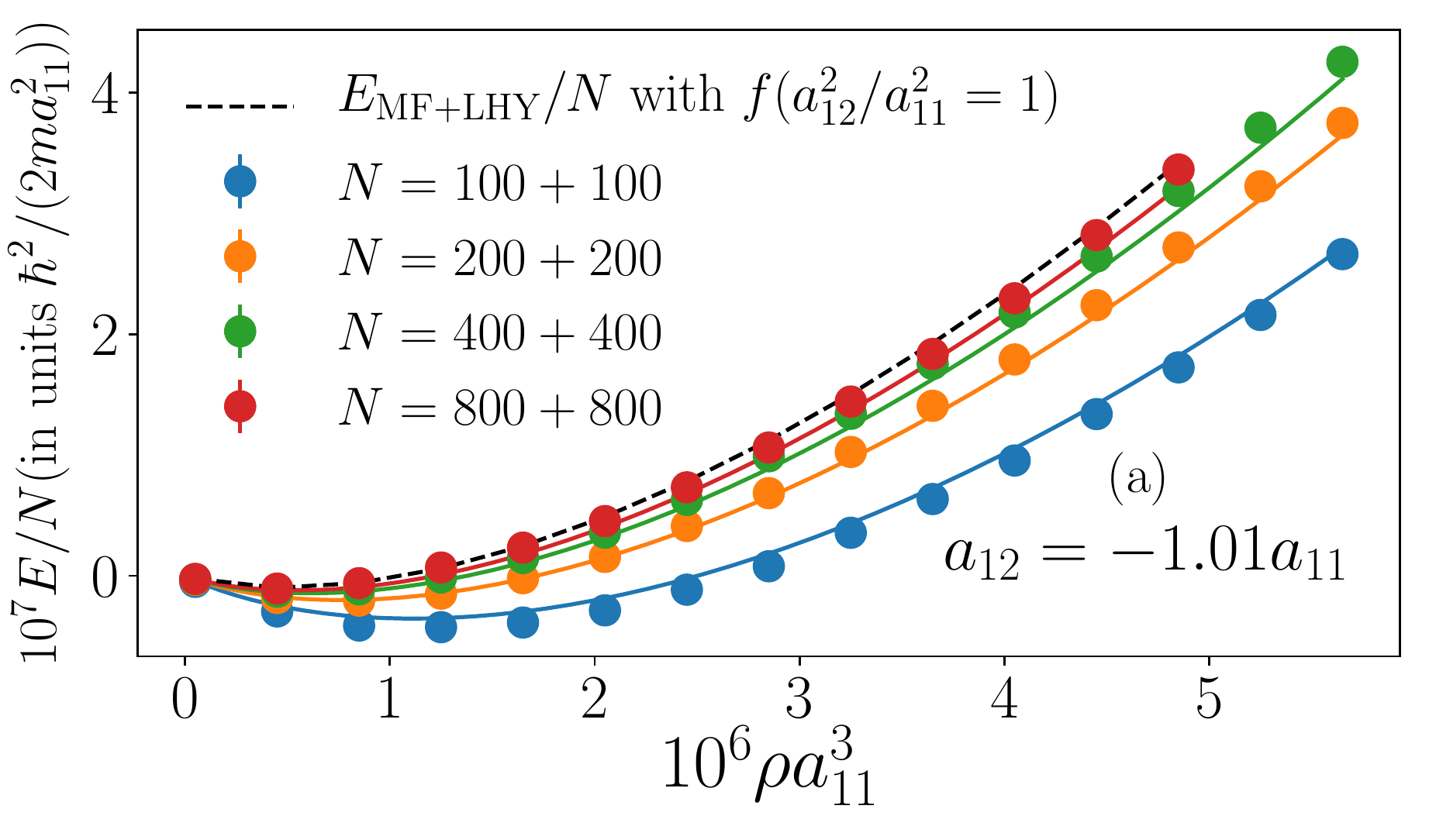}
	\includegraphics[width=0.7\textwidth]{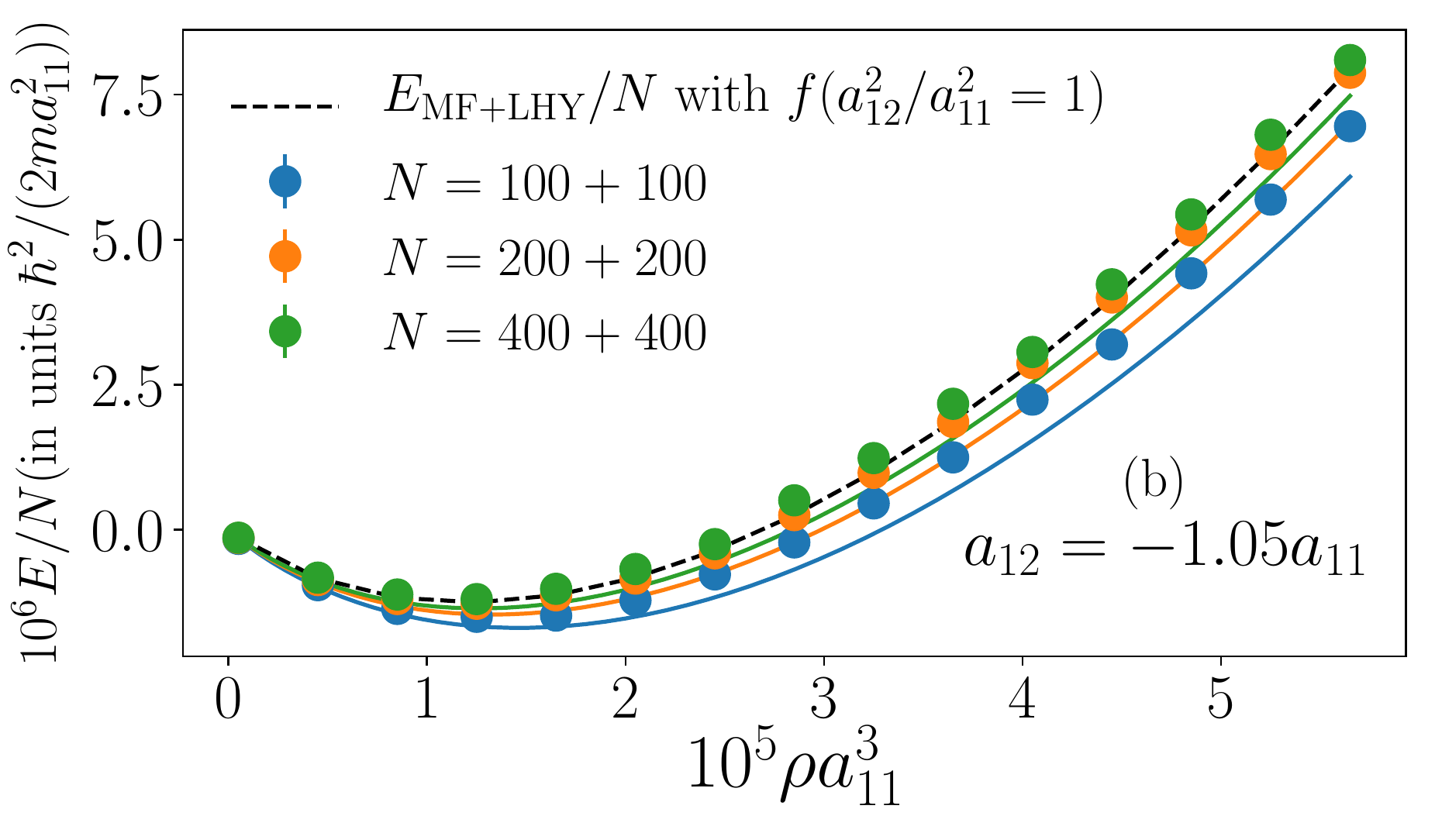}
	\includegraphics[width=0.7\textwidth]{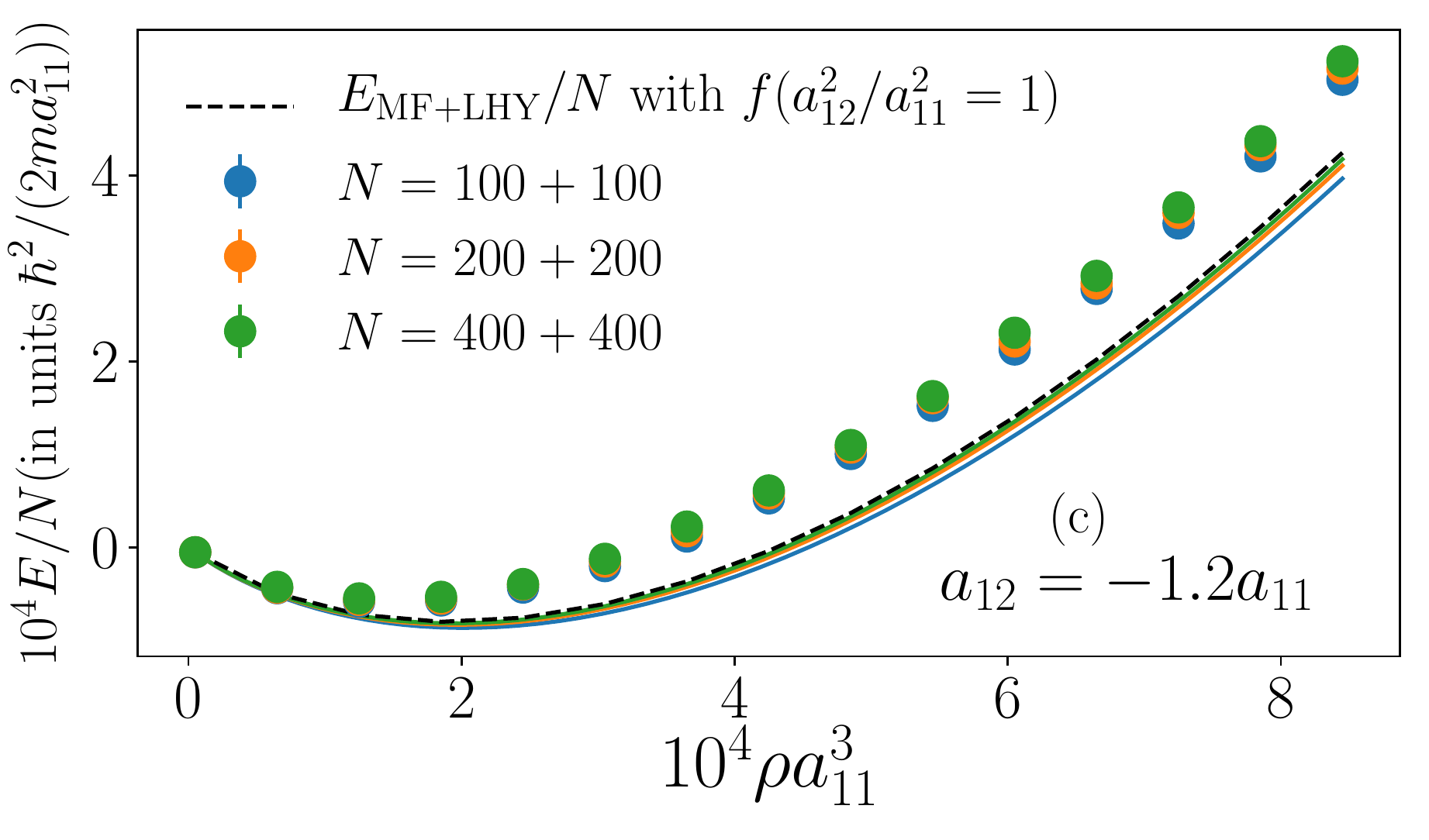}
	\caption[The energy per particle of a symmetric Bose-Bose liquid as a function of the density for $a_{12}= -1.01a_{11}$, $-1.05a_{11}$ and $-1.2a_{11}$]{The energy per particle of a symmetric Bose-Bose liquid as a function of the density for $a_{12}= -1.01a_{11}$, $-1.05a_{11}$ and $-1.2a_{11}$. Different total number of particles $N$ is used to illustrate the finite-size effect. Results are obtained with hard-core of diameter $a_{11}$ for the repulsive intraspecies interaction and a square well potential with diameter $\rho R^3 =10^{-5}$ for interspecies attraction. Full lines are finite-size corrections coming from the mean-field energy Eq. (\ref{eq:mf_corr}) and dashed line is the MF+LHY result.}
	\label{fig:finite_size_effects}        
\end{figure}

To approach the thermodynamic limit, we investigated finite-size effects by 
increasing the number of particles up to the point where energy converged. DMC 
results are presented for the largest particle number used. In 
Fig.~\ref{fig:finite_size_effects}, we show the convergence for three different 
$a_{12}/a_{11}$ values. It is possible to obtain the finite-size correction 
analytically only very close to the mean-field instability limit. In that case, 
the leading correction to mean-field (Hartree) contribution comes from the 
intracomponent number of pairs $N_1 (N_1 - 1)/2$,
\begin{equation}
\dfrac{\Delta E_{\rm MF}(N)}{N} = -\dfrac{\pi}{6}\dfrac{\hbar^2}{ma_{11}^2} 
\dfrac{2\pi}{N} \rho a_{11}^3 \ ,
\label{eq:mf_corr}
\end{equation}
where $N = 2N_1$. This correction is negative, linear with the density and decreases as $N^{-1}$ with the number of particles. 
This correction is also shown in Fig. \ref{fig:finite_size_effects}.

We have also optimized the timestep and population bias to reduce their influence below the statistical noise. Timesteps from 0.1 to 2$\times$10$^{-3}$ $2 m_1 a_{11}^2/\hbar^2$ were used,
with lower values of the timestep being used at higher densities. At low densities for the HCSW model, there
was no difference between calculations with 100 and 1000 walkers, which is a consequence of the good quality of the guiding wave function. In other cases, we used up to 1000 walkers.

\section{Finite-size effects}

\begin{figure}[!htb]
	\centering        
	\includegraphics[width=0.7\textwidth]{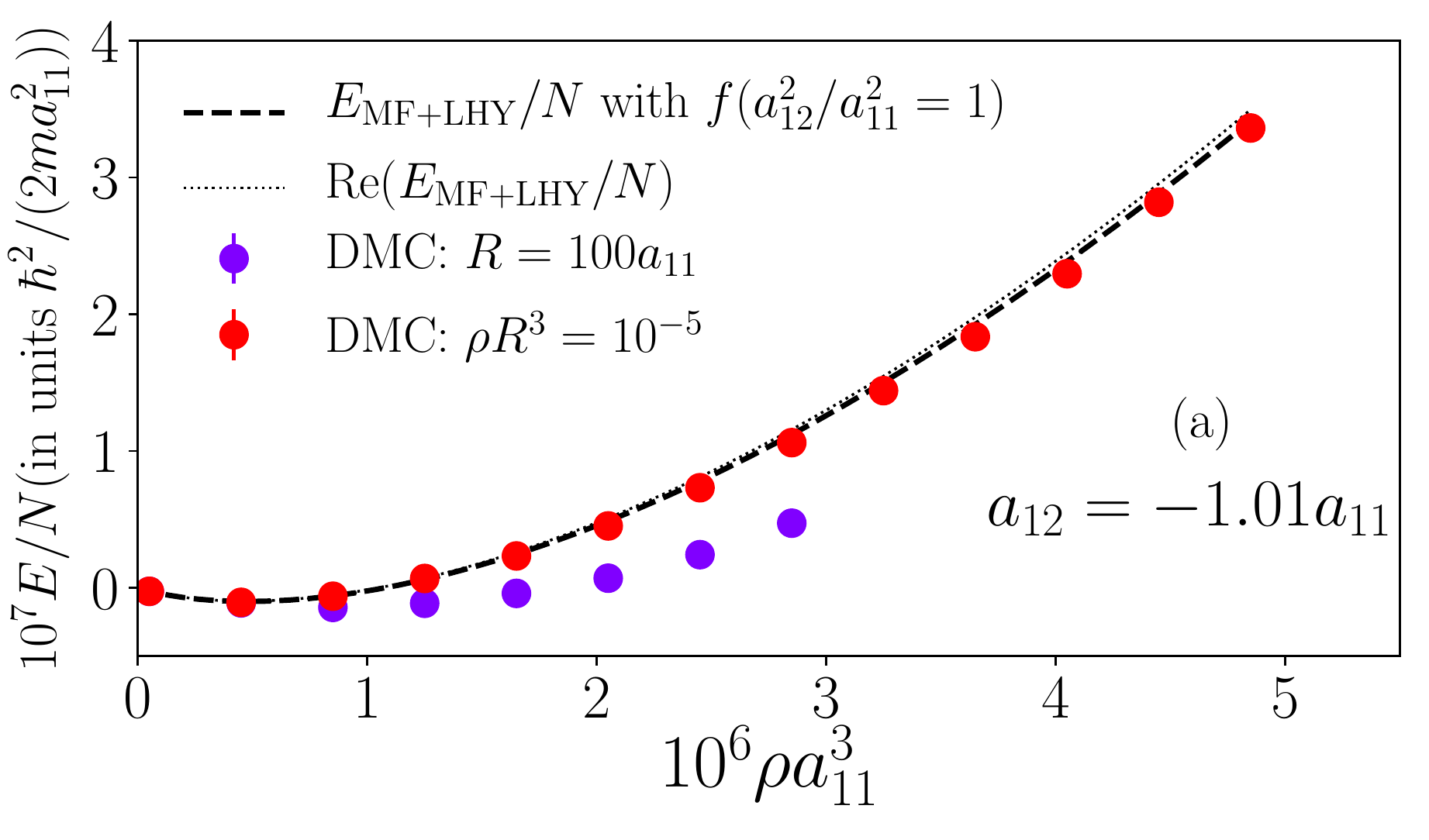}
	\includegraphics[width=0.7\textwidth]{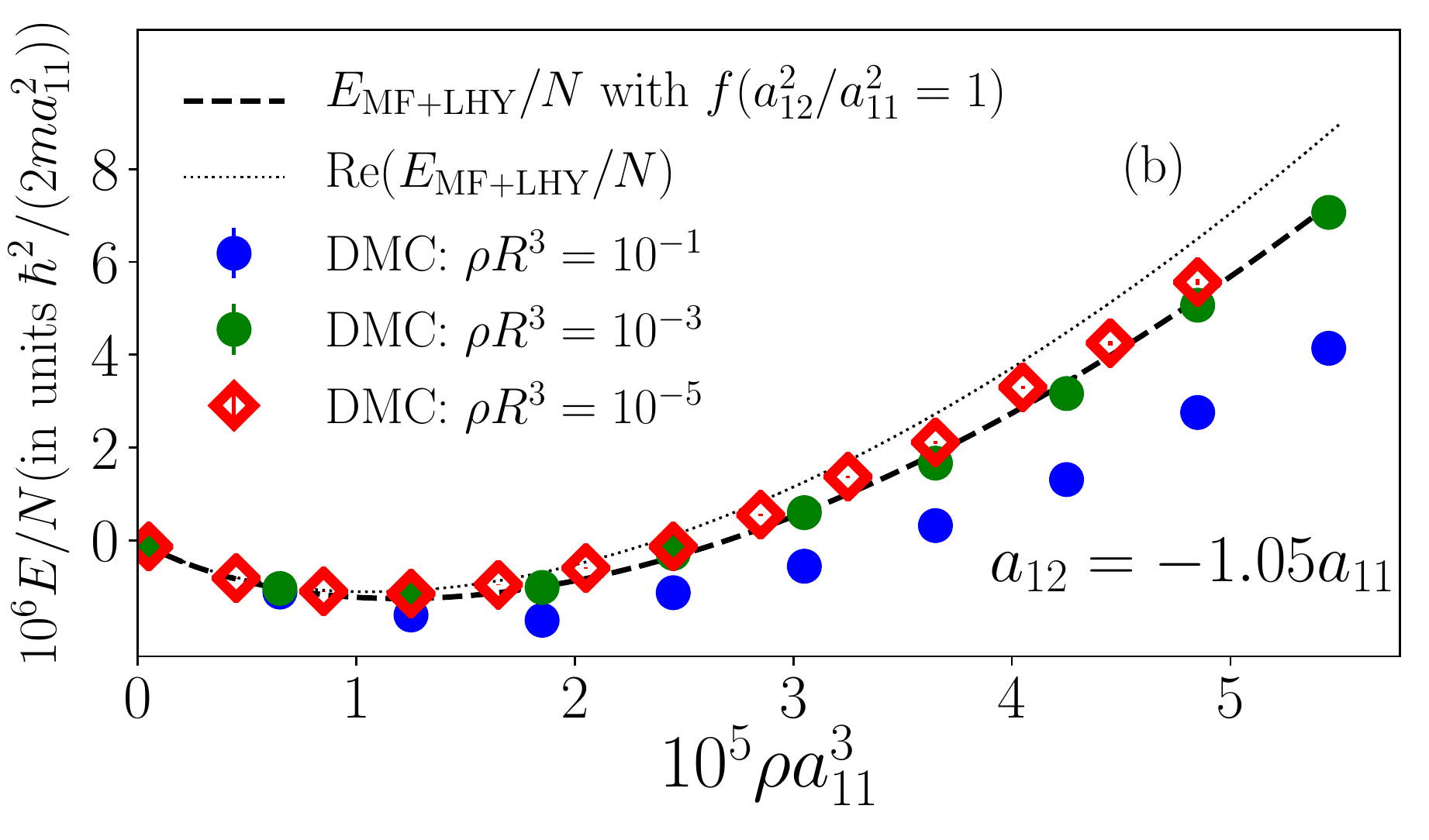}
	\includegraphics[width=0.7\textwidth]{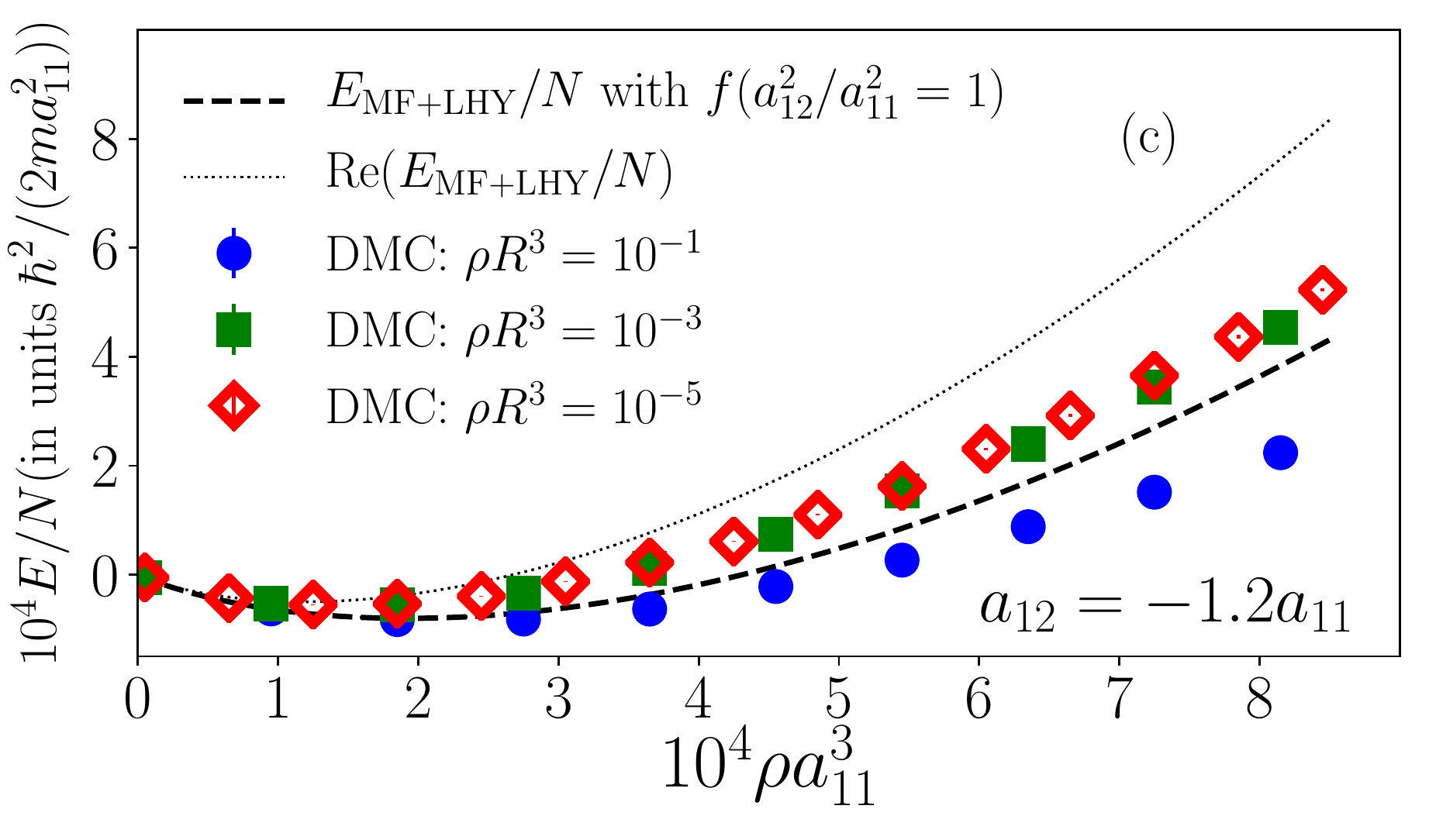}
	\caption[DMC equation of state of the symmetric Bose-Bose liquid mixture for different
	values of $a_{12}$ and different ranges $R$, in comparison with MF+LHY
	theory.]{DMC equation of state of the symmetric Bose-Bose liquid mixture for different
		values of $a_{12}$ and different ranges $R$, in comparison with MF+LHY
		theory.}
	\label{fig:fig1}
\end{figure}
First, we report results obtained using the hard-core model for the 
repulsive interactions and a square-well (SW) potential for the attractive 
ones. 
In Fig.~(\ref{fig:fig1}), we show our results for different values of the 
interspecies scattering length $a_{12}$ and different ranges of the attractive 
well $R$, and compare them to the MF+LHY prediction~\cite{petrov2015quantum}. The equation 
of state in Ref.~\cite{petrov2015quantum} for $m_2 = m_1=m$, $a_{22} = a_{11}$, and $N_2 = 
N_1$ is given by        
\begin{equation}
\frac{E}{N}
= \frac{\hbar^2 \pi (a_{11} + a_{12})}{m} \rho
+
\frac{32\sqrt{2 \pi}}{15}\frac{\hbar^2 a_{11}^{5/2} }{m} f\left(\frac{a_{12}^2}{a_{11}^2}\right) \rho^{3/2}\ ,
\label{eq:petrov}
\end{equation}
with
\begin{equation}
f(x) = (1 + \sqrt{x})^{5/2} + (1- \sqrt{x})^{5/2}.
\end{equation}
Notice that the function $f(x)$ becomes complex for $a_{12} < -a_{11}$
and the presence of the imaginary component reduces the applicability of 
the perturbative theory. If instead the argument is approximated by $x = 
a_{12}^2 / a_{11}^2 = 1$ so that the function $f(x)$ remains real, as it was 
done in Ref.~\cite{petrov2015quantum}, Eq.~(\ref{eq:petrov}) reduces to the following 
form,    
\begin{equation}
\frac{E}{N}
= \frac{\hbar^2 \pi (a_{11} + a_{12})}{m} \rho
+
\frac{256\sqrt{\pi}}{15}\frac{\hbar^2 a_{11}^{5/2} }{m}  \rho^{3/2}    \ .
\label{eq:petrovs}
\end{equation}
shown with a dashed line in Fig.~(\ref{fig:fig1}). We plot as well  the 
energy resulting from taking the real part of $f(x)$~(\ref{eq:petrov}), without 
invoking the approximation $x = 1$. Only very close to the $a_{12}=-a_{11}$ 
limit corresponding to zero equilibrium density, both predictions are nearly the same while 
they clearly differ for finite densities. We report the exact DMC energies in 
Fig.~(\ref{fig:fig1}). The perturbative MF+LHY results are recovered for small 
range $R$ of the square well and $\rho a_{11}^3 \approx 10^{-6}$, see 
Fig.~\ref{fig:fig1}a. However, when $R$ is increased by a large amount (to 
$R=100 a_{11}$) the universality breaks at $\rho R^3 \simeq 10^{-1}$. The 
energies for experimentally relevant densities, $\rho a_{11}^3 \approx 10^{-5}$ 
~\cite{cabrera2018quantum,semeghini2018self}, are reported in Fig.~\ref{fig:fig1}b.
In this case and for larger densities (Fig.~\ref{fig:fig1}c), we observe that the energy depends on the potential range. Furthermore, the two ways of 
writing the perturbative equation of state, given by 
Eqs.~(\ref{eq:petrov},~\ref{eq:petrovs}), differ among themselves but are not 
equal to the obtained DMC equation of state. The latter appears to be independent of $R$ up to approximately $\rho R^{3}=10^{-3}$. Indeed, the difference between the energy per particle $E/N$ calculated at $\rho 
R^{3}=10^{-3}$ and $\rho R^{3}=10^{-5}$ is at most 3 errorbars or 6\% at the highest density and at most 4\% in the minimum.

\section{\label{sec:repulsive_beyond_lhy_symmetric}Repulsive beyond-LHY energy}

\begin{figure}[tb]
	\begin{center}
		\includegraphics[width=0.7\textwidth]{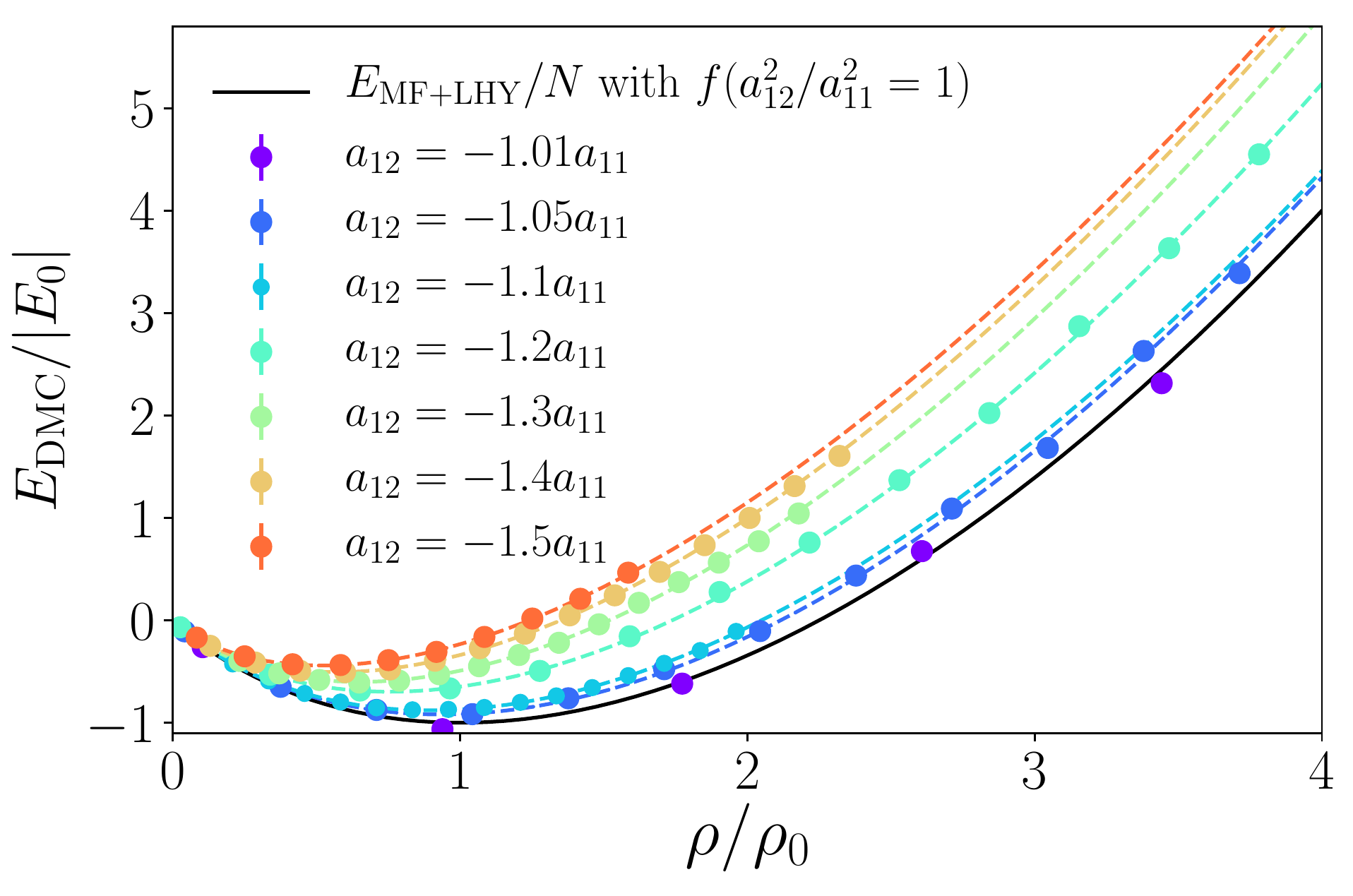}
		\caption[Equations of state of a symmetric Bose-Bose liquid for different $a_{12}/a_{11}$ normalized to the density and the energy at MF+LHY equilibrium point $(\rho_0,E_0)$]{Equations of state of a symmetric Bose-Bose liquid for different $a_{12}/a_{11}$ normalized to the density and the energy at MF+LHY equilibrium point $(\rho_0,E_0)$. Dashed lines show fits to the data in the form of $E/N = \alpha x + \beta x^{\gamma}$ with $x=\rho a_{11}^3$. The range of the square well is $\rho R^3 = 10^{-5}$.}
		\label{fig:fig2}
	\end{center}
\end{figure}

It can be noted that within perturbative theory the energy is a single curve written in units of the equilibrium energy $E_0$ and density $\rho_0$. That is, the equation of state~(\ref{eq:petrovs}) can be conveniently represented as a $(E / E_0, \rho / \rho_0)$ curve,
\begin{equation}
\dfrac{E}{|E_0|} =
-3 \left(\frac{\rho}{\rho_0}\right)
+2 \left(\frac{\rho}{\rho_0}\right)^{3/2}\ ,
\label{eq:petrovs2}
\end{equation}    
with, for the symmetric mixture
\begin{equation}
\rho_0 =  \dfrac{25 \pi  \left(a_{11} + 
	a_{12}\right)^2}{16384 a_{11}^5},
\end{equation}
and
\begin{equation}
E_0/N = -  \dfrac{25 \pi^2 \hbar^2 |a_{11} + 
	a_{12}|^3}{49152 m a_{11}^5}. 
\end{equation}
The DMC equations of state for different scattering lengths are shown in 
Fig.~(\ref{fig:fig2}). The results are obtained for sufficiently small potential 
range, $\rho R^{3}=10^{-5}$, ensuring the universality in terms of the $s$-wave 
scattering length. As already observed in Fig.~(\ref{fig:fig1}), when $|a_{12}| 
\approx a_{11}$, the MF+LHY prediction is recovered. Increasing 
$|a_{12}|/a_{11}$ repulsive contributions to the energy beyond the LHY terms are 
found. 
At the same time, the equilibrium densities become lower compared to the ones 
predicted by Eq.~(\ref{eq:petrovs2}), which was obtained by calculating $f(x)$ 
function at $x=1$. If instead one uses Eq.~(\ref{eq:petrov}) derived by taking 
the real part of $f(x)$, weaker binding is predicted as compared to DMC results. 
Thus, as we can see in Fig.~\ref{fig:fig1}, for small ranges $\rho 
R^{3}=10^{-5}$, the DMC many-body prediction is between Eq.~(\ref{eq:petrov}) 
and (\ref{eq:petrovs2}), but closer to Eq.~(\ref{eq:petrovs2}).

\begin{table}[tb]    
	\caption{Energies (equilibrium and spinodal) and densities for different scattering lengths $a_{12}/a_{11}$ for small ranges $\rho R^3 = 10^{-5}$. Here ``eq'' stands for the minimum from the fit to DMC energy shown in Fig.~\ref{fig:fig2}, ``0'' stands for minimum of perturbative equation of state given by Eq.~(\ref{eq:petrovs2}), spinodal point is denoted by ``sp'' from the fit on DMC data  and ``sp,0'' in case of Eq.~(\ref{eq:petrovs2}). 
	}
	\centering
	\resizebox{\textwidth}{!}{    
		\begin{tabular}{c c  c c c c  c}    
			\hline \hline    \\                    
			$~~\dfrac{a_{12}}{a_{11}}~~$    & $~\dfrac{10^5\rho_{\rm eq}}{a_{11}^{-3}}~$    & $~\dfrac{\rho_{\rm eq}}{\rho_{\rm 0}}~$    & $~\dfrac{10^5\rho_{\rm sp}}{a_{11}^{-3}}~$        & $~\dfrac{\rho_{\rm sp}}{\rho_{\rm sp,0}}~$    & $~\dfrac{10^6\hbar^2E_{\rm eq}}{2 m a_{11}^2N}   ~~$    &  $~~\dfrac{E_{\rm eq}}{E_{\rm 0}}~ $    \\
			\hline \\        
			$-1.05    $&$1.12    $&$0.934    $&$0.715    $&$0.932    $&$-1.15    $&$0.919$\\
			$-1.10    $&$4.28    $&$0.894    $&$2.73     $&$0.888    $&$-8.82    $&$0.879$\\
			$-1.20    $&$14.5    $&$0.754    $&$9.19     $&$0.749    $&$-56.0    $&$0.697$\\
			$-1.30    $&$28.0    $&$0.649    $&$17.7     $&$0.641    $&$-163    $&$0.601$\\
			$-1.40    $&$44.9    $&$0.585    $&$28.3     $&$0.576    $&$-334        $&$0.520$\\
			$-1.50    $&$62.4    $&$0.521    $&$39.3     $&$0.512    $&$-554    $&$0.441$\\
			\hline    
			\label{tab:en}                        
	\end{tabular}    }                                        
\end{table}
The DMC values of the equilibrium energies and densities are reported in 
Table~\ref{tab:en}. They are also compared to predictions from perturbative 
theory given by Eq.~(\ref{eq:petrovs2}). With the increase of $|a_{12}|/a_{11}$ 
the equilibrium and spinodal densities start to depart significantly from the 
MF+LHY values. It is worth noticing again that the MF+LHY equation of state 
becomes complex, and thus unphysical, unless the approximation $f(a_{12}^2 / 
a_{11}^2 = 1)$ is used. Our results show that, even very small (in absolute 
value) negative pressures, can cause spinodal instability. For typical 
experimental parameters $a_{11}=50a_0$~\cite{cabrera2018quantum,semeghini2018self} the uniform 
liquid breaks into droplets when the applied negative pressure is very small, 
from $1.81$pPa for $a_{12}=-1.05 a_{11}$ to $31.3$nPa for $a_{12}=-1.5 a_{11}$.

\begin{figure}[tb]
	\centering
	\includegraphics[width=0.9\linewidth]{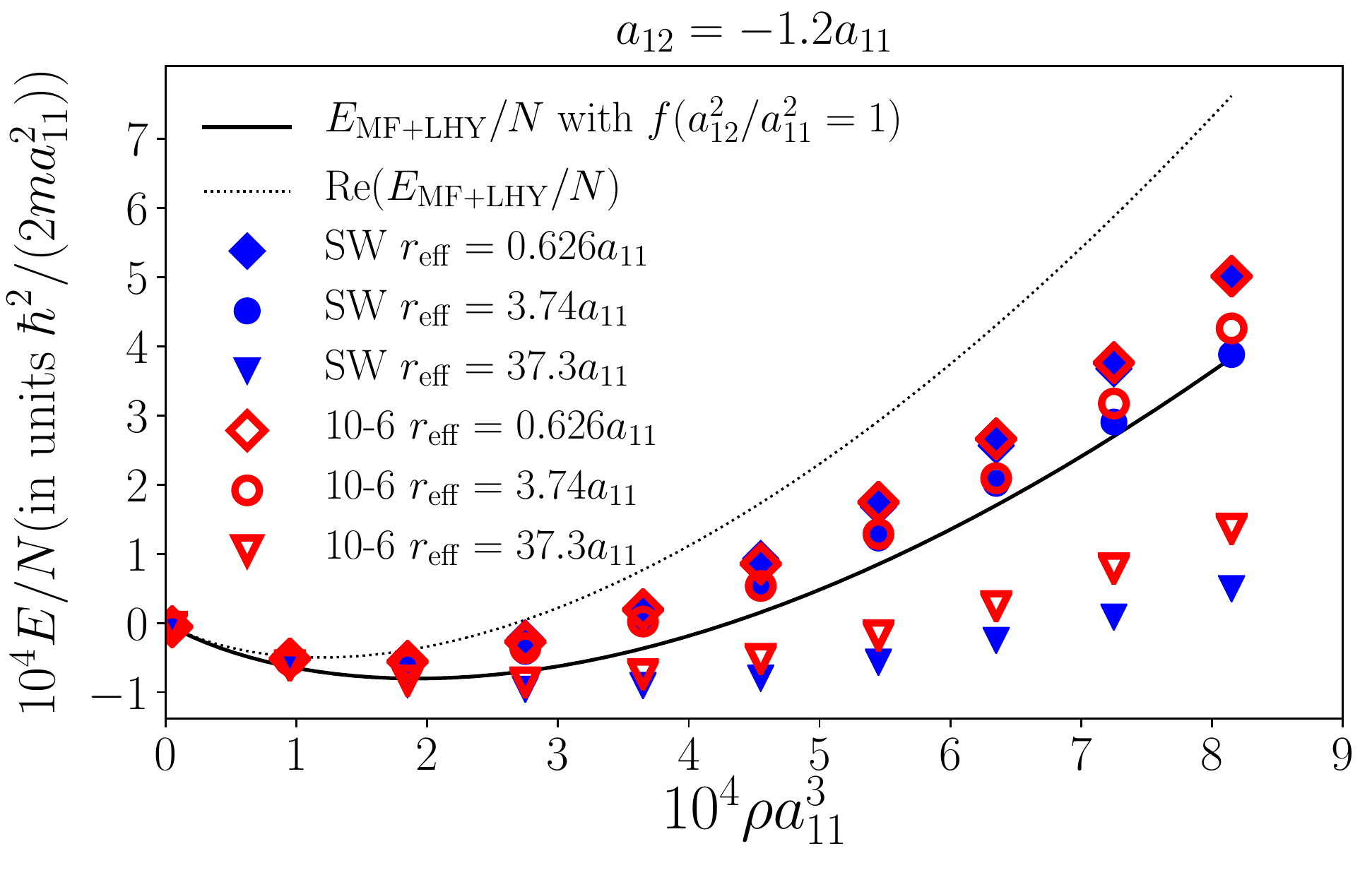}
	\caption{Dependence of the equation of state on the effective range.}
	\label{fig:fig3}
\end{figure}

As can be seen from Fig.~\ref{fig:fig1}, the equation of state loses universality in terms of the scattering length when $\rho R^3 \gtrsim 10^{-3}$. This poses the relevant question of whether by fixing one more parameter, besides the $s$-wave scattering length, it is possible to obtain a universal description. 
To address this question, we performed DMC calculations using the 10-6 model with equivalent values of the $s$-wave scattering lengths and effective range $r_{\rm eff}$ of the attractive interaction. For the repulsive interactions, we fix the range of the 10-6 model potential to $r_0 = 2a_{11}$. In Fig.~\ref{fig:fig3}, we show results for scattering length $a_{12}=-1.2a_{11}$ and three values of the effective range $r_{\rm eff}$. The solid line is for Eq.~(\ref{eq:petrovs}) and the dashed one for the real part of Eq.~(\ref{eq:petrov}). The range of the SW potential is $R/a_{11}$ = 0.531, 2.17, and 9.18 when $r_{\rm eff} / a_{11} = 0.626$, 3.74 and 37.3, respectively. We find that specifying only the scattering length cannot generally fulfill universal results unless the range is sufficiently small, $\rho R^3 \lesssim 10^{-1}$. The interaction potential for a given scattering length predicts different energies and equilibrium densities when different effective ranges are used. Generally, increasing the range lowers the energy and shifts the equilibrium density to larger values. However, if we specify both the scattering length and the effective range, then we observe that the difference between results of two models is always smaller than the difference between results for the same type of model but with a different effective range. In Fig.~\ref{fig:fig3}, the two models with $r_{\rm eff}/a_{11}=0.626$ give, within errorbars, the same energies in the whole density range. Increasing the range, at higher densities, we observe that the two potentials start to give different predictions and that the difference between them grows with the increase in density. Interestingly, even when the effective range is quite large, $r_{\rm  eff}/a_{11} = 37.3$, the relative difference between the models remains lower than 10\%, as long as $\rho R^3 < 0.2$. Increasing the density even further, we would need more parameters beyond $a_{12}$ and $r_{\rm eff}$ to describe the interaction.
The observed dependence on the effective range for $\rho_{eq} R^3>10^{-3}$ is in overall agreement with the recent calculation of unbalanced mixtures~\cite{staudinger2018self} based on the variational hypernetted chain method. It is interesting to notice that the MF+LHY 
equations of state, following Eq.~(\ref{eq:petrovs2}), are closer to our full many-body calculations using rather large values of the effective range. On the other hand, the results using only the real part of 
Eq.~(\ref{eq:petrov}) are above the DMC energies for even the smallest range.

Presuming that the equation of state of the liquid mixture is universal in terms of the scattering length and the effective range for $\rho R^3 \lesssim 10^{-1}$, we use the SW results to deduce the following form for the equation of state    
\begin{equation}
\dfrac{E}{N}=\dfrac{|E_0|}{N}         
\left[
-3\left(\dfrac{\rho}{\rho_0}\right)
+ \beta \left(\dfrac{\rho}{\rho_0}\right)^{\gamma}        
\right] \ ,
\label{eos}
\end{equation}
where $\beta$ and $\gamma$ are functions of $a_{12}/a_{11}$ and $r_{\rm eff}/a_{11}$
\begin{equation}
\beta =  \beta_{01}\dfrac{a_{12}}{a_{11}} + \left(\beta_{10} + \beta_{11}\dfrac{a_{12}}{a_{11}}\right) \dfrac{R(a_{12}, r_{\rm eff})}{a_{11}}
\end{equation}    
\begin{equation}
\gamma = \gamma_{00} + \gamma_{01}\dfrac{a_{12}}{a_{11}}+ \left(\gamma_{10} + \gamma_{11}\dfrac{a_{12}}{a_{11}}\right)\dfrac{R(a_{12}, r_{\rm eff})}{a_{11}}.
\end{equation}
$R$ is the square well diameter associated with the given $a_{12}$ and $r_{\rm eff}>0$. It can be calculated numerically for given $a_{12}$ and $r_{\rm eff}$, and we provide a numerical code in \cite{git_python}. There are 7 free parameters in the model: $\beta_{01}$, $\beta_{10}$, $\beta_{11}$, $\gamma_{00}$, $\gamma_{01}$, $\gamma_{10}$, $\gamma_{11}$.  They are obtained by fitting 18 equations of state  with different $R$ and $a_{12}$. In particular, the chosen $a_{12}/a_{11}$ were -1.01, -1.05, -1.08, -1.1 and -1.2.     The obtained values of the parameters are given in Table~\ref{table:parameters_best}. 

\begin{table}[!htb]
	\caption{Parameters of the equation of state.}
	\label{table:parameters_best}
	\centering
	\resizebox{0.95\textwidth}{!}{    
		\begin{tabular}{  ccccccc}    
			\hline    
			$\beta_{01}$ &    $\beta_{10}$ &     $\beta_{11}$ &     $\gamma_{00}$ &     $\gamma_{01}$& $\gamma_{10}$ &     $\gamma_{11}$  \\ \hline
			$-1.956(3)$ &  ~$0.231(5)$ &   $0.236(5) $  &  ~$1.83(2)$    & ~$0.32(2) $&  $0.030(3) $& $0.030(3) $     \\ \hline
			
	\end{tabular}}
\end{table}

With these values $\beta(a_{12}=-a_{11}, R=0) = 1.956 \pm 0.003$, $\gamma(a_{12}=-a_{11}, R=0) = 1.51 \pm 0.02 $, which is very close to MF+LHY value: $\beta=2$, $\gamma=1.5$. We then verified that this form predicts well the equation of state up to $a_{12}=-1.3 a_{11}$ provided that $R$ is not too large ($\rho R^3<10^{-1}$).

\section{\label{sec:universal_equation_of_state_symmetric}Universal equation of state}

\begin{figure*}[!htb]
	\centering
	\includegraphics[width=0.49\textwidth]{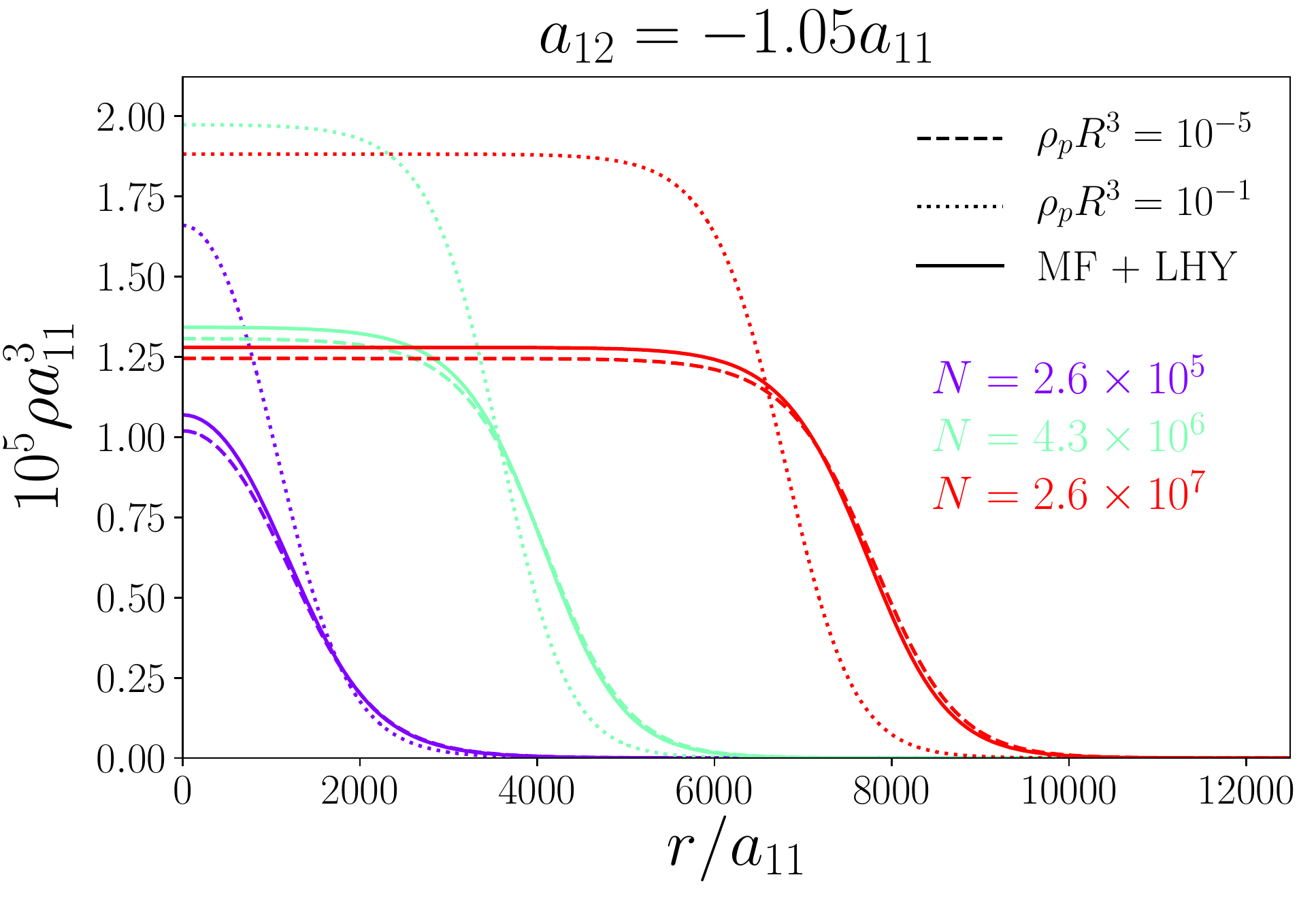}
	\includegraphics[width=0.49\textwidth]{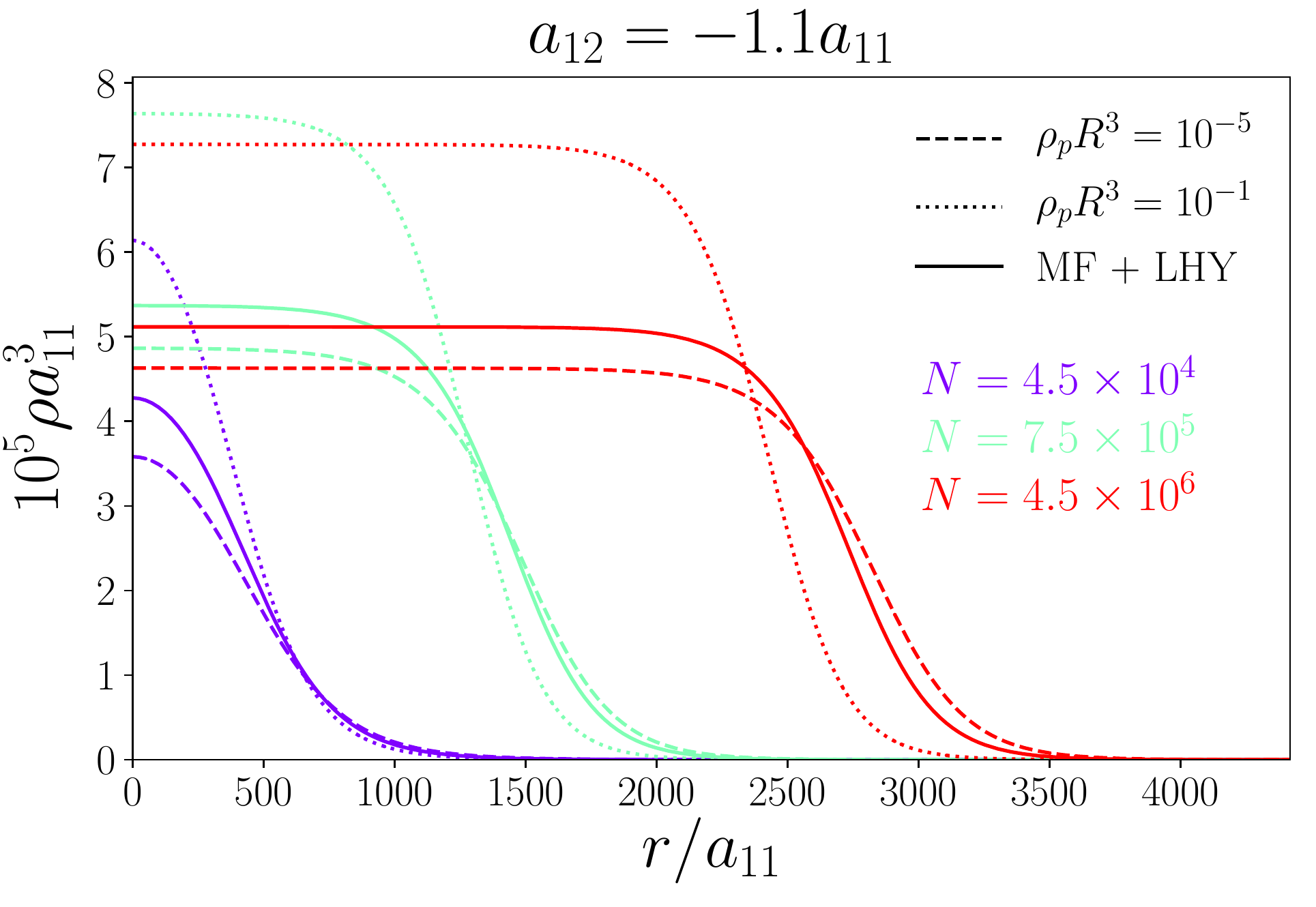}    
	\includegraphics[width=0.49\textwidth]{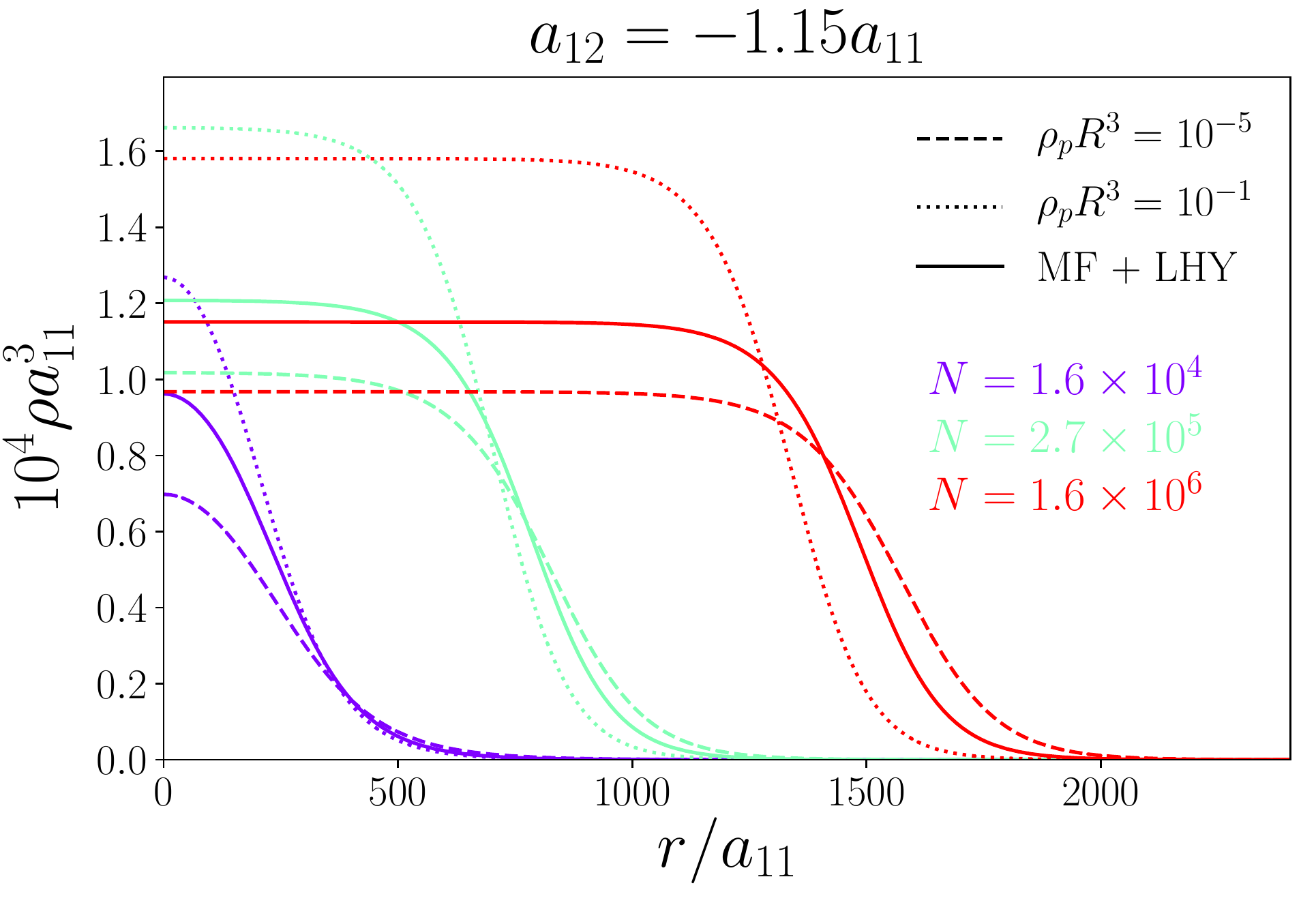}    
	\includegraphics[width=0.49\textwidth]{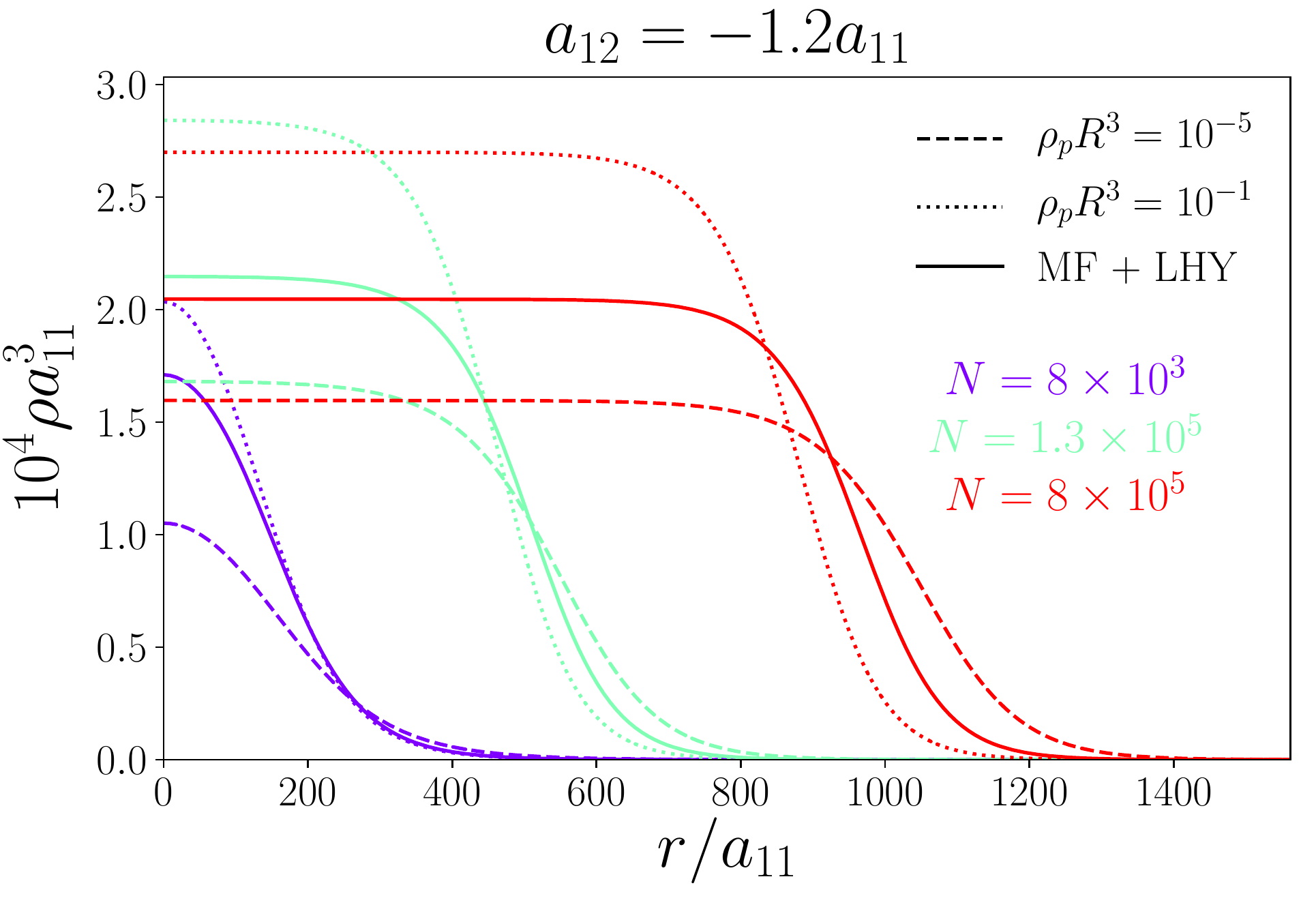}    
	\caption[Density profiles of symmetric drops with different scattering lengths and ranges  compared to MF+LHY predictions]{Density profiles of symmetric drops with different scattering lengths and ranges  compared to MF+LHY predictions. Dashed and dotted lines correspond to LDA calculations using the energy functional (Eq. \ref{eos}), for $\rho_p R^3 = 10^{-5}$ (dashed line) and $\rho_p R^3 = 10^{-1}$ (dotted line), where $\rho_p = 25 \pi  \left(a_{11} + a_{12}\right)^2/(16384 a_{11}^5)$ is equilibrium density from the MF+LHY theory. Full lines correspond to MF+LHY calculation using the approximation $f(a_{12}^2/a_{11}^2 = 1)$. For each $a_{12}$ and $R$, we show profiles for three different values of particle numbers, written below legend, distinguished by color and growing from purple to red. Profiles with the smallest particle number (purple color) are stable ($E<0$) and close to the critical number $N_c = 22.55 \times 96\sqrt{6}/(5\pi^2 |1+a_{12}/a_{11}|^{5/2})$ \cite{petrov2015quantum}. 
	}
	\label{fig:densprof}        
\end{figure*}

\begin{figure}
	\centering
	\includegraphics[width=0.98\linewidth]{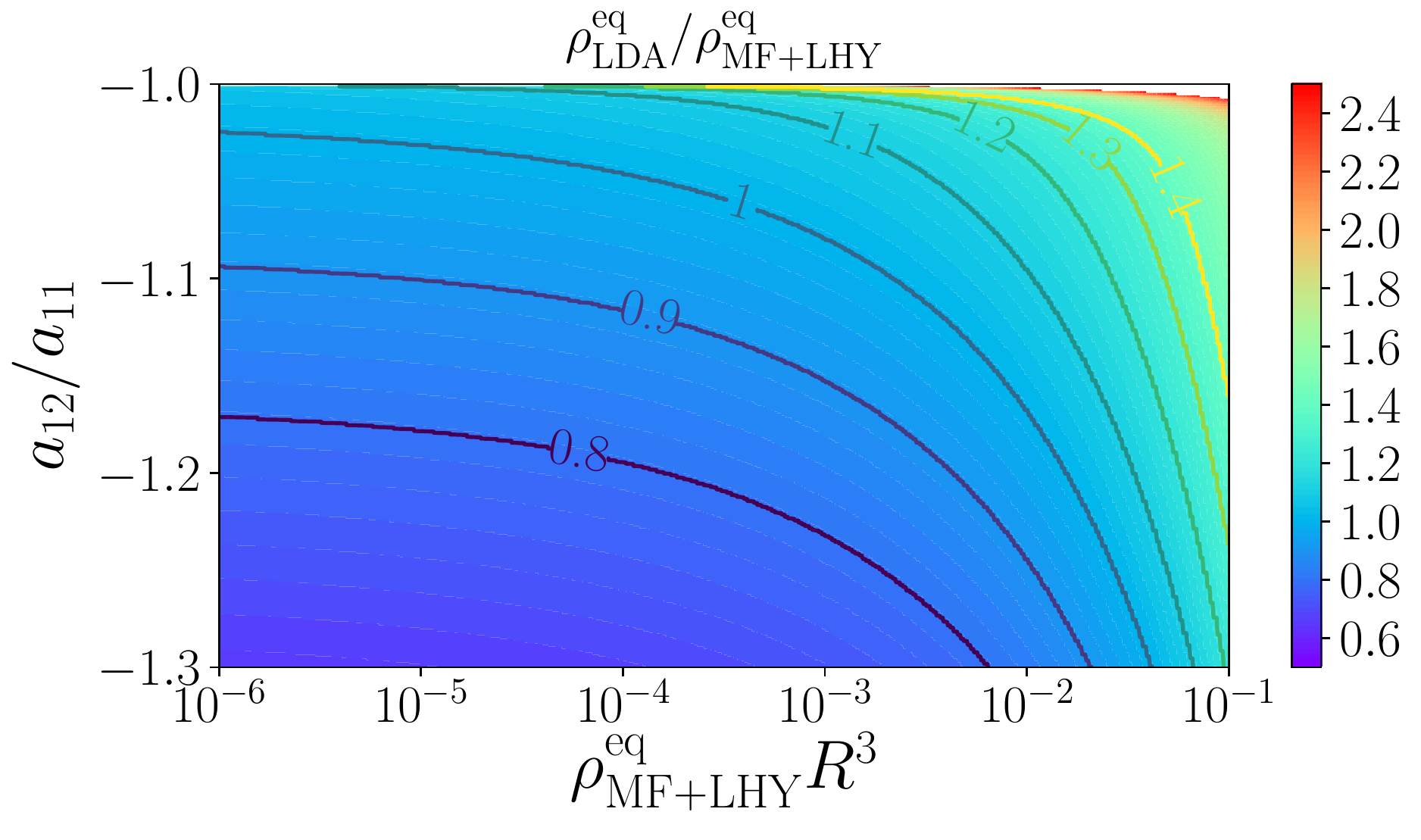}
	\caption{Ratio of equilibrium densities $\rho_{\rm LDA}^{\rm eq} / \rho_{\rm MF+LHY}^{\rm eq}$ vs. scattering length $a_{12}$ and the range $R$. Isolines follow values with the constant $\rho_{\rm LDA}^{\rm eq} / \rho_{\rm MF+LHY}^{\rm eq}$. LDA results are obtained starting from the DMC equation of state. }
	\label{fig:fig4}
\end{figure}

The equation of state~(\ref{eos}) can be used as an energy 
functional~\cite{ancilotto2018self} to calculate density profiles of liquid mixture drops 
within the local density approximation (LDA), following the procedure described in Chapter \ref{ch:overview}. Starting from the DMC equation of state and using LDA we obtain density profiles
of drops for different scattering lengths and effective ranges and compare them with
MF+LHY predictions in Fig. \ref{fig:densprof}. To do so we write the energy functional as
\begin{equation}
\mathcal{E}   =  \dfrac{\hbar^2}{2m}N|\nabla \phi|^2 \nonumber  +\dfrac{25\pi^2 \hbar^2|a_{11}+a_{12}|^3}{49152 m a_{11}^5} \left[-3\dfrac{N^2|\phi|^4}{\rho_0} + \beta \dfrac{(N |\phi|^2)^{\gamma+1} }{\rho_0^{\gamma}}  \right],
\end{equation}
where $N$ is the number of particles and $\phi$ is normalized as
\begin{equation}
\int d^3 r |\phi|^2 = 1.
\end{equation}
Then, we find the stationary solution of the equation of motion
\begin{equation}
i\hbar\dfrac{\partial \phi}{\partial t}
=  -\dfrac{\hbar^2 \nabla^2 \phi}{2m}  + \dfrac{25\pi^2 \hbar^2|a_{11}+a_{12}|^3}{49152 m a_{11}^5} \left(-6 \dfrac{N|\phi|^2}{\rho_0}
+ 
\beta (1+\gamma)\left(\dfrac{N |\phi|^2}{\rho_0}\right)^{\gamma} \right) \phi,
\end{equation}
by propagating it in imaginary time $\tau = i \hbar t $. The results for the equilibrium density as a function of the interspecies scattering length and the square-well range are presented in Fig.~\ref{fig:fig4} and compared to the MF+LHY predictions. For a negligible range $R$, the equilibrium density drops below the MF+LHY prediction as $|a_{12}|/a_{11}$ is increased. The effect of the finite range is to increase the equilibrium density. That is by increasing $R$, the LDA prediction crosses the perturbative result of MF+LHY and goes above.  Overall, by increasing the range and decreasing $|a_{12}|/a_{11}$ (i.e. going in the up-right direction in Fig.~\ref{fig:fig4}) we observe an increase of $\rho_{\rm LDA}^{\rm eq} / \rho_{\rm MF+LHY}^{\rm eq}$.

\section{Summary and conclusion}

We have carried out high-precision DMC calculations of the ground-state equation of state of ultradilute two-component Bose liquids. 
We find out that the use of only the first beyond-MF correction, the LHY term, is accurate only for extremely small densities, and only when the range of the interaction is not very large. In our study, we have used for the range $ R $ the diameter of the square well potential, which has the same scattering length and effective range as the chosen model. If $|a_{12}/a_{11}+1| \le 0.05$ and $\rho R^3 < 
10^{-3}$ one parameter, the $s$-wave scattering length is enough to describe the system, but there is an appreciable difference with respect to MF+LHY. 
Increasing the range, one enters a regime where interaction potentials with the same scattering length and effective range give equivalent results within 10\%, which means that up to $\rho R^3 = 0.1$ we have at hand a  universal equation of state which is a function of two parameters. For even larger values of $\rho R^3$,  additional parameters would need to be specified. The results of scattering calculations of alkali atoms, such as given in Ref.~\cite{flambaum1999analytical,tanzi2018feshbach}, indicate that most likely, the effective ranges are quite far from the zero-range limit. In that case, for obtaining the correct results, one needs a full many-body approach like DMC. Here, we provide a new energy functional based on the best fit of DMC data and use it to calculate the density profiles of realistic drops with LDA.



\chapter{\label{ch:finite_range_effects}Finite-range effects in ultradilute $^{39}$K quantum drops}

\section{Introduction \label{intro} }

In the two labs~\cite{cabrera2018quantum,semeghini2018self}, where the ultradilute Bose-Bose drops have been observed, the Bose-Bose mixture is composed of two hyperfine states of $^{39}$K. In the first experiment by  Cabrera \textit{et al.}~\cite{cabrera2018quantum}, the drops are harmonically confined in one of the directions of space, whereas in the second one by Semeghini \textit{et al.}~\cite{semeghini2018self} the drops are observed in free space. This difference in the setup makes that, in the first case, the drops are not spherical like in the second experiment. The external trap also affects the critical number: the minimum number of atoms required to get a self-bound state. The measured critical numbers differ significantly between the two labs due to the different shapes of the drops, the ones in the confined case being smaller than in the free case. In both works, the experimental results for the critical number are compared with the LHY-extended mean-field (MF+LHY) theory. The agreement between this theory and the drops produced in free space is entirely satisfactory despite the large error bars of the experimental data that hinder a precise comparison. However, in the confined drops of Ref.~\cite{cabrera2018quantum}, where the critical numbers are significantly smaller than in the free case, the theoretical predictions do not match well the experimental data.

Ultradilute liquid drops, which require beyond-mean field corrections to be theoretically understood, offer the perfect benchmark to explore possible effects beyond MF+LHY theory \cite{jorgensen2018dilute} which usually play a minute role in the case of single-component gases~\cite{pethick2008bose,giorgini1999ground}. Indeed, several theoretical studies (see Chapters \ref{chapter:ultradilute_liquid_drops}, \ref{ch:symmetric_liquids}, \ref{ch:finite_range_effects} and Refs. \cite{staudinger2018self, tononi2019zero, tononi2018condensation,salasnich2017nonuniversal}) indicate a strong dependence of the equation of state of the liquid on the details of the interatomic interaction, even at very low densities accessible to the experiment. This essentially means that it is already possible to achieve observations outside the universal regime, in which all the interactions can be expressed in terms of the gas parameter $n a^3$, with $a$ the $s$-wave scattering length. The first correction beyond this universality limit must incorporate the next term in the scattering series, which is the effective range $r_{\rm eff}$~\cite{newton2013scattering, roman1965advanced}, which in fact can be quite large in these drops and alkali atoms in general~\cite{flambaum1999analytical,tanzi2018feshbach}. 

Motivated by experiments with quantum drops, we have investigated the self-bound quantum mixture composed of two hyperfine states of $^{39}$K using nonperturbative quantum Monte Carlo (QMC) methods. Direct QMC simulations, such as those presented in Chapter \ref{chapter:ultradilute_liquid_drops}, of finite particle-number drops, as produced in experiments, would serve as a great test of mean-field theory, but, unfortunately, this is not yet achievable because of the large number of particles in realistic drops ($N>10^4$).  Nevertheless, the problem can be addressed in the Density Functional Theory (DFT) spirit, relying on the Hohenberg-Kohm-Sham 2nd theorem~\cite{hohenberg1964inhomogeneous}, which guarantees that a density functional exists that matches precisely the ground-state solution. To build a functional for the quantum Bose-Bose mixture, we have carried out calculations in bulk conditions using the diffusion Monte Carlo (DMC) method, an exact QMC technique applicable to systems at zero temperature (see Sec. \ref{sec:dmc}). Using that functional, we can access energetics and structure of liquid drops in the same conditions as in the experiment. We focus on the data obtained by Cabrera \textit{et al.}\cite{cabrera2018quantum} in the confined setup since it is in that case where discrepancies between MF+LHY theory were observed. 

The rest of the chapter is organized as follows. In Sec. (\ref{sec:hamiltonian_methods_finite_range_nc}) we introduce the theoretical methods used for the study and discuss how the density functional is built. Sec. (\ref{results_finite_range_mixture}) comprises the bulk liquid results using the available scattering data of the $^{39}$K mixture. The inclusion of the effective range parameters in the interaction model allows for a better agreement with the measured critical numbers. Finally, we summarize the most relevant results here obtained and derive the main conclusions of our work.

\section{\label{sec:hamiltonian_methods_finite_range_nc}Hamiltonian and the methods}

We study a liquid mixture of two hyperfine states of $^{39}$K bosons at zero temperature. The Hamiltonian of the system is
\begin{equation}
H = \sum_{i=1}^{N}-\dfrac{\hbar^2}{2m}\nabla_i^2 + \dfrac{1}{2}   
\sum_{\alpha, \beta=1}^{2} \sum_{i_\alpha, j_\beta=1}^{N_\alpha 
	N_\beta}V^{(\alpha, \beta)}(r_{i_\alpha j_\beta}) \ ,
\label{hamiltonian}
\end{equation}
where $V^{(\alpha, \beta)}(r_{i_\alpha j_\beta})$ is the interatomic potential between species $\alpha$ and $\beta$. The mixture is composed of $N=N_1 + N_2$ atoms, with $N_1$ ($N_2$) bosons of type 1 (2). The potentials are chosen to reproduce the experimental scattering parameters, and we have used different model potentials to investigate the influence of the inclusion of the effective range. The microscopic study has been carried out using  a second-order DMC method~\cite{boronat1994monte}, as described in Chapter \ref{ch:qmc_methods}. In the present case, and similarly as in Chapters \ref{chapter:ultradilute_liquid_drops} and \ref{ch:symmetric_liquids}, we used a trial wavefunction built as a product of Jastrow factors \cite{reatto1967phonons},
\begin{equation}
\label{eq:trial_wf_1}
\Psi(\vec{\mathrm{R}}) = \prod_{i<j}^{N_1} f^{(1,1)}(r_{ij}) 
\prod_{i<j}^{N_2} f^{(2,2)}(r_{ij}) \prod_{i,j}^{N_1, N_2} f^{(1,2)}(r_{ij}) \ ,
\end{equation} 
where the two-particle correlation functions $f(r)$ reproduce the expected behavior at small and large distance between atom pairs  
\begin{equation}
\label{eq:trial_wf_2}
f^{\alpha, \beta}(r)=
\begin{cases}
f_{\rm 2b}(r) &  r < R_0 \\
B\exp(-\frac{C}{r} + \frac{D}{r^2})       ,& R_0 <r < L/2 \\
1      ,& r > L/2 \ , \\
\end{cases}
\end{equation}

thus focusing the sampling where it is physically most likely. The function $f_{\rm 2b}$ is the solution of the two-body problem for a specific interaction model, and it is connected to the long-range phonon wavefunction~\cite{reatto1967phonons} with the corresponding coefficients B, C and D, which are adjusted to match the continuity condition of the wavefunction, its first derivative and the imposing condition of zero derivative at $r=L/2$. $R_0$ is a variational parameter, and $L=(N / \rho)^{1/3}$ is the size of the simulation box. There is a weak dependence of the variational energy on $R_0$. Even though the long-range part of the wavefunction does not contribute significantly to the variational energy, we have kept the phonon part in the two-body correlations in order to smoothly connect the wavefunction to $r=L/2$, where it fullfils the condition $f'(r=L/2) = 0$. Particularly, the value of variational parameter $R_0$ has been kept as $R_0 = 0.9 L/2$ for all the cases. A careful analysis of imaginary time-step dependence and population size bias has been carried out, keeping both well under the statistical error. The time-step dependence is well eliminated for $\Delta \tau = 0.2 \times ma_{11}^2 / \hbar$ and the population bias by using $n_{\rm w} = 100$. Our simulations  are performed in a cubic box with periodic boundary condition, using a number of particles $N$. The thermodynamic limit is achieved by repeating calculations with different particle numbers; we observe that, within our numerical precision, the energy per particle converges at $N\approx 600$ for the range of magnetic fields here considered.

Within density functional theory (DFT), we seek for a many-body wave function built as a product of single-particle orbitals, 
\begin{equation}
\label{eq:dft_wf}
\Psi(\vec{r}_1, \vec{r}_2, \ldots, \vec{r}_N)= 
\prod_{i=1}^{N} \psi(\vec{r}_i).
\end{equation}    
These single-particle wave functions, which in general are time-dependent, are obtained by solving the Schr\"odinger-like equation~\cite{ancilotto2018self},
\begin{equation}
\label{eq_finite_range:time_dep_gp}
i\hbar\dfrac{\partial \psi}{\partial t} = \left(-\dfrac{\hbar^2}{2m} 
\nabla^2 + V_{\rm ext}(\vec{r}) + \dfrac{\partial \mathcal{E}_{\rm 
		int}}{\partial \rho}\right) \psi \ ,
\end{equation}
where $V_{\rm ext}$ is an external potential acting on the system and $\mathcal{E}_{\rm int}$ is an energy per volume term that accounts for the interparticle correlations.  The Eq. (\ref{eq_finite_range:time_dep_gp}) is solved as described in Chapter \ref{ch:overview}.

\subsection{Interatomic model potentials }

For the system under study, $^{39}$K, only two scattering parameters, that is the $s$-wave scattering length and the effective range, are known. In the mixture of $^{39}$K under study, we call the state $\ket{F,m_{\rm F}}=\ket{\downarrow} = \ket{1,0}$ as component 1, and the state $\ket{F,m_{\rm F}}=\ket{\uparrow} = \ket{1,-1}$ as component 2. In order to model the interaction potential with those parameters, we have used three different set of potentials: 

\begin{itemize}
	
	\item[i)] Hard-core interactions (HCSW) with diameter 
	$a_{ii}$, $i=1,2$, for the repulsive intraspecies interaction, and a 
	square-well potential 
	with range $R=a_{11}$ and depth $V_0$ for the interspecies 
	potential. The three potentials reproduce the $s$-wave scattering lengths for the 
	three channels. Mathematical expression for the potential reads
	\begin{equation}
	V_{ii}(r)=
	\begin{cases}
	\infty , &  r < a_{ii} \\
	0  ,&  \text{otherwise,} \\
	\end{cases}
	\end{equation} 
	for $i=1, 2$. The $s$-wave scattering length of this potential corresponds to the diameter of the hard core. Interspecies attraction is modeled by the attractive square well potential
	\begin{equation}
	V_{12}(r)=
	\begin{cases}
	-V_0 , &  r < R_0 \\
	0  ,&  \text{otherwise.} \\
	\end{cases}
	\end{equation} 
	The $s$-wave scattering length and the effective range of the attractive square well can be found in Ref. \cite{pethick2008bose}.
	
	\item[ii)] POT1 stands for a set of potentials which reproduces both the 
	$s$-wave scattering lengths and effective ranges of the three interacting 
	pairs of the $^{39}$K mixture.  To model the interactions, we have chosen a 
	square-well 
	square barrier potential~\cite{jensen2006bcs} for the $11$ channel, a 
	10-6 Lennard-Jones potential~\cite{pade2007exact} for the $22$ channel, and a 
	square-well 
	potential of range $R$ and depth $V_0$~\cite{pethick2008bose} in the $12$ channel. Mathematical expression of these potentials are the following 
	\begin{equation}
	V_{11}(r)=
	\begin{cases}
	-V_0 , &  r < R_0 \\
	V_1  ,&  R_0 < r < R_1 \\
	0  ,&  R_1 > r,  \\
	\end{cases}
	\end{equation} 
	\begin{equation}
	V_{12}(r)=
	\begin{cases}
	-V_0 , &  r < R_0 \\
	0  ,&  \text{else}, \\
	\end{cases}
	\end{equation} 
	\begin{equation}
	V_{22}(r)= V_0 \left[\left(\dfrac{r_0}{r}\right)^{10} - \left(\dfrac{r_0}{r}\right)^{6}\right].
	\end{equation}  
	The effective range of 22 potential was found numerically by solving the two-body scattering problem (see Sec. \ref{subsec:numerov_algorithm} and Ref. \cite{newton2013scattering}), whereas analytical expressions for the effective ranges for 11 and 12 interactions are given in Refs \cite{jensen2006bcs} and \cite{pethick2008bose}, respectively.
	
	\item[iii)] POT2 also reproduce both the $s$-wave scattering lengths and effective ranges, by using a sum of Gaussians in the $11$ channel, a 
	10-6 Lennard-Jones potential in the $12$ channel, and finally a soft-sphere 
	square well in the $22$ channel.  Mathematical expression for these potentials read
	\begin{equation}
	V_{11}(r)= -V_0\exp\left[-\dfrac{r^2}{2r_0^2}\right] + V_1\exp\left[-\dfrac{(r-r_1)^2}{2r_0^2}\right],
	\end{equation} 
	\begin{equation}
	V_{12}(r)= V_0 \left[\left(\dfrac{r_0}{r}\right)^{10} - \left(\dfrac{r_0}{r}\right)^{6}\right],
	\end{equation} 
	\begin{equation}
	V_{22}(r)= 
	\begin{cases}
	V_0 , &  r < R_0 \\
	-V_1  ,&  R_0 < r < R_1 \\
	0  ,&  R_1 > r.  \\
	\end{cases}
	\end{equation} 
	The scattering parameters for the potential in the 11 and 12 channel are found numerically (see Sec. \ref{subsec:numerov_algorithm} and Ref. \cite{newton2013scattering}). Scattering properties of 22 channel can be found using the relations in \cite{jensen2006bcs}.

\end{itemize}

In all cases, the attractive interatomic potential does not support a two-body bound state. We have obtained the $s$-wave scattering length and effective range of the potentials using standard scattering theory (see Sec. \ref{subsec:numerov_algorithm} and Refs. \cite{newton2013scattering, roman1965advanced}).  Note that for a given set of just two scattering parameters, there is an uncountable number of corresponding potentials. We have used HCSW potentials to investigate systems under the limit of zero range. However, this set of potentials cannot reproduce the scattering parameters of $^{39}$K given in Table \ref{table_finite_range:scattering_parameters}. In order to fulfill both scattering conditions, the $s$-wave scattering lengths, and the effective ranges, interaction potentials need to have a more elaborate shape, so we use POT1 and POT2 set of potentials. These particular choices of interaction models are somewhat arbitrary since we have focused only on reproducing two scattering parameters. POT1 and POT2 potentials have different shapes and thus different higher-order scattering parameters. However, they are qualitatively similar, i.e., they share the dominant repulsive or attractive character of the potential, and they have a similar structure. All the potentials we use do not support a two-body bound state.

\begin{table}[]
	\centering
	\caption[Scattering parameters for the $^{39}$K mixture \cite{roy2013test}]{Scattering parameters for the $^{39}$K mixture \cite{roy2013test}, $s$-wave scattering length $a$ and the effective range $r^{\rm eff}$ in units of Bohr radius $a_0$, as a function of the magnetic field $B$.  }
	\label{table_finite_range:scattering_parameters}
	\begin{tabular}{| c | c | c | c | c | c | c |}
		\hline
		$B (G)$ & $a_{11} (a_0)$ & $r_{11}^{\rm eff}(a_0)$ & $a_{22} (a_0)$ & $r_{22}^{\rm eff}(a_0)$ & $a_{12} (a_0)$ & $r_{12}^{\rm eff}(a_0)$ \\
		\hline 
		56.230 & 63.648           & -1158.872                                             & 34.587           & 578.412                                               & -53.435          & 1021.186                                              \\
		56.337 & 66.619           & -1155.270                                             & 34.369           & 588.087                                               & -53.386          & 1022.638                                              \\
		56.395 & 68.307           & -1153.223                                             & 34.252           & 593.275                                               & -53.360          & 1022.617                                              \\
		56.400 & 68.453           & -1153.046                                             & 34.242           & 593.722                                               & -53.358          & 1022.616                                              \\
		56.453 & 70.119           & -1150.858                                             & 34.136           & 599.143                                               & -53.333          & 1023.351                                              \\
		56.511 & 71.972           & -1148.436                                             & 34.020           & 604.953                                               & -53.307          & 1024.121                                              \\
		56.574 & 74.118           & -1145.681                                             & 33.895           & 610.693                                               & -53.278          & 1024.800                                              \\
		56.639 & 76.448           & -1142.642                                             & 33.767           & 616.806                                               & -53.247          & 1025.593    \\
		\hline                                         
	\end{tabular}
\end{table}

\begin{figure}[!htb]
	\centering
	\includegraphics[width=0.49\linewidth]{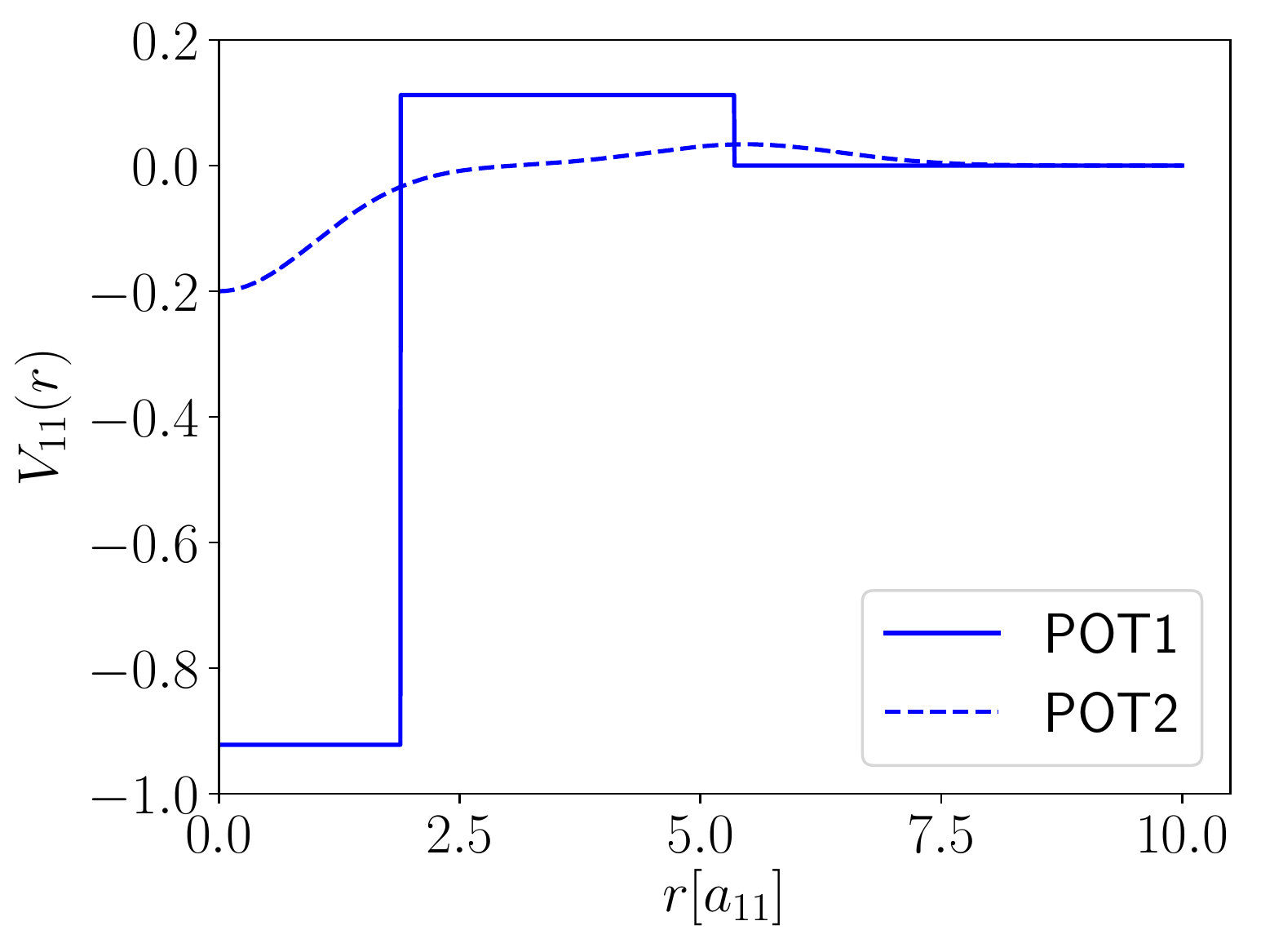}
	\includegraphics[width=0.49\linewidth]{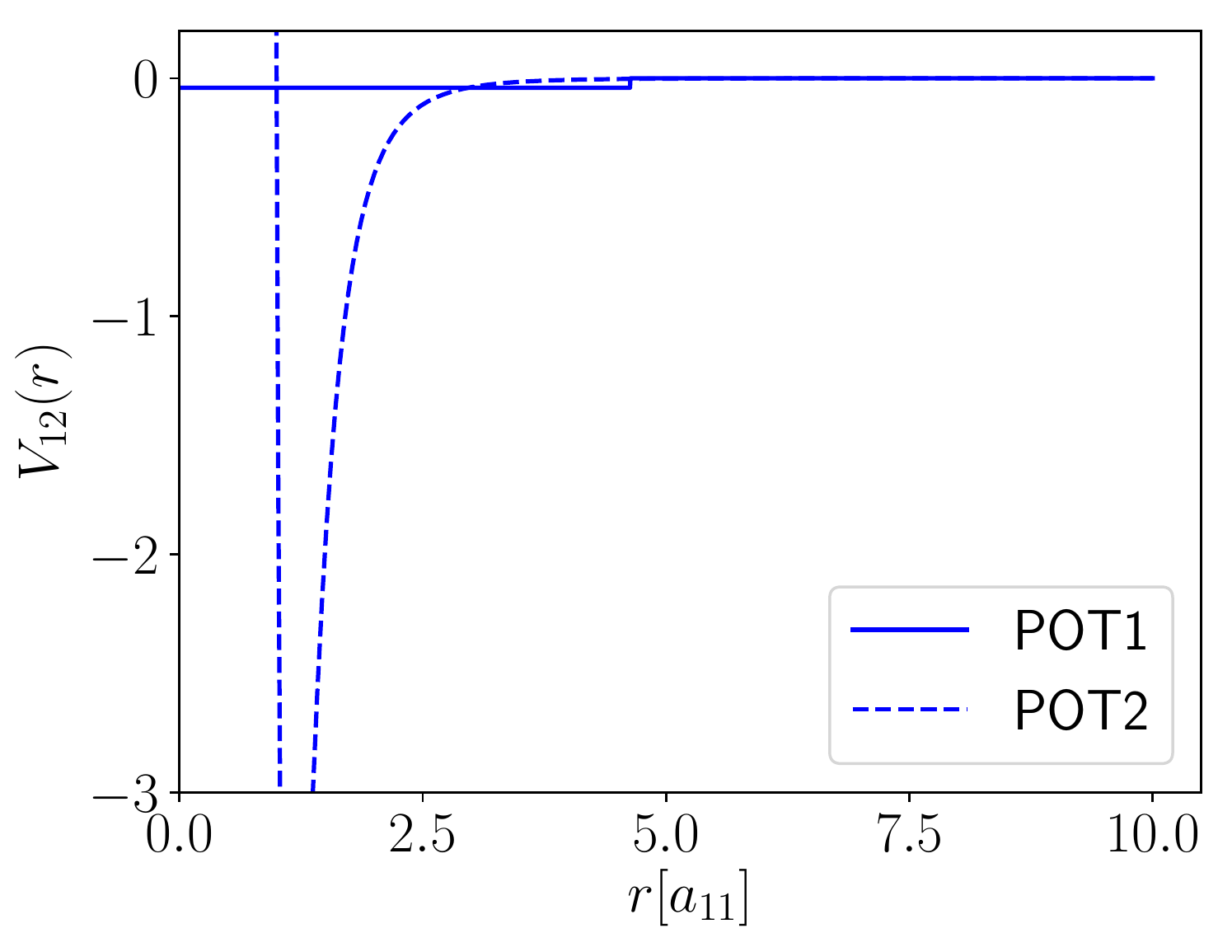}
	\includegraphics[width=0.49\linewidth]{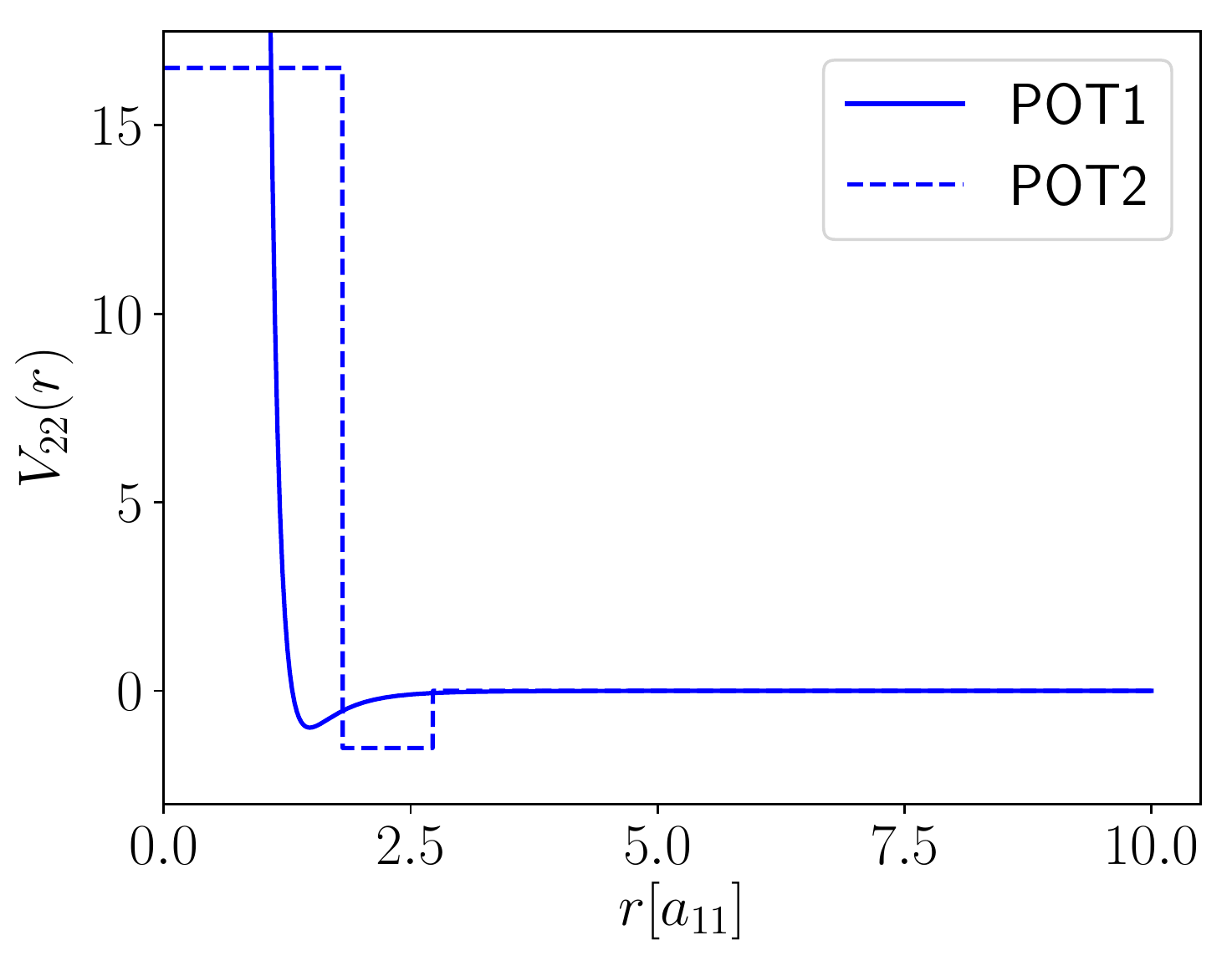}
	\caption[POT1 and POT2 potentials in each of the channel which reproduce the $s$-wave scattering lengths and effective ranges for $^{39}$K mixture at $B=56.337 G$]{POT1 and POT2 potentials in each of the channel which reproduce the $s$-wave scattering lengths and effective ranges for $^{39}$K mixture at $B=56.337 G$.}
\end{figure}

\section{\label{results_finite_range_mixture} Equations of state}

\begin{figure}[tb]
	\centering
	\includegraphics[width=0.9\linewidth]{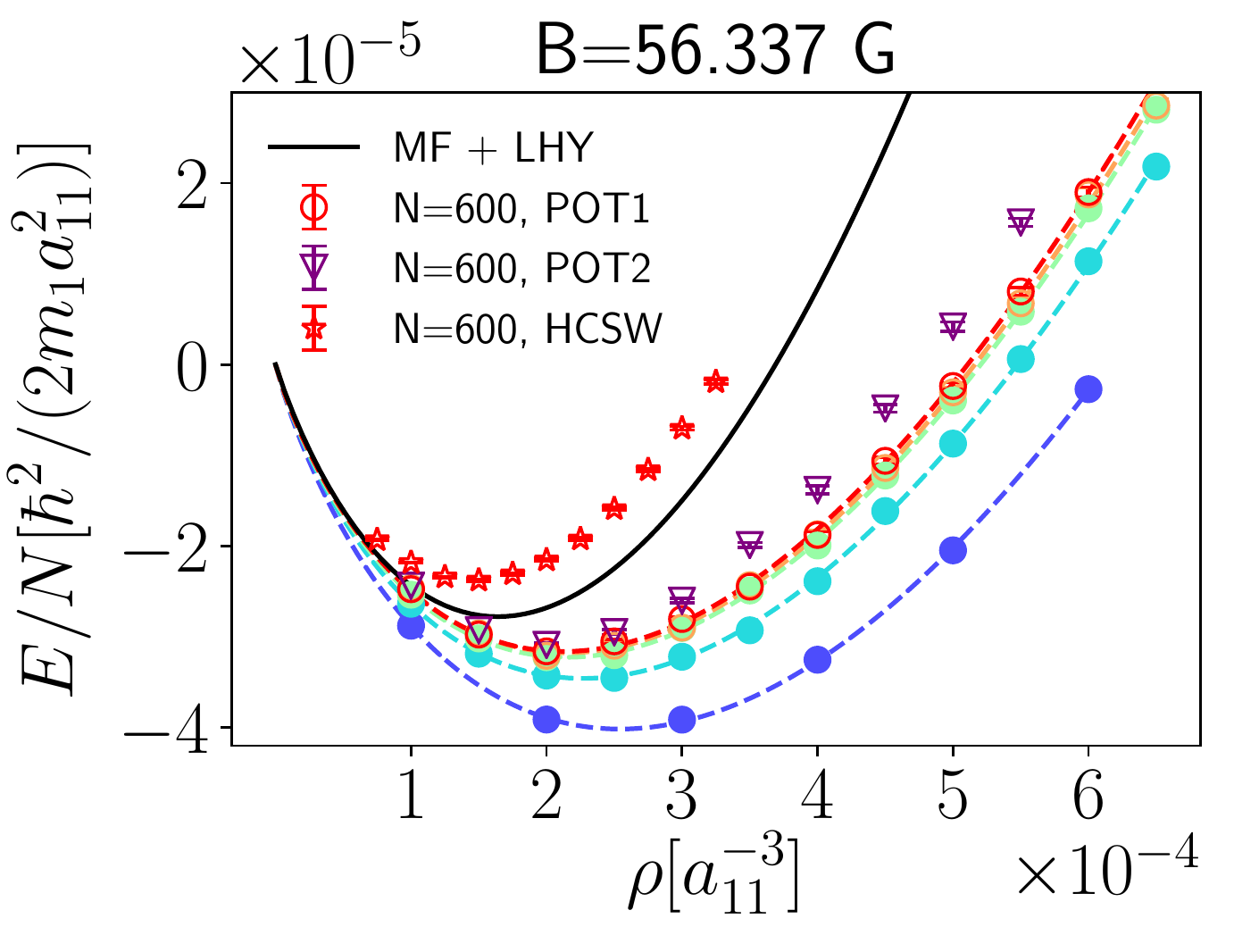}
	\caption[Dependence of the equation of state of $^{39}$K quantum liquid on the effective range for selected potential models, compared with MF+LHY theory.]{Dependence of the equation of state of $^{39}$K quantum liquid on the effective range for selected potential models, compared with MF+LHY theory. Full circles are calculations using POT1, and we illustrate the convergence to negligible finite-size effects starting from $N=100$ (lower points), 
		200, 400, 500 to $N=600$ (upper points). Dashed lines are fits to the DMC data  
		with Eq. (\ref{eq_finite_range:density_functional}).}
	\label{fig:ploteosfinitesizeb56337}
\end{figure}

In order to go beyond the MF+LHY density functional, we have carried out DMC calculations of the bulk liquid.
In Fig.~(\ref{fig:ploteosfinitesizeb56337}), we show the energy per particle of the $^{39}$K mixture as a function of the density, using three different sets of potentials in the Hamiltonian (\ref{hamiltonian}). We compare our DMC results to the MF+LHY theory, which can be compactly written as \cite{petrov2015quantum}
\begin{equation}
\label{eq_finite_range:mflhy_eos}
\dfrac{E/N}{|E_0 / N|} = -3\left(\dfrac{\rho}{\rho_0}\right) + 2 
\left(\dfrac{\rho}{\rho_0}\right)^{3/2} \ ,
\end{equation}
assuming the optimal concentration of particles from mean-field theory, $N_1 / N_2 = \sqrt{a_{22} / a_{11}}$. The energy per particle $E_0 / N$ at the equilibrium density $\rho_0$ of the MF+LHY approximation are
\begin{equation}
\label{eq_finite_range:en_0}
E_0 / N = \dfrac{25 \pi^2 \hbar^2 |a_{12} + \sqrt{a_{11} 
		a_{22}}|^3}{768m a_{22} a_{11} \left(\sqrt{a_{11}} + \sqrt{a_{22}}\right)^6},
\end{equation} 
\begin{equation}
\label{eq_finite_range:rho_0}
\rho_0 a_{11}^3 = \dfrac{25 \pi}{1024} \dfrac{\left(a_{12}/a_{11} + 
	\sqrt{a_{22}/a_{11}}\right)^2}{\left(a_{22}/a_{11}\right)^{3/2}\left(1+\sqrt{a_{
			22}/a_{11}}\right)^4} .
\end{equation}

In Fig.~(\ref{fig:ploteosfinitesizeb56337}), we report DMC results for the equation of state corresponding to a magnetic field $B=56.337$ G, one of the magnetic fields used in experiments. We demonstrate the convergence of energy per particle on the number of particles in the particular case of the POT1 set of potentials. As we can see, the convergence is achieved with $N=600$. We have repeated this analysis for all the potentials, and, in all the magnetic field range explored, we arrive at convergence with similar $N$ values. We have investigated the dependence on the effective range by repeating the calculation using the HCSW and POT2 potentials. As it is clear from  Fig. (\ref{fig:ploteosfinitesizeb56337}), only when both scattering parameters, the $s$-wave scattering length and the effective range, are imposed on the model potentials, we get an approximate universal equation of state, mainly around the equilibrium density. The equation of state so obtained shows a significant and overall decrease of the energy compared to the MF+LHY prediction, with a correction that increases with the density. Instead, using the HCSW potentials, which only fulfill the $s$-wave scattering lengths, the energies obtained are even above the  MF+LHY prediction. Similar behavior has been previously shown to hold in symmetric ($N_1=N_2$) Bose-Bose mixtures (see Chapter \ref{ch:symmetric_liquids}).

\begin{figure}[!htb]
	\centering
	\includegraphics[width=0.9\linewidth]{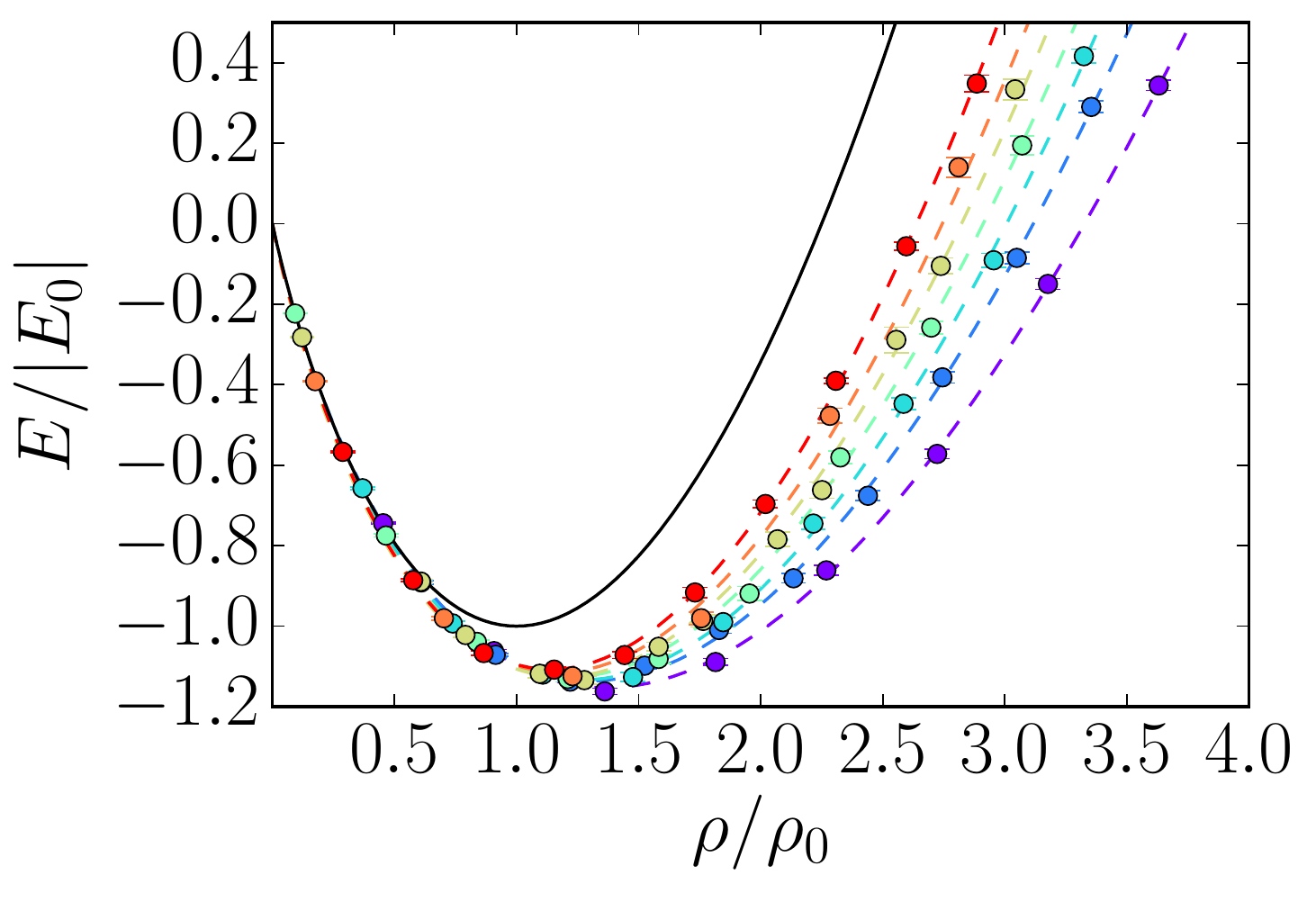}
	\caption[DMC energy per particle for the $^{39}$K liquid as a function of the density (circles), starting from $B=56.230 G$ (lower points) to $B=56.639 G$ (upper points)]{DMC energy per particle for the $^{39}$K liquid as a function of the density (circles), starting from $B=56.230 G$ (lower points) to $B=56.639 G$ (upper points). Energy and density are normalized to $E_0$ and $\rho_0$, given in Eq. \ref{eq_finite_range:en_0} and  \ref{eq_finite_range:rho_0}, respectively. Dashed lines are fits with Eq. (\ref{eq_finite_range:density_functional}). Full line is the MF+LHY theory (Eq. \ref{eq_finite_range:mflhy_eos}), universally expressed in terms of equilibrium density (Eq. \ref{eq_finite_range:rho_0}) and energy (Eq. \ref{eq_finite_range:en_0}).}
	\label{fig:eoscombineddiffbuniversal}
\end{figure}

Equations of state of the bulk mixture, for the seven values of the magnetic field used in the experiments ($B=56.230$ G  to $B=56.639$ G),  are shown in Fig. (\ref{fig:eoscombineddiffbuniversal}). The DMC results are calculated using the model POT1. In all cases, we take the mean-field prediction for the optimal ratio  of partial densities $\rho_1 / \rho_2 = \sqrt{a_{22} / a_{11}}$. We have verified in several cases that this is also the concentration corresponding to the ground state of the system in our DMC calculations, i.e., the one that gives the minimum energy at equilibrium (see Sec. \ref{subsec:effective_mflhy_theory}). The DMC results are compared with the MF+LHY equation of state (Eq. \ref{eq_finite_range:mflhy_eos}). Overall, a reduction of the magnetic field, or equivalently an increase in $|\delta a|=a_{12} + \sqrt{a_{11}a_{12}}$, leads to an increase of the binding energy compared to  the MF+LHY approximation. This happens clearly due to the influence of the large experimental effective range since in the limit of zero range, one would observe overall repulsive beyond-LHY terms (see also Fig.\ref{fig:ploteosfinitesizeb56337} and Chapter \ref{ch:symmetric_liquids}).

DMC energies for the $^{39}$K mixture are well fitted using the functional form 
\begin{equation}
\label{eq_finite_range:density_functional}
E/N = \alpha \rho  + \beta \rho^\gamma,
\end{equation}
as it can be seen in Fig.~(\ref{fig:eoscombineddiffbuniversal}). These equations of state, calculated within the range of magnetic fields used in experiments, are then used in the functional form (\ref{eq_finite_range:time_dep_gp}) with the interacting energy density being
\begin{equation}
	\label{eq:interacting_energy_density}
	\mathcal{E}_{\rm{int}}=\rho \dfrac{E}{N}
\end{equation}
With the new functional, based on our DMC results, we can study the quantum drops with the proper number of particles which is too large for a direct DMC simulation.

\section{\label{critical_atom_number_finite_range}Critical atom number}

\begin{figure}[tb]
	\centering
	\includegraphics[width=0.8\linewidth]{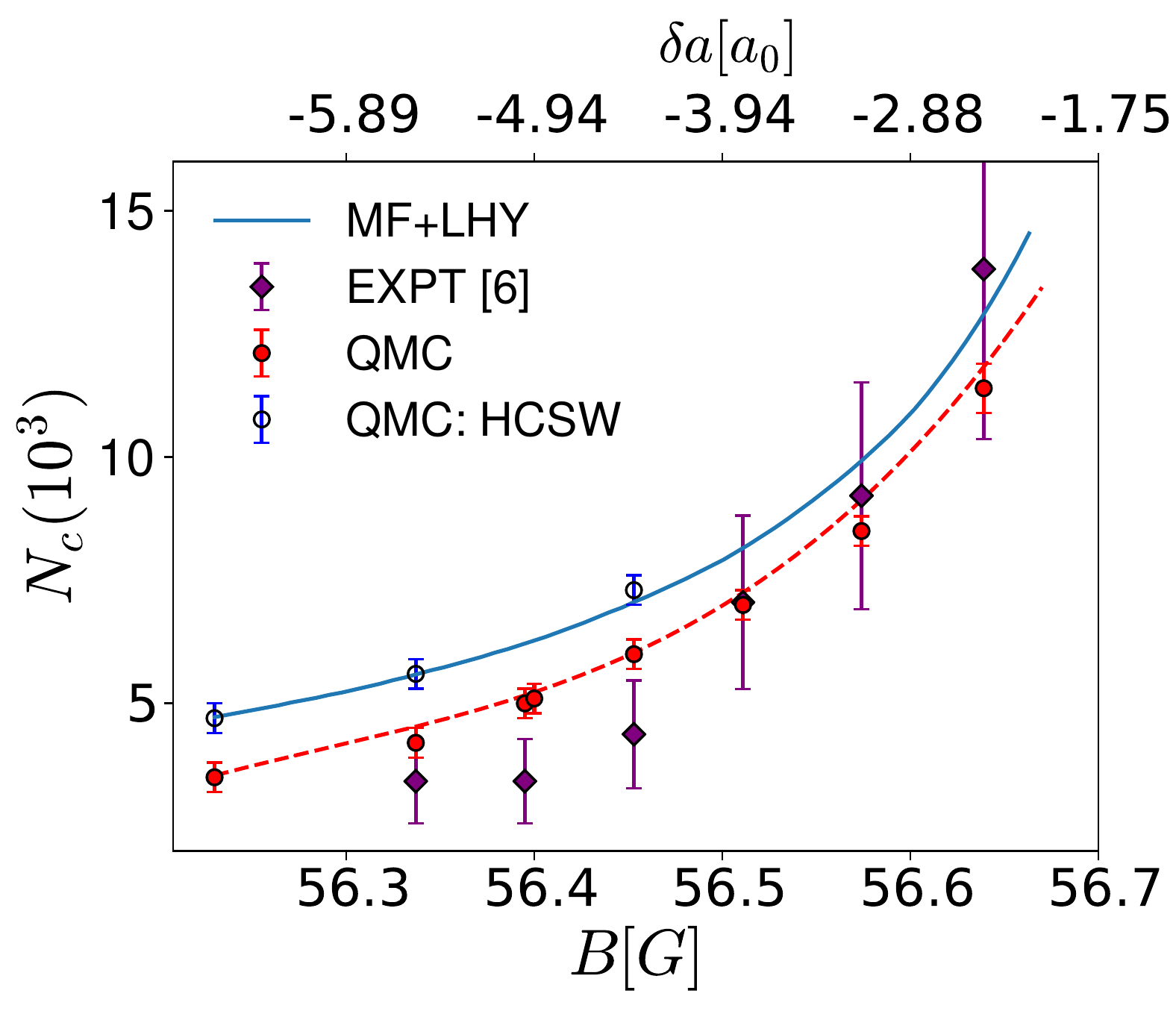}
	\caption[Dependence of the critical atom number of a $^{39}$K droplet on the 
	magnetic field]{Dependence of the critical atom number of a $^{39}$K droplet on the 
		magnetic field. Full circles are predictions  using the QMC functional within DFT with the interaction potentials which reproduce both $a$ and $r_{\rm eff}$. Diamond points are data from the experiment \cite{cabrera2018quantum}. Empty points show the prediction using the QMC functional with the HCSW model potentials. }
	\label{fig:criticalncomparisonwicfo}
\end{figure}

\begin{figure}
	\centering
	\includegraphics[width=0.7\linewidth]{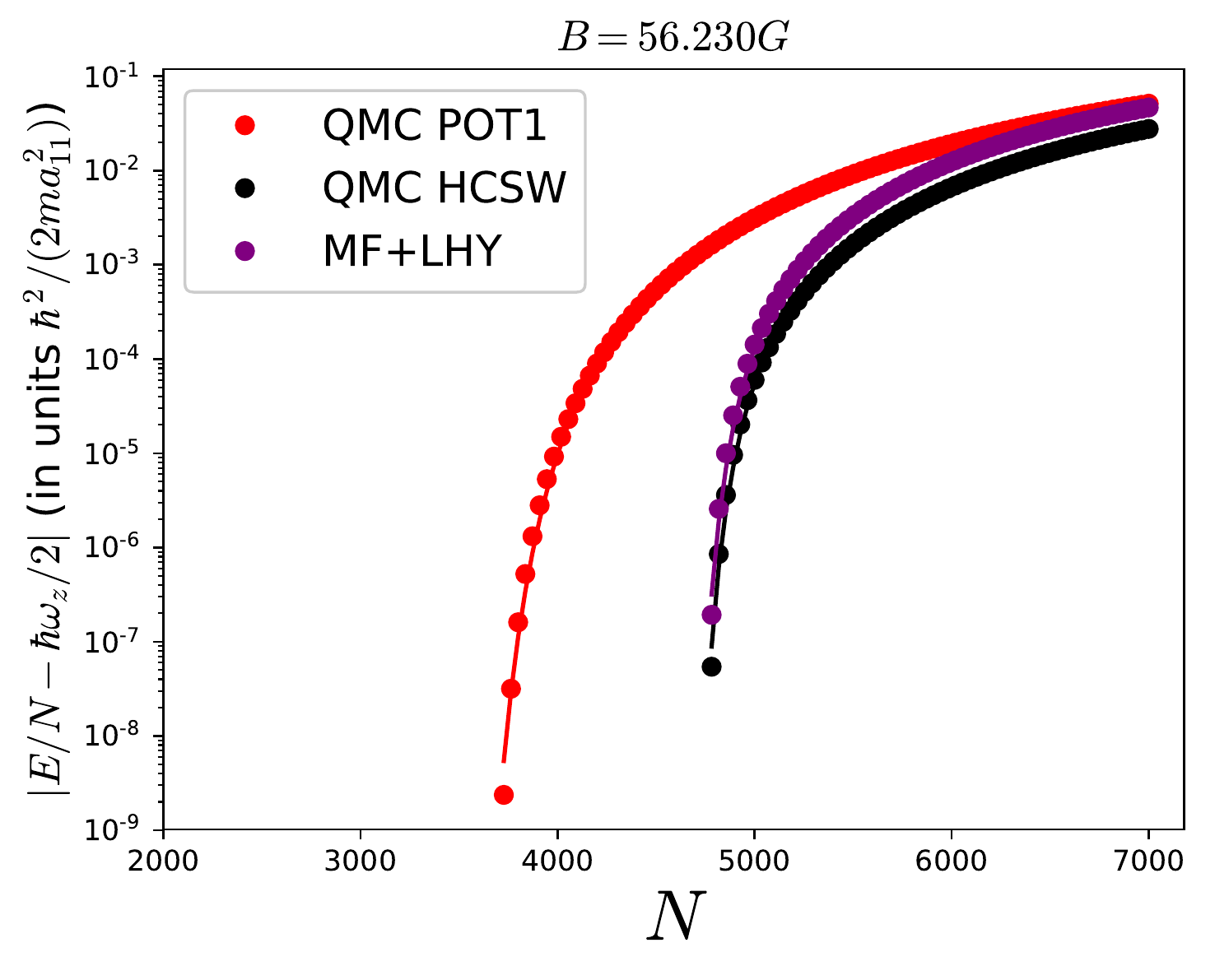}
	\caption[Absolute value of the binding energy with respect to the total atom number, using the scattering parameters corresponding to the magnetic field $B=56.230G$]{Absolute value of the  binding energy with respect to the total atom number, using the scattering parameters corresponding to the magnetic field $B=56.230G$. Points are energies estimated with the density functional Eq. (\ref{eq_finite_range:density_functional}), with the input Gaussian wavefunction. Lines are fits to the energies in a form $\Delta /(N- N_c)^C$, with $\Delta$, $N_c$ and $C$ being the fit parameters. Value of $N_c$ is used as an estimate for the critical atom number. QMC POT1 (QMC HCSW) stands for the QMC-built functional using the POT1 (HCSW) model potentials (see Fig. \ref{fig:ploteosfinitesizeb56337} and Eq. \ref{eq:interacting_energy_density}), whereas the MF+LHY uses functional derived by Petrov (see Eq. \ref{eq_finite_range:mflhy_eos} and Ref. \cite{petrov2015quantum}).}
	\label{fig:enndeterminencb56230}
\end{figure}

Results for the critical atom number $N_c$ at different $B$ are shown  in Fig.~(\ref{fig:criticalncomparisonwicfo}) in comparison with the  experimental results of Ref.~\cite{cabrera2018quantum}. To make the comparison reliable, we have included the same transversal confinement as in the experiment. In particular, theoretical predictions are obtained within DFT, using a Gaussian ansatz normalized to total particle number $N$
\begin{equation}
    \phi= \dfrac{\sqrt{N}}{\pi^{3/4} \sigma_r \sqrt{\sigma_z}} \exp\left(-r^2 / (2\sigma_r^2) -z^2 /(2\sigma_z^2)\right),
\end{equation}
where we take $\sigma_r$ and $\sigma_z$  to be variational parameters. To obtain the critical atom number, we have calculated the binding energy per particle (in absolute value) $E/N - \hbar \omega_z / 2$, with $\omega_z = \hbar / (m a_{\rm ho}^2)$, as a function of the total atom number. By extrapolating the fit to the point where binding energy is equal to zero, we obtain the critical atom number. This is illustrated for $B=56.230 G$ at Fig. (\ref{fig:enndeterminencb56230}).

When the equation of state of the bulk takes into account the effective range of all the pairs, we observe an overall decrease of $N_c$ with respect to the MF+LHY prediction. Interestingly, if we use the HCSW model potentials, with essentially zero range, our results are on top of the  MF+LHY line (see the points at $B=56.23$ G, $B=56.337$ G and $B=56.453$ G in Fig.~\ref{fig:criticalncomparisonwicfo}). The observed decrease of $N_c$ leads our theoretical prediction closer to the experimental data in a significant amount and all the $\delta a$ range, clearly showing the significant influence of the effective range on the $N_c$ values.

Experiments on quantum droplets were performed either in the harmonic trap \cite{cabrera2018quantum} or in a free-drop setup \cite{semeghini2018self}. Predictions of $N_c$ for these two geometries are given in Table (\ref{tab:nc_icfo}) and (\ref{tab:nc_florence}), using MF+LHY and QMC functionals. The absolute difference of predicted $N_c$ values between the two functionals are about 1000 atoms. On the other hand, relative difference is much higher in the harmonically-trapped system because the introduction of an external trap significantly reduces the $N_c$. 
\begin{table}[tb]
	\centering
	\caption{Critical atom number to form a droplet in a harmonic trap $V_z = \frac{1}{2}m\omega_z^2 z^2$, where $a_{\rm ho} = \sqrt{\hbar / (m\omega_z)} = 0.639 \mu m$ is the same value as in the experiment \cite{cabrera2018quantum}. $\mathcal{E}_r$ stands for $\left({N_c^{\rm QMC} - N_c^{\rm MFLHY}}\right)/{N_c^{\rm MFLHY}}$.}
	\label{tab:nc_icfo} 
	\begin{tabular}{ c | c | c | c | c | c}
		$B (G)$ & $N_c^{\rm QMC}$ & $N_c^{\rm MFLHY}$ &  $N_c^{\rm ICFO}$\cite{cabrera2018quantum} & $\mathcal{E}_r$ & $N_c^{\rm QMC} - N_c^{\rm MFLHY}$  \\[+0.3em] \hline
		56.23  & 3500  & 4650  &  -     & -0.25 & -1150 \\
		56.337 & 4200  & 5570  & 3420  & -0.25 & -1370 \\
		56.395 & 5000  & 6200  & 3421  & -0.19 & -1200 \\
		56.4   & 5100  & 6250  &  -     & -0.18 & -1150 \\
		56.453 & 6000  & 7000  & 4373  & -0.14 & -1000 \\
		56.511 & 7000  & 8050  & 7052  & -0.13 & -1050 \\
		56.574 & 8500  & 9800  & 9217  & -0.13 & -1300 \\
		56.639 & 11300 & 12700 & 13819 & -0.11 & -1400    \\
	\end{tabular}
\end{table}

\begin{table}[tb]
	\centering
	\caption{Critical atom number for spherical free drops~\cite{semeghini2018self}. $\mathcal{E}_r$ stands for $\left({N_c^{\rm QMC} - N_c^{\rm MFLHY}}\right)/{N_c^{\rm MFLHY}}$.}
	\label{tab:nc_florence}
	\begin{tabular}{ c | c | c | c | c }	
		
		$B (G)$ & $N_c^{\rm QMC}$ & $N_c^{\rm MFLHY}$ & $\mathcal{E}_r$ & $N_c^{\rm QMC} - N_c^{\rm MFLHY}$  \\
		\hline 
		56.23  & 16000  & 15800  & 0.01  & 200   \\
		56.337 & 24600  & 24900  & -0.01 & -300  \\
		56.395 & 32700  & 33900  & -0.04 & -1200 \\
		56.4   & 35300  & 35500  & -0.01 & -200  \\
		56.453 & 47200  & 47700  & -0.01 & -500  \\
		56.511 & 69100  & 70600  & -0.02 & -1500 \\
		56.574 & 114000 & 119000 & -0.04 & -5000 \\
		56.639 & 230000 & 236000 & -0.03 & -6000    \\
	\end{tabular}
\end{table}

\section{\label{size_39K_droplet}Size of a $^{39}$K droplet}

\begin{figure}
	\centering
	\includegraphics[width=0.8\linewidth]{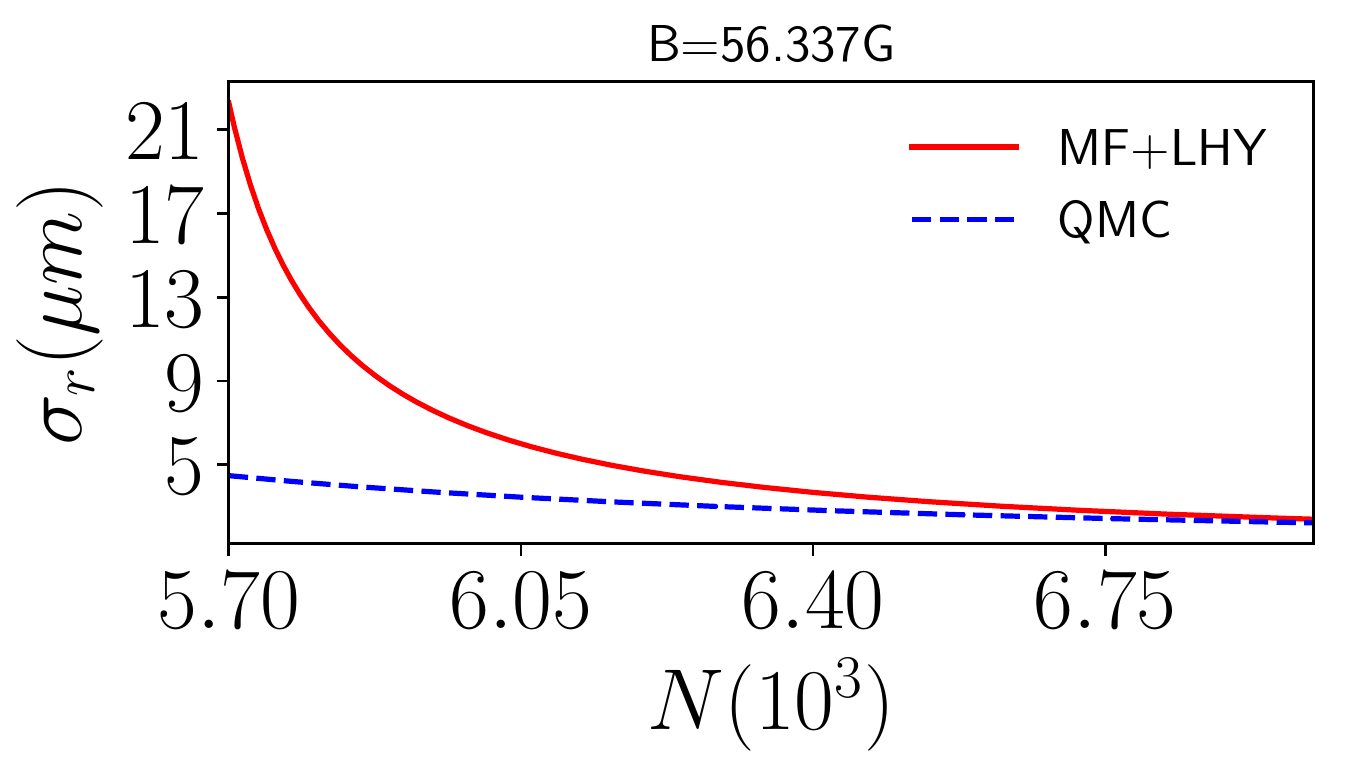}
	\caption[Dependence of the radial size $\sigma_r$ of the $^{39}$K drop on the 
	number of particles.]{Dependence of the radial size $\sigma_r$ of the $^{39}$K drop on the 
		number of particles. The size is obtained from the variational ansatz, since 
		close 
		to the critical atom number the density profile in the radial direction is well 
		approximated by a Gaussian. In both functionals, it is assumed that the 
		relative 
		concentration is optimal $N_2 / N_1 = \sqrt{a_{11} / a_{22}}$. QMC functional includes the correct finite-range $r_{\rm eff}$ through POT1 set of potentials, Fig.  \ref{fig:ploteosfinitesizeb56337}.}
	\label{fig:sigmarvsncompareqmcmflhy}
\end{figure}

A second observable measured in experiments is the size of the drops. Close to the critical atom number, the density profile of a drop can change drastically depending on the functional. We illustrate this effect in Fig.~(\ref{fig:sigmarvsncompareqmcmflhy}) for a magnetic field $B=56.337$ G. In the figure, we show the dependence of the radial size on the number of particles, with the same harmonic confinement strength as in one of the experiments~\cite{cabrera2018quantum}. We observe a substantial difference between the MF+LHY and QMC functional results, mainly when $N$ approaches the critical number $N_c$.

\begin{figure}[tb]
\centering
\includegraphics[width=0.7\linewidth]{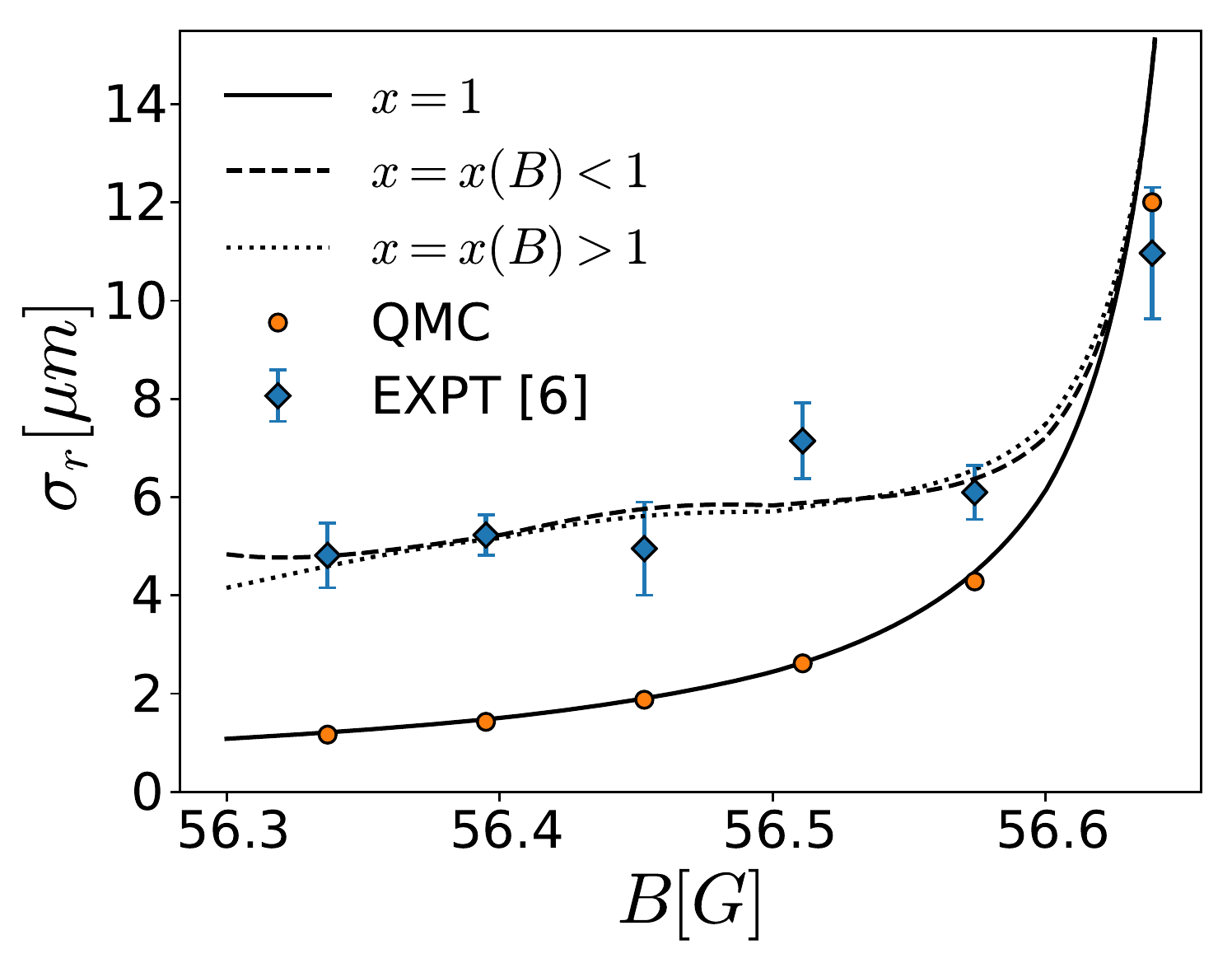}
\caption[Dependence of the radial size of a $N=15000$ $^{39}$K drop on the 
external magnetic field, or equivalently the residual $s$-wave scattering length.]{Dependence of the radial size of a $N=15000$ $^{39}$K drop on the 
	external magnetic field, or equivalently the residual $s$-wave scattering length. 
	Lines are predictions under MF+LHY theory; full line is a prediction with 
	$x=1$, 
	dashed and dotted lines are fits of experimental sizes using a parameter $x$.}
\label{fig:sigmarcompwicfoxofb}
\end{figure}

The radial size of a $N=15000$ drop for different magnetic field values was reported in Ref.~\cite{cabrera2018quantum}. In Fig.~(\ref{fig:sigmarcompwicfoxofb}), we compare the experimental values with different theoretical predictions. We observe a slight reduction in size using QMC functionals, compared to MF+LHY theory, which is a consequence of the stronger binding produced by the inclusion of finite range interactions. Since the experimental data go in the opposite direction, it means that drop size can not be explained solely with the non-zero effective range. One possible explanation for this clear disagreement could be a deviation from the optimal relative number of particles, which can occur in non-equilibrated drops or when one of the components has a large three-body recombination coefficient. In section (\ref{subsec:effective_mflhy_theory}) we further explore the theory of non-equilibrated droplets and introduce the $x$ parameter.

\subsection{\label{subsec:effective_mflhy_theory}Effective MF+LHY theory for $x\neq 1$ liquids   }

\begin{figure}
\centering
\includegraphics[width=0.9\linewidth]{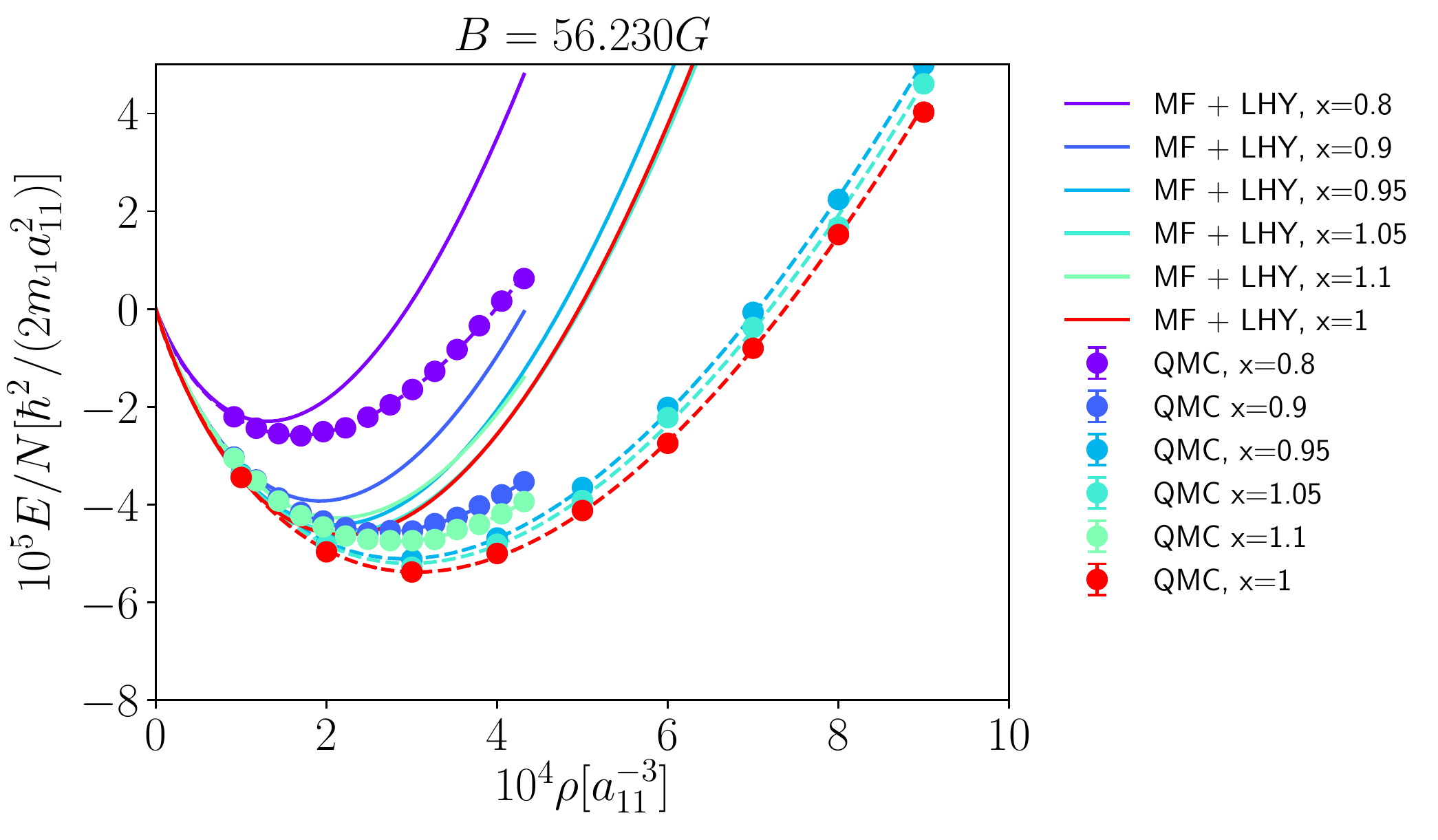}
\caption{Energy per particle as a function of total density $\rho$. Different colors differentiate population ratio throught the parameter $x = (\rho_2 / \rho_1) \sqrt{g_{11} / g_{22}} $, full lines are MF+LHY predictions (see Eq. \ref{eq:mf_diffx} and \ref{eq:lhy_diffx}), points are QMC results, and dashed lines are the fits to the DMC energies.}
\label{fig:ploteosb56230wx}
\end{figure}

We explain the discrepancy in size of the droplets due to the non-optimal ratio between particles, namely $N_2 / N_1 \neq \sqrt{a_{11} / a_{22}}$. For simplicity, let us consider a homogeneous metastable liquid with the ratio of partial densities $\rho_1$ and $\rho_2$ being
\begin{equation}
\label{eq:definition_x}
\dfrac{\rho_2}{\rho_1}
=
x 
\sqrt{\dfrac{g_{11}}{g_{22}}}.
\end{equation}
Mean field optimal particle ratio is fulfilled for $x=1$,  i.e., the concentration corresponding to the ground-state of the system.  and a deviation of $x$ from one implies an excess of one of the component in the liquid: when $x<1$ ($x > 1$), there is an excess of species 1 (2).  MF and LHY terms read
\begin{equation}
\label{eq:mf_diffx}
\dfrac{E_{\rm MF}}{V} 
=
\dfrac{\hbar^2}{ma_{11}^2}\dfrac{2\pi \left(1 + x^2 + 2x\dfrac{a_{12}/a_{11}}{\sqrt{a_{22}/a_{11}}}\right)}{\left(1+x/\sqrt{a_{22}/a_{11}}\right)^2}(\rho a_{11}^3)^2,
\end{equation}
\begin{equation}
\label{eq:lhy_diffx}
\dfrac{E_{\rm LHY}}{V}
=\dfrac{\hbar^2}{ma_{11}^2}
\dfrac{256\sqrt{\pi}}{15}
\left(\dfrac{1 + x\sqrt{a_{22}/a_{11}}}{1+\dfrac{x}{\sqrt{a_{22}/a_{11}}}}\right)^{5/2} \left(\rho a_{11}^3\right)^{5/2},
\end{equation}    
where $\rho = \rho_1 + \rho_2$ is the total density. Note that this energy functional keeps the $x$ parameter fixed. This is in contrast with the ground state of the two-component MF+LHY theory, which is fulfilled when $x=1$. Parameter $x$ thus plays a role of how much the liquid is metastable. On Fig. (\ref{fig:ploteosb56230wx}) we compare the equations of state with various non-optimal ratios, obtained with DMC. We see that finite-range effects are of second-order compared to the influence of non-optimal particle ratio.

\begin{figure}
\centering
\includegraphics[width=0.7\linewidth]{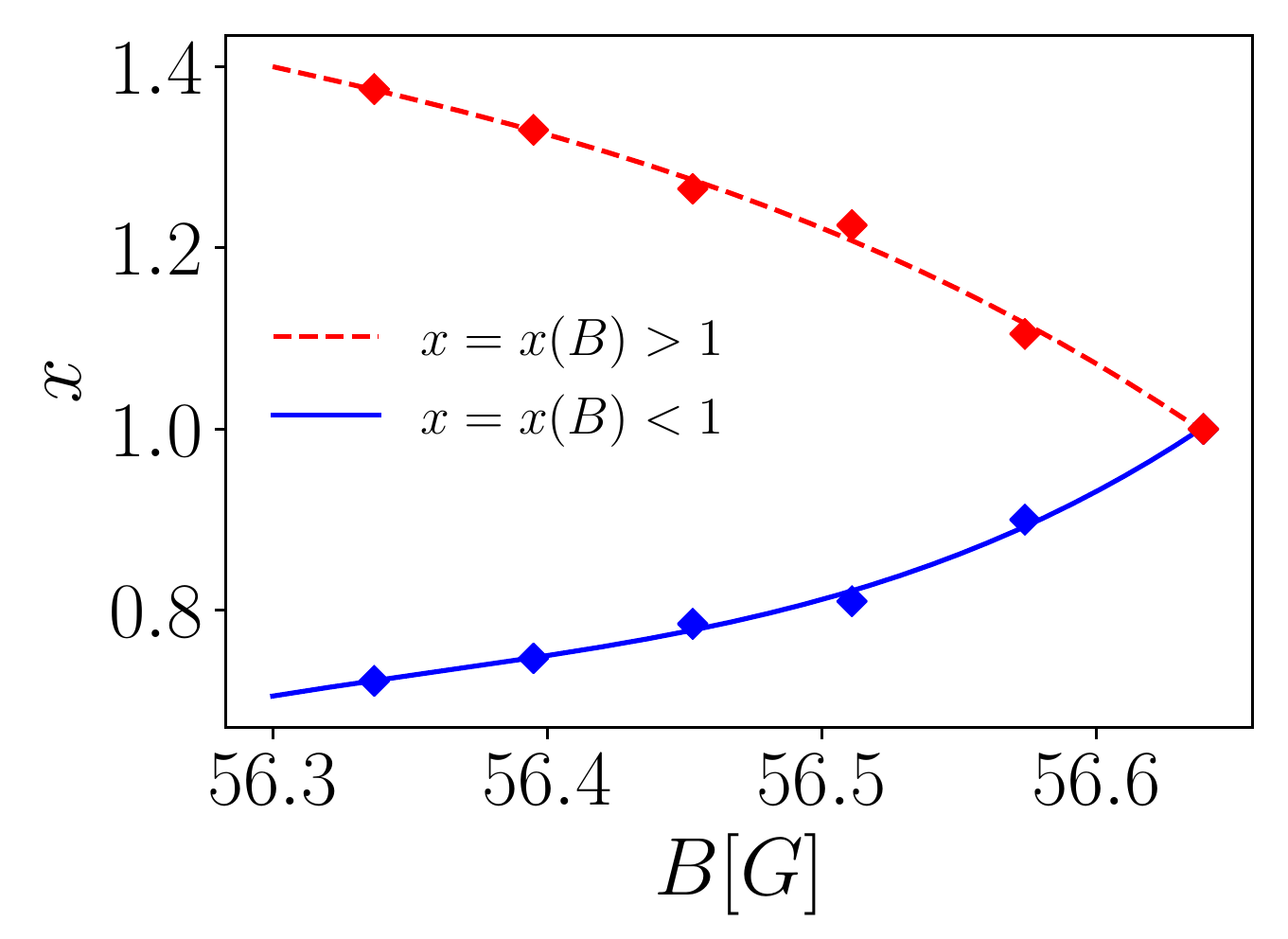}
\caption[Values of $x=N_2 / N_1 \sqrt{a_{22} / a_{11}}$ which reproduce 
the experimental size of a N=15000 $^{39}$K drop (Fig. \ref{fig:sigmarcompwicfoxofb}) 
within the MF+LHY theory, as a function of the magnetic field.]{Values of $x=N_2 / N_1 \sqrt{a_{22} / a_{11}}$ which reproduce 
	the experimental size of a N=15000 $^{39}$K drop (Fig. \ref{fig:sigmarcompwicfoxofb}) 
	within the MF+LHY theory, as a function of the magnetic field. Points are the 
	values which reproduce the size, and lines are power-law fits of $x$ as a 
	function of the magnetic field $B$. Note that two solutions exist since the 
	choice of naming each component is twofold.}
\label{fig:xofbfitcombined}
\end{figure}

We have investigated the behavior of both the MF+LHY and QMC functionals under variations in $x$, and both predict a decrease in the drop size proportional to the deviation from $x=1$. Using the MF+LHY functional, we have obtained the $x$ values that fit the experimental size for every $B$ (Fig.~\ref{fig:sigmarcompwicfoxofb}). We report the result of this analysis in Fig.~(\ref{fig:xofbfitcombined}); notice that there is symmetry on $x \leftrightarrow 1/x$, equivalent to relabeling the components as $1\leftrightarrow2$, and so only its absolute deviation from one is important. This result clearly shows the sensitive dependence of drop structural properties on the relative atom number.

\begin{figure}[tb]
\centering
\includegraphics[width=1.\linewidth]{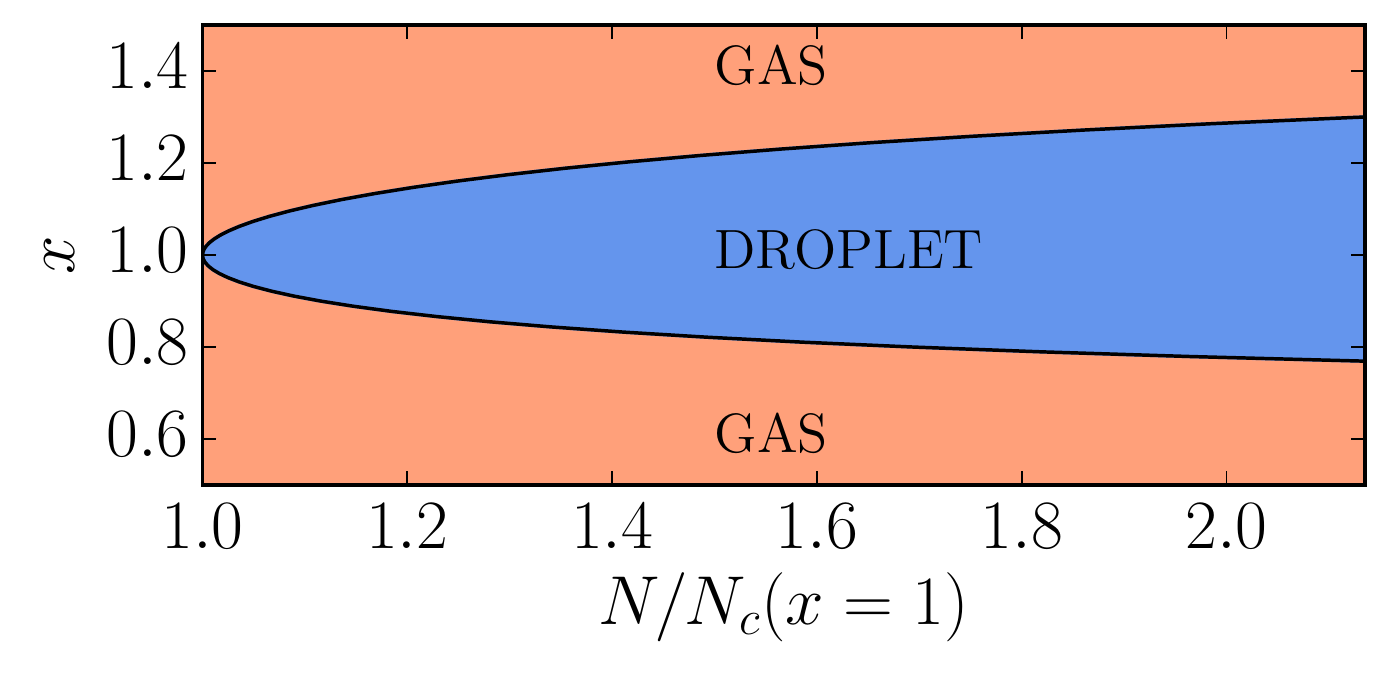}
\caption[Phase diagram of $^{39}$K at $B=56.230$G.]{Phase diagram of $^{39}$K at $B=56.230$G using MF+LHY theory, 
	spanned with $x = N_2 / N_1 \sqrt{a_{22} / a_{11}}$ and the total particle number 
	$N$, normalized with the critical atom number $N_c$ evaluated at $x=1$ 
	\cite{petrov2015quantum}.}
\label{fig:xnc}
\end{figure}

As we can see in Fig.~\ref{fig:xofbfitcombined}, the value for $x$ becomes 1 (optimal value) when the drop composed of 15000 particles is studied at the highest magnetic field. This can be understood if we observe that the critical number for this magnetic field matches approximately this number of atoms (see Fig. \ref{fig:criticalncomparisonwicfo}). When the number of atoms of a drop is larger than the critical number (lower $B$ in Fig.~\ref{fig:sigmarcompwicfoxofb}) $x$ departs from one. This can be better understood if one calculates the drop phase diagram as a function of $x$. The result is plotted in Fig.~\ref{fig:xnc}. As the number of particles is approaching the critical one, the range of possible values of $x$, which supports a drop state, is reducing. This is a supporting fact that drops close to the critical atom number observed in the experiment fulfill the condition $x=1$. On the other hand, there is an increasing range of relative particle concentrations for which a drop can emerge as the number of particles increases.

\begin{figure}[H]
\centering
\includegraphics[width=0.7\linewidth]{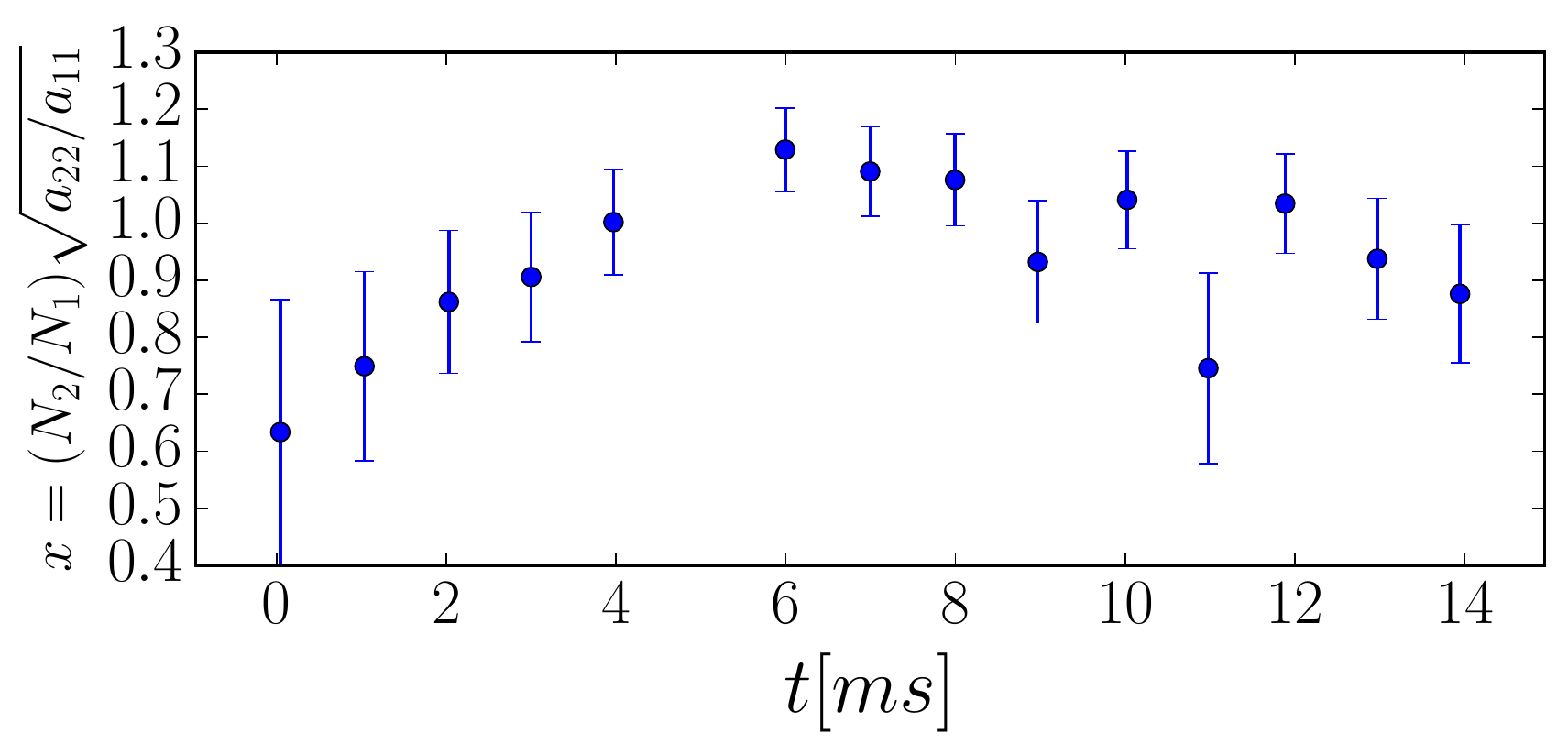}
\caption[Evolution of $x=(N_2 / N_1) \sqrt{a_{22} / a_{11}}$ in time]{Evolution of $x=(N_2 / N_1) \sqrt{a_{22} / a_{11}}$ in time as reported in Fig 2C in \cite{semeghini2018self}.}
\label{fig:xevolution}
\end{figure}

The phenomenological density functional with $x \neq 1$ is motivated by the experimental observations in which $x$ is always reported to deviate from $x=1$. For example, in Fig. (\ref{fig:xevolution}) we show the time evolution of $x$ as reported in Ref. \cite{semeghini2018self}, and it can be observed that $x$ can deviate even 30\% from 1, which is still within the values we report in Fig. (\ref{fig:xofbfitcombined}). Two important processes describe the relaxation of the drop to the true ground state $x=1$: (i) evaporation of the excess component from the drop and (ii) three-body recombination. When a droplet is created with $x \neq 1$, during relaxation time, the particles from excess component evaporate out of the droplet. With three-body losses being present in these drops, there is a possibility that these drops do not reach the ground state $x=1$ and are observed in the metastable regime. This indication is further grounded by noting that both the MF+LHY and QMC functionals predict much smaller sizes than experimentally observed.

We have also compared the size of spherical droplets in a vacuum, which is a geometry used in the experiment~\cite{semeghini2018self}. The qualitative behavior of the droplet size with respect to $x$ is very similar to the one we present for the drops in harmonic traps~\cite{cabrera2018quantum}.

\begin{figure}[H]
\centering
\includegraphics[width=0.7\linewidth]{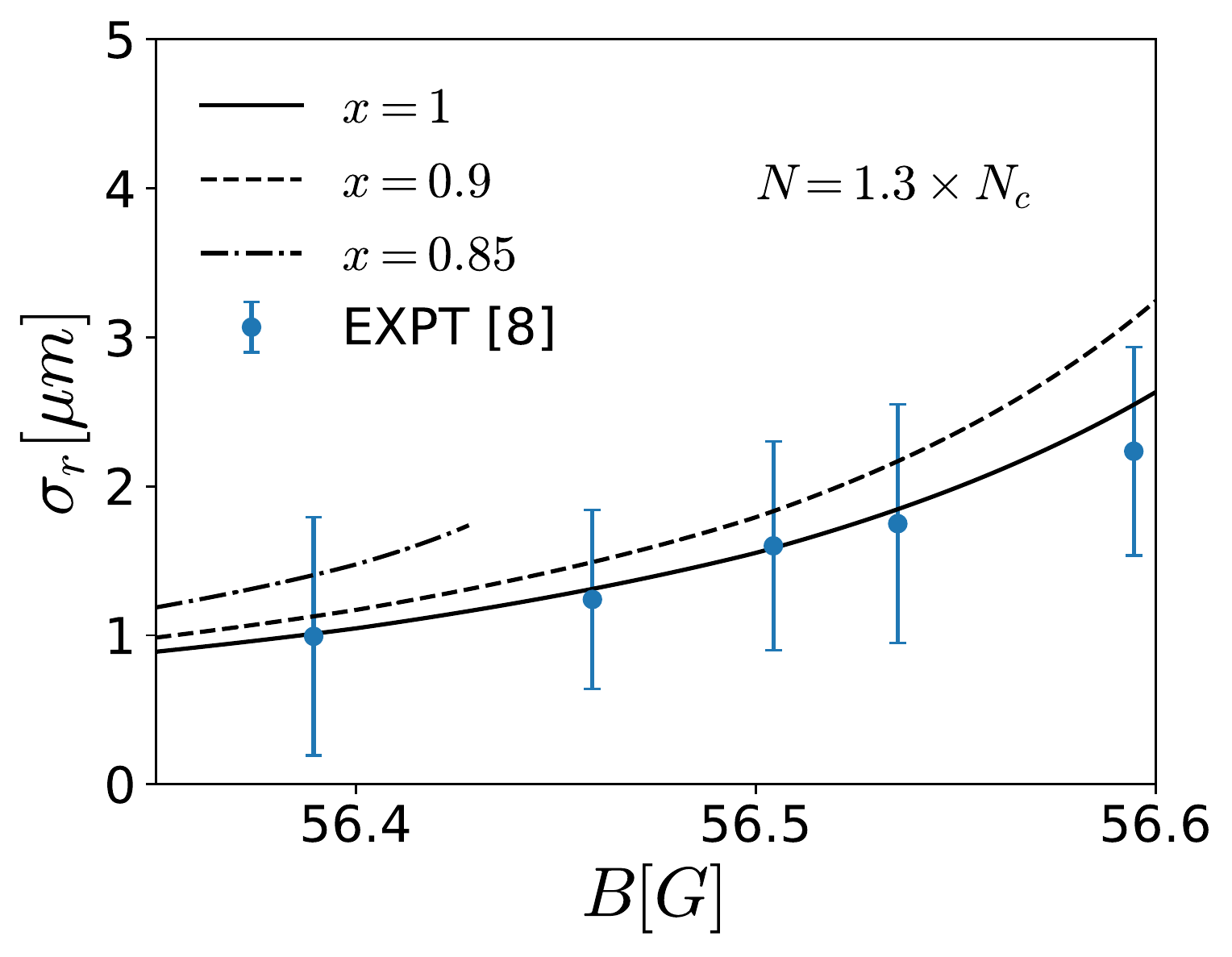}
\caption{Evolution of size $R=\sigma_r/\sqrt{2}$ as a function of magnetic field $B$, for a spherical droplet in a vacuum. $\sigma_r$ is defined as the distance at which the density falls to $e^{-1}$ of the peak density.}
\label{fig:sigmarcompwflorencexofb}
\end{figure}

\section{Summary and discussion}

An experiment in Ref.~\cite{cabrera2018quantum} showed significant disagreement between the measured data and the MF+LHY perturbative approach. In order to determine the possible origin of these discrepancies, we have pursued a beyond  MF+LHY theory, which explicitly incorporates the finite range of the interaction. To this end, we have carried out DMC calculations of the bulk liquid to estimate its equation of state accurately. We have observed that the inclusion in the model potentials of both the $s$-wave scattering length and the effective range produces a rather good universal equation of state in terms of this pair of parameters. This is in agreement with the results from Chapter \ref{ch:symmetric_liquids}, where a symmetric mixture with $a_{11}=a_{22}$ and $m_1 = m_2$ was studied. Excluding the effective range, significant differences are obtained from these universal results. This relevant result points to the loss of universality in terms of the gas parameter in the study of these dilute liquid drops.

Introducing the DMC equation of state into the new functional, following the steps which are standard in other fields, such as DFT in liquid Helium~\cite{barranco2006helium}, we derive a new functional that allows for an accurate study of the most relevant properties of the drops. In particular, we observe that the inclusion of finite range effects reduces the critical atom number in all the magnetic field ranges, approaching the experimental values significantly. On the other hand, our QMC functional cannot explain the apparent discrepancy between theory and experiment about the size of the drops. We attribute this difference to the dramatic effect on the size that small shifts on the value of $x$ produce. Our analysis provides a reasonable explanation of this feature: above the critical atom number, the window of stability of the drops increases from the single point $x=1$ to a range of values that, in absolute terms, grow with the number of particles. With the appropriate choice of $x$, one can obtain agreement with the experiment.

The drops produced in the different setup of Ref.~\cite{semeghini2018self} are spherical since all magnetic confinement is removed. The corresponding critical numbers, in this case, are larger than in the confined setup~\cite{cabrera2018quantum}, and MF+LHY theory accounts reasonably well for the observed features. We have applied our formalism to this case, and the corrections are not zero but relatively less important than in the case analyzed here.

	

\chapter{\label{chapter:breathing_modes}  Finite range effects on the excitation modes of a $^{39}$K quantum droplet}

\section{Introduction}

	In Chapter \ref{ch:finite_range_effects} we have performed diffusion Monte Carlo (DMC) calculations 
	\cite{boronat1994monte,giorgini1999ground} using model potentials that reproduce both scattering parameters, obtaining the equation of state 
	for a $^{39}$K mixture in the homogeneous liquid phase.  We concluded that  one could
	reproduce the  critical atom number determined  in the experiment \cite{cabrera2018quantum} only for the model potentials which incorporate 
	the correct effective range. This critical number is a static property  of the  quantum droplet at equilibrium. Besides a good knowledge of the
	equilibrium properties of  a quantum many-body system, determining the excitation spectrum is essential to unveil its microscopic structure. 
	
	In the present Chapter, we present a study of the monopole and quadrupole excitation spectrum of a $^{39}$K quantum droplet using 
	the QMC functional introduced in Chapter \ref{ch:finite_range_effects}, which correctly describes the inner part of large drops, constituting 
	an extension to the MF+LHY theory. The excitation spectrum of these droplets has already been calculated within the MF+LHY 
	approach \cite{petrov2015quantum,jorgensen2018dilute}. Our goal is to make visible  the appearance of any  beyond-LHY effect arising from 
	the inclusion of the effective range in the interaction potentials.

	This Chapter is organized as follows. We build in Sec. \ref{sec_quad:functional_form} the QMC density functional, in the local density approximation (LDA), and
	compare it with the MF+LHY approach, which can be expressed in a similar form. In Sec. \ref{sec_quad:method}, we give details on the application of the density 
	functional method, static and dynamic,  to the obtainment of the ground state and excitation spectrum of quantum droplets.
	In Sec. \ref{sec_quad:results}, we report the results of the monopole and quadrupole frequencies obtained with the QMC functional and compare them 
	with the MF+LHY predictions. Finally, a conclusion is in  Sec. \ref{sec_quad:summary}.

	\section{\label{sec_quad:functional_form} The QMC Density Functional}
	
	We shall consider  $^{39}$K mixtures at   
	the optimal relative atom concentration yielded by the mean-field theory, namely $N_1 / N_2 = \sqrt{a_{22} / a_{11}}$ \cite{petrov2015quantum}. 
	For these mixtures, in Chapter \ref{ch:finite_range_effects} we have  shown that the energy per atom in the QMC approach can be  accurately written as    
	\begin{equation}
	\dfrac{E}{N} = \alpha \rho + \beta \rho^{\gamma} \ ,
	\label{eq_quad:eos} 
	\end{equation}
	where $\rho$ is the total atom number density. The parameters $\alpha$, $\beta$, and $\gamma$ 
	have been determined by fits to the DMC results for the model potentials satisfying the s-wave scattering length and effective range, given in Table \ref{table_finite_range:scattering_parameters}. The QMC approach does not yield a universal  expression for $E/N$,
	as it depends on the value of the applied $B$. 
	For the optimal concentration,  the MF+LHY energy per particle can be cast in a similar expression 
	\begin{equation}
	\label{eq_quadropole:mflhy_eos}
	\dfrac{E/N}{|E_0|/ N} = -3\left(\dfrac{\rho}{\rho_0}\right) + 2 
	\left(\dfrac{\rho}{\rho_0}\right)^{3/2} \ ,
	\end{equation}
	where $E_0/N$ and $\rho_0$ are the energy per atom and atom density at equilibrium given in equations \ref{eq_finite_range:en_0} and \ref{eq_finite_range:rho_0}, respectively.  MF+LHY theory 
	is thus universal if it is expressed in terms of $\rho_0$ and $E_0$. According to this theory,  the droplet properties do not change separately 
	on $N$ and $a_{ij}$ but rather combined through
	\begin{equation}
	\label{eq_quadropole:ntilde_def}
	\dfrac{N}{\tilde{N}} = \dfrac{3\sqrt{6}}{5\pi^2} \dfrac{\left(1+\sqrt{a_{22}/a_{11}}\right)^5}
	{\left|a_{12} / a_{11} + \sqrt{a_{22} / a_{11}}\right|^{5/2}} \ ,
	\end{equation}
	where $\tilde{N}$ is a dimensionless parameter~\cite{petrov2015quantum}. Additionally, the healing length corresponding to the mixture is
	\begin{equation}
	\label{eq_quadropole:xi_mflhy}
	\dfrac{\xi}{a_{11}} = 
	\dfrac{8\sqrt{6}}{5\pi} \sqrt{\dfrac{a_{22}}{a_{11}}} \dfrac{(1 + \sqrt{a_{22} / a_{11}})^3}{\left|a_{12}/a_{11} + \sqrt{a_{22} / a_{11}}\right|^{3/2}} \ .
	\end{equation}

	\begin{table}
	    \centering
		\caption{Parameters of the QMC energy per atom calculated at several magnetic fields $B$, 
			assuming $\rho_1/\rho_2 = \sqrt{a_{22} / a_{11}}$, satisfying the s-wave scattering length $a$ and effective range $r^{\rm eff}$ 
			given in Table \ref{table_finite_range:scattering_parameters}. $\alpha$ is  in $\hbar^2 a_{11}^2 / (2m)$ units, $\beta$ is in 
			$\hbar^2 a_{11}^{3\gamma - 2} / (2m)$ units,
			$m$ being the mass of a $^{39}$K atom, and $\gamma$ is dimensionless.}
		\label{table_quadropole:eos_params}
		\begin{tabular}{c  | c | c | c }
			\hline
			$B({\rm G})$  &  $\alpha$ & $\beta$ & $\gamma$ \\ \hline
			56.230  & -0.812     &  5.974  &  1.276    \\ 
			56.453    & -0.423        &  8.550    &  1.373 \\ 
			56.639    & -0.203       &  12.152    & 1.440 \\  
		\end{tabular}
	\end{table}

	\begin{figure}
		\centering
		\includegraphics[width=0.8\linewidth]{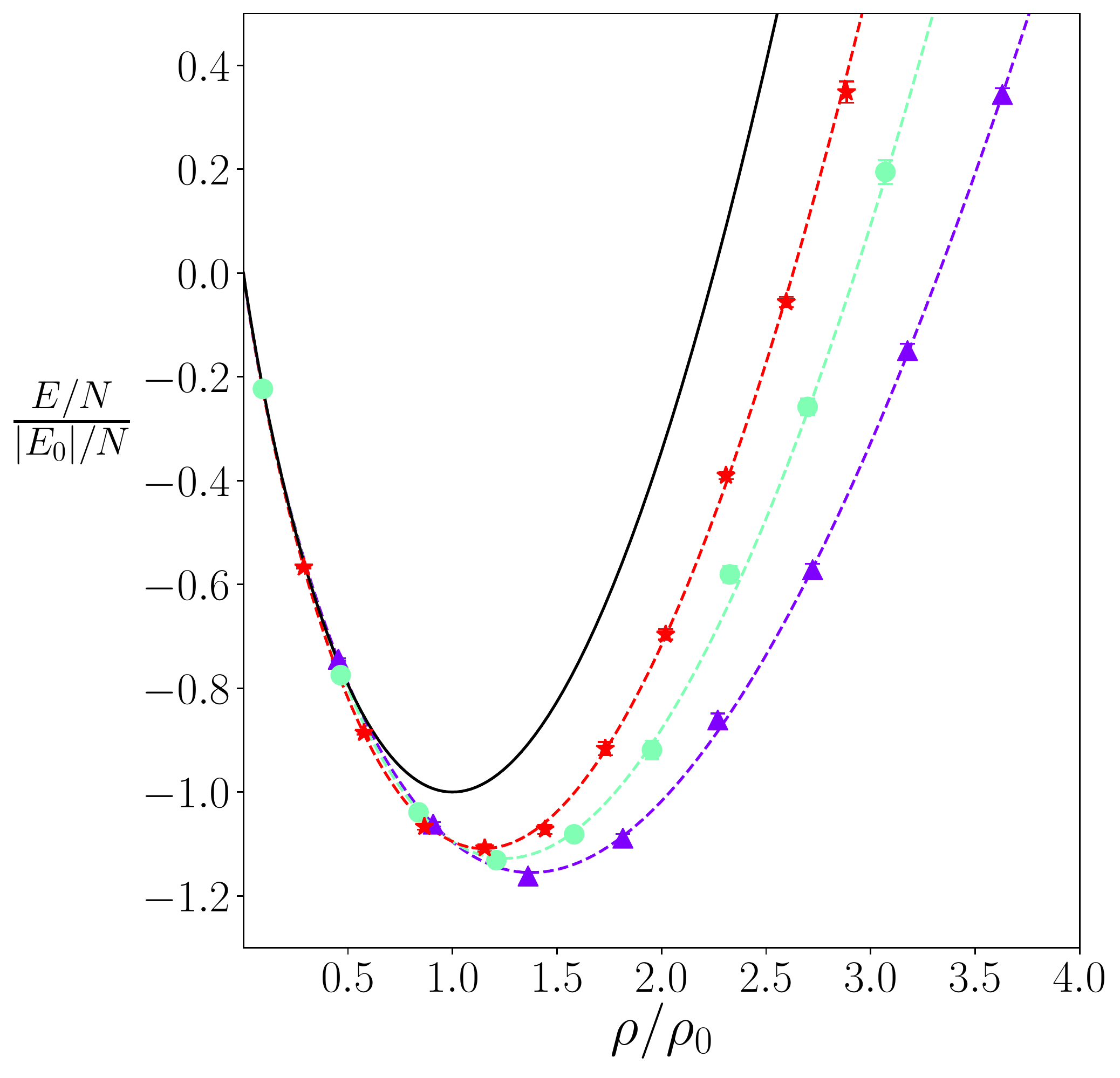}
		\caption{DMC energy per particle as a function of the density. From bottom (purple triangles) to top (red stars), the results correspond to
			magnetic fields  $B$=56.23, 56.453 and 56.639 G.
			Calculations were performed for the mean-field optimal  ratio $\rho_2 / \rho_1 = \sqrt{{a_{11} / a_{22}}}$. The energy per 
			atom  and atom density 
			are normalized to the  $|E_0|/N$ and $\rho_0$ MF+LHY values obtained  from Eqs. (\ref{eq_finite_range:en_0}) and (\ref{eq_finite_range:rho_0}), respectively. 
			The dashed lines are fits in the form $E/N = \alpha \rho + \beta \rho^\gamma$. The black solid line corresponds to the MF+LHY theory, 
			Eq. (\ref{eq_quadropole:mflhy_eos}).} 
		\label{fig_quad:eoscombineddiffbuniversal}
	\end{figure}	
	The energy per atom Eq. (\ref{eq_quad:eos}) allows one to readily introduce, within LDA,  a density functional whose interacting part is 
	\begin{equation}
	\mathcal{E}_{\rm int} = \rho \frac{E}{N} = \alpha \rho^2 + \beta\rho^{\gamma + 1} \ .
	\end{equation}
	A similar expression     holds in the MF+LHY approach.    
	In the homogeneous phase, one may easily obtain the pressure 
	\begin{equation}
	p(\rho) = \rho^2 \dfrac{\partial}{\partial \rho}\left(\dfrac{E}{N}\right) = \alpha \rho^2 + \beta \gamma \rho^{\gamma+1}
	\end{equation}
	and incompressibility
	\begin{equation}
	\kappa(\rho) =\rho \frac{\partial p}{\partial \rho},
	\end{equation}
	which can be written as 
	\begin{equation}
	\kappa(\rho) =\rho^2 \frac{\partial^2 \mathcal{E}_{\rm int}}{\partial \rho^2}=
	\rho^2\left\{ 2 \frac{\partial}{\partial \rho} \left( \frac{E}{N}\right) + \rho \frac{\partial^2}{\partial \rho^2} \left(\frac{E}{N}\right) \right\}.
	\label{eq_quadropole:K}
	\end{equation}
	
\begin{figure}
		\centering
		\includegraphics[width=0.6\linewidth]{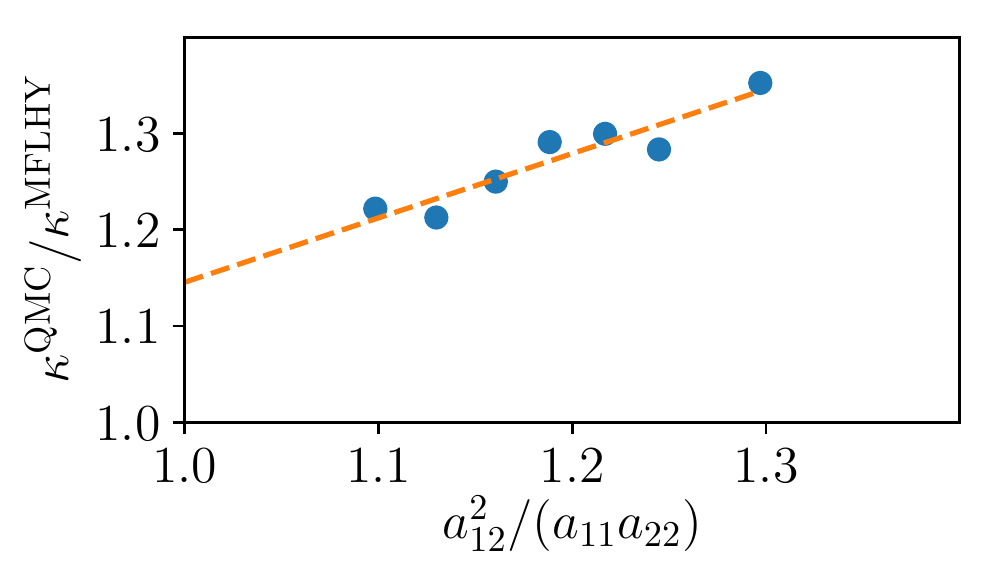}
		\caption{DMC  over MF+LHY incompressibility  ratio at equilibrium for the magnetic fields  considered in Chapter \ref{ch:finite_range_effects}. The dashed line is a linear fit to the points.
		}
		\label{fig:KQMC-o-MF}
	\end{figure}

\begin{figure}
		\centering
		\includegraphics[width=0.6\linewidth]{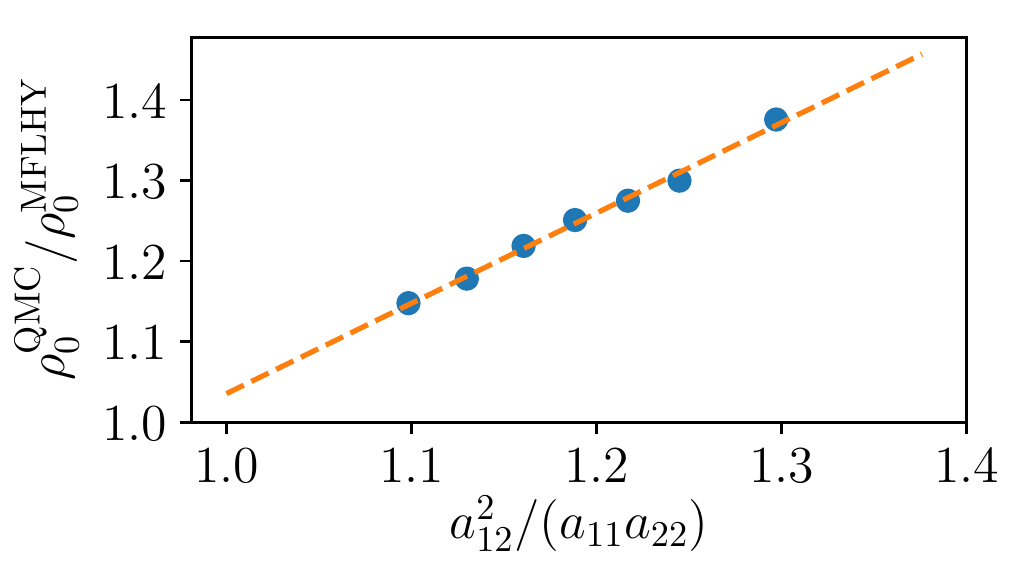}
		\caption{DMC  over MF+LHY equilibrium density ratio for the magnetic fields  considered in Chapter \ref{ch:finite_range_effects}. The dashed line is a linear fit to the points.
		}
		\label{fig:rhoQMC-o-MF}
	\end{figure}

	Figure  \ref{fig_quad:eoscombineddiffbuniversal} shows the  DMC energy per atom as a function of the density for selected values of the magnetic field, together
	with the result for the  MF+LHY theory. It is worth noticing the rather different equations of state yielded by the QMC functional and MF+LHY approaches. 
	The QMC approach yields a substantially larger equilibrium density and more binding. The QMC incompressibility is also larger,  as can be 
	seen in Fig. \ref{fig:KQMC-o-MF};
	at first sight, this seems to be in contradiction with the results in Fig.  \ref{fig_quad:eoscombineddiffbuniversal}, which clearly indicate that the curvature of the 
	$E/N$ vs $\rho$ curve at equilibrium ($\partial (E/N)/\partial \rho=0$ point) is smaller for the QMC functionals than for the MF+LHY approach. 
	However, this is compensated by the larger QMC value of the atom density at equilibrium, see Eq. (\ref{eq_quadropole:K}) and  Fig. \ref{fig:rhoQMC-o-MF},
	where we  show the ratio of QMC and MF+LHY equilibrium densities. Besides its importance for a quantitative description of the
	monopole droplet oscillations addressed here, inaccurate incompressibility may affect the description of processes  where the liquid-like
	properties of quantum droplets play a substantial role, as {\it e.g.} droplet-droplet collisions  \cite{ferioli2019collisions}. 
	
	\begin{table}
	    \centering
		\caption{Surface tension of a $^{39}$K Bose-Bose mixture at the MF+LHY optimal mixture composition 
			in $10^{-8} \times \hbar^2 / (m a_{11}^4)$ units.}
		\label{table_quadropole:surface_tension}
		\begin{tabular}{c  | c | c  }
			\hline
			$B({\rm G})$  &  $\sigma_{\rm MF+LHY}$ & $\sigma_{\rm QMC}$ \\ \hline
			56.230  &  35.1   &  48.8    \\ 
			56.453    &    9.31     &  12.2 \\ 
			56.639    &    1.21    & 1.46     \\ 
		\end{tabular}
	\end{table}

	Another fundamental property of  the liquid is the surface tension $\sigma$ of the free-surface. The origin of the definition of surface tension is as follows. Let us consider a semi-infinite system having a surface in the $z$-direction, with the translational invariance in $xy$ coordinates, and let the conditions of the density profile be $\rho(z\rightarrow -\infty) = \rho_0$ and $\rho(z\rightarrow\infty ) =0$, as in Ref. \cite{stringari1985surface}. Then, the surface tension is defined as the grand potential per unit surface
	\begin{equation}
		\label{eq:surface_tension_definition}
		\sigma = \dfrac{E - \mu N}{S} = \int_{-\infty}^{+\infty} dz \left\{  \mathcal{E}(\rho) - \mu \rho\right\},
	\end{equation}
	where $S$ is unit surface, $\mu$ is the chemical potential evaluated at the equilibrium density, and  $\mathcal{E}$ is energy density
	\begin{equation}
		\label{eq:quadropole_energy_density_surface}
		\mathcal{E} = \alpha \rho^2 + \beta \rho^{\gamma + 1} + \dfrac{\hbar^2}{2m} \dfrac{1}{4} \dfrac{(\nabla \rho)^2}{\rho},
	\end{equation}
	where the last term in Eq. (\ref{eq:quadropole_energy_density_surface}) is the kinetic energy density. Minimizing the energy density, or equivalently the surface tension, leads to the condition
	\begin{equation}
		\dfrac{\delta \mathcal{E}}{\delta \rho} = \mu,
	\end{equation}
	which gives the equation for the surface density profile along $z$
	\begin{equation}
		2\alpha \rho + \beta (1 + \gamma ) \rho^\gamma + \dfrac{\hbar^2}{8m} \left(\dfrac{\rho'^2}{\rho^2} -2 \dfrac{\rho''}{\rho}\right) = \mu.
	\end{equation}
	Multiplying by $\rho'$, integration yields
	\begin{equation}
		\label{eq:drdz_surface}
		\dfrac{d\rho }{dz} = - \left(\dfrac{\alpha\rho^{3} + \beta\rho^{\gamma + 2} - \mu \rho^2}{\frac{\hbar^2}{8m}}\right)^{1/2}.
	\end{equation}
	Since $(\nabla \rho)^2 = (d\rho / dz)^2$, the equations (\ref{eq:surface_tension_definition}), (\ref{eq:quadropole_energy_density_surface}) and (\ref{eq:drdz_surface}) can be combined for the simple quadrature formula of the surface tension
	\begin{equation}
	\label{eq_quadropole:surface_tension_formula}
	\sigma = 2 \int_{0}^{\rho_0} d\rho \left[\left(\dfrac{\hbar^2}{8m} \right) \left(\alpha \rho + \beta \rho^{\gamma} - \mu\right)\right]^{1/2}.
	\end{equation}

	The surface tension of several QMC functionals, i.e. functionals corresponding to different magnetic fields, is given in Table \ref{table_quadropole:surface_tension}. As can be seen, QMC functionals yield consistently 
	higher values of the surface tension than the MF+LHY approach.	
	Within MF+LHY, the surface tension  can be written in terms of the equilibrium density (\ref{eq_finite_range:rho_0}) and healing length (\ref{eq_quadropole:xi_mflhy})  \cite{petrov2015quantum}
	\begin{equation}
	    \sigma_{\rm MF+LHY} =  \dfrac{ 3(1 + \sqrt{3}) \rho_0 \hbar^2}{35 m \xi }   .
	\end{equation}

	\section{\label{sec_quad:method} The LDA-DFT approach}
	
	\subsection{Statics}
	
	Once $\mathcal{E}_{\rm int}[\rho]$ has been obtained, we have used density functional theory (DFT) to address the static and dynamic 
	properties of $^{39}$K droplets similarly as for superfluid $^4$He droplets \cite{ancilotto2017density}.    
	Within DFT, the energy of the quantum droplet  at the optimal composition mixture is written as a functional of the atom density $\rho({\mathbf r})$ as
	\begin{equation}
	E[\rho] = T[\rho] + E_c[\rho] =
	\frac{\hbar^2}{2m} \int d {\mathbf r} |\nabla \Psi({\mathbf r})|^2 +  \int d{\mathbf r} \,{\cal E}_{\rm int}[\rho],
	\label{eq_quad:eq1}
	\end{equation}
	where the first term  is the kinetic energy, and the effective wavefunction $\Psi({\mathbf r})$ of the droplet  is  related to the
	atom density as $\rho({\mathbf r})= |\Psi({\mathbf r})|^2$.  
	The equilibrium configuration is obtained by solving the Euler-Lagrange  equation arising 
	from the functional minimization of Eq. (\ref{eq_quad:eq1})
	\begin{equation}
	\left\{-\frac{\hbar^2}{2m} \nabla^2 + \frac{\partial {\cal E}_{int}}{\partial \rho}  \right\}\Psi 
	\equiv {\cal H}[\rho] \,\Psi  = \mu \Psi,
	\label{eq_quad:eq3}
	\end{equation}
	where $\mu$ is the chemical potential corresponding to the number of $^{39}$K atoms in the droplet, $N = \int d{\bf r}|\Psi({\bf r})|^2$. The time-dependent  version of  Eq. (\ref{eq_quad:eq3}) is obtained minimizing the action and adopts the form
	\begin{equation}
	i \hbar \frac{\partial}{\partial t}  \Psi({\mathbf r},t) = {\cal H}[\rho] \,\Psi({\mathbf r},t).
	\label{eq_quad:eq4}
	\end{equation}
	We have implemented a three-dimensional numerical solver based 
	on the Trotter decomposition of the time-evolution 
	operator with second-order accuracy  in the time-step $\Delta t$ ~\cite{chin2009any}. Within this scheme, it is possible to obtain both the ground 
	state and the dynamical evolution, as described in Sec. \ref{numerical_solution_gpe_equations}. It is known that in real time, the 
	Trotter decomposition may be unstable for different combinations of the time 
	and space steps used in the discretization \cite{chin2007higher}. To tackle 
	this problem, we have carefully chosen the time step that ensures that the 
	dynamic evolution is stable during the total propagation time.
	
	\begin{figure}
		\centering
		\includegraphics[width=0.7\linewidth]{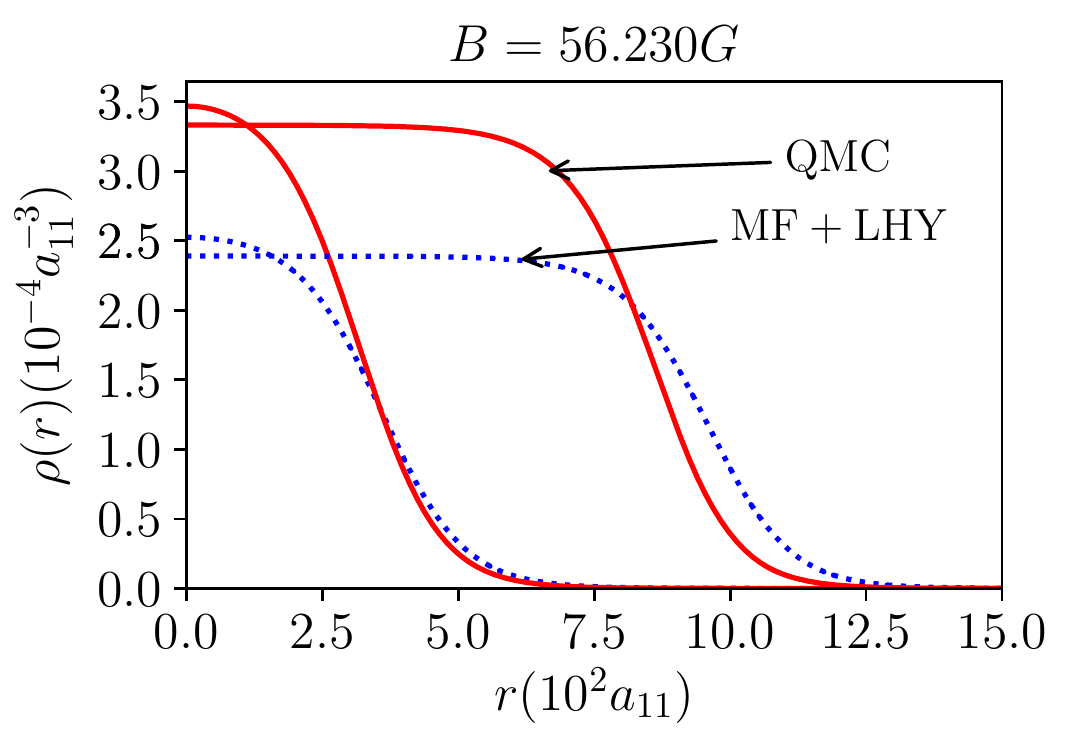}
		\caption{	Density profiles of two $^{39}$K quantum droplets corresponding to a small  $(\tilde{N} -\tilde{N}_c)^{1/4}=3$, and to a large droplet
			$(\tilde{N} -\tilde{N}_c)^{1/4}=6$, where $\tilde{N}_c=18.65$ is the critical number below which the droplet becomes unstable within the MF+LHY theory \cite{petrov2015quantum}. Solid lines, QMC functional; dotted lines, MF+LHY approach.		}
		\label{fig:density-profiles}
	\end{figure}    
	
	Figure \ref{fig:density-profiles} shows the density profile of two droplets, one corresponding to a small gaussian-like droplet and the other to a large saturated one. They have
	been obtained within the QMC ($B=56.230$ G) functional and MF+LHY methods. The sizeable difference between the 
	profiles yielded by both approaches reflects the different value of their equilibrium densities, see Fig. \ref{fig:rhoQMC-o-MF}.
	
	\subsection{Real-time dynamics and excitation spectrum}

	The multipole excitation spectrum of a quantum droplet can be obtained {\it e.g.} by solving the equations
	obtained linearizing Eq. (\ref{eq_quad:eq4}) \cite{dalfovo1999theory,petrov2015quantum,baillie2017collective}. We have used an equivalent method based on 
	the Fourier analysis of the real-time oscillatory response of the droplet to an appropriated external field \cite{stringari1979damping,pi1986time}.    
	The method, which we outline now, bears clear similarities with the experimental procedure to access to some excited states of confined 
	Bose-Einstein condensates (BEC) \cite{jin1996collective,altmeyer2007precision}. For monopole oscillations, our method is similar to that used in Ref. \cite{ferioli2020dynamical}.
	
	A droplet at the equilibrium, whose ground-state effective wavefunction $\Psi({\mathbf r})$ is obtained by solving the DFT Eq. (\ref{eq_quad:eq3}), is displaced from it by the action of a static  external  one-body field $Q$ whose intensity is controlled by a parameter $\lambda$. The new equilibrium wavefunction $\Psi'({\mathbf r})$ is
	determined by solving Eq. (\ref{eq_quad:eq3}) for the constrained Hamiltonian ${\cal H}'$
	\begin{equation}
	{\cal H} \rightarrow {\cal H}' = {\cal H} + \lambda Q .
	\label{eq16}
	\end{equation}
	If $\lambda$ is  small enough so that $\lambda Q$ is a perturbation and linear response theory applies, switching off $Q$  and letting 
	$\Psi'({\mathbf r})$ evolve in time according to Eq. (\ref{eq_quad:eq4}), $\langle Q(t) \rangle$ will oscillate around the equilibrium value
	$Q_{eq}=\langle \Psi({\mathbf r}) | Q | \Psi({\mathbf r})\rangle$.  Fourier analyzing $\langle Q(t) \rangle$, one gets the non-normalized strength function  corresponding to the excitation operator $Q$, which displays peaks at the frequency values corresponding to the excitation modes of the droplet. Specific values of $\lambda$ that we use are in the range from  $\lambda = 10^{-13}$ to $10^{-15}$ for the monopole modes, and $\lambda = 10^{-15}$ to $10^{-17}$ for the quadropole modes, with $\lambda$ being measured in $\hbar^2 / (2m a_{11}^4)$ units, and the smaller values corresponding to larger magnetic fields, i.e. less correlated drops.
	
	\section{\label{sec_quad:results} Results}
	
	We have used as excitation fields the monopole $Q_0$ and quadrupole $Q_2$ operators 
	\begin{eqnarray}
	Q_{\rm 0} & = & \sum_{i}^N  r_i^2 \\
	Q_{\rm 2} & = & \sum_{i}^N \left(r_i^2 - 3z_i^2\right) 
	\end{eqnarray}    
	which allows one to obtain the $\ell=0$ and 2 multipole strengths. The $\ell=0$ case corresponds to pure radial oscillations of the
	droplet and for this reason it is called ``breathing'' mode. In a pure hydrodynamical approach, its frequency is determined by the 
	incompressibility of the liquid  and the radius of the droplet \cite{bohigas1979sum,pitaevskii2016bose}.
	
\begin{figure}[H]
		\centering
		\includegraphics[width=\linewidth]{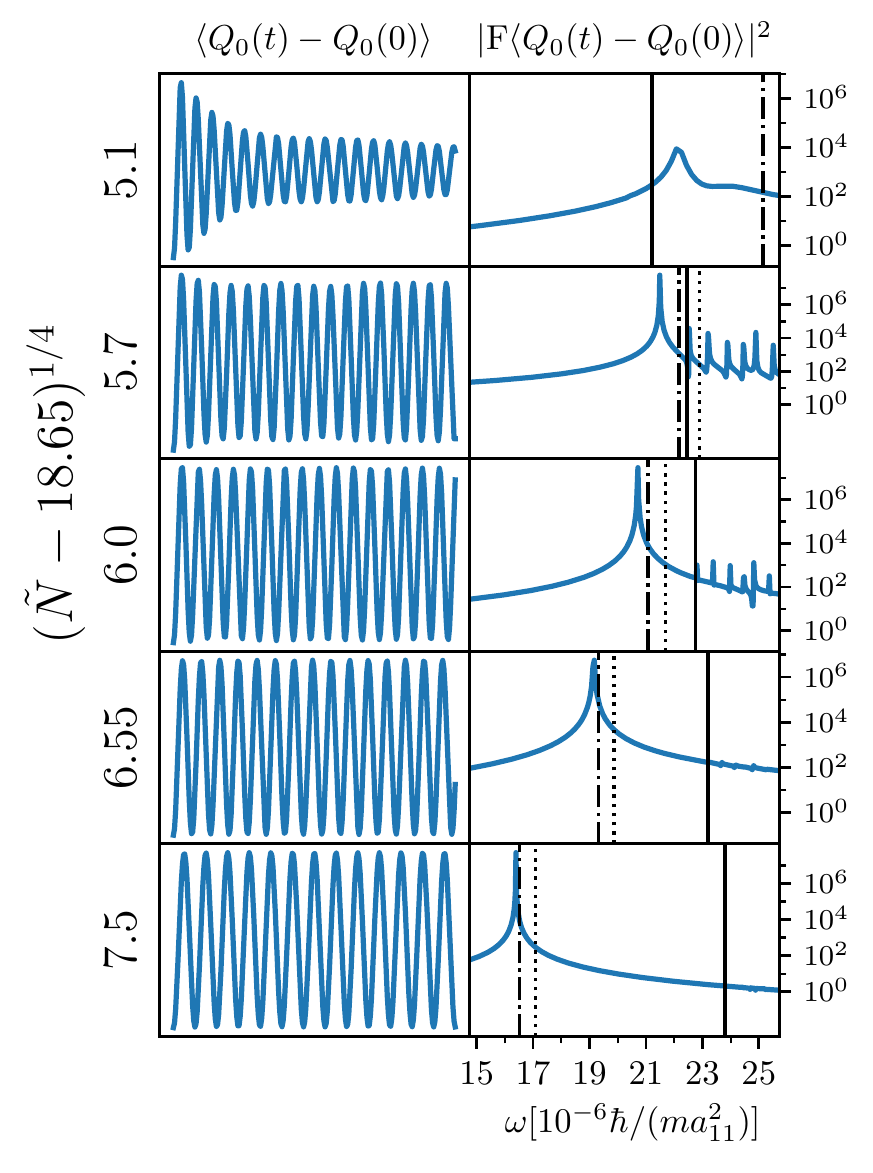}    
		\caption{ 
			Time evolution of the monopole moment $\langle Q_0(t) \rangle$  
			and  strength function (right) for $^{39}$K quantum droplets of different sizes
			obtained  using the QMC functional at $B = 56.230$ G. 
			In the right panels, the vertical solid line corresponds to
			the frequency $|\mu| / \hbar$ corresponding to the atom emission energy $|\mu|$, and the dotted and dash-dotted lines to the $E_3/\hbar$ and $E_1/\hbar$ frequencies, obtained by the sum rules in Eq. (\ref{eq_quadropole:monopole_E3}) and (\ref{eq_quadropole:monopole_E1}), respectively.
		}
		\label{fig:monopole}
	\end{figure}

	We have propagated the excited state  $\Psi'({\mathbf r})$ for a very long period of time,  storing  $\langle Q(t) \rangle$ and Fourier analyzing it.
	Fig. \ref{fig:monopole} (left)  shows $\langle Q_0(t) \rangle$ for $^{39}$K quantum droplets of different sizes. We choose the same scale of particle numbers (x-axis) as in Ref. \cite{petrov2015quantum}, as the monopole frequency $\omega_0$ close to the instability point $\tilde{N}_c = 18.65$ is directly proportional to $(\tilde{N} - \tilde{N}_c)^{1/4}$ \cite{petrov2015quantum}. Whereas a harmonic behavior 
	is clearly visible for the largest droplets, as corresponding to a single-mode excitation, for small droplets the radial oscillations are damped 
	and display different oscillatory behaviors (beats), anticipating
	the presence of several modes in the monopole strength, as the Fourier analysis of the signal unveils.
	
	Figure  \ref{fig:monopole} (right) displays the monopole strength function in logarithmic scale as a function of the excitation frequency. The solid vertical line represents the frequency $|\mu| / \hbar$ corresponding to the atom emission threshold, i.e. the absolute value of the atom chemical potential, $|\mu|$. It can be seen that for $(\tilde{N} -18.65)^{1/4} =5.1$  the strength  is in the continuum frequency region above $|\mu|/\hbar$.  
	Hence,  self-bound small $^{39}$K droplets, monopolarly excited, have excited states (resonances) that may decay by atom emission \cite{petrov2015quantum,ferioli2020dynamical}.
	This decay does not imply that the droplet breaks apart; it just loses the energy deposited into it by emitting a number of atoms, in a way  
	similar to the decay of  some states appearing in the atomic nucleus, the so-called ``giant resonances'' \cite{bohigas1979sum}.
	We want to stress that the multipole strength is not normalized, as it depends on the value of the arbitrary small parameter $\lambda$. However, 
	the relative intensity of the peaks for a given droplet is properly accounted for in this approach.
	
\begin{figure}[H]
		\centering
		\includegraphics[width=0.9\linewidth]{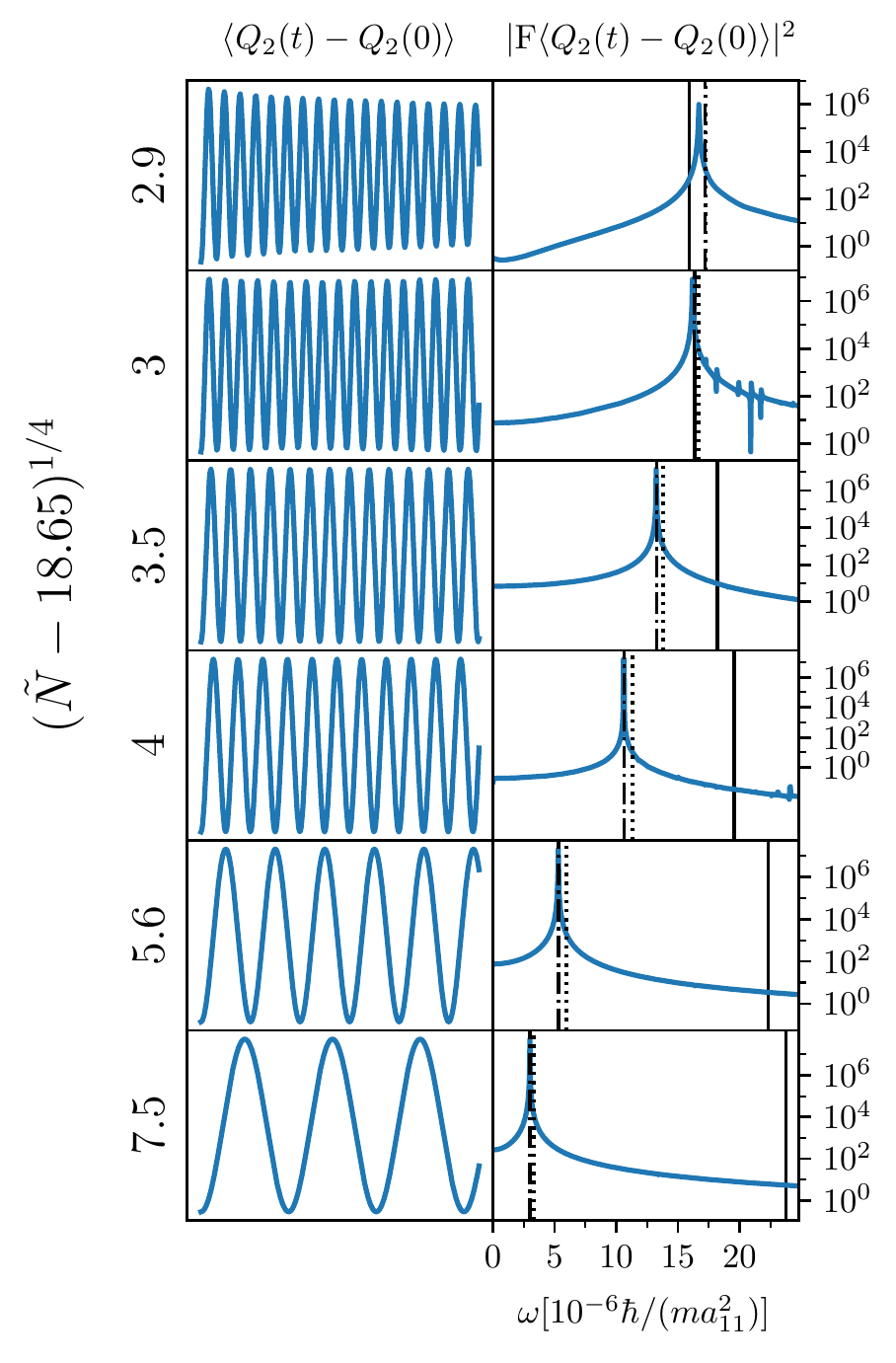}    
		\caption{  
			Time evolution of the quadrupole moment $\langle Q_2(t) \rangle$  
			and  strength function (right) for $^{39}$K quantum droplets of different sizes
			obtained  using the QMC functional at $B = 56.230$ G. 
			In the right panels, the vertical solid line corresponds to
			the frequency $|\mu| /\hbar$ corresponding to the atom emission energy $|\mu|$, and the dotted and dash-dotted lines to the $E_3/\hbar$ and $E_1/\hbar$ frequencies,  obtained by the sum rules in Eq. (\ref{eq_quadropole:quadropole_E3}) and (\ref{eq_quadropole:quadropole_E1}), respectively.			
		}
		\label{fig:quadrupole}
	\end{figure}

	A similar analysis for the quadrupole mode is  presented in Fig. \ref{fig:quadrupole}.
	In this case, we have found a more harmonic behavior for $\langle Q_2(t) \rangle$, 
	and therefore the quadrupole strength function  is dominated by one single peak.
	
	Figures  \ref{fig:monopole} and \ref{fig:quadrupole} show an interesting evolution of the strength function from the continuum to the discrete
	part of the frequency spectrum as the number of atoms in the droplet increases. For small $N$ values, but still corresponding to self-bound
	quantum droplets, the spectrum is dominated by  a broad resonance that may  decay by atom emission. The   
	$\langle Q (t) \rangle$ oscillations are damped, and when several resonances	are present (monopole case), distinct beats appear 
	in the oscillations.
	
	This remarkable evolution of the monopole and quadrupole  spectrum  has also been found for $^3$He 
	and $^4$He droplets \cite{serra1991collective,barranco1994response}. In the $^4$He case, it has been  experimentally confirmed by detecting 
	``magic'' atom numbers in the size distribution of $^4$He droplets which correspond to especially stable droplets  \cite{bruhl2004diffraction}. The
	magic numbers occur at the threshold sizes for which the excitation modes of the droplet, as calculated by the diffusion Monte Carlo method, 
	are stabilized when they pass below the atom emission energy. This constituted the  
	first experimental confirmation for the energy levels of $^4$He droplets. On the other hand,  in  confined BECs, the energy of the breathing mode is
	obtained by direct analysis of the radial oscillations of the atom cloud \cite{pitaevskii2016bose}.

\begin{figure}[H]
		\centering
		\includegraphics[width=0.6\linewidth]{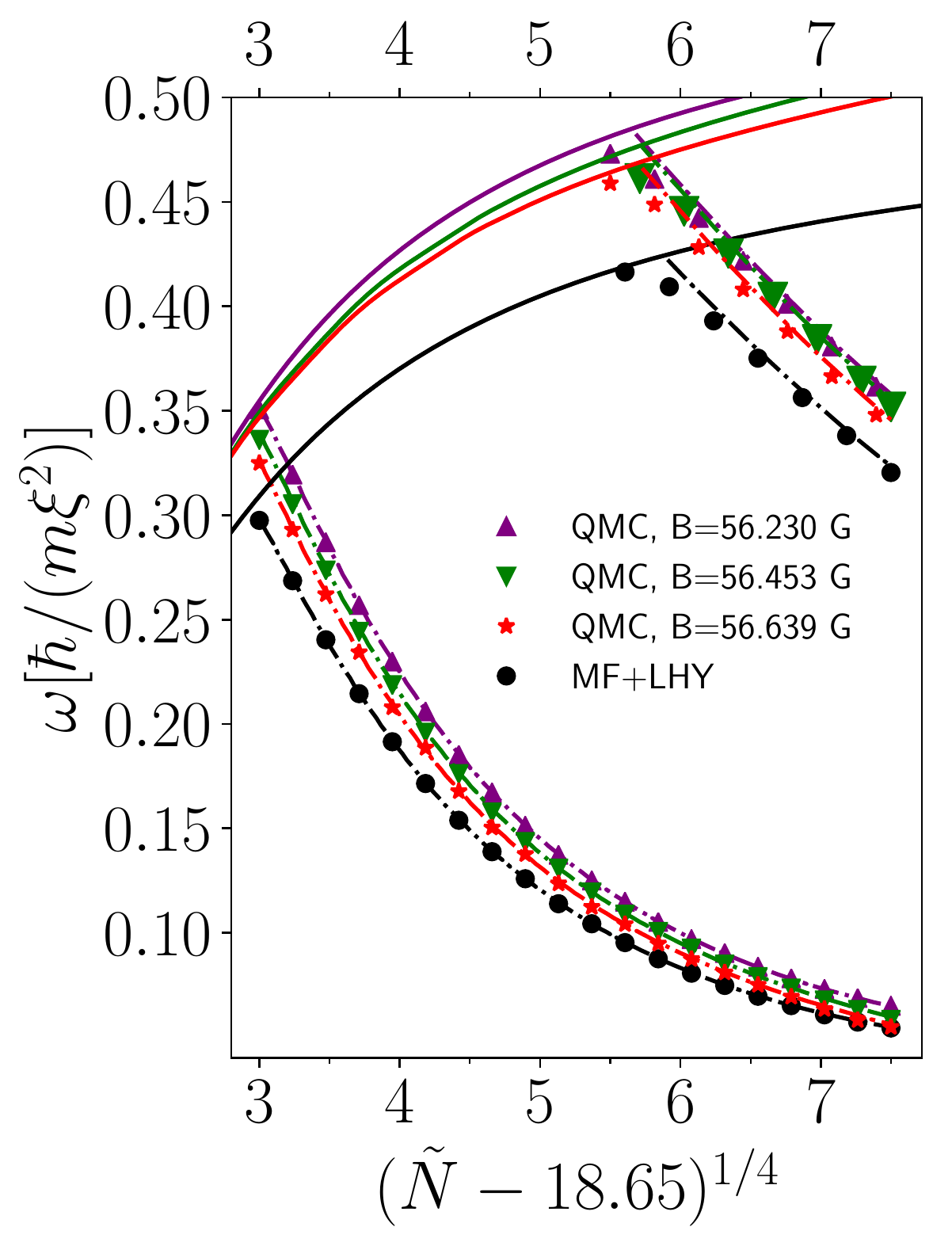}    
		\caption{Breathing (upper points) and quadrupole (lower points) frequencies as a function of the total atom number in  units of $\tilde{N}$. 
			Points are the results obtained from QMC and MF+LHY TDDFT calculations, and dashed lines
			are the $E_1/\hbar$ frequencies from the sum-rule approach (Eqs. (\ref{eq_quadropole:monopole_E1}) and (\ref{eq_quadropole:quadropole_E1})). Full lines represent the frequency corresponding to the absolute value of the droplet chemical potential $| \mu|$, corresponding to the legend from top to bottom
		}
		\label{fig:breathingmodecompareqmcmflhydiffb}
	\end{figure}

	We show in Fig. \ref{fig:breathingmodecompareqmcmflhydiffb}
	the breathing and quadrupole frequencies, corresponding to the more intense peaks, as a function of the number of atoms obtained with the QMC
	functional and the MF+LHY approach. For the latter, our results are in full 
	agreement with both those obtained using the
	Bogoliubov-de Gennes method \cite{petrov2015quantum}, which is fully 
	equivalent to ours, and with the ones of Ref. \cite{ferioli2020dynamical}. 
	The results are plotted in the universal units of the MF+LHY theory. 
	We find that the QMC functional predicts systematically larger monopole and quadrupole frequencies in all the range of particle numbers we have studied. 
	Additionally, as we change the magnetic field, i.e. the  scattering parameters, QMC predictions do not fall on the same curve, 
	meaning that the QMC functional breaks the MF+LHY universality.
	
	When the multipole strength is concentrated in a single narrow peak, it is possible to estimate the peak frequency using the sum rules approach
	\cite{bohigas1979sum,pitaevskii2016bose}. 
	Sum rules are energy moments of the strength function that, for some excitation operators,  can be written as 
	compact expressions involving expectation values on the ground state configuration. For the multipole operators considered here,
	two such sum rules are the linear-energy $m_1$ and cubic-energy $m_3$ sum rules. The inverse-energy sum rule $m_{-1}$ can be obtained from a 
	constrained calculation involving the Hamiltonian ${\cal H}'$ of Eq. (\ref{eq16}). Once determined, these three sum rules may be
	used to define two average energies $E_1 =\sqrt{m_1/m_{-1}}$  and  $E_3 =\sqrt{m_3/m_1}$ expecting, {\it bona fide}, that they are good estimates of
	the peak energy.
	
	For the monopole and quadrupole modes, the $E_1$ energies are \cite{bohigas1979sum}         
	\begin{equation}
	\label{eq_quadropole:monopole_E1}
	E_1 (\ell=0)= 
	\sqrt{- \dfrac{4 \hbar^2}{m} \,      \dfrac{\ave{r^2}}{\left( \partial \ave{Q_0} / \partial \lambda \right) \rvert_{\lambda = 0}}} 
	\end{equation}
	and
	\begin{equation}
	\label{eq_quadropole:quadropole_E1}
	E_1 (\ell=2) = \sqrt{-\dfrac{8 \hbar^2}{m} \, \dfrac{\ave{r^2}}{\left( \partial \ave{Q_2} / \partial \lambda  \right) \rvert_{\lambda = 0}}} \; ,
	\end{equation}
	with $\lambda$ being the parameter in the constrained Hamiltonian ${\cal H}'$, Eq.(\ref{eq16}),
	and $\ave{r^2} = \displaystyle \int d\vec{r}  \rho(r) r^2 / N$ evaluated at $\lambda = 0$.
	The frequencies corresponding to these energies are drawn in Figs. \ref{fig:monopole} and \ref{fig:quadrupole} as vertical dash-dotted lines.
	Except for small droplets, for which the monopole strength is very fragmented, one
	can see that they are good  estimates of the peak frequency.
	
	Closed expressions for the $E_3$ averages can be easiliy obtained for the monopole and the quadrupole modes \cite{bohigas1979sum,pitaevskii2016bose}. For the sake of completeness, we present the result obtained for the QMC functional. 
	
	Defining 
	\begin{eqnarray}
	E_{\alpha} &=& \alpha \int  d\mathbf{r} \rho^2(\mathbf{r})  
	\nonumber
	\\
	E_{\beta} &=& \beta \int  d\mathbf{r} \rho^{\gamma +1}(\mathbf{r})
	\nonumber
	\\
	\langle T \rangle & =& \frac{\hbar^2}{2m} \int d {\mathbf r} |\nabla \Psi({\mathbf r})|^2  \; ,
	\label{eq222}    
	\end{eqnarray}
	where $\Psi({\mathbf r})$  and $\rho(\mathbf{r})$  are those of the equilibrium configuration, we have 
	\begin{equation}
	\label{eq_quadropole:monopole_E3}
	E_3 (\ell=0) = \left[\frac{\hbar^2}{N m \langle r^2 \rangle}\right]^{1/2} [4  \langle T \rangle + 9 (E_{\alpha} +  \gamma^2 E_{\beta})]^{1/2}
	\end{equation}
	\begin{equation}
	\label{eq_quadropole:quadropole_E3}
	E_3 (\ell=2) = \left[\frac{\hbar^2}{N m \langle r^2 \rangle}\right]^{1/2} [4  \langle T \rangle]^{1/2}\; .
	\end{equation}
	We have $E_3 (\ell=2) < E_3 (\ell=0)$. The $\omega_3=E_3/\hbar$ frequencies are shown in Figs. \ref{fig:monopole} and  \ref{fig:quadrupole}
	as vertical dotted lines. It can be seen 
	that even when the strength is concentrated in a single peak, $\omega_3$ is a worse estimate of the peak frequency than  $\omega_1=E_1/\hbar$.
	This is likely so because $m_3$ gets contributions from the high energy part of the spectrum. At variance, since contributions to $m_{-1}$ 
	mainly come from the low energy part of the spectrum, $\omega_1$ is better suited for estimating the  peak frequency.
	
\begin{figure}[H]
		\centering
		\includegraphics[width=0.7\linewidth]{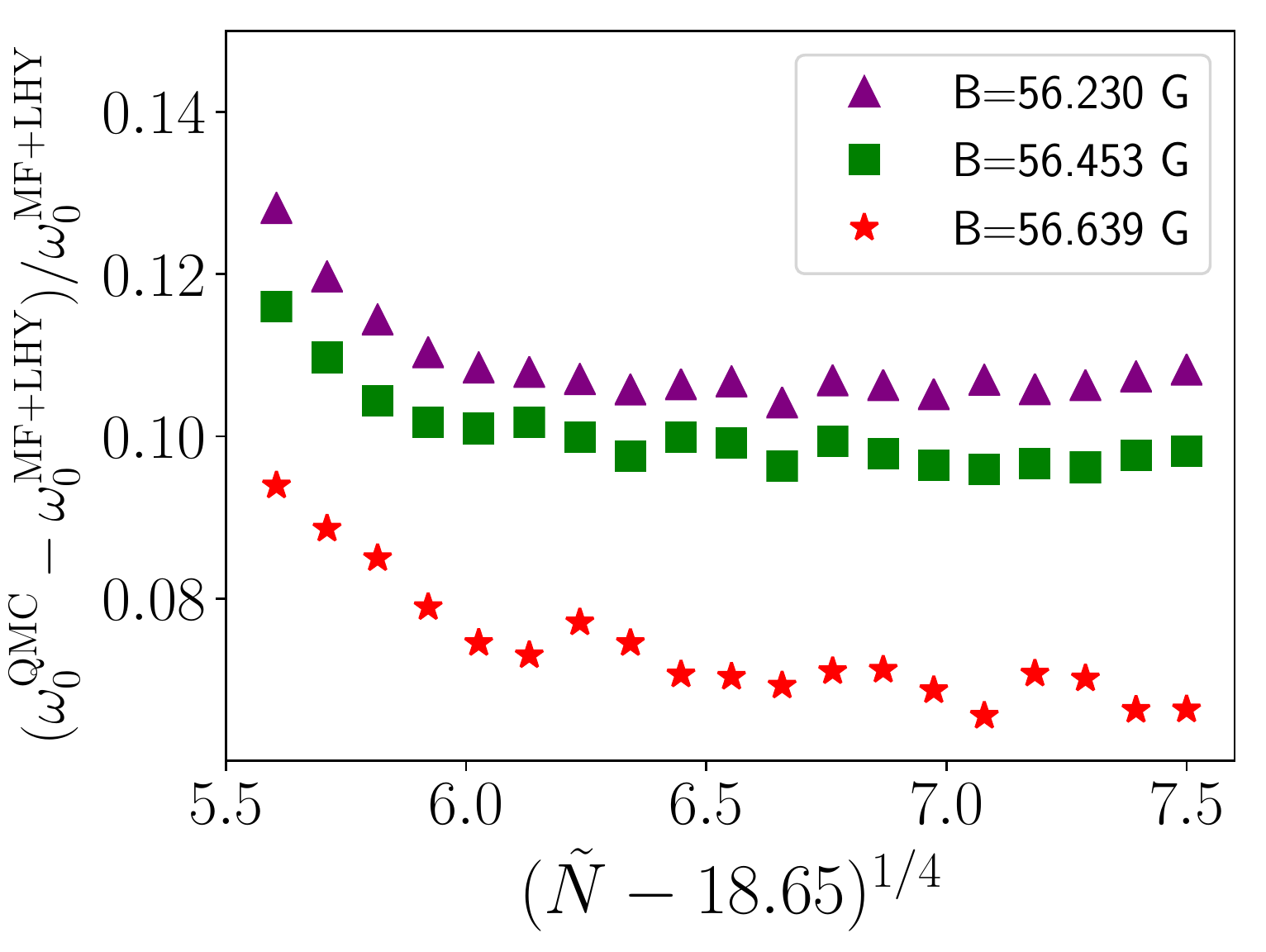}
		\includegraphics[width=0.7\linewidth]{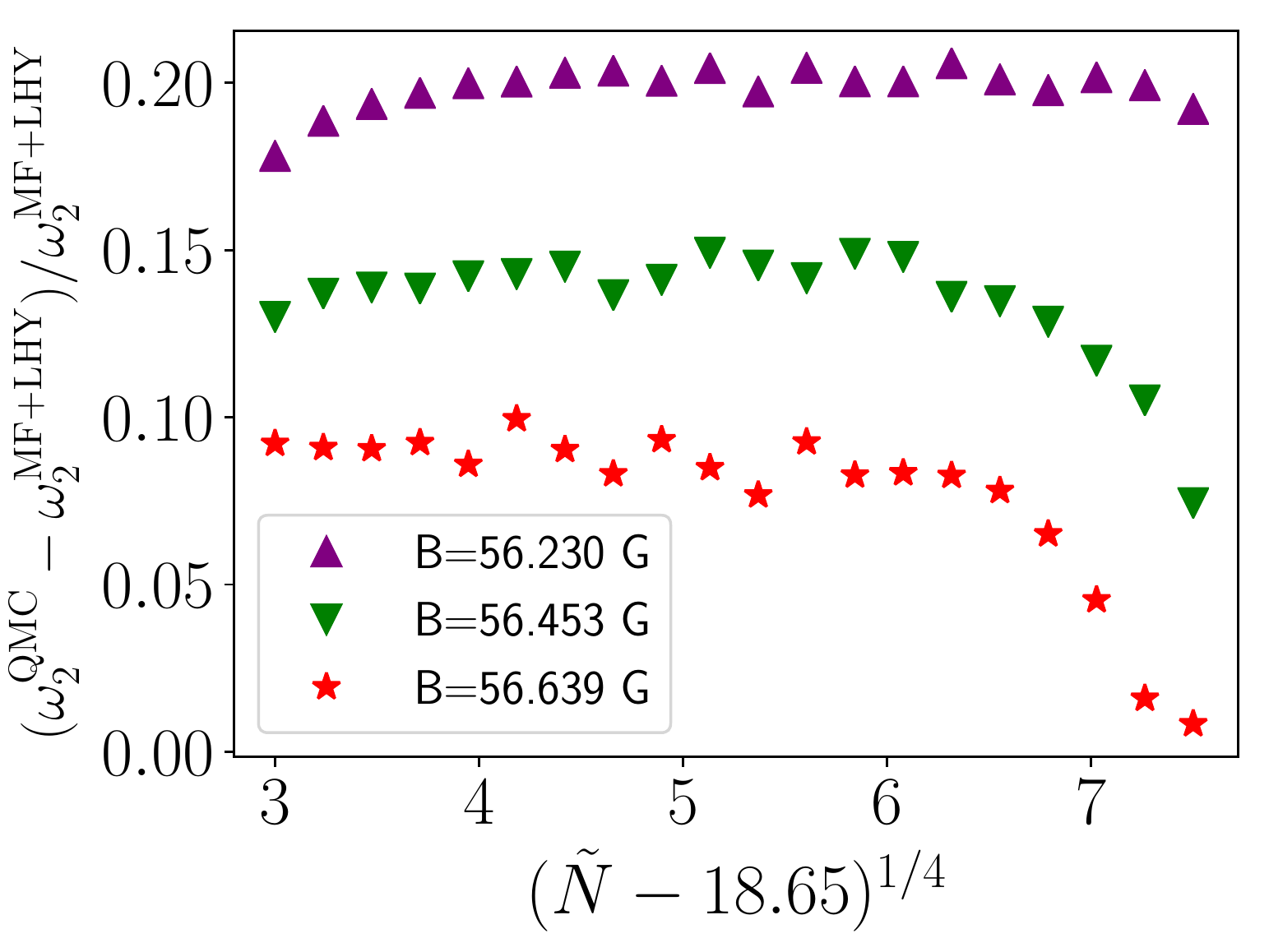}
		\caption{Relative frequency difference  between QMC and MF+LHY  TDDFT calculations for quadrupole (bottom figure) and monopole
			modes (top figure) as a function of the total atom number in  units of $\tilde{N}$.}
		\label{fig:relative_diff}
	\end{figure}

	The relative differences between the MF+LHY theory and the QMC functional for
	monopole and quadrupole frequencies  are presented in Fig. \ref{fig:relative_diff}. As the magnetic field 
	increases, the droplet is more correlated and differences of even $20\%$ can be observed.

\begin{figure}[H]
		\centering
		\includegraphics[width=\linewidth]{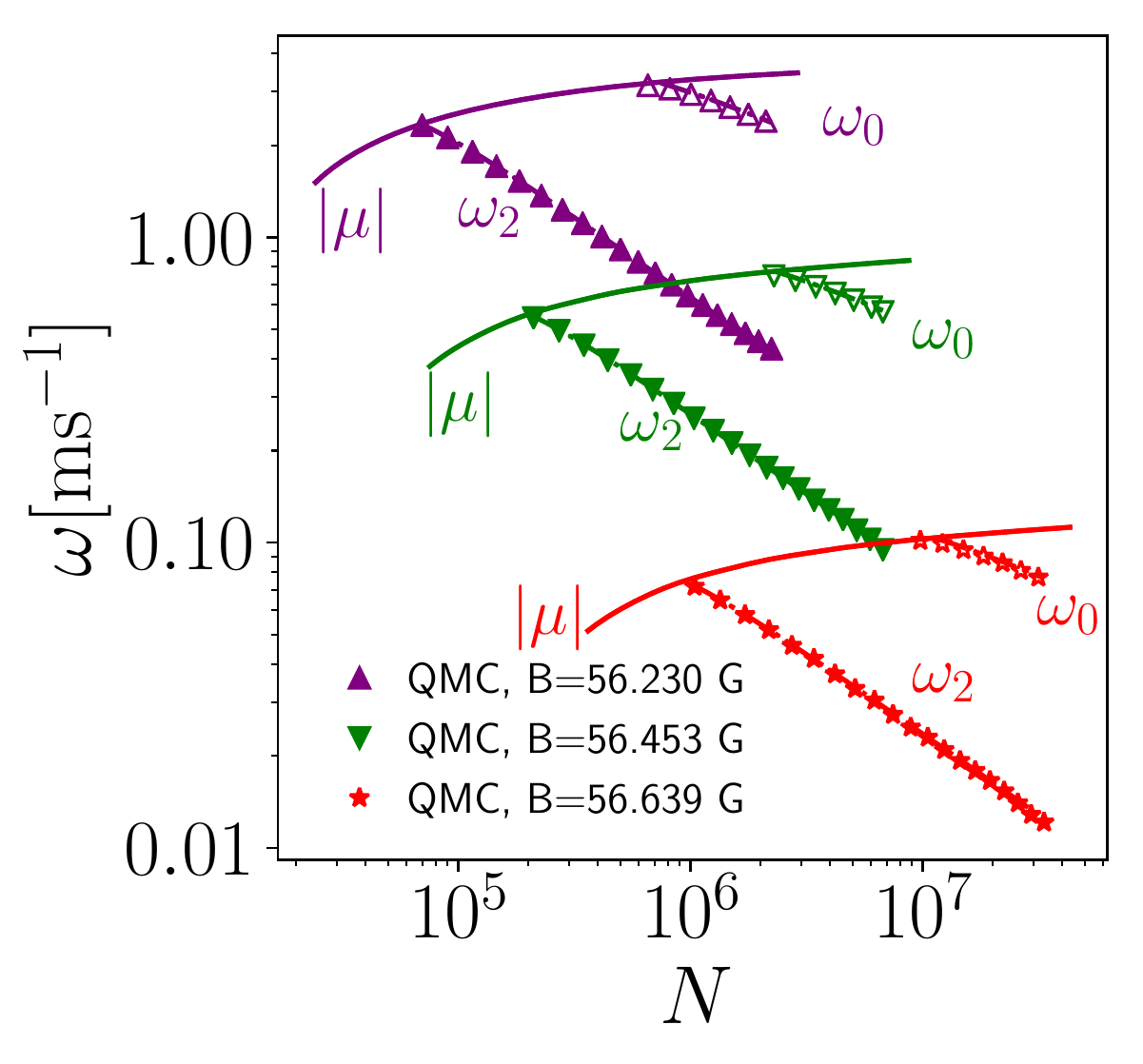}
		\caption{Predictions of the frequency corresponding to the absolute value of the chemical potential $|\mu|$, breathing frequency $\omega_0$ and quadropole frequency $\omega_2$ as a function of total atom number, using the QMC functionals. Dashed lines are $\omega_1  = E_1 / \hbar$ frequencies.}
		\label{fig:bothmodescompareqmcmflhydiffbrealisticunits}
	\end{figure}

	We finally compare in more detail the frequencies obtained with the QMC and MF+LHY functionals at  $B = 56.230$ G 
	for $\tilde{N} = 100$ and $\tilde{N} =1010$, which correspond to $N = 7 \times 10^4$ and $N= 7.1\times 10^5$, 
	respectively. Although it might require rather large droplets to observe neat breathing oscillations, systems with $\tilde{N}>100$, for which clean quadrupole modes show up (see Fig. \ref{fig:breathingmodecompareqmcmflhydiffb}),
	are already accessible in experiments \cite{cabrera2018quantum,semeghini2018self,ferioli2019collisions,derrico2019observation}. 
	For $N=7 \times 10^4$,   the quadrupole frequencies are
	$\omega_2^{\rm QMC} = 2323 \,{\rm Hz}$ and $\omega_2^{\rm MF+LHY} = 1972 \, {\rm Hz}$, i.e. oscillation 
	periods $\tau_2^{\rm QMC} = 2.70 \, {\rm ms}$ and $\tau_2^{\rm MF+LHY} = 3.19 \, {\rm ms}$. A similar comparison can 
	be made for the monopole frequency; for $N = 7.1 \times 10^5$,  the frequencies are
	$\omega_0^{\rm QMC} = 3114 \,{\rm Hz}$ and $\omega_0^{\rm MF+LHY} = 2755 \,{\rm Hz}$, and the 
	oscillation periods are $\tau_0^{\rm QMC} = 2.02 \,{\rm ms}$, and $\tau_0^{\rm MF+LHY} = 2.28 \,{\rm ms}$. In Fig. (\ref{fig:bothmodescompareqmcmflhydiffbrealisticunits}), we report our results for the breathing and quadrupole modes in not-reduced units to facilitate future comparisons with experiments.

	\section{\label{sec_quad:summary} Summary and outlook}
	
	Using a new QMC-based density functional which properly incorporates
	finite-range effects, we have determined the monopole and quadrupole excitation modes of $^{39}$K quantum droplets at 
	the optimal MF+LHY mixture composition.  Comparing with the results obtained within the MF+LHY approach, we have found that finite-range effects have a 
	detectable influence on the excitation spectrum, whose study  may thus be a promising way to explore physics beyond the LHY correction.
	
	We have shown  that introducing the QMC functional into the usual DFT methodology can easily be done, as only  minor changes need to be made in the (many)  existing Gross-Pitaevskii numerical 
	solvers~\cite{antoine2015gpelab,wittek2015extended,schloss2018gpue}.
	This opens the door to using better functionals --based on including  quantum  effects beyond mean-field--	in the current applications of the extended Gross-Pitaevskii approach  \cite{astrakharchik2018dynamics,ferioli2019collisions}.



\chapter{\label{chapther:repulsive_mixtures} Harmonically trapped Bose-Bose mixtures with repulsive interactions}

\section{\label{sec_repulsive:introduction}Introduction}

Before the adventure of BEC gases, the study of Bose-Bose mixtures was purely academic 
since there was not such a stable system in Nature. Nevertheless, its 
stability was deeply studied in connection with the stable $^3$He-$^4$He 
mixture at low $^3$He concentration. In particular, it was proved that the 
Bose-Bose mixtures of isotopic Helium are always unstable and phase separate 
and that only the right consideration of the Fermi nature of $^3$He atoms could 
account for its finite miscibility~\cite{massey1970theory,miller1978theory,kurten1982stability,chakraborty1982structure}. 

The very low density of the BEC gases makes feasible a theoretical description 
where atomic interactions are modeled by a single parameter, the $s$-wave 
scattering length. The miscibility of bulk Bose-Bose mixtures can be easily 
derived within mean-field theory (see Sec. \ref{sec:stability_bose_mixture} and references \cite{ho1996binary,pu1998properties,ohberg1999stability}) and the predictions of 
this approximation 
account well for the observed properties in different experiments. However, 
quite recently this simple argument has been questioned when the mixtures are  
harmonically trapped and a new parameter, based on the shape of the density 
profiles, has been suggested~\cite{lee2016phase}. The theoretical descriptions rely on 
the 
Gross-Pitaevskii (GP) equation whose range of applicability has been assumed to 
fit in the relevant experimental setups. Recently, it has been proved that if 
the interspecies interaction is attractive, instead of repulsive, the mixture 
can be stable due to the Lee-Huang-Yang correction which cancels the 
mean-field collapse~\cite{petrov2015quantum}, resulting in a self-bound 
(liquid) system.

In the present chapter, we report results of harmonically trapped repulsive 
Bose-Bose mixtures in different interaction regimes and with different masses 
for the constituents. Our approach is microscopic and relies on the use of 
quantum Monte Carlo methods able to solve exactly a given many-body Hamiltonian 
for Bose systems (within some statistical noise). The number of particles of 
our system is much smaller than the typical values used in GP calculations due 
to the complexity of our approach but this allows for an accurate study 
of effects 
going beyond the mean-field treatment. The numerical simulations have been 
carried out 
for different combinations of the interaction strengths, covering the mixed and 
phase separated regimes. In agreement with previous GP results, the miscibility 
rule derived for homogeneous gases fails to describe some of the results.

The rest of this chapter is organized as follows. In the next Section we introduce 
the theoretical method used for the study. Sec. III contains the density 
profiles corresponding to the points of the phase diagram here analyzed, both 
in miscible and phase-separated regimes. 
In Sec. IV, we study the scaling in terms of the GP interaction strength. Sec. 
V discusses the universality of our results by changing the model potential.
Finally, Sec. VI reports the main conclusions of our work.

\section{\label{sec_repulsive:method}Model and the methods}

We study a mixture of two kind of bosons with masses $m_1$ and $m_2$, 
harmonically confined and at zero temperature. The Hamiltonian of the system is
\begin{eqnarray}
H & = &  -\frac{\hbar^2}{2} \sum_{\alpha=1}^{2} \sum_{i=1}^{N_{\alpha}} 
\frac{\nabla_i^2}{m_{\alpha}} +
\frac{1}{2}\sum_{\alpha, \beta=1}^{2} \sum_{i_{\alpha}, 
	j_{\beta=1}}^{N_{\alpha}, N_{\beta}} V^{(\alpha, \beta)} 
(r_{i_{\alpha}j_{\beta}}) \nonumber \\ 
& &		+		\sum_{\alpha=1}^{2} \sum_{i}^{N} V_{\rm 
	ext}^{(\alpha)} (\mathbf{r}_{i})
\label{hamiltonian_repulsive_mixture} \ .
\end{eqnarray}
The mixture is composed of $N=N_1+N_2$ particles, with $N_1$ and $N_2$ bosons 
of type 1 and 2, respectively. The interaction between particles is modeled by 
the potentials $V^{(\alpha, \beta)} (r_{i_{\alpha}j_{\beta}})$ and the 
confining potential is a standard harmonic term, with frequencies that can be 
different for each species,
\begin{equation}
V_{\rm ext}^{(\alpha)} (\mathbf{r})
=   \frac{1}{2} m_{\alpha} \omega_{\alpha}^2 r^2 \ .
\label{harmonic}
\end{equation}

The many-body problem is solved by means of the diffusion Monte Carlo method 
(DMC), presented in Chapter \ref{ch:qmc_methods}. In the present problem, we have chosen a Jastrow model for the trial 
wave function,
\begin{eqnarray}
\Psi(\mathbf{R}) & = &
\prod_{1=i<j}^{N_1} f^{(1,1)}(r_{ij})
\prod_{1=i<j}^{N_2}f^{(2,2)}(r_{ij}) \prod_{i,j=1}^{N_1, 
	N_2}f^{(1,2)}(r_{ij}) \nonumber \\
& & \times \prod_{i=1}^{N_1} h^{(1)}(r_i)
\prod_{i=1}^{N_2} h^{(2)}(r_i) \ ,
\label{trialw}
\end{eqnarray}
with $\mathbf{R}=\{\mathbf{r}_1,\ldots,\mathbf{r}_N\}$, $f^{(\alpha,\beta)}(r)$ 
the two-body Jastrow factors accounting for the pair interactions between 
$\alpha$ and $\beta$ type of atoms, and $h^{(\alpha)}(r)$ the one-body terms 
related to the external harmonic potential. 
Systematic errors, which are the time-step and population-size
biases, were investigated and reduced below
the statistical noise; we used about 100 walkers and a time step of around $10^{-3} m_1 a_{11}^2 / \hbar$.

As one of the objectives of our work is to estimate the validity regime for a 
mean-field approach we have also studied the problem by solving the 
Gross-Pitaevskii (GP) equations. In this case, one assumes contact interactions 
between particles,
\begin{equation}
V^{(\alpha, \beta)} (r_{i_{\alpha}j_{\beta}})
=
g_{\alpha \beta} \, \delta(|\mathbf{r}_{i_{\alpha}} - 
\mathbf{r}_{j_{\beta}}|) \ ,
\label{contact}
\end{equation}
with strengths
\begin{equation}
g_{\alpha, \beta} = \frac{2 \pi \hbar^2 a_{\alpha 
		\beta}}{\mu_{\alpha\beta}} \ ,
\label{gfactors}
\end{equation}
with $\mu_{\alpha\beta}^{-1}=m_{\alpha}^{-1}+m_{\beta}^{-1}$ the reduced mass and $a_{\alpha 
	\beta}$ the $s$-wave scattering length of the two-body interaction between 
$\alpha$ and $\beta$ particles. With the Hartree \textit{ansatz},
\begin{equation}
\Psi(\mathbf{R})
=
\prod_{i=1}^{N_1}
\phi_{1}(\mathbf{r}_i,t)
\prod_{j=1}^{N_2}
\phi_{2}(\mathbf{r}_j,t)
\label{hartree}            
\end{equation}
one obtains the coupled GP equations for the mixture~\cite{ho1996binary},
\begin{eqnarray}
i \hbar \frac{\partial \phi_{1} (\mathbf{r},t)}{\partial t}
& = &
\left(
-\frac{\hbar^2}{2 m_{1}} \nabla^2
+
V_{\rm ext}^{(1)}(\mathbf{r})
+
g_{11} |\phi_{1}(\mathbf{r},t)|^2  \right.  \nonumber \\
& &    +       g_{12} |\phi_{2}(\mathbf{r},t)|^2
\bigg) 
\phi_{1} (\mathbf{r},t) \ ,
\label{gp1}
\end{eqnarray}
\begin{eqnarray}
i \hbar \frac{\partial \phi_{2} (\mathbf{r},t)}{\partial t}
& = &
\left(
-\frac{\hbar^2}{2 m_{2}} \nabla^2
+
V_{\rm ext}^{(2)}(\mathbf{r})
+
g_{22} |\phi_{2}(\mathbf{r},t)|^2 \right. \nonumber \\
& &	+
g_{12} |\phi_{1}(\mathbf{r},t)|^2
\bigg) 
\phi_{2} (\mathbf{r},t) \ .
\label{gp2}
\end{eqnarray}
We have solved the GP equations (\ref{gp1},\ref{gp2}) by imaginary-time 
propagation using a 4th order Runge-Kutta method.

\section{\label{sec_repulsive:phases}Results }

We have explored the phase space of the Bose-Bose mixture using the DMC method 
and, in all cases, we have compared DMC with GP results in the 
same conditions. We 
drive our attention to the density profiles of both species since our main goal 
is to determine if the systems are miscible or phase separated. The results 
contained in this Section have been obtained by using hard-core potentials 
between the different particles,
\begin{equation}
V^{(\alpha\beta)}(r) = 
\begin{cases}
\infty  , &  r \leq a_{\alpha \beta}   \\
0 	  ,&  r > a_{\alpha \beta} \\
\end{cases}
\ ,
\label{hardcore}
\end{equation}
with $a_{\alpha \beta}$ the radius of the hard sphere which 
coincides with 
its $s$-wave scattering length. The Jastrow factor in the trial wave function 
(\ref{trialw}) is chosen as the two-body scattering solution, 
$f^{(\alpha,\beta)}(r)=1-a_{\alpha \beta}/r$, and the one-body term corresponds 
to the exact single-particle ground-state wave function, 
$h^{(\alpha)}(r)=\exp(-r^2/(2a_\alpha^2))$. The length $a_\alpha$ is optimized 
variationally but, even for the strongest interaction regime here studied, its 
value is at most $~10$\% larger than the one of the non-interacting system, 
$l_{\text{ho},\alpha}=\sqrt{\hbar/(m_\alpha \omega_\alpha)}$. The influence of 
internal parameters of the DMC calculations, such as the number of walkers and the
imaginary-time step, is analyzed and the reported results are converged with 
respect to them. The density profiles that we report are derived using the mixed 
estimation since we have checked that the correction introduced by using pure 
estimators~\cite{casulleras1995unbiased} is at the same level as the typical statistical noise.

\begin{figure}[]
	\centering	
	\includegraphics[width=\linewidth]{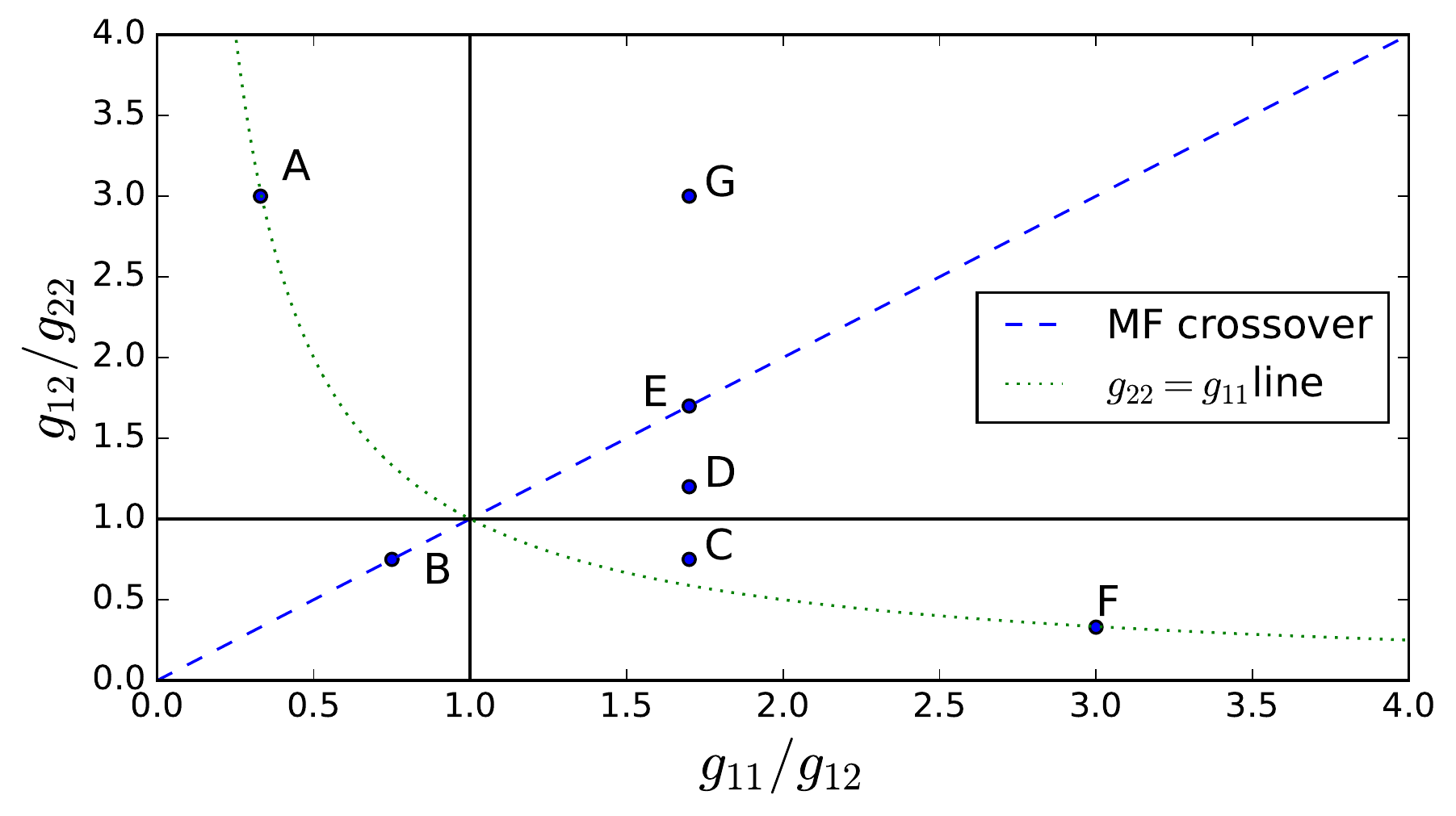}
	\caption{Representation of the phase space for the mixture as a function of the 
		interaction strengths $g_{\alpha \beta}$. The points correspond to the cases 
		studied, with coordinates given in Table 
		\ref{table1}. The mean-field theory for 
		homogeneous system predicts separation (mixing) for all the points above (below) 
		the mean-field critical line (dashed line). The dotted (green) line stands for 
		points where $g_{11}=g_{22}$.}
	\label{fig1}
\end{figure}

As in Ref.~\cite{lee2016phase}, we plot the phase space in terms of the adimensional 
variables 
$g_{12}/g_{22}$ and $g_{11}/g_{12}$. In Fig. \ref{fig1} we plot it showing the 
different regimes, with the line $g_{12}^2=g_{11} g_{22}$ standing for the 
critical line separating miscibility and phase separation, using the mean-field 
criterion.
In the figure, we plot the points which we have studied; they are selected to 
cover the most interesting areas of the phase space. In Table \ref{table1}, we 
report the specific coordinates of the interaction strengths.

\begin{table}[]
	\centering
	\begin{tabular}{c c c c}
		\hline \hline  \rule{0pt}{11pt}
		$\text{Label}$ & $g_{12}/g_{22}$ & $g_{11}/g_{12}$ & $\Delta$     \\ 
		\hline
		$\text{A}$  & $3.0$ & $0.33$ & $-0.89$  \\
		$\text{B}$  & $0.75$ & $0.75$ & 0  \\
		$\text{C}$  & $0.75$ & $1.7$ & $1.27$ \\
		$\text{D}$  & $1.2$ & $1.7$ & $0.42$ \\
		$\text{E}$  & $1.7$ & $1.7$ & 0 \\
		$\text{F}$  & $0.33$ & $3$ & 8 \\
		$\text{G}$  & $3$ & $1.7$ & -0.43 \\
		\hline \hline
	\end{tabular}	
	\caption{Representative phase space points analyzed in our study.  
		The values of $\Delta$ are obtained from Eq. (\ref{delta}).}
	\label{table1}	
\end{table}

As usual in the study of Bose-Bose mixtures we define an adimensional  
parameter $\Delta$,
\begin{equation}
\Delta = \frac{g_{11} g_{22}}{g_{12}^2} - 1 \ ,
\label{delta}
\end{equation} 	
which classifies the regimes of phase separation ($\Delta < 0$) and miscibility 
($\Delta > 0$) according to the mean-field treatment of bulk mixtures. When 
$\Delta=0$ we are on the critical line separating both regimes (dashed line in 
Fig. \ref{fig1}). In the results reported below, we also calculate the 
parameter $\Delta n$ defined as~\cite{lee2016phase}
\begin{equation}
\Delta n = \frac{\rho_1(0)}{\max \rho_1(r)} - \frac{\rho_2(0)}{\max 
	\rho_2(r)} \ ,
\label{deltan}
\end{equation}
which compares the value of the density profiles $\rho_\alpha(r)$ at the 
origin $r=0$ with its maximum value. Then, $\Delta n \simeq 0$ when the peaks 
of both density profiles coincide to be at the origin (mixed state) or when 
the ratio of the central and maximum density is the same for both species, which 
occurs when two species of the same mass separate to two blobs.  
For other types of phase separation $|\Delta n| > 0$.

The density profiles reported in this Section have been obtained with a total 
number of particles $N=200$ and considering a balanced mixture, i.e., 
$N_1=N_2$. We have repeated some calculations considering  
$N_1/N_2=\sqrt{g_{22}/g_{11}}$, which is the optimal balance from the 
mean-field theory \cite{petrov2015quantum}, 
and the differences in energy with respect to $N_1=N_2$ are at most 10\%. 
Therefore, we 
concentrate on the balanced mixture. The role of the confining frequencies 
$\omega_\alpha$ is a bit more relevant since we have observed changes in the 
results that can reach the 30\%. In general, the energies are lower when the 
frequencies obey the rule $m_1 \omega_1^2= m_2 \omega_2^2$ which corresponds to 
applying the same harmonic confinement for both species. Therefore, the results 
presented below correspond always to this choice.

\subsection{\label{mixed}$\mathbf{\Delta>0}$}

\begin{figure}[!h]
	\centering
	\includegraphics[width=0.9\linewidth]{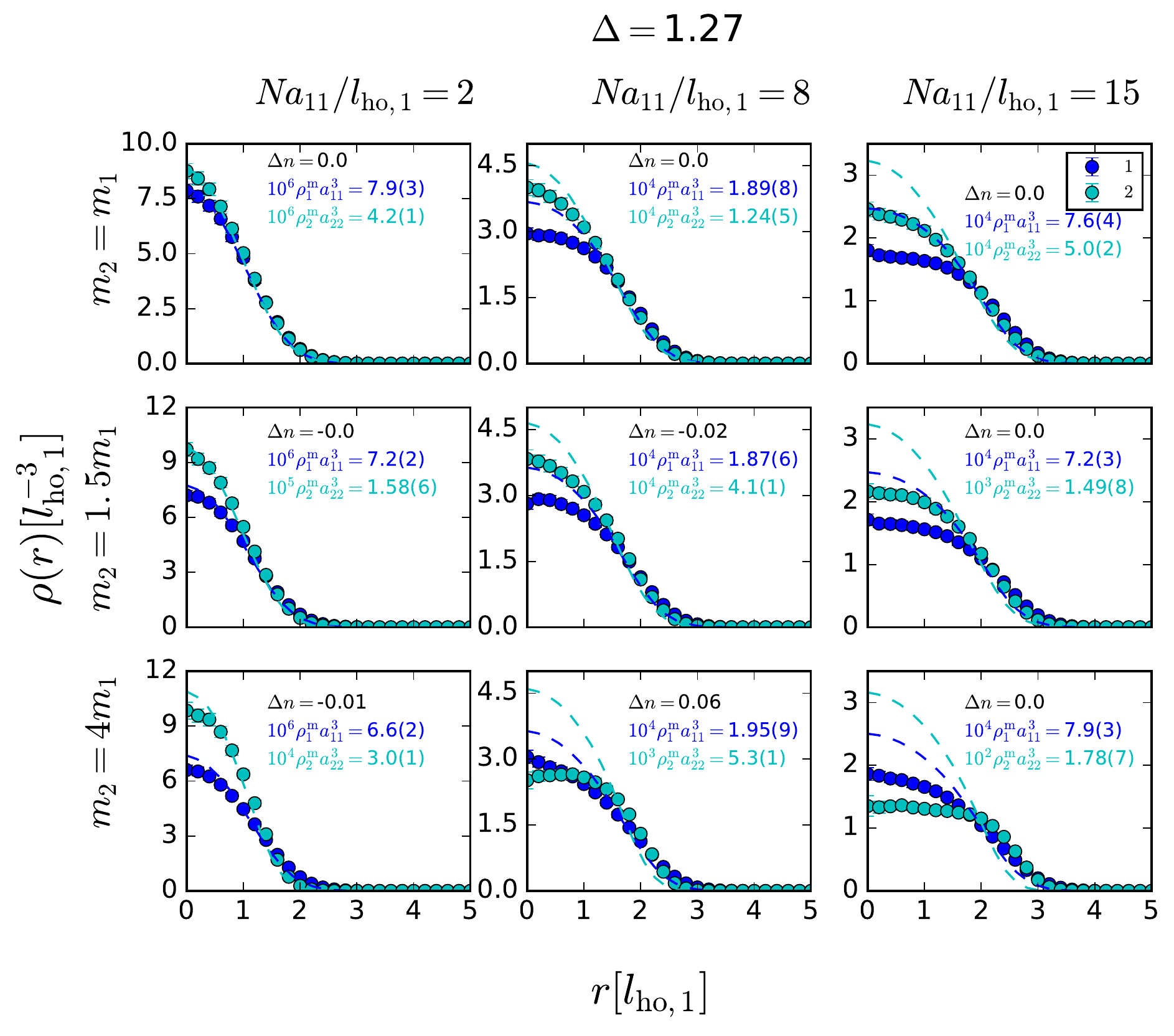}
	\caption{Point C of the phase space. The points correspond to the DMC results 
		and the lines to the solution of the GP equations for the same system.}
	\label{pointc}
\end{figure}

When $\Delta>0$ the mean-field criterion predicts
mixing between the 
two species. We have explored the confined system in three different points of 
the phase space. We start with point C, with 
$\Delta=1.27$ and the $g_{\alpha \beta}$ values reported in Table \ref{table1}. 
In Fig. \ref{pointc}, we show the density profiles of both species. In all 
cases, we use as unit length the harmonic oscillator length 
$l_{\text{ho},1}$ of species 1. In the nine subfigures of 
Fig. \ref{pointc}, 
going from left to right we increase the parameter  $N a_{11}/ 
l_{\text{ho},1}$, and from top to bottom we increase the mass of species 2. 
The parameter  $N a_{11}/l_{\text{ho},1}$ is chosen because it is the 
mean-field scaling variable contained in the GP equation. As we keep the total 
number of particles $N$ fixed, increasing that parameter means to increase  the 
scattering length of the 11 interaction and thus making $g_{11}$ larger. As 
the coordinates in the phase space are fixed at the point C, increasing 
$g_{11}$ implies that also the other strengths $g_{12}$ and $g_{22}$ 
increase. Therefore, moving to the right in the panels of Fig. \ref{pointc} 
means an increase of both the interspecies and intraspecies interaction. Moving 
down in the panels, for a fixed value $N a_{11}/l_{\text{ho},1}$, means 
an 
increase of the mass of species 2 and therefore an increase of the scattering 
lengths $a_{12}$ and $a_{22}$ because the couplings $g_{12}$ and $g_{22}$ are 
kept constant.

The density profiles shown in Fig. \ref{pointc} show in all cases a mixed 
state, with $\Delta n \simeq 0$ (\ref{deltan}). In the leftmost column, when 
the interaction is very soft, we appreciate that $\rho^{(\alpha)}(r)$ are 
basically Gaussians, following the shape of the non-interacting gas. When $m_2$ 
grows, they are still Gaussians but slightly different in shape because the 
frequencies are different. The Gaussian profiles disappear progressively moving 
to the right due to the increase of interactions. The comparison between DMC 
and GP shows agreement when the interaction is low and they clearly depart when 
$N a_{11}/l_{\text{ho}}^{(1)}$ grows. The DMC profiles show the emergence of a 
plateau close to $r=0$, in significant contrast with the GP prediction.

\begin{figure}[!h]
	\centering
	\includegraphics[width=0.9\linewidth]{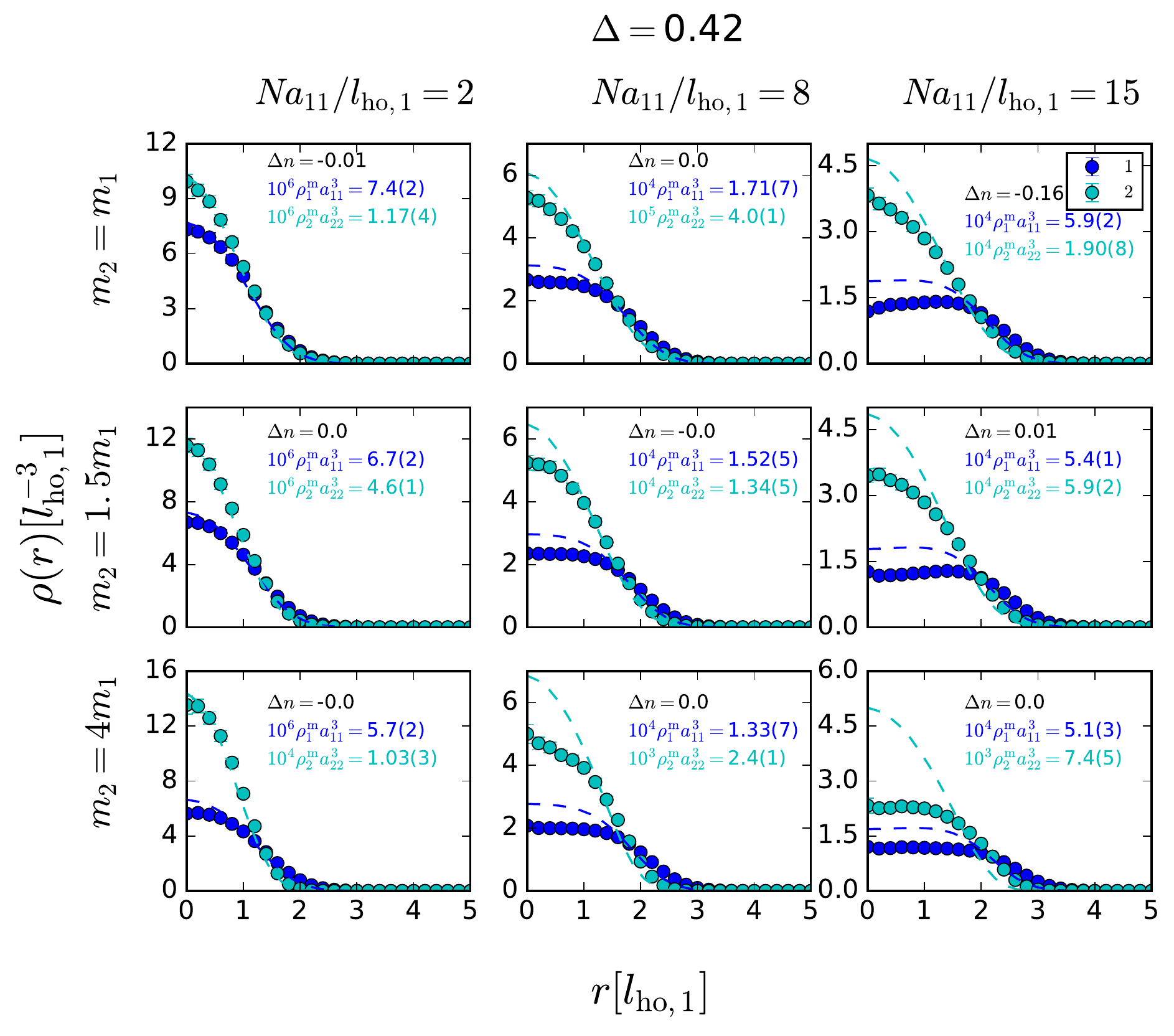}
	\caption{Point D of the phase space. The points correspond to the DMC results 
		and the lines to the solution of the GP equations for the same system.}
	\label{pointd}
\end{figure}
\begin{figure}[!h]
	\centering
	\includegraphics[width=0.9\linewidth]{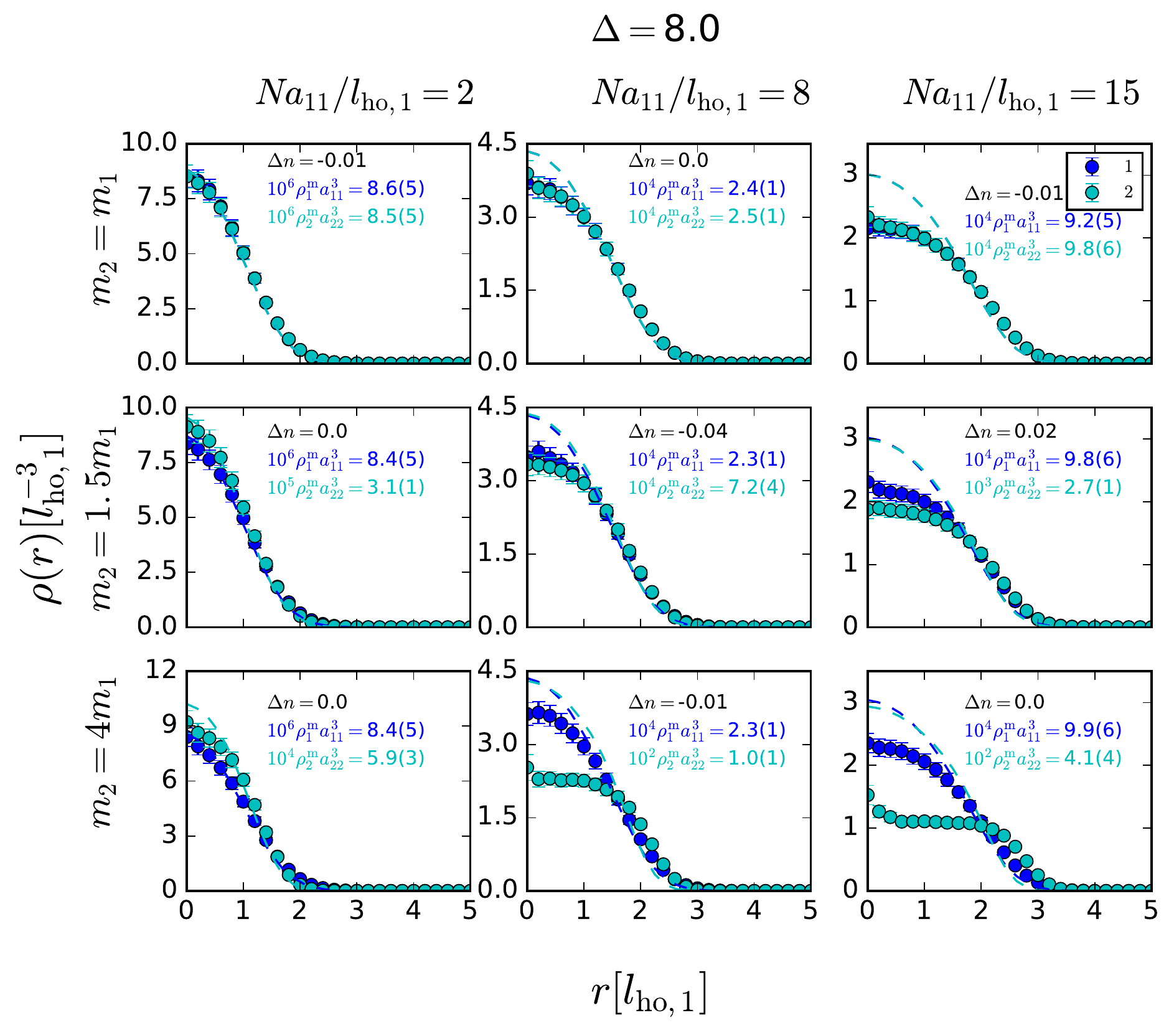}
	\caption{Point F of the phase space. The points correspond to the DMC results 
		and the lines to the solution of the GP equations for the same system.}
	\label{pointf}
\end{figure}

Density profiles for points D ($\Delta= 0.42$) and F ($\Delta=8$) are reported 
in Figs. \ref{pointd} and \ref{pointf}, respectively. The results are 
qualitatively similar to the ones of point C, showing mixing in both cases. 
Point F is deeply located in the mixed part of the phase space (large $\Delta$ 
value) and the agreement with GP is, in this case, quite satisfactory except 
quantitatively  when the mass difference is large and the strength of the 
interaction increases. Point $D$, 
with a small $\Delta$ value, shows slightly more significant  departures 
from GP 
predictions that again increase when the difference in mass between both 
species increases.

\subsection{\label{critical}$\mathbf{\Delta=0}$}

We have studied two points (B and E, see Table \ref{table1}) of the phase 
space which are illustrative of the $\Delta=0$ case. Assuming mean-field theory,
this corresponds to the critical line for mixing in bulk mixtures.

\begin{figure}[!h]
	\centering
	\includegraphics[width=0.8\linewidth]{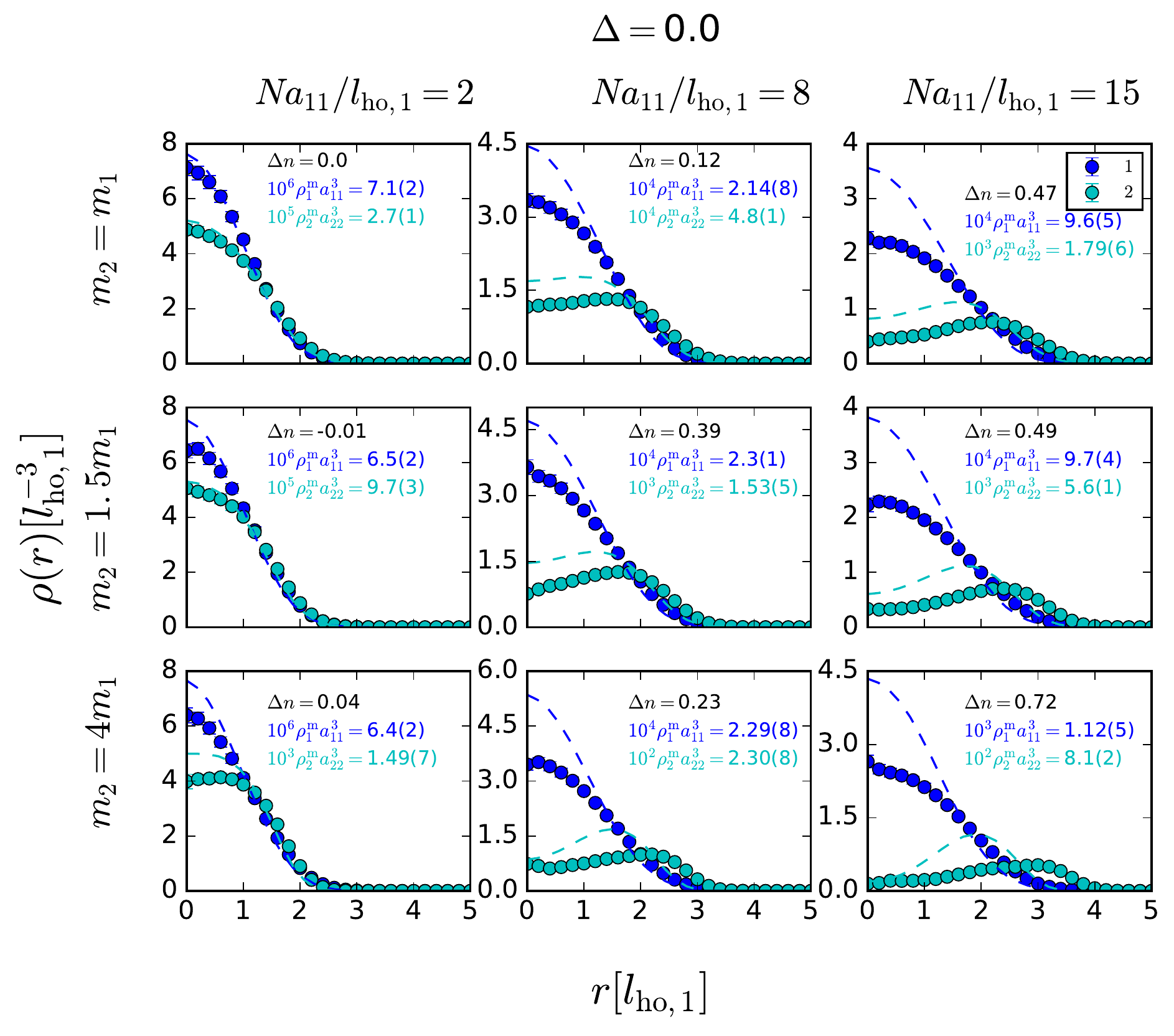}
	\caption{Point B of the phase space. The points correspond to the DMC results 
		and the lines to the solution of the GP equations for the same system.}
	\label{pointb}
\end{figure}

In point B, we have 
$g_{12}/g_{22} = 0.75$ and $g_{11}/g_{12} = 0.75$. With these values,  
$g_{22} > g_{12} > g_{11}$, and thus one expects that species 2 goes out of 
the trap center because it is more repulsive than species 1. This effect is 
emphasized when $m_2>m_1$ because the relation between scattering lengths has 
an additional factor $m_2/m_1$. In Fig. \ref{pointb}, we can see the evolution 
of the density profiles as a function of the interaction strength and mass 
ratio. When the system is only weakly interacting (left column) one appreciates 
Gaussian profiles that coincide with GP predictions. The situation changes when 
the interaction grows (second and third columns) as we can see that the two 
systems start to phase separate, a feature that is measured by the positive 
value of the factor $\Delta n$ (\ref{deltan}). If the difference in mass is 
enlarged, bottom panels, one can see that the phase separation is even more 
clear. In this case, the heaviest component (2) is manifestly going out of the 
center and thus it surrounds the core,  mainly occupied by the species 1. 
Comparison with GP shows that there is a qualitative agreement with DMC but 
quantitatively GP is rather inaccurate, specially when the mass ratio is 
$m_2/m_1=4$.

\begin{figure}[!h]
	\centering
	\includegraphics[width=0.8\linewidth]{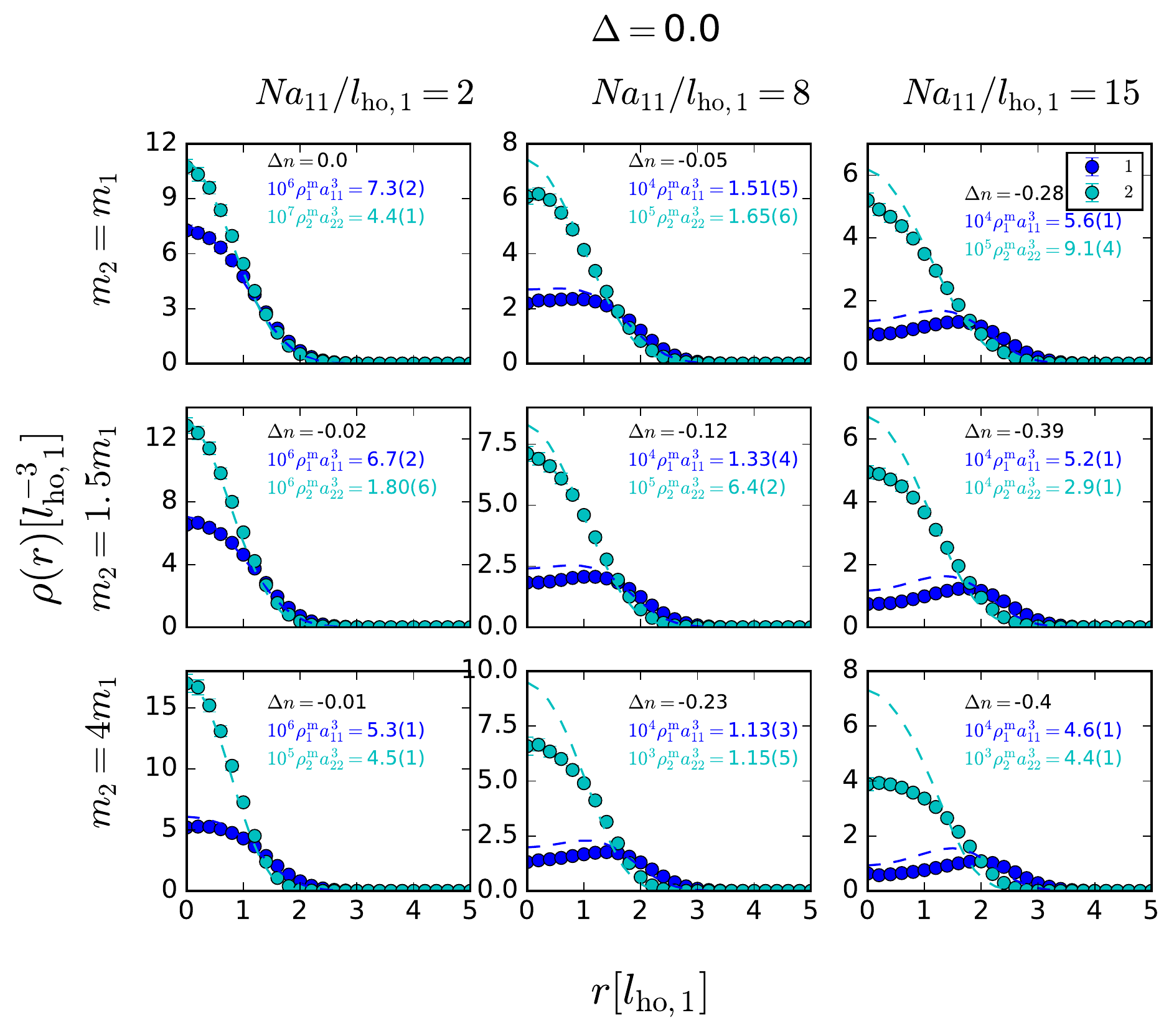}
	\caption{Point E of the phase space. The points correspond to the DMC results 
		and the lines to the solution of the GP equations for the same system.}
	\label{pointe}
\end{figure}

In point E, we are still in the critical line $\Delta=0$ but now the relation 
of interaction strengths is inverted with respect to point B. That is, 
$g_{22} < g_{12}<g_{11}$. Therefore, one now expects  species 2  occupying 
the center and species 1 moving to the external part of the trap. This is 
reflected in the negative values of the parameter $\Delta n$ which are reported 
for every panel in Fig. \ref{pointe}. The increase of the factor $m_2/m_1$ goes 
in reverse direction and slightly compensates the increase in $a_{22}$. 
However, it is shown not to be large enough to change the description. 
Comparing with Fig. \ref{pointb}, the phase separation is not complete because 
one can see that there is always a finite fraction of species 1 close to the 
center.

\subsection{\label{phaseseparated}$\mathbf{\Delta<0}$}

In this subsection, we move to points of the phase space where phase separation 
is expected. We have studied two representative points of the phase space, 
points A and G (see Table \ref{table1}). 

\begin{figure}[H]
	\centering
	\includegraphics[width=0.7\linewidth]{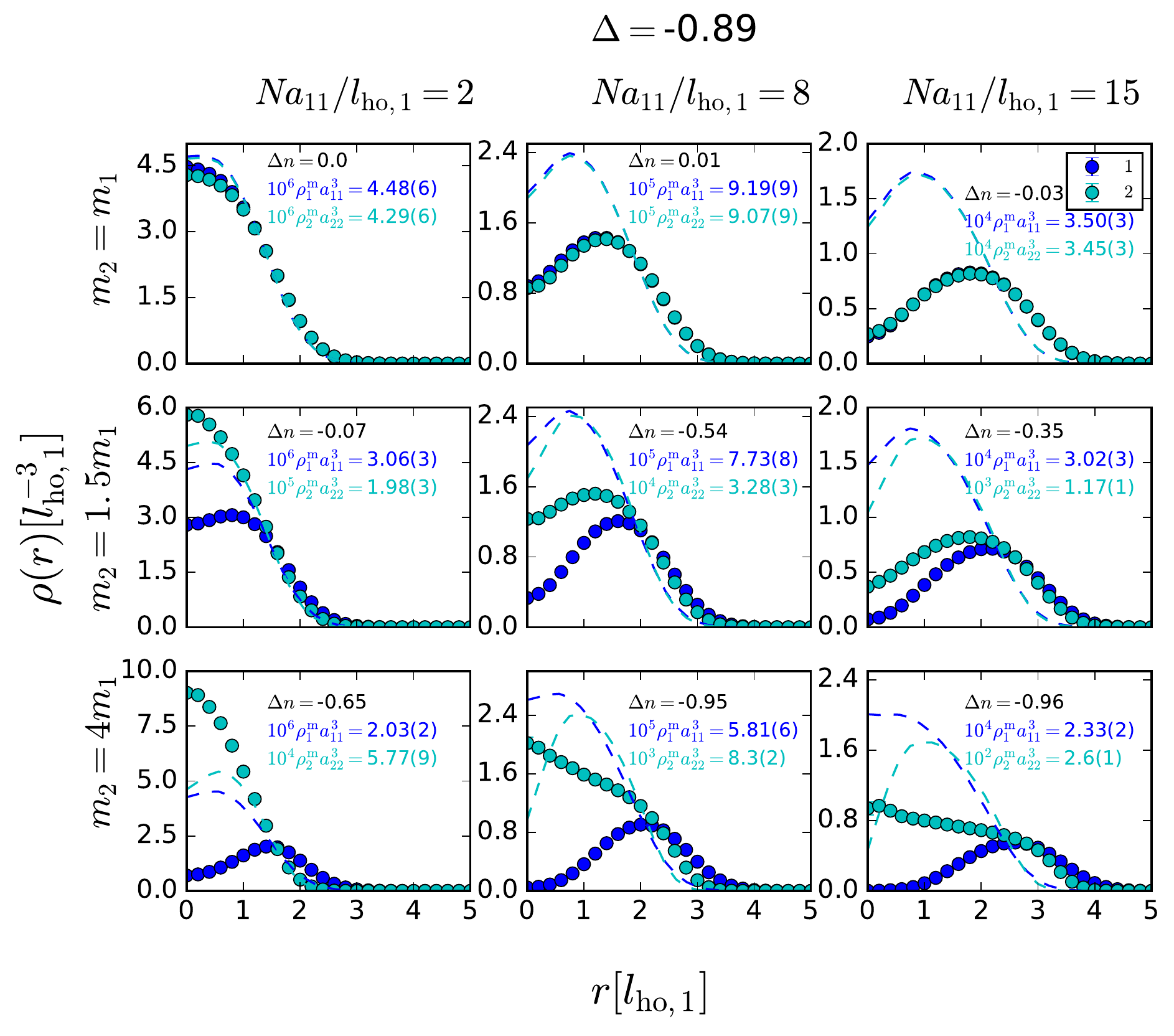}
	\caption{Point A of the phase space. The points correspond to 
		the DMC results and the lines to the solution of the GP equations for the same 
		system.}
	\label{pointa}
\end{figure}

\begin{figure}[H]
	\centering
	\includegraphics[width=0.6\linewidth, 
	angle=270]{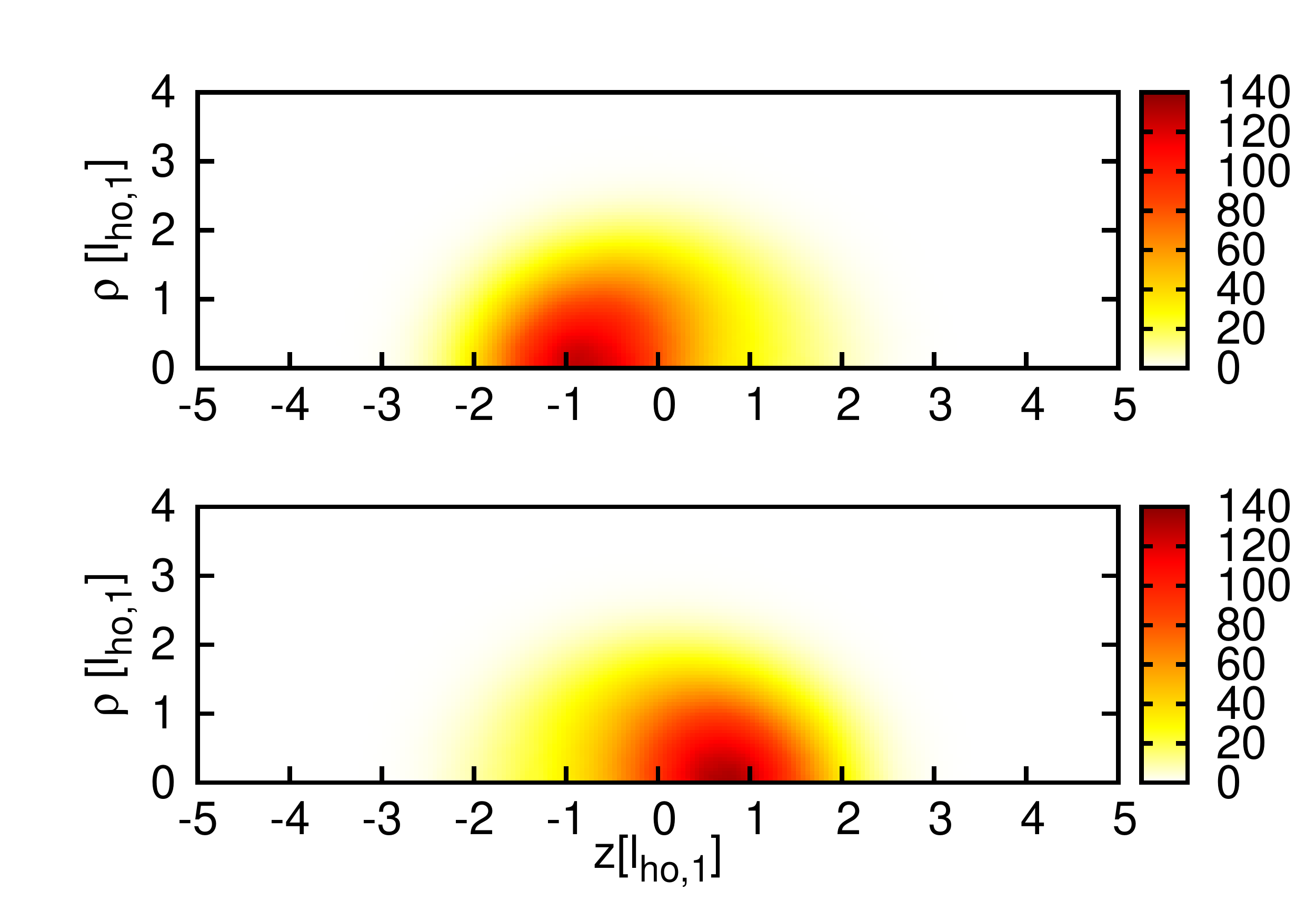}
	\caption{Point A of the phase space for $m_1=m_2$ and $N 
		a_{11}/l_{\text{ho},1} = 2$. $z$-axis corresponds to the line going through the 
		centers of mass of the two components, while $\rho$ corresponds to the distance 
		of a particle from that line. Top and bottom panels stand for the distribution 
		of species 1 and 2, respectively.}
	\label{splot1}
\end{figure}

\begin{figure}[H]
	\centering
	\includegraphics[width=0.6\linewidth, 
	angle=270]{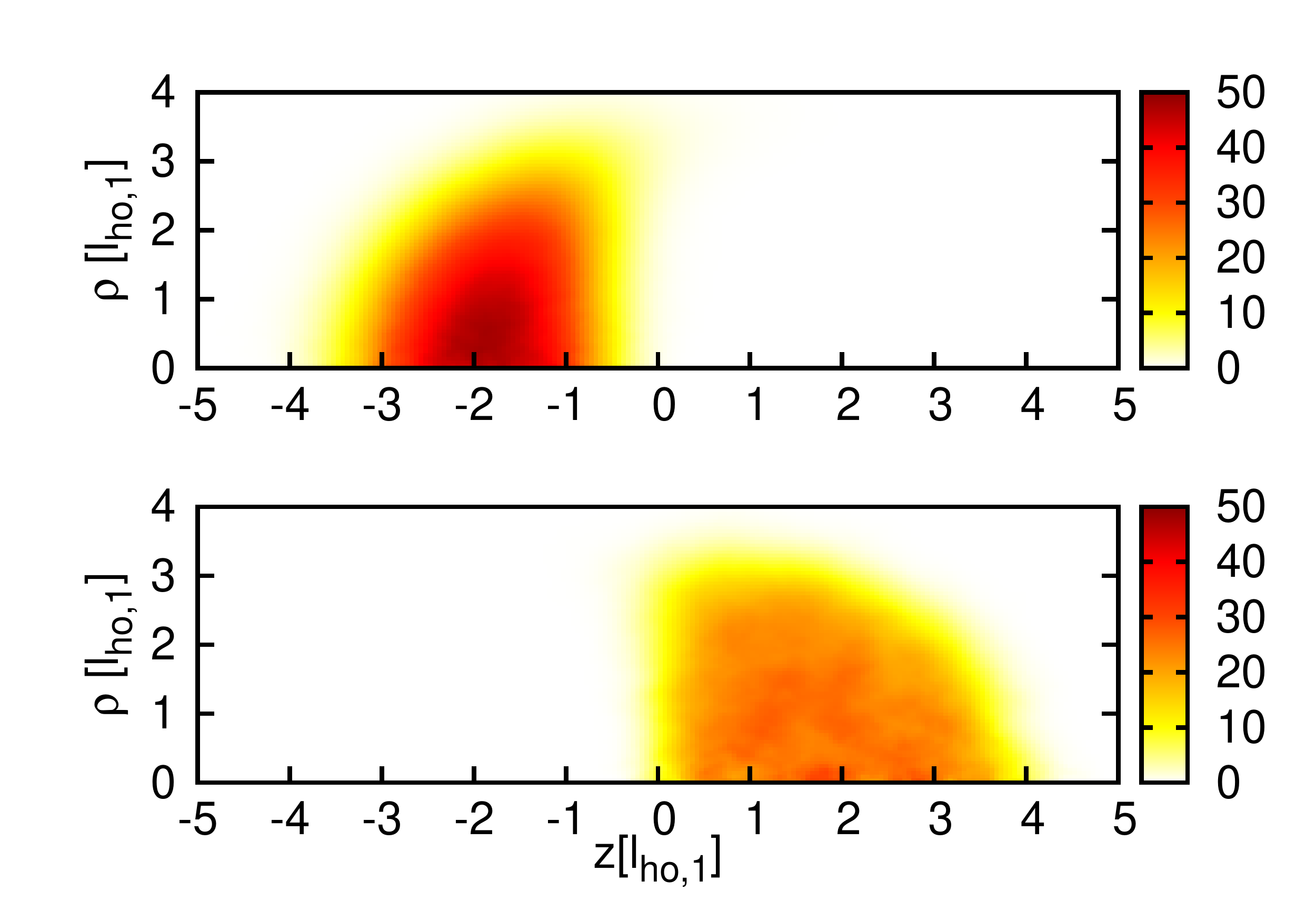}
	\caption{Same as Fig. \ref{splot1} (point A) for $m_2=4m_1$ and $N 
		a_{11}/l_{\text{ho},1} = 15$.}
	\label{splot2}
\end{figure}

In point A, $g_{12}/g_{22} = 3$ and $g_{11}/g_{12} = 1/3$ producing 
$\Delta=-0.89$. The relation of strengths is now 
$g_{11}=g_{22} < g_{12}$. By going from left to right in the panels of Fig. 
\ref{pointa} one can see that the mixture phase separates when the interaction 
between atoms is more important than the one-body confining harmonic 
potential. When the masses of both species are equal, one identifies a phase 
separation in form of two symmetric separated blobs, similar to what one 
would observe in 
a bulk system. This is also consistent with $\Delta n = 0$. Again, the 
mean-field prediction becomes quantitatively worse, as the interaction 
strength and the difference in mass between the two components increase. 
However, although it is not visible from the radial profiles, in all cases of 
point A there is at least partial phase separation in two blobs. In order to 
show this, we have calculated the $P(z,\rho)$ distribution, where  the 
$z$-direction is defined as a line passing through the two centers of mass, 
with the second component being in the positive $z$-direction. $z=0$ is the 
geometric center between the two centers of mass. The second variable, $\rho$, 
is just the distance of a single particle from this line. Results are 
normalized such that $\int 2 \pi \rho d\rho dz P(z, \rho) = N/2$. The results in 
the two illustrative cases are presented in Fig. \ref{splot1} and \ref{splot2}. 
In the case of equal masses and $N a_{11}/l_{\text{ho},1} = 2$ (Fig. 
\ref{splot1}) although both species overlap significantly, the maxima of 
their probability distributions are clearly separated. In fact the average 
distance of their centers of mass is about $0.4 l_{\text{ho},1}$. Increasing  
$N a_{11}/l_{\text{ho},1}$ the overlap between the two species decreases and a 
clear two-blob structure becomes visible, with the average distance between the 
two centers of mass becoming $3 l_{\text{ho},1}$. The increase of the mass 
difference of the two species also favors their separation. The extreme case of 
$m_2 = 4m_1$ and  $N a_{11}/l_{\text{ho},1} = 15$ is shown in Fig. \ref{splot2}. 
We observe that the two species are clearly separated in two blobs and more 
spread in both the $\rho$ and $z$ directions than in Fig. \ref{splot1}, due to 
the  larger repulsive interspecies interaction. DMC predicts the more 
massive component to be closer to the center of the trap, unlike GP. $|\Delta 
n|>0$ is quite large, but remarkably has opposite sign in GP and DMC.

\begin{figure}[H]
	\centering
	\includegraphics[width=0.6\linewidth, 
	angle=270]{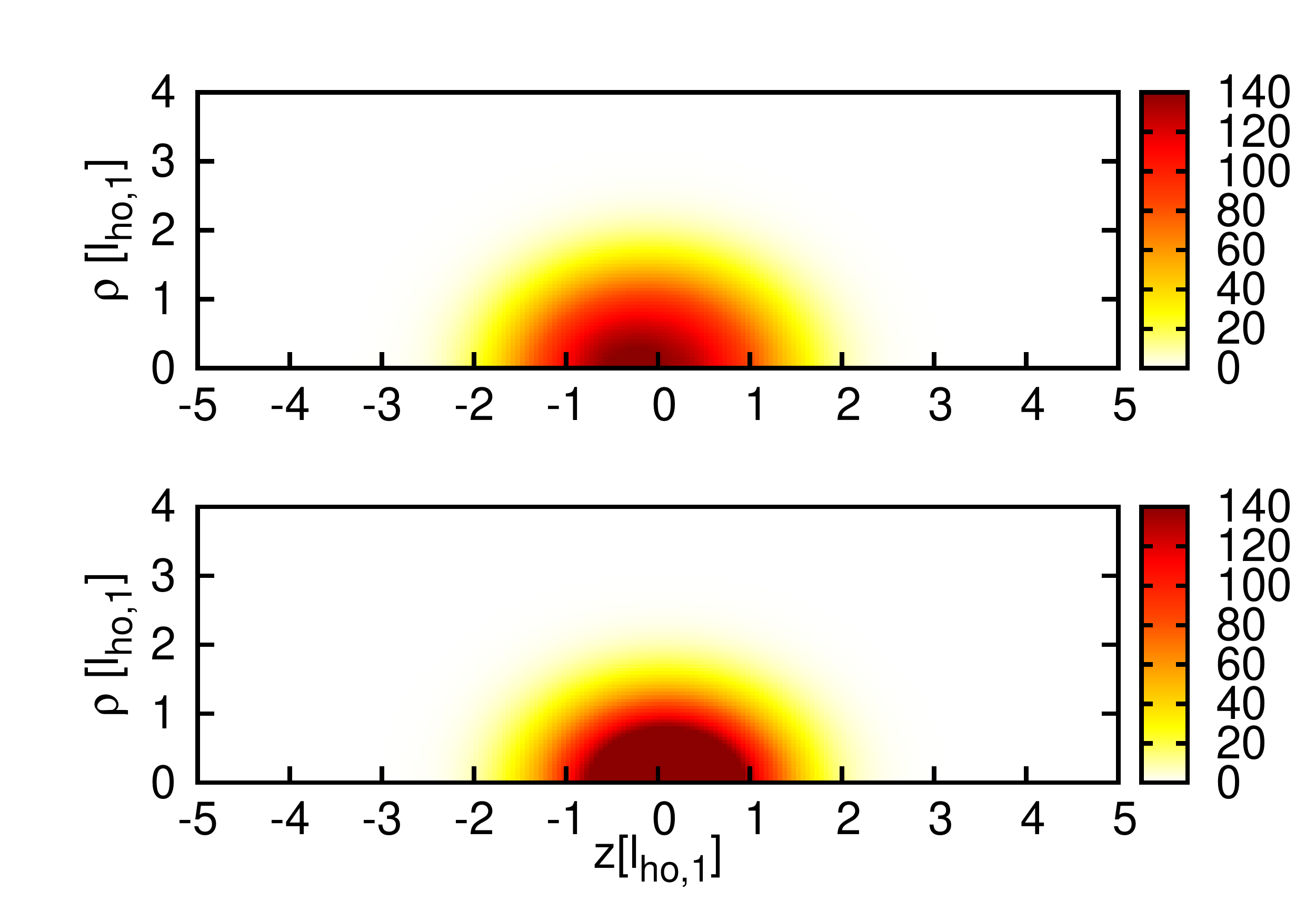}
	\caption{Same as Fig. \ref{splot1} (point G) for $m_2=m_1$ and $N 
		a_{11}/l_{\text{ho},1} = 2$.}
	\label{splot_g1}
\end{figure}
\begin{figure}[H]
	\centering
	\includegraphics[width=0.6\linewidth, 
	angle=270]{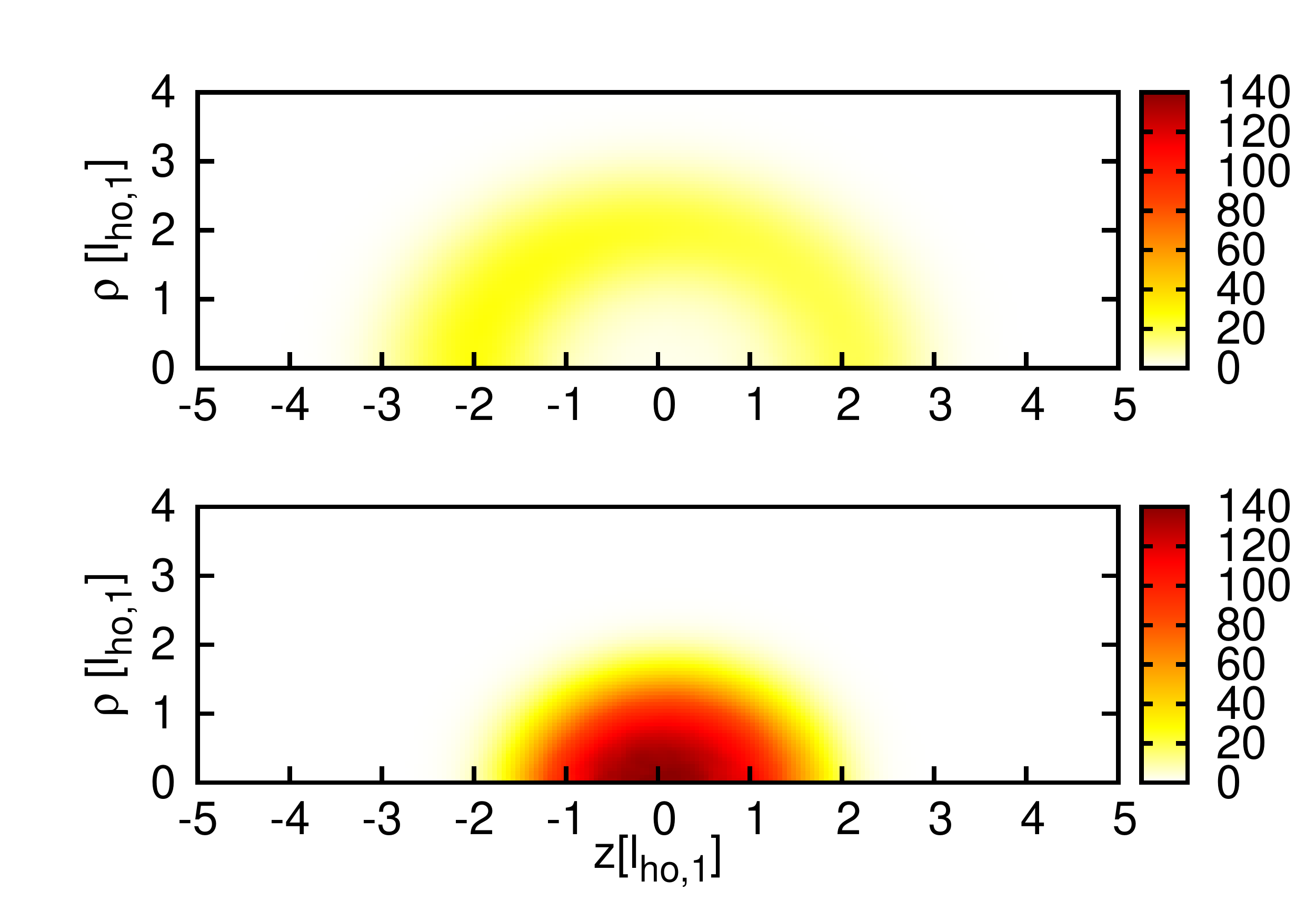}
	\caption{Same as Fig. \ref{splot1} (point G) for $m_2=4m_1$ and $N 
		a_{11}/l_{\text{ho},1} = 15$.}
	\label{splot_g2}
\end{figure}

In the last point (G), one expects phase separation because  
$\Delta=-0.43$. The 
relation of strengths is now $g_{11} > 
g_{12} > g_{22}$. We clearly observe a 
phase separated system for medium and large interactions and a mixed one when 
$Na_{11}/l_{ho,1}=2$. The lighter particle moves progressively out of the 
center and finally, when the relation of masses is large, it surrounds 
completely the heavier one, which occupies the center of the trap. Contrarily 
to point A, here the GP description is in nice agreement with the DMC data even 
when the difference in masses is large.

\begin{figure}[H]
	\centering
	\includegraphics[width=0.9\linewidth]{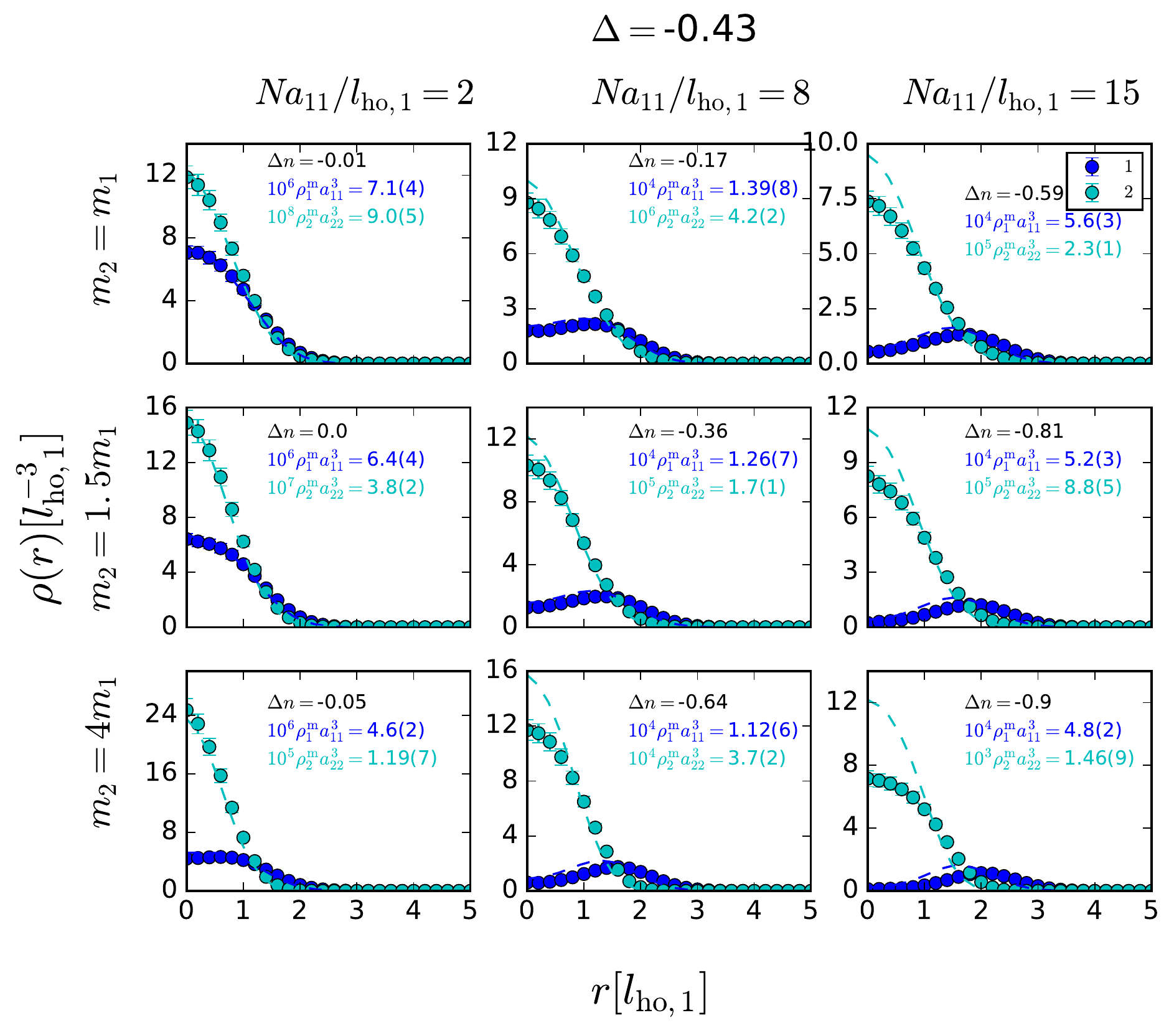}
	\caption{Point G of the phase space. The points correspond to 
		the DMC results 
		and the lines to the solution of the GP equations for the same system.}
	\label{pointg}
\end{figure}

\section{\label{scaling}Scaling with the interaction parameter ${\bf N a_{11}/l_{\textbf{ho},1}}$ }

In the previous Section, we have plotted the density profiles considering the 
parameter $N a_{11}/l_{\textbf{ho},1}$ as a scaling parameter to determine the strength of 
the interactions. For a given value of this adimensional parameter we have then 
changed the relation of masses between the two species. This parameter has been 
taken from the GP equation for a single Bose gas harmonically confined where 
it is proved to be the \textit{only}  input parameter of the calculations.

This universality  emerges from the Gross-Pitaevskii equations (Eqs. 
\ref{gp1} and \ref{gp2}) when they are written in  
length, energy, and time scales given by $l_{\rm ho,1} = \sqrt{\hbar / (m_1 
\omega_1)}$, $\hbar^2/(m_1 l_{\rm ho,1}^2)$ and $\tau =m_1 l_{\rm ho,1}^2 / \hbar $, 
respectively
\begin{eqnarray}
i  \frac{\partial \tilde{\phi}_{1} (\tilde{\mathbf{r}},\tilde{t})}{\partial \tilde{t}}
& = &
\left(
-\frac{\tilde{\nabla}^2}{2}
+
\dfrac{1}{2}\tilde{r}^2
+
\dfrac{4 \pi N_1 a_{11}}{l_{\rm ho, 1}} |\tilde{\phi}_{1}(\tilde{\mathbf{r}},\tilde{t})|^2  \right.  \nonumber \\
& &    +       \dfrac{4 \pi N_2 a_{12}}{l_{\rm ho, 1}} |\tilde{\phi}_{2}(\tilde{\mathbf{r}},\tilde{t})|^2
\bigg) 
\tilde{\phi}_{1} (\tilde{\mathbf{r}},\tilde{t}) \ ,
\label{gp1_red}
\end{eqnarray}
\begin{eqnarray}
i  \frac{\partial \tilde{\phi}_{2} (\tilde{\mathbf{r}},\tilde{t})}{\partial \tilde{t}}
& = &
\left(
-\frac{\tilde{\nabla}^2}{2}
+
\dfrac{m_2}{m_1}\dfrac{\omega_2^2}{\omega_1^2} \dfrac{1}{2}\tilde{r}^2
+
\dfrac{4 \pi N_2 a_{22}}{l_{\rm ho, 1}} |\tilde{\phi}_{2}(\tilde{\mathbf{r}},\tilde{t})|^2  \right.  \nonumber \\
& &    +       \dfrac{4 \pi N_1 a_{12}}{l_{\rm ho, 1}} |\tilde{\phi}_{1}(\tilde{\mathbf{r}},\tilde{t})|^2
\bigg) 
\tilde{\phi}_{2} (\tilde{\mathbf{r}},\tilde{t}) \ ,
\label{gp2_red}
\end{eqnarray}
where $\tilde{\phi}_1$ and $\tilde{\phi}_2$ are normalized according to $\int 
d^3\tilde{r} |\tilde{\phi}_i|^2 = 1$ ($i=1,2$). Notice that the universality is 
recovered when fixing the ratios $g_{12} / g_{22}$, $g_{11} / g_{12}$ and $N_1 
a_{11}/l_{\rm ho, 1}$ since in our study $N_1=N_2$, with $\omega_1$ and $\omega_2$ kept fixed.

In this section, we check if this universality also holds true when performing DMC calculations, both in 
regimes where DMC results and GP ones essentially coincide and in others where 
we have observed significant discrepancies.

We have chosen two illustrative cases of both situations. In particular, 
points A and E of the phase space (see Table \ref{table1}). In both cases we 
have changed independently $N$ and $a_{11}$ in such a way to keep the GP 
parameter as equal. To this end, we have used a system with $N=100+100$ and a 
smaller one, composed by half the number of particles $N=50+50$.

\begin{figure}
	\centering
	\includegraphics[width=\linewidth]
	{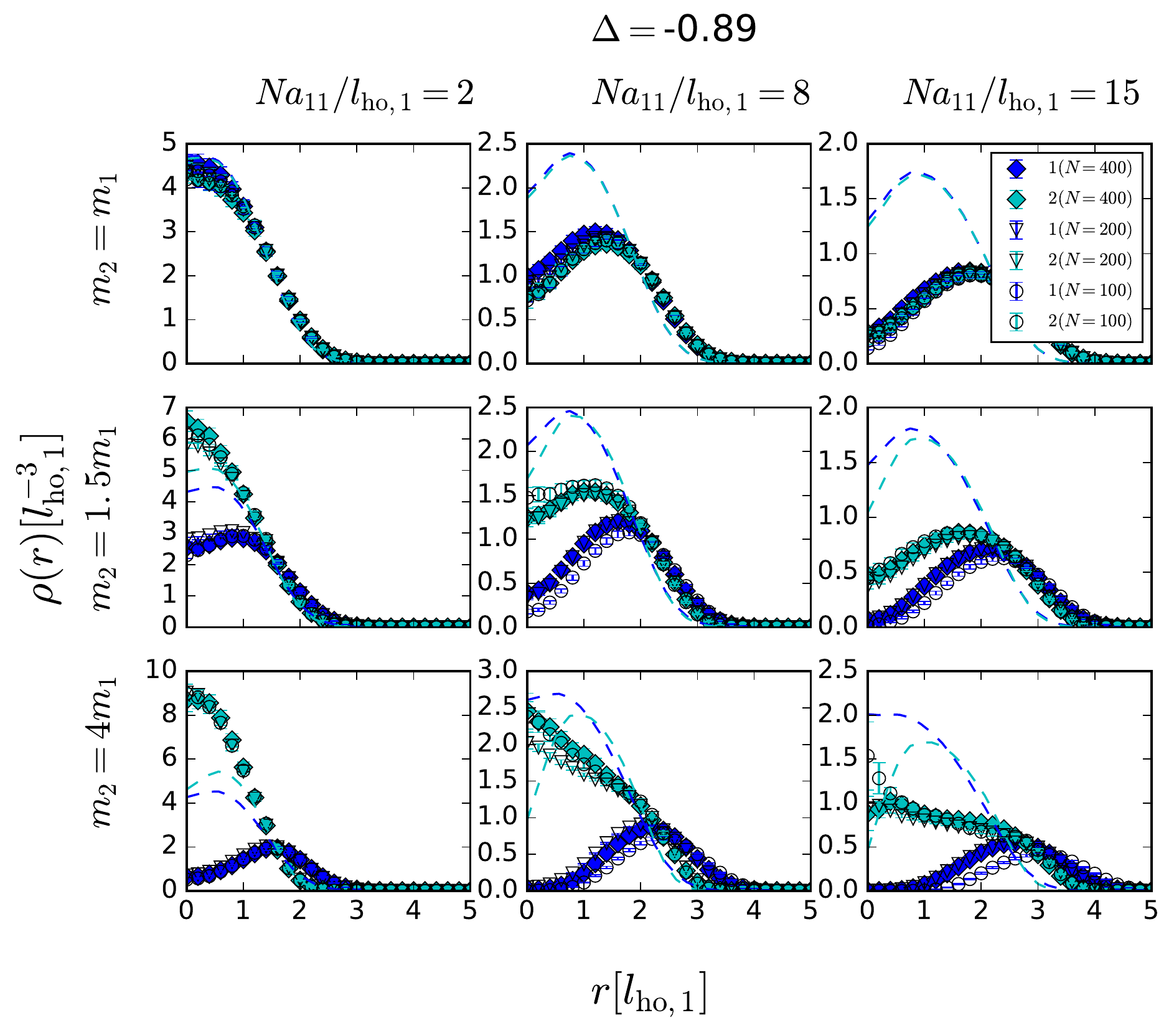} 
	\caption{Scaling on the GP parameter in point A of the phase space. Solid and 
		open points stand for results with $N=200$ and $N=100$, respectively.
	}
	\label{pointascal}
\end{figure}

In Fig. \ref{pointascal}, we report the results of this analysis for point A. 
The results of the density profiles are all normalized to sum up to 200 in order 
to make the comparison easier. As we commented in the previous Section, point A 
is the one where we have observed the largest departures from the GP results. 
The figures shows excellent agreement when the interaction is low and some 
discrepancies when the GP parameter grows. However, the effect is not dramatic 
and affects only the heaviest species. It is remarkable that even in 
situations like the ones of Fig. \ref{pointascal},  where GP strongly 
departs from DMC, 
one can still observe a very reasonable scaling with the GP interaction 
parameter.

\begin{figure}[H]
	\centering
	\includegraphics[width=\linewidth]
	{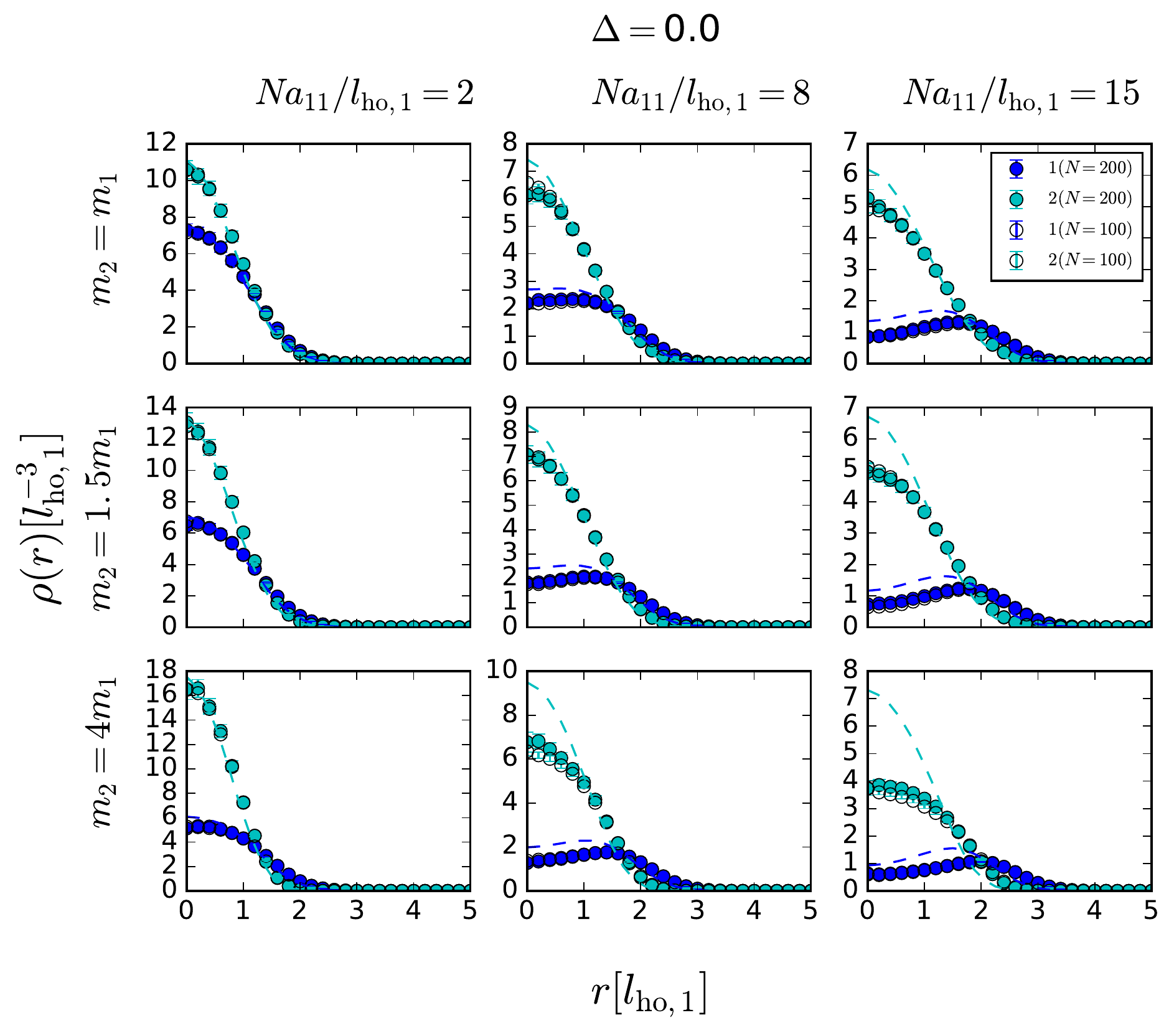} 
	\caption{Scaling on the GP parameter in point E of the phase space. Solid and 
		open points stand for results with $N=200$ and $N=100$, respectively.
	}
	\label{pointescal}
\end{figure}

Point E was one of the points where the agreement between GP and DMC was 
better. In Fig. \ref{pointescal}, we study the the dependence of the density 
profile results on the GP parameter as a scaling factor. In this case, the 
agreement is practically perfect because the discrepancies are just of the 
order of the error bars.

\section{\label{universality} Universality test}
A relevant point in our numerical simulations is the influence of the model 
potential on the results. Universality in these terms means that the 
interaction can be fully described by a single parameter, the $s$-wave 
scattering length, as it corresponds to a very dilute system. In all the 
previous results we have used a hard-core model for the interactions between 
the atoms (\ref{hardcore}). In this Section, we compare these results with 
other ones obtained with a 10-6 potential,
\begin{equation}
V^{(\alpha\beta)}(r)=\frac{\hbar^2}{2 \mu_{\alpha\beta}}\, V_0 \left[ 
\left(\frac{r_0}{r}\right)^{10}-\left(\frac{r_0}{r}\right)^6 \right] \ ,
\label{lenjones}
\end{equation}
whose $s$-wave scattering length is analytically known~\cite{pade2007exact}. We fix the 
parameter 
$r_0=2 a_{\alpha \beta}$ for all cases and modify the strength $V_0$ to 
reproduce the desired scattering length. In Eq. (\ref{lenjones}), the 
parameter $\mu_{\alpha\beta}$ is the reduced mass, $\mu_{\alpha\beta}=m_\alpha m_\beta/(m_\alpha+m_\beta)$.

Our analysis has been performed for $m_2=1.5 m_1$, $N=100+100$ particles, and 
considering equal harmonic frequencies $\omega_1=\omega_2$. Two interaction 
strengths have been used, $Na_{11}/l_{\text{ho}, 1}=8$ and 15. As in the 
previous 
Section, we have studied points E and A of the phase space, i.e., those 
characteristic of agreement and disagreement with GP.

\begin{figure}[!h]
	\centering
	\includegraphics[width=\linewidth]
	{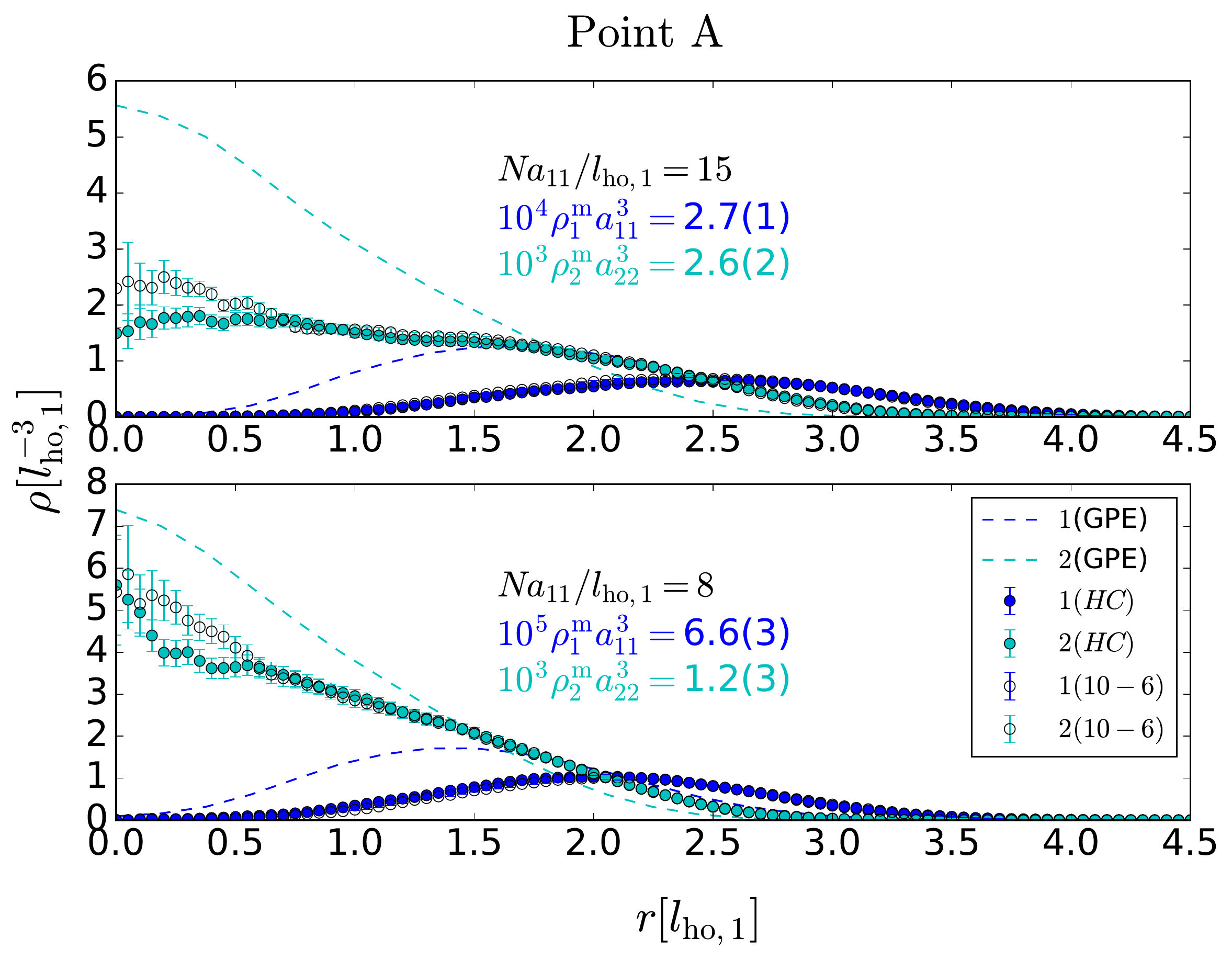} 
	\caption{DMC results for two models are compared to GP results for two values of GP parameter in point A .
	}
	\label{aunivers}
\end{figure}

In Fig. \ref{aunivers}, we show the density profiles of point A, obtained using 
the two model potentials normalized in the same way. We can see an overall 
agreement between both results with only some differences in the estimation of 
the density in the center. Close to $r=0$ the statistical fluctuations are 
bigger due to the normalization in a small volume. This feature is always 
present but we observe that these fluctuations are larger in the case of the 
10-6 potential (\ref{lenjones}). 

\begin{figure}[!h]
	\centering
	\includegraphics[width=\linewidth]
	{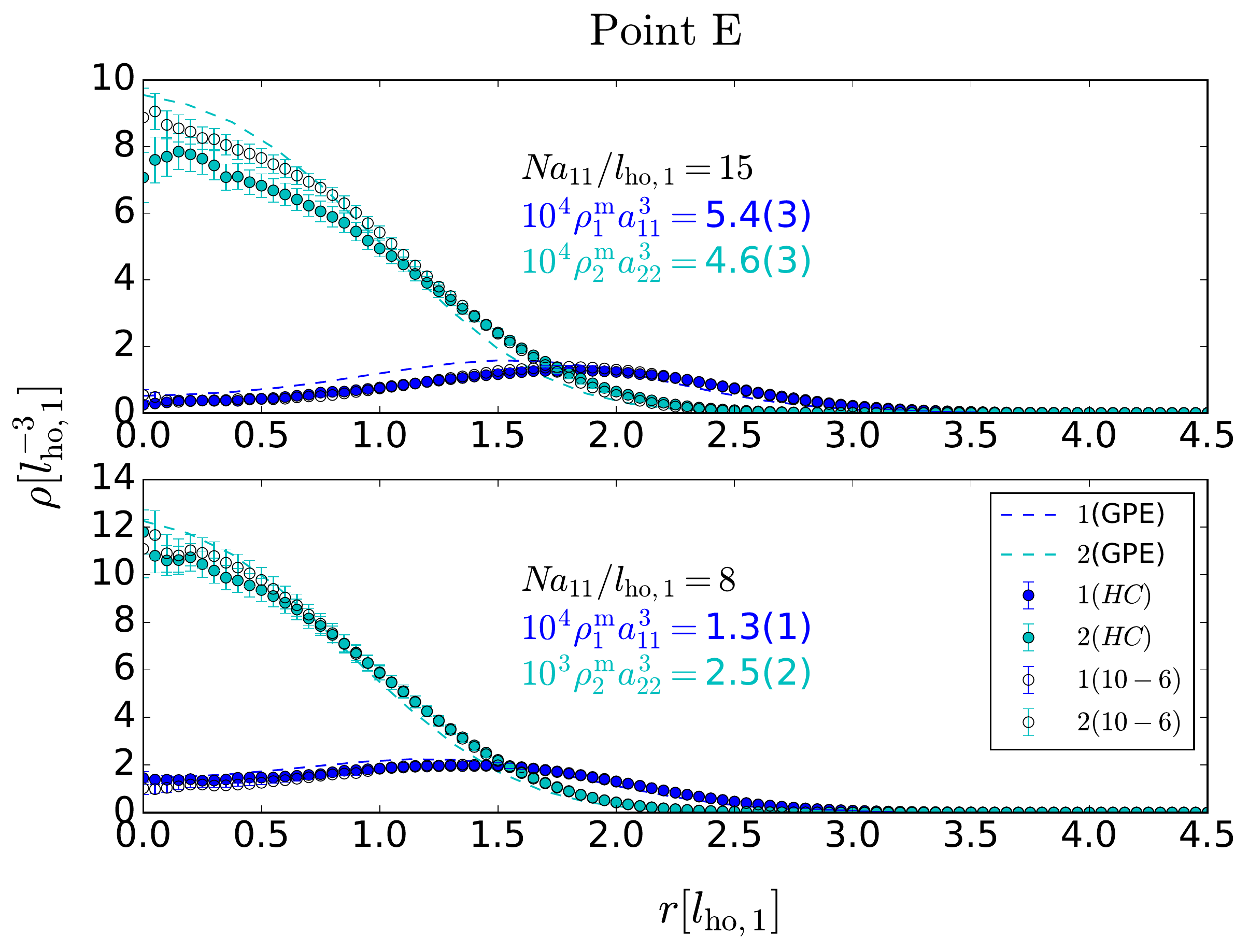} 
	\caption{DMC results for two models are compared to GP results for two values of GP parameter in point E.
	}
	\label{eunivers}
\end{figure}

In Fig. \ref{eunivers}, we analyzed the same for point E in 
which we are closer to an effective mean-field description. The comparison also 
shows a good agreement between results obtained for both potentials, with some 
differences in the inner core of the trap. 

We have verified in other points of the phase space the universality of our 
results and the conclusion is that this is maintained with respect to the 
character of the system (miscible or phase separated) and also the overall 
shape of the density profiles. The influence of the model potential is at the 
scale of our statistical fluctuations, except close to $r=0$ where small 
differences are observed in some cases.

\section{\label{discuss} Summary and discussion}
Using the diffusion Monte Carlo method, able to provide exact results for 
bosonic systems,we have explored the phase space of an harmonically confined 
Bose-Bose mixture at zero temperature. Our results are compared in all cases 
with a mean-field Gross-Pitaevskii calculation in the same conditions. As 
expected, our DMC results agree better with the GP ones when the strength of 
interactions is small and worsens progressively when that grows. In spite of 
the fact that the prediction for miscibility or phase separation is coincident 
in both cases, it is also true that the density profiles can be 
rather different. 

Our results are by construction exact and go beyond mean-field, showing the 
limits of this approach. A systematic trend in which GP fails is the dependence 
on the mass of the two species. When the asymmetry in the masses grows GP 
becomes clearly wrong in some cases. That could be argued to be an effect of 
the interaction model used in DMC but our results contradict this point, within 
the range of interactions here explored.

The tunability of atomic interactions and the possibility of using atoms with 
different mass ratios makes this system specially rich. The phase space shows 
regimes of miscibility and, interestingly, two different situations for phase 
separation, two blobs or in-out spherical separation, depending in the 
relation of masses and interaction strength. It would be very interesting to 
produce in the lab smaller systems, with hundreds of atoms, to check the 
departure of the physics from the mean-field GP treatment. In this way, one can 
start to enter into the realm of fully quantum many-body physics.



\chapter{Summary and outlook}

In this Thesis, we performed numerical studies of different bosonic mixtures with essentially four techniques of various range of applicability. 
We built numerical codes specialized to the study of quantum mixtures, which can stimulate further research, as we will make them available in open access repositories. 

Firstly, we used a Variational Monte Carlo method, which, as the name suggests, is a variational method aimed at solving the many-body Schr\"odinger equation. Throughout the Thesis, we investigated various forms of the trial wave functions and reported those which worked the best for each system when combined with the Diffusion Monte Carlo method. 
Secondly, we applied the Diffusion Monte Carlo technique, an exact method suitable for the study of static properties of bosonic systems. It is restricted to zero temperature, making it ideal for the study of ultracold systems where neglecting thermal effects is an excellent first approximation. Diffusion Monte Carlo is very computationally demanding, thus making it somewhat limited to systems with at most 1000 atoms. 
This led us to resort to the final set of methods applicable to large systems, which opened access to time-dependent phenomena as well, the bosonic mean-field theory and density functional theory. We used Gross-Pitaevskii equation, arising from mean-field theory. In the regime where the Gross-Pitaevskii equation fails, we built an underlying density functional by relying on the Quantum Monte Carlo data.

The results presented in this Thesis are relevant both to theoretical and experimental activities in quantum Bose mixtures. We benchmarked the existing mean-field theories, extended them where they are expected to fail, and explained the physical mechanism behind the numerical results. The main body of the Thesis is the study of ultradilute bosonic droplets, where we presented numerical evidence of the necessity to go beyond the LHY-extended mean-field theory, as we discovered both the repulsive and attractive contribution to the energy, depending on the value of the effective range of the interaction. Part of these findings are already manifested in experimentally obtained quantum droplets in $^{39}$K. We believe that next-generation experiments with quantum droplets will rely on our findings. 

The last topic of this Thesis is the study of repulsive mixtures, where we aimed at understanding the repulsive part of the bosonic mixture phase diagram. These findings will allow one to connect the Quantum Monte Carlo results with future experimental findings in mesoscopic quantum systems, where the description needs to go beyond the mean-field.

Following is the summary and outlook for the subtopics of this Thesis.

\vspace{0.1cm}
{\large \bf Properties of finite-size symmetric quantum drops}
\vspace{0.1cm}

Properties of symmetric finite-size quantum drops using essentially the exact DMC method were presented. For simplicity, we restricted to symmetric Bose mixtures, i.e., the ones with $a_{11} = a_{22}$. We studied the critical atom number, which is the minimum total atom number required to have a self-bound drop, for a wide range of interparticle attractive strength $a_{12} / a_{11}$, where the Petrov's theory cannot be applied. There we already observed the repulsive beyond-LHY contributions which seem to be a general feature in a Bose mixture with short-ranged interactions \cite{ota2020beyond,hu2020consistent}. We predicted on density profiles which feature the density saturation, for two values of the attractive $s$-wave scattering length $a_{12}/a_{11}$. These are in agreement with our predictions of the full equation of state of a symmetric liquid, i.e., the total energy per particle as a function of the density, for three values of $a_{12}/a_{11}$. Finally, by fitting the energies of finite-size drops within a liquid drop model, we obtained the surface tension. A further investigation of small-$N$ quantum drops could lead to the improved density-functional which generalizes the Petrov's approach, and which could reproduce the QMC results summarized here.


\vspace{0.1cm}
{\large \bf Beyond-LHY contributions to the energy in quantum liquids}
\vspace{0.1cm}

The bothersome feature of the LHY-extended mean-field theory is the appearance of imaginary contributions to the energy of homogeneous mixtures, coming from the LHY term \cite{petrov2015quantum,hu2020consistent}. We extensively benchmarked and studied the energy per particle in the thermodynamic limit by means of DMC calculations, which provided exact estimations of the energy. We discovered that there exist substantial beyond-LHY contributions in a symmetric $a_{11} = a_{22}$ liquid, which have a repulsive nature, and whose contribution increases when entering in a more correlated, dense regime. We predicted that the aforementioned beyond-LHY terms also depend on the effective range, which is another coefficient in the expansion of the $l=0$ phase shift. We concluded that the inclusion of the effective range leads to a decrease in energy, a feature which appears both in symmetric liquids and in a liquid of $^{39}$K atoms. By performing many, in particular 18 calculations of the equations of state of symmetric liquids, we produced a QMC-built density functional which also incorporates the effects of finite-range. The outlook of presented results is very wide: we believe that the correct inclusion of the effective range is crucial in the description of next-generation quantum droplet experiments. Additionally, our results could be used as a benchmark of beyond-LHY theories, such as those in Refs. \cite{hu2020consistent,ota2020beyond}.


\vspace{0.1cm}
{\large \bf Finite-range effects on the static and dynamic properties of $^{39}$K drops}
\vspace{0.1cm}

We thoroughly studied a quantum droplet in a mixture of $^{39}$K, which was first experimentally realized by the Cabrera et. al \cite{cabrera2018quantum} and by Semeghini et. al. \cite{semeghini2018self}. For this mixture, there are known values of the effective ranges, on top of the $s$-wave scattering lengths. We provided the QMC-built density functionals for various values of the applied magnetic field, or alternatively the scattering parameters. Based on QMC energies, we devised a density functional which leads to the prediction of a substantially lower critical atom number, observed in \cite{cabrera2018quantum}. We outlined the MF+LHY theory for non-optimal composition of particles in a drop, which we used to explain the discrepancy in the predictions of observed size in the experiment \cite{cabrera2018quantum}. Non-optimal composition might occur due to strong three-body losses, in which case finite-range effects would be of second-order of importance. The discrepancy in size might also occur due to thermal effects \cite{wang2020thermal}, which is an outlook for a future investigation.

We presented a study of the thermodynamic properties of a QMC-built density functional in a $^{39}$K liquid. We predicted frequencies of excitation modes in a spherical self-bound $^{39}$K quantum droplet, in particular the quadropole and the monopole modes. The significant differences between the results of QMC and MF+LHY functionals for the excitation spectrum indicates that finite-range effects could show up in other dynamical problems as well. In particular, in  droplet-droplet collisions \cite{ferioli2019collisions}, where the actual value of the incompressibility might play a relevant role. A reliable functional might also be useful to
study quantum droplet aspects that are currently  under study for superfluid $^4$He droplets, such as the appearance of quantum turbulence and of bulk and surface vorticity
in  droplets merging, the equilibrium phase diagram of rotating quantum  droplets, and the
merging of vortex-hosting quantum droplets \cite{ancilotto2018spinning,escartin2019vorticity,oconnell2020angular}.  These aspects are at present under investigation. Further improvements in the building of a more accurate QMC functional should  consider the inclusion of surface tension effects others that those arising from the quantum kinetic energy term  \cite{marin2005free}.

\vspace{0.1cm}
{\large \bf Phase diagram and universality in repulsive Bose-Bose mixtures}
\vspace{0.1cm}

We presented an extensive study of the phase diagram of  harmonically trapped Bose-Bose mixtures at zero temperature where all the interactions are repulsive. We systematically explored the phase diagram of the system and compared the density profiles obtained by DMC with Gross-Pitaevskii solutions.
We found good qualitative agreement between the DMC and Gross-Pitaevskii equations when the interactions are weak and mass asymmetry is small. However, the differences magnify when the asymmetry between masses and interaction strength increases. Our study was limited to zero temperature, and a future outlook would be the proper inclusion of thermal fluctuations, which proved to trigger a phase separation in the homogeneous system even when the mean-field zero temperature theory predicts mixing \cite{ota2019magnetic}.

We showed that the shape of DMC-predicted density profiles remained the same, with just the norm changing accordingly, when the calculations were performed
for $N = 200$ and 400 atoms, provided that $N_1 a_{11} /l_{\rm ho,1}$ ,
$N_1 a_{12} /l_{\rm ho,1}$, $N_2 a_{12} /l_{\rm ho,1}$ and $N_2 a_{22} /l_{\rm ho,1}$ were kept fixed. Thus, the universality in Gross-Pitaevskii equations seems to be valid even in the strongly correlated regime, which allows for the comparison of ab-initio microscopic calculations with experimental results.

We provided numerical evidence for the universality of our results, meaning that the interaction can be fully described by a single parameter, the $s$-wave scattering length, as it corresponds to a very dilute system. This is relevant as our density profiles differ from the Gross-Pitaevskii solutions. Since the next correction to the mean-field, the LHY term, is imaginary for $g_{12}^2 > g_{11} g_{22}$, our results might prove useful as a benchmark of more elaborate many-body theory of repulsive mixtures.


	\begin{spacing}{0.9}
		
		
		\bibliographystyle{ieeetr}

		\cleardoublepage
		\bibliography{References/refs.bib} 

		

	\end{spacing}

	
	
	\begin{appendices} 
		
\chapter{\label{appendix:local_energy}Derivation of quantum force and local energy in VMC and DMC}

Crucial elements for the propagation of particle coordinates during the course of DMC and VMC methods are the quantum force 
\begin{equation}
	\vecF(\vecR) = 2\dfrac{\nabla_{\vecR} \psi_{\rm T}(\vecR)}{\psi_{\rm T}(\vecR)},
\end{equation}
and the local energy
\begin{equation}
	E_L(\vecR) = -\dfrac{\hbar^2}{2m} \dfrac{\nabla_{\vecR}^2 \psi_{\rm T}(\vecR)}{\psi_{\rm T}(\vecR)} + V(\vecR),
\end{equation}
where $V(\vecR) = \sum_{i}^N V_{\rm ext}(\vec{r}_i) + \sum_{i<j} V_{\rm pair}(|\vec{r}_i - \vec{r}_j|)$. Since we work directly in the coordinate representation, potential energy evaluation is trivial. In this Appendix the explicit expression for these quantities are derived under the assumption of the trial wavefunction in the form
\begin{equation}
	\psi_{\rm T}(\vecR) = \prod_{i}^{N} f_{\rm 1b} (\vec{r}_i) \prod_{i<j}^{N} f_{\rm 2b} (|\vec{r}_i - \vec{r}_j|).
\end{equation}
Let us write the wavefunction as
\begin{equation}
	\psi_{\rm T}(\vecR) = \psi_1(\vecR) \psi_2(\vecR),
\end{equation}
where
\begin{equation}
	\psi_1(\vecR) = \prod_{i}^{N} f_{\rm 1b}(\vec{r}_i),
\end{equation}
and
\begin{equation}
	\psi_2(\vecR) = \prod_{i<j}^{N} f_{\rm 2b} (|\vec{r}_i - \vec{r}_j|).
\end{equation}
Quantum force is therefore
\begin{equation}
	\vecF(\vecR) = 2\dfrac{\nabla_{\vecR} \psi_{\rm T}(\vecR)}{\psi_{\rm T}(\vecR)} = 2\dfrac{\nabla_{\vecR} \psi_1(\vecR)}{\psi_1(\vecR)} + 2\dfrac{\nabla_{\vecR} \psi_2(\vecR)}{\psi_2(\vecR)} = \vecF_1(\vecR) + \vecF_2(\vecR),
\end{equation}
where the force acting on the particle $i$ is given by
\begin{eqnarray}
	\vecF_1(\vec{r}_i) & =&   2\dfrac{\nabla_{\vec{r}_i} \psi_1(\vec{r}_i)}{\psi_1(\vec{r}_i)} \\
	& = & \dfrac{2}{ \psi_{\rm 1b}(\vec{r}_i)} \sum_{k=1}^{3} \hat{e}_k  \dfrac{\partial}{\partial x_k^{(i)}}   \psi_{\rm 1b}(\vec{r}_s)
\end{eqnarray}
and 
\begin{eqnarray}
	\vecF_2(\vec{r}_i) & = &  2 \sum_{k=1}^3 \hat{e}_k  \sum_{j = 1 \neq i}^N  \dfrac{1}{f(r_{ij})} \dfrac{\partial f(r_{ij})}{\partial r_{ij}}   \dfrac{x_k^{(i)} - x_k^{(j)}}{r_{ij}}.
\end{eqnarray}
Calculation of $\vecF_2(\vec{r})$ can be utilized such that iterations go only through $N (N - 1) / 2$ atom pairs. Finally, kinetic part can be calculated by invoking formula
\begin{equation}
	\dfrac{\nabla^2 \psi}{\psi} = \dfrac{1}{2}\nabla \vecF + \dfrac{1}{4}\vecF^2,
\end{equation}
where the gradient for the particle $i$ is calculated as
\begin{equation}
	\nabla_{\vec{r}_i} \vecF(\vec{r}_i) = G_1(\vec{r}_i) + G_2(\vec{r}_i),
\end{equation}
where
\begin{equation}
	G_1(\vec{r}_i) = 2 \sum_{k=1}^3 \dfrac{\partial }{\partial x_k^{(i)}} \left\{ \dfrac{1}{\psi_{\rm 1b}} \dfrac{\partial \psi_{\rm 1b}(\vec{r}_i)}{\partial x_k^{(i)}}   \right\},
\end{equation}
\begin{equation}
	G_2(\vec{r}_i) = 2 \sum_{j=1 \neq i}^{N} \left\{ \dfrac{\partial }{\partial r_{ij}} \left(\dfrac{1}{f(r_{ij})} \dfrac{\partial f(r_{ij})}{\partial r_{ij}}\right) + \dfrac{2}{f({r}_{ij})} \dfrac{\partial f(r_{ij})}{\partial r_{ij}}  \right\},
\end{equation}
where again the calculation of $G_2$ can be done by looping only through each atom pair.

\chapter{Curriculum vitae and related publications}

\hspace{5ex} I, Viktor Cikojević, was born on 29 January 1993 in Split, Croatia. In 2011, I graduated from the 3rd (mathematical) gymnasium in Split. In 2016, I graduated from the University of Split, Faculty of Natural Sciences, with a degree in Computational Physics. Since December 2016 I am a research and teaching assistant at Physics Department within the Faculty of Natural Sciences, University of Split.

 \hspace{5ex} During my doctoral studies, I contributed to a number of publications related to this work.
 These include the following:

\begin{itemize}[noitemsep]

	\item \underline{Cikojević, V.}, Marki{\'c}, L. V., Pi, M., Barranco, M. \& Boronat, J.. "Towards a quantum Monte Carlo–based density functional including finite-range effects: Excitation modes of a  $^{39}$K quantum droplet", Phys. Rev. A {\bf 102}, 033335 (2020)

	\item \underline{Cikojević, V.}, Marki{\'c}, L. V. \& Boronat, J.. "Finite-range effects in ultradilute quantum drops." New J. Phys  {\bf 22}, 053045 (2020).

	\item \underline{Cikojević V.} , Marki{\'c}, L. V., Astrakharchik G. E. \&  Boronat J.  "Universality in ultradilute liquid Bose-Bose mixtures". Phys. Rev. A {\bf 99}, 023618 (2019)

	\item \underline{Cikojević, V.}, Marki{\'c}, L. V. \& Boronat, J. (2018). "Harmonically trapped Bose-Bose mixtures: a Quantum Monte Carlo study." New J. Phys \textbf{20}, 085002 (2018).

	\item \underline{Cikojević, V.}, Dželalija, K., Stipanović, P., Markić, L. V., \& Boronat, J, "Ultradilute quantum liquid drops." Phys. Rev. B, \textbf{97}, 140502(R)  (2018).

\end{itemize}

\vspace{0.5cm}
	Other papers to which the author contributed:

\begin{itemize}

\item Ra\'ul Bomb\'in, \underline{Viktor Cikojevi\'c,} Juan S\'anchez-Baena, Jordi Boronat, "Finite-range effects in the two-dimensional repulsive Fermi polaron", 	arXiv:2008.10510 (2020)
	
\item K. Dželalija, \underline{V. Cikojević}, J. Boronat, L. Vranješ Markić, "Trapped Bose-Bose mixtures at finite temperature: a quantum Monte Carlo approach", Phys. Rev. A {\bf 102}, 063304 (2020)

\end{itemize}

	\end{appendices}

\end{document}